\documentclass[english,a4paper,headsepline]{scrartcl}
\usepackage[cp850]{inputenc}    
\usepackage{babel}
\usepackage{amsmath2000}
\usepackage{fancybox}
\usepackage{epsfig}
\newcommand{\E}{\ensuremath{\mathrm{e}}}

\newcommand{\1}{\ensuremath{\boldsymbol{1}}}

\newcommand{\bra}[1]{\ensuremath{\langle #1|}}
\newcommand{\ket}[1]{\ensuremath{|#1\rangle}}

\newcommand{\braket}[2]{\ensuremath{\langle #1|#2\rangle}}

\begin{document}
\thispagestyle{empty}
\begin{flushright}
MS-TP-01-4

June 2002
\end{flushright}
\vspace*{-15mm}
\epsfig{file=Logo_f.ps,width=7.5cm}
\vspace*{1.2cm}
\begin{center}
\LARGE{
Field-Theory Amplitudes as Resurgent Functions}\\[5ex]
\Large{M. Stingl}\\
\Large{Institut für Theoretische Physik, Universität Münster}\\
\Large{Wilhelm-Klemm-Str. 9, D-48149 Münster, Germany}\\
\Large{e-mail: stingl@uni-muenster.de}
\end{center}
\vspace*{0.5cm}
\begin{center}
{\bf Abstract}
\end{center}

\qquad
A series of informal seminars at graduate-student level
on the subject of coupling dependence in quantum field theory, with an 
elementary introduction to the notion of resurgent function that forms the 
appropriate framework for the coupling dependence of strictly renormalizable 
theories. While most of the discussion is pedagogical, there are also a
few things for the expert: we demonstrate, by studying a model spectral 
integral, that an amplitude may possess both the t'Hooft set of singularities 
in the coupling-constant plane, and Borel-plane singularities of the
infrared-renormalon type in its perturbative part, and yet be uniquely 
reconstructible from its resurgent symbol, the appropriate generalization of 
the semiconvergent perturbation expansion. In the same model we demonstrate 
the virtues of a quasi-perturbative expansion, obtained by resummation of the 
resurgent symbol in its nonperturbative direction, and which in contrast to 
the perturbative one is Borel summable. On the basis of this expansion,
we discuss a systematic approximation method for the reconstruction of 
correlation functions with the resurgent coupling dependence typical of 
strictly renormalizable and asymptotically free theories.

\thispagestyle{empty}
\tableofcontents
\newpage
\setcounter{page}{1}
\renewcommand{\thesection}{0}
\section{Generalities on Coupling Dependence}

\qquad
We know everything that is knowable about a quantum field theory (QFT)
once we know the full set of its Euclidean correlation functions,
\renewcommand{\theequation}{0.1}
\begin{equation}
\label{xcorr}
G_{N}\,(x_{1}\,...\,x_{N})\,=\,\braket{0|\varphi\,(x_{1})\,\varphi\,(x_{2})\,...\,
\varphi\,(x_{N})}{0}\,\,,
\end{equation}
for all  configurations $N$ of elementary fields, or
equivalently their Fourier transforms defined by
\renewcommand{\theequation}{0.2}
\begin{equation}
\label{pcorr}
\begin{array}{l}
\displaystyle{
\int d^{D}\,x_{1}\,...\,d^{D}\,x_{N}\,\E^{i\,(k_{1}\,\cdot\,
x_{1}\,+\,...\,+\,k_{N}\,\cdot\,x_{N})}\,G_{N}\,(x_{1}\,...\,x_{N})}\\
\mbox{}\\
=\,(2\,\pi)^{D}\,\delta^{D}\,(k_{1}\,+\,k_{2}\,+\,...\,+\,k_{N})\,
\tilde{G}_{N}\,(k_{1}\,...\,k_{N})
\end{array}
\end{equation}
at Euclidean momenta $k_{i}$. Here we denote by $D$ the Euclidean
spacetime dimension, which we will allow to deviate from the realistic
value of $D\,=\,4$, both for the purpose of looking at lower-dimensional
model theories $(D\,=\,3,\,2,\,1)$ and for dealing with ultraviolet
divergence in $D\,=\,4$ through dimensional regularization
($D\,=\,4\,-\,2\,\varepsilon$). For simplicity we mostly consider,
instead of (\ref{pcorr}), the momentum-space, Euclidean, proper vertex
functions $\Gamma_{N}$, that is, the connected, amputated, and
one-particle irreducible functions, since these are the simplest
building blocks from which the full $\tilde{G}_{N}$ can be pieced
together purely algebraically. Their full set of functional dependences
is
\renewcommand{\theequation}{0.3}
\begin{equation}
\label{depends}
\Gamma_{N}\,=\,\Gamma_{N}\,(k_{1},\,k_{2},\,...\,k_{N};\,g^{2}\,(\mu),\,
m\,(\mu),\,\xi\,(\mu)\,...\,;\,\mu)_{R}\,\,,
\end{equation}
where the momenta obey the constraint $\Sigma\,k_{i}\,=\,0$ of
total-momentum conservation, and $\mu$ denotes the arbitrary
renormalization mass scale introduced by renormalization in a scheme $R$
not usually indicated explicitly -- here, a
dimensional-regularization-plus-minimal-subtraction scheme. Moreover,
$g^{2}\,(\mu),\,m\,(\mu),\,\xi\,(\mu)$ ... are the renormalized
coupling constant, renormalized elementary masses, renormalized
gauge-fixing parameter ... of the theory, defined at the "sliding
scale" $\mu$ \cite{WEI} and therefore "running" with that scale,
since upon changing the arbitrary $\mu$ they all need to be readjusted
to keep the observable quantities constant.

\qquad
In these seminars our focus will not, as usual, be on the momentum
dependence of the functions (\ref{depends}), but rather on their
{\em dependence on the coupling constant} $g^{2}\,(\mu)$.
(Renormalization-Group
(RG) analysis shows that these two are in fact related, and we are
going to use this relation in places, but the link is really tight and
restrictive only for two-point functions depending on a single squared
Euclidean momentum $k^{2}$, since for higher correlation functions the
RG governs only the behavior under uniform scaling of {\em all} momentum
arguments but says little about individual momentum dependences.) We are
less interested here in the dependence on other parameters such as
$m\,(\mu),\,\xi\,(\mu)$, and therefore will mostly get rid of those
by considering massless theories ($m\,=\,0$), adopting special choices
of
gauge fixing (such as $\xi\,=\,0$), etc. This, of course, is not to say
that the other dependences may not be of paramount interest in dealing
with specific problems, but the interaction strength $g^{2}\,(\mu)$,
which makes for a nontrivial theory in the first place, does play a
distinguished role, of which these talks are intended to provide a
glimpse.

\qquad
To anyone with a basic knowledge of QFT, the full $g^{2}\,(\mu)$
dependence of amplitudes generated by a realistic, interacting,
renormalized QFT would at first sight appear to be of unfathomable
complexity,
nothing short of a horrible mess. It is therefore surprising that for a
class of interacting theories with mild ultraviolet problems, the
so-called superrenormalizable theories (essentially, theories in which
the mass dimension $[g^{2}]$ of $g^{2}\,(\mu)$ is positive, and which
therefore mostly live in spacetime dimensions $D\,\leq\,3$) this full
coupling dependence can {\em in principle} be recovered from (though it
is not given explicitly by) the most straightforward and unsophisticated
ansatz conceivable, that of a "perturbative" power-series expansion
about the free-field situation $g^{2}\,=\,0$:
\renewcommand{\theequation}{0.4}
\begin{equation}
\label{pertex}
\Gamma^{\text{pert}}_{N}\,(\{k_{i}\};\,\alpha\,(\mu);\,\mu)\,=\,
\sum\limits_{p\,=\,0}^{\infty}\,\alpha^{p}\,\Gamma^{(p)\,\text{pert}}_{N}\,(
\{k_{i}\};\,\mu)\,\,,\qquad
\alpha\,=\,\frac{g^{2}\,(\mu)}{4\,\pi}\,\,,
\end{equation}
where the zeroth-order terms, the "bare vertices"
$\Gamma^{(0)\,\text{pert}}_{N}$, are just the ordinary Feynman rules to
be read off from the action of the theory. This situation will be
briefly sketched and commented upon in chapter 1 below. We recall that
an expansion of type (\ref{pertex}) is straightforward to generate
formally, either by expanding the path-integral representation of
correlation functions (see (\ref{pathint}) below) in powers of the
interaction term of the Lagrangian ("integral" approach), or
alternatively by iterating the coupled equations of motion of the vertex
functions around their zeroth-order terms in $\alpha$, which again are
the $\Gamma^{(0)\,\text{pert}}_{N}$ of (\ref{pertex}) ("differential"
approach). Either route leads to the familiar representation of the p-th
order term $\Gamma^{(p)\,\text{pert}}_{N}$ as a sum of Feynman diagrams
with the external-legs configuration $N$ and $p$ successive momentum
integrations ("loops"). The main practical obstacle at present to
actually evaluating terms of higher perturbative order $p\,(p\,\geq\,3$,
say) is the explosive growth with $p$ of the number of diagrams
contributing that will soon come to haunt us in the form of eq.
(\ref{largep}) below, but for the superrenormalizable theories, which
enjoy the property that overall convergence of loop diagrams in the
ultraviolet {\em improves} with increasing $p$, the time is probably not
far away when computer programs, following one of the above two routes,
will furnish us with $\Gamma^{(p)\,\text{pert}}_{N}$´s for all
configurations $N$ of practical interest at the orders $p$ (in the
range of $10^{1}$ to $10^{2}$) that are nowadays routinely computed in
series expansions for statistical-mechanics models.

\qquad
That expansion (\ref{pertex}) is at best a first step in analyzing
coupling dependence has to do with the fact that, in the
superrenormalizable and in more complex theories alike, the series turns
out to be badly divergent. Dyson \cite{DYS5} as early as 1952 was the
first to show, by an ingenious physical argument, that in the case of
quantum electrodynamics (QED), the correlation functions must possess
some kind of non-analyticity at real negative $\alpha$, and that
consequently the radius of convergence in the complex $\alpha$ plane for
expansion (\ref{pertex}) is zero. Later work by many authors, work which
received a particular boost from the instanton results of Lipatov
\cite{LIP}, established techniques for calculating the leading
large-$p$ behavior of the coefficients $\Gamma^{(p)\,\text{pert}}_{N}$,
with the results always fitting into the generic form
\renewcommand{\theequation}{0.5}
\begin{equation}
\label{largep}
\Gamma^{(p)\,\text{pert}}_{N}\quad \longrightarrow\quad
c\,p^{B}\,A^{-p}\,p!\,\left[1\,+\,{\cal{O}}\,\left(\frac{1}{p}\right)\right]\qquad
(p\,\to\,\infty)\,\,,
\end{equation}
where $c$ carries the momentum dependence, and $A,\,b$ are real
constants. (The large-$p$ calculations form an interesting subject in
themselves, into which these seminars cannot really enter, so I must
refer you to the editor´s introduction and the reprinted primary sources
in the edited volume \cite{LGZ} if you wish to pursue this special
topic.) In particular, the {\em instanton contribution} to large-$p$
behavior, which is calculated along the "integral" route mentioned
above by expanding the path-integral representation semi-classically
around a finite-action solution to the Euclidean classical equations of
motion ("instanton"), has the constant $A$ equal to that finite
action. That this instanton saturation does catch the full large-$p$
behavior is, as far as I know, currently a hope. Indeed in the strictly
renormalizable theories additional contributions have been identified in
perturbation theory, which we shall have occasion to mention in sect.
2.4; they also follow the pattern (\ref{largep}) but with different
constants $A,\,b,\,c$. In the absence of a complete classification of
possible contributions to large-$p$ behavior, the mathematically minded
person would not regard the divergence of the $\alpha$ expansion as
proven, but since the pattern of eq. (\ref{largep}) has resurfaced in
all large-$p$ studies performed so far, we shall assume here
pragmatically that up to linear combination with possibly different sets
of constants, {\em but with the p! factor common to all terms}, this
relation does give the large-$p$ behavior. The factorial growth, which
in the superrenormalizable theories simply reflects the fast increase in
the number of combinatorial possibilities of drawing a $p$-loop Feynman
diagram, precludes convergence of the series (\ref{largep}) at any
$\alpha$.

\qquad
It is then obvious that for coupling dependence, expansion
(\ref{largep})
is not simply "the" answer. To make some use of the information
contained in it, one essentially has three options. If a regime exists
where the running coupling $\alpha\,(\mu)$ is small, one may exploit the
fact that (\ref{pertex}) with (\ref{largep}), its divergent character
notwithstanding, possesses the property of {\em semiconvergence}
(section 1.1) to halt the expansion at a finite $p\,=\,p_{\text{max}}$
and use the resulting polynomial approximant directly as an estimate
for the function $\Gamma_{N}$ -- an estimate whose accuracy, while
limited in principle, may in practice be quite satisfactory. This is
what Poincaré, in chapter VIII of his celebrated {\em Nouvelles Méthodes
de la Mécanique Céleste} \cite{POIN} where he introduced the modern
concept of semiconvergent series, referred to as {\em la règle des
astronomes} -- and "les astronomes" were, of course, the perturbative
physicists of his day, the practitioners of celestial mechanics
calculating planetary-orbit perturbations.
This is also
what is routinely done in the electroweak theory, where the gauge
couplings are indeed small in the currently accessible range of scales,
$\mu\,\stackrel{<}{\sim}\,v\,\approx\,250$ GeV (the electroweak
scale).

\qquad
In regimes where $\alpha\,(\mu)$ is not small enough to provide good
semiconvergence, one may try to adduce physical arguments to justify,
for particular kinematical situations, an estimation of $\Gamma_{N}$ by
"partial resummation", i.\,e., the {\em exact summation to all orders
of certain subseries} of (\ref{pertex}) whose terms one has reason to
believe will dominate the particular phenomenon. In most cases the
series summed are subseries of nonzero convergence radius -- geometric,
exponential, or Bessel-function series -- which stay clear of the $p!$
growth problem. Examples are the "leading-logarithms" geometric
resummations one performs (as discussed in sect. 2.1 below) in order to
trade the coupling $\alpha\,(\mu)$ for a coupling
$\overline{\alpha}\,(Q^{2})$ running with some momentum variable $Q^{2}$
(one of the $k_{i}^{2}$ or $k_{i}\,\cdot\,k_{j}$ or a combination of
these) of
$\Gamma_{N}$,
or calculations of bound states like positronium by solving a
Bethe-Salpeter equation with low-order irreducible kernel, which amounts
to summing an integral-kernel-valued geometric series in $\alpha$. Such
summations are useful and (as in the bound-state problem) sometimes
necessary to extend the range of applicability of low-order perturbative
information, and to the extent that the resummed function often has
properties qualitatively different from those of the finite polynomial
approximants, they may be called "nonperturbative". One should however
keep in mind, particularly when looking beyond the superrenormalizable
theories, that they will not recover terms that are invisible in
principle in the semi-convergent perturbative expansion, and in the
context of these seminars I will therefore refer to them as "weakly"
nonperturbative.

\qquad
On the other hand if one is interested in the more theoretical question
of whether, or to what extent, the series determines the "true", exact
function $\Gamma_{N}$, one must look for uniqueness properties
justifying the application of what mathematicians call limiting
procedures, or summability methods -- the fine art of sensibly assigning
a value to a divergent series. Such uniqueness results exist for the
superrenormalizable theories and will be sketched in sect. 1.2; sect.
1.3 will then go on to discuss the most frequently used technique, the
Borel-Laplace transform, of actually performing the reconstruction of
$\Gamma_{N}$. In this context, sect. 1.3 will also take a look at the
analytic-continuation step required for that transform -- apparently a
purely technical problem, but one that will guide us in more central,
analogous problems arising further down the road in sect. 3.1. -- All of
the material of chapter 1 is "old", and most of it in fact was old
already when physicists became interested in it in the late 1960´s,
dating back as it does to the work of mathematicians of the late 19th
and early 20th centuries, but I believe it is of timeless interest and
value, and deserves to be retaught on occasion.

\qquad
The situation is markedly different in the realistic field theories in
$D\,=\,4$ -- in particular, the asymptotically free gauge theories
central to present-day particle physics -- that are "strictly" or
"marginally" renormalizable but not superrenormalizable (chapter 2).
There, the very definition of the theory involves a crucial new element,
a {\em divergent coupling renormalization}.
(More precisely, the tri- and quadrilinear {\em interaction monomials}
in the Lagrangian of the theory, which in the superrenormalizable cases
underwent at most finite rescalings, now demand their own
ultraviolet-divergent renormalizations, but here we will continue to
somewhat loosely refer to just the coupling renormalization that is at
the core of these).
If $\overline{g}_{0}\,(\varepsilon)$ denotes the bare coupling
originally
appearing in the interaction terms of the Lagrangian, with mass
dimension
$[\overline{g}_{0}]\,=\,\varepsilon\,=\,\frac{1}{2}\,(4\,-\,D)$ in
$D\,=\,4\,-\,2\,\varepsilon$ dimensions, then its relation to the
dimensionless and renormalized coupling $g\,(\mu)$ that parametrizes the
functions (\ref{depends}) must in these theories be written as
\renewcommand{\theequation}{0.6}
\begin{equation}
\label{recoup}
\overline{g}^{2}_{0}\,(\varepsilon)\,=\,g^{2}\,(\mu)\,\mu^{2\,\varepsilon}\,Z_{\alpha}\,(g^{2}\,(\mu),\,\varepsilon)\,\,,
\end{equation}
where the coupling-renormalization constant, $Z_{\alpha}$, is
ultraviolet divergent as $\varepsilon\,\to\,0$ at any finite order in
$g^{2}\,(\mu)$ except for the zeroth order,
$Z_{\alpha}\,(0,\,\varepsilon)\,=\,1$. This is in contrast to the
superrenormalizable theories where $Z_{\alpha}$ is ultraviolet finite
term by term and can always be transformed to 1 by a {\em finite}
rescaling of the elementary fields $\varphi\,(x)$ (and, possibly, of
$m\,(\mu)$). You have become accustomed (or so I presume) to the idea
that the renormalization process is every bit as important in
determining the dynamics of a QFT as the essentially classical
Lagrangian, and you will therefore be prepared to hear that this new
ingredient draws a fine but fateful line between the superrenormalizable
and the strictly renormalizable field theories, a line signalling in
fact a deep and fundamental divide. (I confess that doubts sometimes
beset me as to whether the authors of the myriad scholarly papers
studying superrenormalizable models in order to learn something for the
realistic theory have always fully realized what they were up against.)
As we shall recall in sect. 2.1, this new feature directly translates
into the emergence of an "RG-invariant mass scale", $\Lambda$, that is
non-analytic at $\alpha\,=\,0$ in such a way as to permit no
power-series expansion in $\alpha$ at all, not even in the
semiconvergent sense \cite{GNV}. In such a theory, the very act of
writing expansion (\ref{pertex}) already implies that one has lost an
important class of terms in the correlation or vertex functions, and
although certain traces of the loss remain, it is essentially
irretrievable. (The terms lost will become important as soon as one or
more of the Lorentz-invariant momentum arguments $k_{i}\,\cdot\,k_{j}$
in our functions are not much larger than $\Lambda^{2}$.) To specify the
minimal information required for a reconstruction of such a function one
then needs, in place of (\ref{pertex}), a more detailed formal construct
called a general resurgent symbol, or a resurgent double expansion,
which
still has a "perturbative direction" proceeding in powers of $\alpha$,
but now also features a "nonperturbative direction" providing for a
systematic approximation of the new, strongly nonanalytic $\alpha$
dependence. As discussed in sect. 2.3, the reconstruction (through an
extension of the Borel-Laplace process) then takes place in the
framework of what might be called, in relation to the summable
semi-convergent series, the next larger, mathematically well-delineated
class of functions, known to mathematicians as the {\em resurgent
functions}. In sect. 2.2 we sketch how an old acquaintance, the
operator-product expansion (OPE) of Wilson \cite{WIL} in the case of an
asymptotically free theory, can be recognized as a resurgent double
expansion, and provides hints as to the special form this expansion will
take in an asymptotically free situation. Section 2.4, in which I hope
you will find something new, examines a simple model of a
coupling-resurgent function, close enough in form to simple field-theory
amplitudes to claim relevance for our subject, but still fully tractable
by analytic means. This example will (I hope) convince you of the
importance of treating a resurgent function as a whole: you will see how
its analytic properties seem to sabotage the Laplace
reconstruction from the perturbative expansion (\ref{pertex}), but how
in fact the reconstruction is entirely well defined when the strongly
nonperturbative parts, whose separation is structurally averse to the
function´s resurgent character, are fully taken into account.

\qquad
The final chapter 3 is more special. It attempts to explain the basic
ideas (while glossing over much technical detail) of a particular
strategy for dealing with the one question to which the present
mathematics of resurgent functions does not seem to contain a ready-made
answer, and on which the physicist, for the time being, is therefore
left to his own heuristic devices. This is the question of how to obtain
some systematic formal representation of a resurgent amplitude (from
which resurgent analysis proper may start) when the double formal series
is not satisfactory for this purpose but needs a certain amount of
analytic continuation to begin with. This need arises from the simple
fact that the basic equations of motion generating the resurgent
amplitudes of QFT are not, as in almost all cases studied in
mathematical theory, ordinary differential, integro-differential, or
difference equations {\em with respect to the coupling},
but a rather complex, hierarchically
coupled, and renormalizably divergent system of integral equations
{\em with respect
to the momentum (or coordinate) arguments} of correlation functions,
equations into which the coupling enters only as a parameter. Actually,
the emergence of functions resurgent with respect to a parameter --
rather than the variable acted upon by a differential operator -- from
differential equations has been studied in mathematics under the name of
{\em quantum resurgence}, and properly speaking it is this kind of
resurgence that will concern us below. But there is an extra
difficulty here, arising from the fact that the nonperturbative coupling
dependence enters as a {\em dimensionful} parameter, the mass scale
$\Lambda$, and is therefore inextricably linked with the dependence on
the momentum variables $k_{i}$: in the integral equations, the loop
integrals extend down to the smallest momenta $|k_{i}|$, and in order to
get started at all, we need to have something to insert for the
integrands that is valid also at $|k_{i}|\,\ll\,\Lambda$, whereas
resurgent symbols of the OPE variety only furnish asymptotic
representations for $|k_{i}|\,\gg\,\Lambda$. We are therefore going to
need, from the very start, at least some partial continuation (namely,
with respect to the $\Lambda$ variable) of the symbols.
Section 3.1 outlines one practical form of
such a partial continuation, drawing on what we may have learned in
sect.
1.3, which can be cast in the suggestive form of {\em generalized
Feynman rules}, objects of zeroth order ($p\,=\,0$) in the
"perturbative
direction" but providing a systematic continuation-through-resummation
in the "nonperturbative direction". Section 3.2 sketches the special
self-consistency mechanism by which the generalized Feynman rules can
establish themselves in the hierarchical vertex equations, and mentions
some nontrivial consequences of the fact that this mechanism is tied to
the renormalizable loop divergences of those equations. Section 3.3 is a
brief outlook on some of the features of the coupling-resurgent
iterative solution that can then in principle be generated from the
generalized Feynman rules -- an outlook certain to be incomplete, as
many other such features probably remain to be explored. My goal here
will have been attained if you come away from this with a realization
that the {\em two-step nature} of this reconstruction process is not due
to the shortcomings of the physicist´s heuristic patchwork, but is
firmly rooted in the mathematical nature of resurgent functions.

\qquad
Despite their title, these talks are not lectures on mathematics.
Although we are going to quote a few mathematical theorems without
proofs and indulge in a certain amount of name-dropping, we will mostly
proceed, as physicists do more often than not, by plausibility, formal
calculation, and simple examples. The reason is lack of expertise,
certainly not lack of enthusiasm, on the part of your speaker. I do feel
obliged, however, to draw your attention to the fact that there already
exists a substantial body of mathematical research on the fascinating
objects called resurgent functions. (I learned what little I hope to
have understood from selected chapters of the books \cite{CNP} and
\cite{STSH}.) A large part of this research builds on the fundamental
contribution of one person, the French mathematician Jean Écalle, who in
{\em Les fonctions résurgentes} (1981 - 85), a three-volume work
totalling almost 600 pages \cite{ECAL}, first singled out resurgent
functions as a remarkable and unexplored phenomenon within classical
complex analysis, studied comprehensively their structural aspects --
the algebras and subalgebras, auto- and homomorphisms, dual-space and
fiber-bundle structures engendered by them -- that are of premier
interest
to the mathematician, introduced, in the form of the so-called alien
derivatives ({\em dérivations étrangères}) and of the associated alien
differential equations, a whole array of powerful new tools and concepts
for characterizing the peculiar
continuation behavior of resurgent functions around singularities of
their Borel transforms, and developed on this basis a new classification
of ordinary differential equations of truly remarkable scope and depth,
where large classes of equations related by quite general analytic
mappings can now be characterized by a single "complete set of
holomorphic
invariants". In these seminars, because of the limitations imposed by
time and by my insufficient mathematical preparation, I am going to use
only a few of the very simplest and most qualitative concepts from
this
beautiful edifice, but even these few will, I am confident, demonstrate
to you one hallmark property of Écalle´s theory: its power to structure,
to clarify, and to provide larger perspective on, what previously
may have looked like heaps of isolated and poorly understood results.

\renewcommand{\thesection}{1}
\section{Superrenormalizable Theories: Summability}
\subsection{Semiconvergent or Asymptotic Series}

\qquad
We are accustomed to dealing with functions $f\,(z)$ of a complex
variable $z$ analytic around $z\,=\,0$ and therefore having a
power-series expansion in $z$ within a circle of convergence around the
origin. Semiconvergent, or asymptotic, expansions at the origin usually
arise in situations where $z\,=\,0$ is not only a point of
non-analyticity of $f\,(z)$, but is in fact a member of a whole set of
non-analyticity points -- an endpoint of a branch cut, say, or an
accumulation point of a sequence of poles, etc. -- that reach up to the
origin from somewhere in the complex plane, and the analyticity
situation, not only globally but already locally around $z\,=\,0$, is
therefore "anisotropic" to begin with: the residual domain of
analyticity, starting at $z\,=\,0$, is usually concentrated along some
direction, or sector of directions, in the $z$ plane. In our context,
where $z$ is the coupling $g^{2}$ or $\alpha$, the positive real axis is
distinguished among possible directions of this kind: it represents the
{\em physical} values of the coupling, and if a field-theory amplitude
were singular or two-valued at points on this axis, the underlying QFT
would quite simply be nonexistent as a physically acceptable theory.
We will therefore always assume that our functions $f\,(z)$ are analytic
in at least some narrow region ${\cal{R}}$ along the positive real axis
(mathematical theory, to be sure, deals with much more general
situations). This, for example, may simply be a strip of constant width
along the axis (Fig. 1.1 (a)), or a sector of opening angle $\gamma$
with apex at the origin (Fig. 1.1 (b)), or, to mention a seemingly
exotic case that is nevertheless of relevance for QFT, a region bounded
by circular arcs tangent to the positive real axis at $z\,=\,0$ (Fig.
1.1 (c)). The "sectorial" analyticity of Fig. 1.1 (b) is the standard
situation postulated in practically all mathematical treatments, and
indeed often made part of the very definition of asymptotic series, but
we should keep in mind that physics may force other kinds of domains on
us. However, they all have in common that they do not allow for a finite
circle of analyticity around the point $z\,=\,0$.

\qquad
A formal power series in $z$, i.\,e., an object of the form
\renewcommand{\theequation}{1.1}
\begin{equation}
\label{formser}
\sum\limits_{n\,=\,0}^{\infty}\,a_{n}\,z^{n}
\end{equation}
with real or complex coefficients $a_{n}$ and with nothing being said as
yet about its possible value, is called a {\em semiconvergent or
asymptotic expansion} for our function $f\,(z)$ if for all $z$ in the
analyticity region ${\cal{R}}$ the error estimate
\renewcommand{\theequation}{1.2}
\begin{equation}
\label{error}
\left|f\,(z)\,-\,\sum\limits_{n\,=\,0}^{N}\,a_{n}\,z^{n}\right|\,\leq\,C_{N\,+\,1}\,|z|^{N\,+\,1}
\end{equation}

\begin{center}
\epsfig{file=fig11.eps,width=12.5cm}\\
$\begin{array}{ll}
{\bf Fig. 1.1}\quad
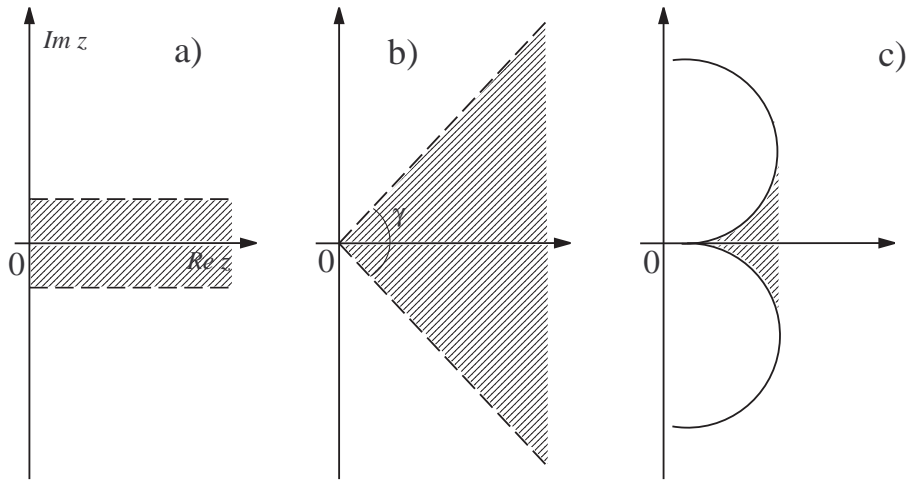
& \text{Examples of regions {\cal{R}} of analyticity}\\
& \text{including the positive real axis}
\end{array}$
\end{center}

is valid for all nonnegative integers $N$, with constants $C_{N\,+\,1}$
such that the error bound eventually diverges for large $N$ everywhere
in ${\cal{R}}$:
\renewcommand{\theequation}{1.3}
\begin{equation}
\label{boundiv}
C_{N\,+\,1}\,|z|^{N\,+\,1}\quad \longrightarrow\quad \infty\qquad
(N\,\to\,\infty)\,\,.
\end{equation}
One expresses this situation by writing
\renewcommand{\theequation}{1.4}
\begin{equation}
\label{asyto}
f\,(z)\,\sim\,\sum\limits_{n\,=\,0}^{\infty}\,a_{n}\,z^{n}\,\,,
\end{equation}
with the $\sim$ sign being read as "has the asymptotic expansion".
In contrast to the case of a convergent series, (\ref{error}) does not
guarantee that the error on its l.h.s. can be made arbitrarily small at
all points in ${\cal{R}}$ by going to sufficiently large $N$. The
expansion may nevertheless by useful for obtaining estimates of $f\,(z)$
at small $z$ in ${\cal{R}}$: for $|z|\,\ll\,1$,
the exponential decrease
of $|z|^{N}$ with $N$ may, for the first few $N$´s, win over the growth
of the $C_{N\,+\,1}$, with the divergence (\ref{boundiv}) making itself
felt only at larger $N$´s, and if one halts the expansion "in time",
one may come away with an error that is tolerable for the purpose at
hand. Such an error pattern, illustrated in Fig. 1.2, is behind all
"pragmatic" applications of asymptotic expansions, where one strictly
avoids asking questions about a possible meaning or value of the series
(\ref{formser}) to all orders.

\begin{center}
\epsfig{file=fig12.eps,width=14cm}\\
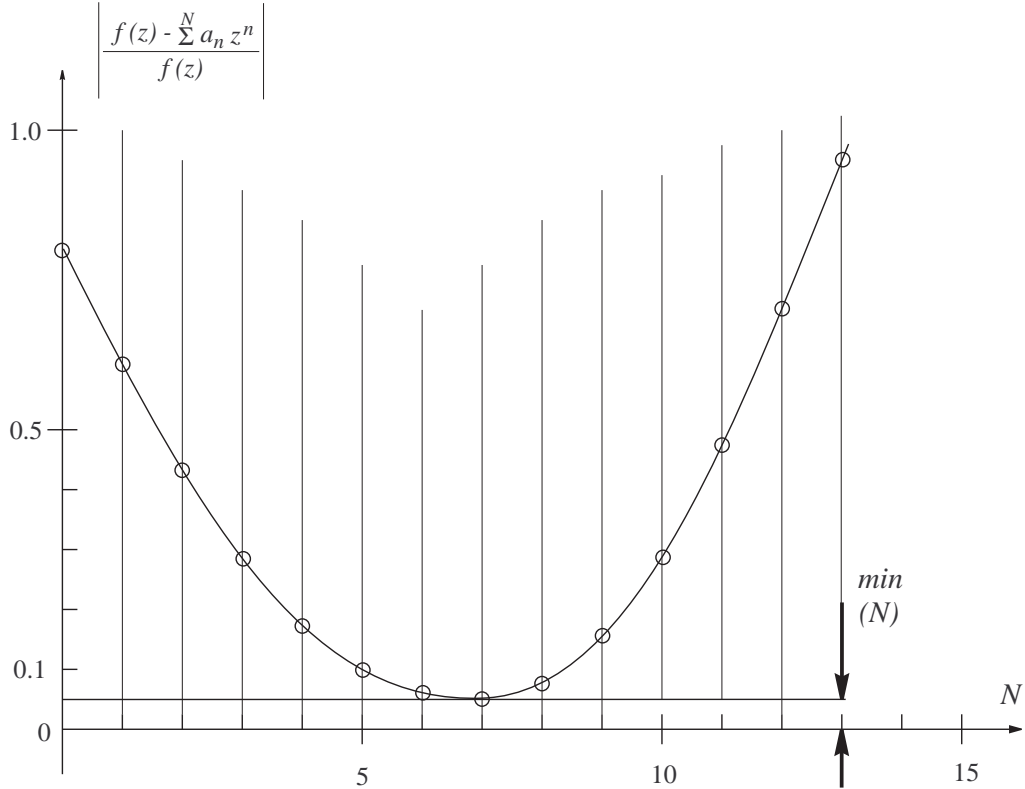
{\bf Fig. 1.2}\quad Typical error behavior of semiconvergent expansion
\end{center}

\qquad
This pragmatic handling of asymptotic series is facilitated by some
simple formal properties: two formal series semiconvergent to functions
$f_{1}\,(z)$ and $f_{2}\,(z)$ in the same region ${\cal{R}}$ can be
linearly combined, with complex coefficients $\zeta_{1}$ and
$\zeta_{2}$, and multiplied to give formal series that are
semiconvergent to the corresponding linear-combination and product
functions, $\zeta_{1}\,f_{1}\,+\,\zeta_{2}\,f_{2}$ and
$f_{1}\,\cdot\,f_{2}$, respectively. A series (\ref{formser}) asymptotic
to $f\,(z)$ may be integrated termwise from $z\,=\,0$ out to some
$z\,\in\,{\cal{R}}$ to give a series asymptotic to a primitive of $f$,
\renewcommand{\theequation}{1.5}
\begin{equation}
\label{serint}
\int\limits_{0}^{z}\,f\,(w)\,d\,w\,\sim\,
\sum\limits_{n\,=\,1}^{\infty}\,\frac{a_{n\,-\,1}}{n}\,z^{n}\,\,.
\end{equation}
On the other hand the series produced by termwise differentiation,
\renewcommand{\theequation}{1.6}
\begin{equation}
\label{diffser}
\sum\limits_{n\,=\,0}^{\infty}\,(n\,+\,1)\,a_{n\,+\,1}\,z^{n}\,\,,
\end{equation}
can be trusted to be
semiconvergent to the derivative function $f^{'}\,(z)$ in
${\cal{R}}$ only if one knows in advance that this derivative exists, is
continuous, and does possess an asymptotic expansion.

\qquad
Can a function have more than one asymptotic expansion? It is easy to
see that the answer is no \cite{WIGH}: divide the estimate (\ref{error})
by
$|z|^{N}$ so that one factor of $|z|$ remains on the r.\,h.\,s., and let
$z\,\to\,0$ from within ${\cal{R}}$ or, as we will put it in view of the
special nature of our regions ${\cal{R}}$, "from the right". You get
\renewcommand{\theequation}{1.7}
\begin{equation}
\frac{f\,(z)\,-\,\sum\limits_{n\,=\,0}^{N}\,a_{n}\,z^{n}}{z^{N}}\quad
\longrightarrow\quad 0\quad \text{as}\quad
z\,\to\,0+\qquad ({\rm all}\,\,\,N)\,\,,
\end{equation}
a property often used as an alternative {\em definition} of
semiconvergence. Spelling this out for successive $N$´s one has
\renewcommand{\theequation}{1.8}
\begin{equation}
\begin{array}{lcl}
\Bigl[f\,(z)\,-\,a_{0}\Bigr]_{z\,\to\,0+}\,=\,0 & \Rightarrow &
a_{0}\,=\,f\,(0+)\,\,,\\
\mbox{}\\
\displaystyle{
\left[\frac{f\,(z)\,-\,a_{0}}{z}\,-\,a_{1}\right]_{z\,\to\,0+}\,=\,0 } &
\Rightarrow & a_{1}\,=\,f^{'}\,(0+)\,\,,\\
\mbox{}\\
\displaystyle{
\left[\frac{\frac{f\,(z)\,-\,a_{0}}{z}\,-\,a_{1}}{z}\,-\,a_{2}\right]_{z\,\to\,0+}\,=\,0}
& \Rightarrow & a_{2}\,=\,\frac{f^{''}\,(0+)}{2!}\,\,,
\end{array}
\end{equation}
etc. That is, if a function $f\,(z)$, analytic in a region ${\cal{R}}$
as specified above, has an asymptotic expansion at all, then that
expansion is identical with its Taylor series "from the right" at
$z\,=\,0$:
\renewcommand{\theequation}{1.9}
\begin{equation}
\label{atayl}
a_{n}\,=\,\frac{1}{n!}\,\left(\frac{d^{n}\,f}{d\,z^{n}}\right)_{z\,\to\,0+}\qquad
(n\,=\,0,\1,\,2,\,...\,)\,\,.
\end{equation}
Moreover, by writing the Taylor expansion of $f\,(z)$ with remainder
term at $z\,\in\,{\cal{R}}$, one obtains the
integral representation
\renewcommand{\theequation}{1.10}
\begin{equation}
\label{remtayl}
f\,(z)\,-\sum\limits_{n\,=\,0}^{N}\,a_{n}\,z^{n}\,=\,\left[
\frac{1}{N!}\,\int\limits_{0}^{1}\,d\,t\
(1\,-\,t)^{N}\,f^{(N\,+\,1)}\,(t\,z)\right]\,z^{N\,+\,1}
\end{equation}
of the error term. If the derivatives $|f^{(N\,+\,1)}|$ are all bounded
throughout ${\cal{R}}$ by $N$-dependent constants, then so is the
bracketed integral, and the bound gives back the
constants $C_{N\,+\,1}$ of (\ref{error}). This point is important -- for
the bound of (\ref{error}) to hold, it is not sufficient to know the
large-$n$ behavior of the Taylor coefficients at $z\,=\,0+$ (such as
relation
(\ref{largep})). We also need some boundedness properties of the
function $f$ and its derivatives {\em away from zero}, which the formal
series alone does not supply.

\qquad
Importantly, the statement converse to this result is not true: a given
formal series (\ref{formser}) can be a semiconvergent expansion for more
than one function $f\,(z)$ analytic in ${\cal{R}}$, and in fact to
infinitely many such functions. Indeed if $f_{1}\,(z)$ and $f_{2}\,(z)$
differ by a function $g\,(z)$ analytic in ${\cal{R}}$ and having all its
derivatives vanishing as $z\,\to\,0+$,
\renewcommand{\theequation}{1.11}
\begin{equation}
\label{evanesc}
f_{2}\,(z)\,-\,f_{1}\,(z)\,=\,g\,(z)\,\,;\qquad
\left(\frac{d^{n}\,g}{d\,z^{n}}\right)_{z\,\to\,0+}\,=\,0\quad ({\rm
all}\,\,\,n)\,\,,
\end{equation}
then both have the same asymptotic expansion. Such a $g\,(z)$ would
vanish identically if it were to be analytic in a region around
$z\,=\,0$, but with our much weaker condition of "anisotropic"
analyticity in a region ${\cal{R}}$, it is quite possible for $g\,(z)$
to be nonzero. From (\ref{error}) we can immediately deduce an
alternative characterization of such functions,
\renewcommand{\theequation}{1.12}
\begin{equation}
\label{evapow}
|g\,(z)|\,\leq\,D_{N\,+\,1}\,|z|^{N\,+\,1}\qquad
(z\,\in\,{\cal{R}},\,\,{\rm all}\,\,\,N)\,\,,
\end{equation}
with constants $D_{N}$. That is, they "vanish faster than any power of
$z$" as $z\,\to\,0+$ in ${\cal{R}}$. Of course, the two functions
$f_{1}$ and
$f_{2}$ differing by $g$ have different constants $C_{N}$ (namely,
$C_{N\,+\,1}^{(2)}\,=\,C_{N\,+\,1}^{(1)}\,+\,D_{N\,+\,1}$) in their
error estimates (\ref{error}), so these constants are characteristic of
the {\em individual functions}, not of their asymptotic expansion.

\qquad
In order to pin down more precisely this class of functions which "fail
to show up" in an asymptotic expansion, and also to estimate the
minimal error observed in Fig. 1.2, we need to adopt more specific
hypotheses about the constants $C_{N}$ in (\ref{error}). In view of what
we said in the introduction about the large-$p$ pattern (\ref{largep})
in QFT, it seems natural to assume from now on that
\renewcommand{\theequation}{1.13}
\begin{equation}
\label{gevrey}
C_{N}^{(f)}\,=\,c_{(f)}\,N^{B}\,|A|^{-N}\,N!\qquad
(N\,\geq\,1)\,\,,
\end{equation}
(although as already emphasized this does {\em not follow} from
(\ref{largep}) alone).
Here only the positive constant $c_{(f)}$ depends on the individual
function $f\,(z)$, while $B$ and $A$ are to be the same for all
functions
having the same asymptotic expansion (this is not essential but
simplifies calculations). This form clearly has the property
(\ref{boundiv}). It would obtain, for example, if in (\ref{remtayl}) we
had the estimate
\renewcommand{\theequation}{1.14}
\begin{equation}
\label{derbnd}
|f^{(N)}\,(z)|\,\leq\,c_{(f)}\,N^{B}\,|A|^{-N}\,(N!)^{2}\,\,,\qquad
z\,\in\,{\cal{R}}\,\,.
\end{equation}
The case $B\,=\,0$ is known as {\em functions of Gevrey
class 1} in the mathematical literature. With this assumption, a
function $g\,(z)$ as in (\ref{evanesc}) satisfies
\renewcommand{\theequation}{1.15}
\begin{equation}
\label{gevbnd}
|g\,(z)|\,\leq\,c_{(g)}\,N^{B}\,\left|\frac{z}{A}\right|^{N}\,N!\qquad
(N\,\geq\,1)\,\,.
\end{equation}
We can find the lowest among this infinite sequence of bounds by
minimizing the r. h. s. with respect to $N$ at fixed $|z|$. For this
purpose, we temporarily treat $N$ as a continuous variable, writing
$N!\,=\,N\,\Gamma\,(N)$. The condition for (\ref{gevbnd}) to be minimal
at $N\,=\,N_{m}$ then turns out to be
\renewcommand{\theequation}{1.16}
\begin{equation}
\label{mincond}
\ln\,\left|\frac{z}{A}\right|\,+\,\psi\,(N_{m})\,+\,\frac{B\,+\,1}{N_{m}}\,=\,0\,\,,
\end{equation}
where $\psi\,(x)$ is the logarithmic derivative of the gamma function.
There is no closed-form solution to this equation, but we are primarily
interested in small-$z$ behavior, $|z|\,\ll\,|A|$, where we expect
$N_{m}\,\gg\,1$, so we may use
$\psi\,(x)\,=\,\ln\,x\,+\,{\cal{O}}\,(x^{-2})$ \cite{GRY}, the Stirling
expansion applied to $\psi$, to write
\renewcommand{\theequation}{1.17}
\begin{equation}
\label{minlead}
\ln\,\left(N_{m}\,\left|\frac{z}{A}\right|\right)\,+\,
\frac{B\,+\,1}{N_{m}}\,+\,{\cal{O}}\,\left(\frac{1}{N^{2}_{m}}\right)\,=\,0\,\,.
\end{equation}
The leading-order solution
obviously is $N_{m}\,\approx\,\left|\frac{A}{z}\right|$, so we can write
$N_{m}$ as a Laurent expansion in $\frac{z}{A}$ and obtain
\renewcommand{\theequation}{1.18}
\begin{equation}
\label{ennmin}
N_{m}\,=\,\left|\frac{A}{z}\right|\,-\,(B\,+\,1)\,+\,{\cal{O}}\,\left(\left|\frac{z}{A}\right|\right)\,\,.
\end{equation}
Plugging this into the r. h. s. of (\ref{gevbnd}) while using for $N!$
once more the Stirling formula, we obtain the lowest bound on
$g\,(z)$ in the form of a small-$z$ expansion,
\renewcommand{\theequation}{1.19}
\begin{equation}
\label{bndmin}
|g\,(z)|\,\leq\,\left(c_{(g)}\,\sqrt{2\,\pi}\,|A|^{B^{'}}\right)\,|z|^{-B^{'}}\,\E^{-\frac{|A|}{|z|}}\,
[1\,+\,{\cal{O}}\,(|z|)]\,\,,
\end{equation}
where $B^{'}\,=\,B\,+\,\frac{1}{2}$, and where the last factor in square
brackets is a power series in $|z|$ arising from the higher corrections
in both eq. (\ref{ennmin}) and the Stirling formula. (Of course, unless
the $N_{m}$ of (\ref{ennmin}) happens to be an integer, the actual
minimum will occur at $[N_{m}]$ or $[N_{m}]\,+\,1$, but in the case
$N_{m}\,\gg\,1$ of primary interest here, this leads only to a minor
readjustment of the constants in (\ref{bndmin}) \cite{STSH}). We will
refer to functions satisfying a bound of this type as {\em exponentially
suppressed} (remember that even if $B^{'}\,>\,0$, the function
$\exp\,\left(-\frac{|A|}{|z|}\right)$ vanishes so fast as $|z|\,\to\,0$
as to suppress arbitrary negative powers). Thus if a function $g\,(z)$,
analytic in ${\cal{R}}$, has vanishing asymptotic expansion and in
addition satisfies a bound of the special form (\ref{gevbnd}), then our
earlier statement "vanishes faster than any power" can be replaced by
the more precise statement of exponential suppression. It is clear that
the same result (\ref{bndmin}), with $c_{(g)}$ replaced by $c_{(f)}$,
also holds as a minimum-error estimate for the semiconvergent
approximation of a function $f\,(z)$ having a {\em nonvanishing}
semiconvergent expansion, and in addition obeying the special condition
(\ref{gevrey}) on its error-bounding constants $C_{N}$.

\qquad
On the other hand if $g\,(z)$ is known to be of the form
\renewcommand{\theequation}{1.20}
\begin{equation}
\label{exsupp}
g\,(z)\,=\,d_{(g)}\,z^{-B^{'}}\,\E^{-\frac{A^{'}}{z}}\,[1\,+\,
{\cal{O}}\,(z)]\qquad (A^{'}\,>\,0)
\end{equation}
with real constants $A^{'},\,B^{'},\,d_{(g)}$, then all its derivatives
contain the factor $\exp\,(-A^{'}\,/\,z)$ and therefore vanish as
$z\,\to\,0$ from the right, no matter what $B^{'}$ and the bracketed
power series are, and so it at least has vanishing asymptotic
expansion. That in addition it also falls into the class of eq.
(\ref{gevbnd}) can be proved {\em generally} only for the "sectorial"
analyticity situation of Fig. 1.1 (b) with opening angle
$\gamma\,<\,\pi$, where one may write
\renewcommand{\theequation}{1.21}
\begin{equation}
\label{sectex}
\left|\E^{-\frac{A^{'}}{z}}\right|\,\leq\,\E^{-\frac{|A|}{|z|}}\quad
\text{with} \quad
|A|\,=\,A^{'}\,\cos\,\left(\frac{1}{2}\,\gamma\right)\,\,,
\end{equation}
but it may be true in individual other cases, and in any case the weaker
statement is quite sufficient for our purposes.

\qquad
A very interesting observation in this context is that the
square-bracketed power series in
(\ref{bndmin}), arising as it does from the semiconvergent Stirling
approximations for $\psi\,(N)$ and $\Gamma\,(N)$, is itself
semiconvergent, and that semiconvergent series are also admissible for
the bracketed series in (\ref{exsupp}), without changing our conclusion.
Thus within the class of functions characterized by (\ref{gevrey}), the
subclass of "escaping" functions that are left undetermined by a
semiconvergent series are typically of the form of an
exponential-suppression factor (possibly preceded by an arbitrary power
$z^{-B^{'}}$ that does no harm) {\em times another power series which
may itself be semiconvergent}. This observation is one of the possible
heuristic starting points for the more general theory discussed in
chapter 2.

\qquad
Of course, life would be much simpler if we could get rid of the
undetermined part altogether when trying to reconstruct a function from
its asymptotic expansion. In such cases the series is referred to as
{\em summable}: there is then a {\em unique} function $f\,(z)$ having
this series as its small-$z$ expansion. It is to such a situation
that we turn next.

\subsection{Nevanlinna-Sokal Analyticity and Application to QFT}

\qquad
A {\em sufficient} condition that has been widely
invoked for unique reconstruction of a function from its
asymptotic expansion, in both mathematics and physics contexts, is
based on a theorem found by Phragmén and Lindelöf already
at the end of the 19th century. It again refers to the "sectorial"
analyticity situation of Fig. 1.1 (b), but this time with an opening
angle $\gamma$ larger than $\pi$ (see Fig. 1.3):
\renewcommand{\theequation}{1.22}
\begin{equation}
\label{phrag}
\left\|\quad
{\parbox{10cm}{
{\bf Phragmén-Lindelöf Theorem:} {\em A function $g\,(z)$ analytic
in a}
{\em sector ${\cal{S}}$ of radius $R$ and angle $\gamma>\pi$ with
apex at the origin, and}
{\em fulfilling the condition of exponential suppression
(\ref{bndmin}) throughout}
{\em ${\cal{S}}$, must vanish identically.}}}
\quad\right\|
\end{equation}

\begin{center}
\epsfig{file=fig13.eps,width=11cm}\\
$\begin{array}{ll}
{\bf Fig. 1.3}\quad &
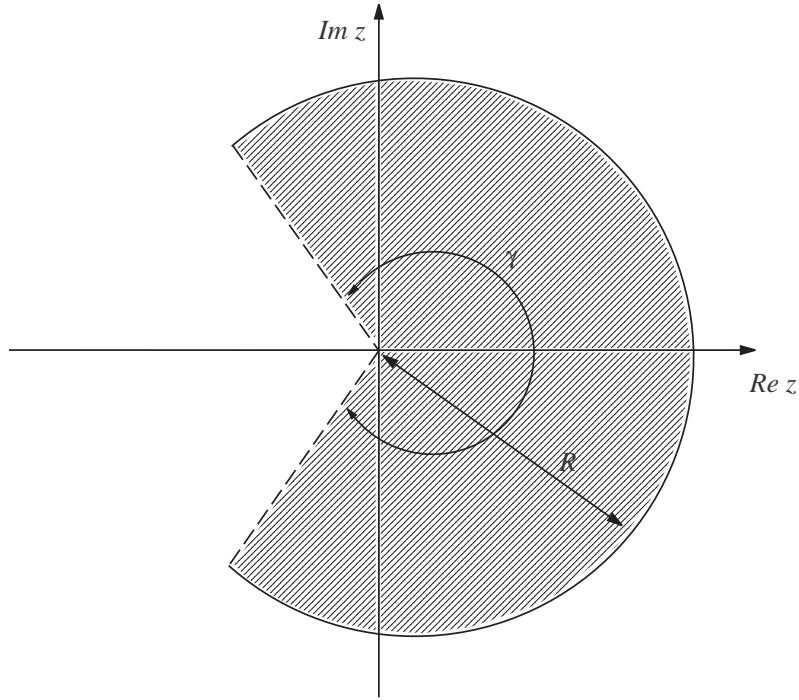
\text{Sectorial region of analyticity of opening angle}
>\,\pi\\
& \text{in the Phragmén-Lindelöf theorem}
\end{array}$
\end{center}
\qquad
I refer you to, for example, ref. \cite{STSH} for a proof of this
theorem, which is based on an application of the maximum principle of
complex-function theory. Note that a finite radius $R$ of the sector is
quite sufficient for this theorem; the essential point is the large
angle of the analyticity sector.

\qquad An immediate application of this theorem is to a function
$f\,(z)$ analytic in such a sector of wide opening and having an
asymptotic expansion (\ref{error}) there, with error-bounding constants
$C_{N}$ given by (\ref{gevrey}). We have seen in the last section that
for such a function the "escaping" functions $g\,(z)$, by which it
differs from all other functions having the same asymptotic expansion in
the same analyticity domain, obey the bound (\ref{bndmin}). We therefore
have the
\renewcommand{\theequation}{1.23}
\begin{equation}
\label{watson}
\left\|\quad
{\parbox{10cm}{
{\bf Uniqueness Theorem I}: {\em If a function $f\,(z)$ is analytic
in a sector
${\cal{S}}$ of opening angle $\gamma\,>\,\pi$, and
possesses in this sector an
asymptotic expansion (\ref{formser}) with error-bounding
constants given by
(\ref{gevrey}), then it is the only function
with this asymptotic expansion
in ${\cal{S}}$.}}}
\quad \right\|
\end{equation}
This application was made in 1912 by G. N. Watson \cite{WAT}, who
combined it with a proof that one particular method of reconstructing
the unique $f\,(z)$ from the series -- the Borel-Laplace process we are
going to discuss in the next section -- works under the same hypothesis.
His result, known as {\em Watson´s theorem}, became the favorite
summability provider for more than half a century, right through to the
first summability proofs for (superrenormalizable) QFT´s in the
mid-1970´s. It is a quirk of mathematical history that nobody seems to
have been irritated by the fact that the Laplace-transform
representation
established by Watson -- our eq. (\ref{laplace}) below -- did not
recover $f\,(z)$ in its full Phragmén-Lindelöf analyticity domain of
Fig. 1.3 used as input to the proof, but only in a much smaller
subdomain -- the hatched circle of Fig. 1.4. Didn´t this look
suspiciously as if analyticity in that smaller domain should have been
sufficient in the first place?

\qquad
Sokal \cite{SOK} in 1979 proved  precisely this, and also redressed an
injustice of history by pointing out that the result had in fact been
given as early as 1918 by the Finnish mathematician F. Nevanlinna
\cite{NEV}, though in a journal perhaps not readily accessible to many
researchers. We may state a first segment of this result as our
\renewcommand{\theequation}{1.24}
\begin{equation}
\label{unique}
\left\|\quad
{\parbox{10cm}{
{\bf Uniqueness Theorem II}:
{\em If a function f\,(z) is (i) analytic in the
circular region ${\cal{C}}_{R}$ given by\\
\mbox{}\\
\hspace*{2cm}\text{$\displaystyle{
Re\,\frac{1}{z}\,>\,\frac{1}{R}\quad \Leftrightarrow\quad
\left|z\,-\,\frac{1}{2}\,R\right|\,<\,\frac{1}{2}\,R}$}\\
\mbox{}\\
(R real and positive), and (ii) possesses in this region the
semiconvergent expansion (\ref{formser}) with "factorial" error
bound
(\ref{gevrey}), then it
is the only function with this
asymptotic expansion in ${\cal{C}}_{R}$. In
particular, with $R\,\to\,\infty$, analyti\-city in the half-plane
$|\arg\,z|\,<\,\frac{\pi}{2}$ is a
sufficient form of (i).}}}
\quad \right\|
\end{equation}

\begin{center}
\epsfig{file=fig16.eps,width=10cm}\\
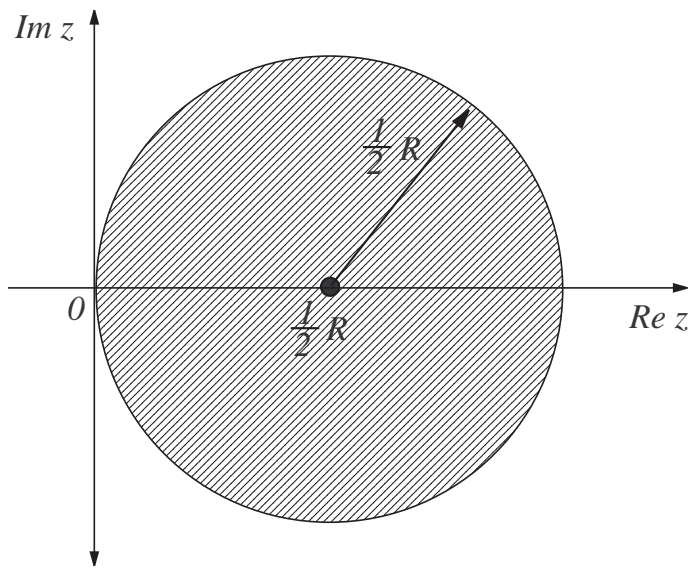
{\bf Fig. 1.4}\quad The circular region ${\cal{C}}_{R}$ in theorem
(\ref{unique}).
\end{center}

\qquad
The second, more technical segment of the Nevanlinna-Sokal result, which
again concerns the Borel-Laplace reconstruction, will be stated and
discussed as theorem (\ref{sokal}) below.

\qquad
Our two examples of sufficient summability conditions, theorems
(\ref{watson}) and (\ref{unique}), create (when applicable) a situation
which, if you think about it, is quite remarkable: under their
hypotheses a {\em divergent} power series, initially a purely formal
object, is never\-theless found to be associated with, and thus in a
sense
to determine, a unique analytic function, in much the same way as a
convergent expansion does. They are therefore sometimes \cite{STSH}
referred to as {\em quasi-analyticity principles}. What makes them work
is the potent mixture of analyticity and strong error bounding they
postulate for the associated function in suitable domains -- I only need
to drop the catchword of Liouville´s theorem in complex analysis in
order to remind you of how powerful that kind of combination can be. Yet
it is prudent to remember that both theorems spell out {\em sufficient},
not necessary, conditions for unique reconstruction. Uniqueness may
still hold in special cases not fitting into their assumptions, and
section 2.4 will indeed treat you to an entirely nontrivial example of
this.

\qquad
For the physicist interested in QFT, the most important applications of
these theorems undoubtedly have been the {\em summability results for
the coupling dependence of correlation functions in superrenormalizable
theories}, beginning in 1975 with a proof \cite{EMS} for the $\phi^{4}$
theory -- a neutral scalar field with quartic self-interaction -- in
$D\,=\,2$ Euclidean dimensions, and later extended along similar lines
to $\overline{\psi}\,\phi\,\psi$ (Yukawa) theory in $D\,=\,2$ and, most
significantly, to the superrenormalizable $\phi^{4}$ theory in
$D\,=\,3$, the workhorse of the modern theory of phase transitions and
critical exponents \cite{MAG}. (It is widely believed that analogous
summability properties also hold for other superrenormalizable theories,
and in particular for abelian and non-abelian gauge theories in
$D\,=\,2$ and $D\,=\,3$, though for some reason these do not seem to
have been analyzed with the same degree of mathematical rigor.) These
proofs involve a good deal of heavy mathematical technicality, and I
cannot even begin to do them justice here, but let me try to give you at
least a superficial glance at their main ideas. The starting point is
always the Euclidean path-integral representation of correlation
functions: taking the $\phi^{4}$ theories as an example, we have
\renewcommand{\theequation}{1.25}
\begin{equation}
\label{pathint}
G_{N}\,(x_{1}\,...\,x_{N};\,g^{2})\,=\,
\frac{\int\,{\cal{D}}\,\mu\,[\phi]\,\phi\,(x_{1})\,...\,\phi\,(x_{N})\,\E^{-S_{I}\,[\phi]}}
{\int\,{\cal{D}}\,\mu\,[\phi]\,\E^{-S_{I}\,[\phi]}}\,\,,
\end{equation}
where
\renewcommand{\theequation}{1.26}
\begin{equation}
\label{selfint}
S_{I}\,[\phi]\,=\,\frac{g^{2}}{4!}\,\int\,d^{D}\,x\,\phi\,(x)^{4}\qquad
(D\,=\,2,\,3)\,\,,
\end{equation}
and where the Gaussian functional measure, ${\cal{D}}\,\mu\,[\phi]$,
incorporates the bilinear part of the Euclidean action. If the integrals
were ordinary multi-dimensional integrals, it would be straightforward
to conclude they are analytic functions of $g^{2}$ in at least the
half-plane $|\arg\,(g^{2})|\,<\,\frac{\pi}{2}$, i. e.
$Re\,(g^{2})\,>\,0$, since the integrands are obviously analytic and the
integrations strongly convergent in that region. In fact the
integrations are infinite-dimensional and, for the axiomatic field
theorist, therefore call for an exact definition of the Gaussian
measures ${\cal{D}}\,\mu\,[\phi]$, and since ultraviolet renormalization
in these theories can be dealt with at the level of one- and two-point
functions, that definition will also have to account for
renormalization. Conveniently, in the years before 1975, the measures
${\cal{D}}\,\mu\,[\phi]$ for the prototype superrenormalizable theories
had already been studied exhaustively, and it had been found that the
(small finite number of) {\em renormalization counterterms suggested by
perturbation theory} were in fact sufficient for such a construction.
Therefore these measures were already under good enough control
mathematically so the analyticity argument could be carried over without
difficulty.

\qquad
To establish the second pillar of summability, the factorial error bound
of eq. (\ref{gevrey}) in the analyticity domain $Re\,(g^{2})\,>\,0$, the
successive derivatives with respect to $g^{2}$ of expression
(\ref{pathint}) were analyzed with the help of the cluster-expansion
technique of statistical mechanics, which in particular allows division
of the denominators into numerators -- the "removal of vacuum bubbles"
in physics language -- to be made explicit, and were estimated using
suitable correlation-function inequalities also established by earlier
constructive work. The result typically was an estimate of the form of
eq.
(\ref{derbnd}) above,
\renewcommand{\theequation}{1.27}
\begin{equation}
\label{derest}
\left|\frac{\partial^{n}}{\partial\,(g^{2})^{n}}\,G_{N}\,
(\,...\,;\,g^{2})\right|\,\leq\,C_{1}\,(C_{2})^{n}\,(n!)^{2}\qquad
(Re\,g^{2}\,>\,0)\,\,,
\end{equation}
which when used in the remainder term (\ref{remtayl}) of the Taylor
expansion around $g^{2}\,=\,0$ yields the desired bound.

\qquad
An inconspicuous but nevertheless crucial precondition for the success
of these proofs is that in eq. (\ref{selfint}) we did not need to
distinguish between a bare coupling $g_{0}$ that appears in the
interaction Lagrangian, and a renormalized coupling $g$ that serves as
the perturbative expansion parameter. Indeed as remarked above, in the
superrenormalizable theories the two differ at most by a {\em finite}
rescaling, so one may expand directly in the $g^{2}$ of eq.
(\ref{selfint}) without loss of generality, since a finite rescaling
merely changes the meaning of the constant $|A|$ in eq. (\ref{gevrey}).
In the strictly renormalizable theories this simple connection is lost:
between the Lagrangian $g_{0}$ and the expansion parameter of the
perturbation series, there is now the abyss of an
ultraviolet-divergent renormalization (\ref{recoup}), which can no more
be absorbed into a rigorous construction of a {\em Gaussian} measure
${\cal{D}}\,\mu\,[\phi]$. That this simplicity is lost is presumably the
main reason for the continuing lack, after more than 25 years, of
comparable results for realistic theories.

\qquad
The superrenormalizable QFT´s are as close as one can come to a
perturbation theorist´s paradise: in these theories the perturbation
series, while divergent, never\-theless "contains the whole truth" in
that it has associated with it {\em unique} correlation functions
without escaping remainders. Conversely, in such theories there is in
principle nothing that cannot be represented by, and recovered from, a
semiconvergent expansion in the coupling. The term "nonperturbative"
in these models can at best have a rather restricted meaning: it can
refer only to what in the introduction we dubbed {\em weakly
nonperturbative} phenomena, whose calculation requires some all-orders
resummation of their perturbation series. This makes both for the
special
charm of these theories and for their limited usefulness when trying to
learn something for the realistic, marginallly renormalizable cases.

\qquad
Before moving on to the technicalities of the most widely used
reconstruction procedure, it may be useful to emphasize that once you
have verified some sufficient condition for uniqueness of $f\,(z)$, it
is in principle a secondary matter what method you use to get hold of
that function. You may simply guess it if you are clever. Among more
systematic methods, the choice depends largely on the pattern of
divergence in your given series, but even for the factorially divergent
series of type (\ref{largep}) one usually faces in perturbative QFT, the
Borel-Laplace procedure discussed below has no monopoly. To mention
just one alternative, many functions having asymptotic expansions with
this divergence pattern can be reconstructed by the method of Padé
approximants \cite{BAK}: one approximates $f\,(z)$ by a rational
function with numerator and denominator degrees and $N$ respectively,
\renewcommand{\theequation}{1.28}
\begin{equation}
\label{fpade}
f\,(z)\,\approx\,f_{M,\,N}\,(z)\,=\,
\frac{b_{0}\,+\,b_{1}\,z\,+\,b_{2}\,z^{2}\,+\,...\,
b_{M}\,z^{M}}{1\,+\,c_{1}\,z\,+\,c_{2}\,z^{2}\,+\,...\,c_{N}\,z^{N}}\,\,,
\end{equation}
and adjusts the $M\,+\,N\,+\,1$ coefficients of this function so as to
make its first $M\,+\,N\,+\,1$ Taylor coefficients at $z\,=\,0$ agree
with the coefficients $a_{0}\,...\,a_{M\,+\,N}$ of the given divergent
series (\ref{formser}). Of course, this method is easiest to apply if
you have some advance information about the large-$z$ behaviour of
$f\,(z)$ to guide you in the choice of the relative degree $M\,-\,N$.
For example, if $f\,(z)$ is known to fit into the general form
\renewcommand{\theequation}{1.29}
\begin{equation}
\label{fstielt}
f\,(z)\,=\,\int\limits_{0}^{\infty}\,d\,x\,\frac{\sigma\,(x)}{z\,+\,x}\,\,,\qquad
\sigma\,(x)\,\geq\,0\,\,,
\end{equation}
i. e. to admit a spectral representation with a cut on the negative
real $z$ axis (so that in fact we have the luxury of Phragmén-Lindelöf
analyticity in a "sector" with $R\,=\,\infty$ and angle just
infinitesimally smaller than $2\,\pi$), and if the coefficients in its
formal small-$z$ expansion
\renewcommand{\theequation}{1.30}
\begin{equation}
\label{astielt}
a_{n}\,=\,(-)^{n}\,\int\limits_{0}^{\infty}\,d\,x\,
\frac{\sigma\,(x)}{x^{n\,+\,1}}\,\,,\qquad n\,=\,0,\,1,\,2,\,...
\end{equation}
{\em while forming a factorially divergent sequence}, are individually
finite, then the "diagonal" Padé approximants with $N\,=\,M$,
$f_{(N,\,N)}\,(z)$, are known \cite{BAK} to converge to $f$ uniformly on
compact domains in the cut plane. (If (\ref{astielt}) exists also
for $n\,=\,-1$, so that $f\,(z)\,\to\,-a_{-1}\,/\,z$ for
$|z|\,\to\,\infty$ outside the cut, then the $N\,=\,M\,+\,1$
approximants, of course, will converge better). Such functions are known
as {\em Stieltjes functions}, and their Padé reconstruction has been
used successfully to sum perturbation series in the $\phi^{4}$ theory at
$D\,=\,1$ (the theory of the anharmonic oscillator) \cite{LOE}. However,
even
in the absence of such detailed information, the method is often applied
heuristically because of its simplicity, generality, and relative
robustness. It does have its drawbacks -- when applied directly to a
factorially divergent series, its convergence in low orders $M\,+\,N$
may be erratic, and for this reason it is more frequently used as an
analytic-continuation engine {\em within} the Borel-Laplace method, to
which we turn in our next section. However, in principle it can stand on
its own feet.

\subsection{The Borel-Laplace Summation Process}

\qquad
In the case of a summable formal power series
\renewcommand{\theequation}{1.31}
\begin{equation}
\label{fasymp}
f_{f}\,(z)\,=\,\sum\limits_{n\,=\,0}^{\infty}\,a_{n}\,z^{n}\,\,,
\end{equation}
the most widely used method of recovering from the series the function
$f\,(z)$ uniquely associated with it is the venerable Borel-Laplace (BL)
process \cite{BOR}. It requires some mild additional
restriction on that function, but it is capable of handling the
"factorial" coefficient growth
\renewcommand{\theequation}{1.32}
\begin{equation}
\label{largen}
a_{n}\quad \longrightarrow \quad
c\,A^{-n}\,\Gamma\,(n\,+\,B\,+\,1)\qquad
(n\,\to\,\infty)\,\,,
\end{equation}
which is essentially the same as that of (\ref{largep}) since
$\Gamma\,(n\,+\,B\,+\,1)\,\to\,(n\,+\,1)^{B}\,\Gamma\,(n\,+\,1)$ for
$n\,\to\,\infty$ \cite{GRY}. It is therefore suitable for summing the
perturbative expansions of superrenormalizable QFT.

\qquad
The first step is to map $f\,(z)$ by an invertible integral
transformation ${\cal{B}}$ onto a function $b\,(s)$ of another complex
variable $s$, the {\em Borel transform} of $f$:
\renewcommand{\theequation}{1.33}
\begin{equation}
\label{borel}
b\,(s)\,=\,{\cal{B}}\,[f;\,s):\,=\,\frac{1}{2\,\pi\,i}\,\int\limits_{\rho
- i\infty}^{\rho +
i\infty}\,d\,\zeta\,e^{s\,\zeta}\,\tilde{f}\,(\zeta)\,\,,
\end{equation}
where
\renewcommand{\theequation}{1.34}
\begin{equation}
\label{ftilde}
\tilde{f}\,(\zeta)\,=\,\frac{1}{\zeta}\,f\,\left(\frac{1}{\zeta}\right)\,\,.
\end{equation}
(This peculiar-looking prescription is adopted here to conform with the
convention used below in eq. (\ref{laplace}) for the Laplace transform.)
The notation ${\cal{B}}\,[f;\,s)$ is used for a quantity that is both a
functional of the function $f$ and an ordinary function of the variable
$s$. The path of integration in the $\zeta$ plane, $Re\,\zeta\,=\,\rho$,
is chosen such that all singularities of $\tilde{f}\,(\zeta)$ remain to
the left of it.

\begin{center}
\epsfig{file=fig15.eps,width=14cm}\\
$\begin{array}{ccl}
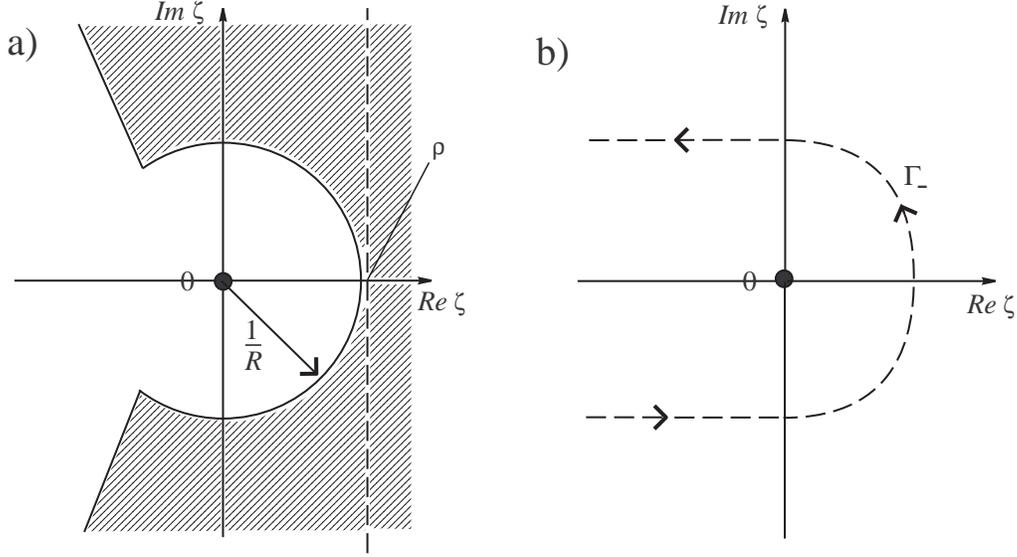
{\bf Fig. 1.5}\quad & \text{(a)} & \text{Region of analyticity under
Phragmén-Lindelöf conditions}\\
&  &
\text{and integration contour in the Borel transform.}\\
\mbox{}\\
 & \text{(b)} & \text{Deformation of integration contour into}\,\,\,
\Gamma_{-}
\end{array}$
\end{center}

In the situation where uniqueness of $f$ is
guaranteed by its fulfilling the conditions of theorem (\ref{watson}),
the domain of analyticity of $\tilde{f}\,(\zeta)$ is the image of the
"wide-angle-sector" of Fig. 1.3 under the inversion
$\zeta\,=\,1\,/\,z$, and is therefore the shaded region in Fig. 1.5 (a),
while under the hypothesis of uniqueness theorem II, (\ref{unique}), it
is simply the half-plane $Re\,\zeta\,>\,\frac{1}{R}$, so in both cases
we should choose
\renewcommand{\theequation}{1.35}
\begin{equation}
\label{rho}
\frac{1}{R}\,<\,\rho\,<\,\infty\,\,.
\end{equation}

From eq. (\ref{borel}), it is clear that the growth restriction we must
impose on our function $f\,(z)$ is modest: $f\,(z)$ should be such that
its associated $\tilde{f}\,(\zeta)$, eq. (\ref{ftilde}), possesses a
Fourier transform with respect to the imaginary part
$\sigma\,=\,Im\,\zeta$ of its argument.

The transformation can next be applied, termwise, to the series
(\ref{fasymp}), giving a series called the {\em formal} Borel transform,
\renewcommand{\theequation}{1.36}
\begin{equation}
\label{borser}
b_{f}\,(s)\,=\,{\cal{B}}\,[f_{f};\,s)\,=\,\sum\limits_{n\,=\,0}^{\infty}\,a_{n}\,\left\{
\frac{1}{2\,\pi\,i}\,\int\limits_{s(\rho - i\infty)}^{s(\rho + i\infty)}
d\,\eta\,e^{\eta}\,\eta^{-(n\,+\,1)}\right\}\,s^{n}\,\,,
\end{equation}
where $\eta\,=\,s\,\rho$, and if the integrals in both (\ref{borel}) and
(\ref{borser}) exist, this series is in turn semiconvergent to $b\,(s)$.
The integration path is $Re\,\eta\,=\,s\,\rho$ for real and positive $s$
and is "tilted" for complex $s$, but since the only singularity of the
integrand is now a pole at $\eta\,=\,0$, it can always be deformed, by
bending its ends over to the left, into the contour $\Gamma_{-}$ of Fig.
1.5 (b). The resulting integral is Hankel´s integral representation
\cite{GRY} of the inverse gamma function,
\renewcommand{\theequation}{1.37}
\begin{equation}
\label{hankel}
\frac{1}{2\,\pi\,i}\,\int\limits_{\Gamma_{-}}\,d\,\eta\
e^{\eta}\,\eta^{-\alpha}\,=\,\frac{1}{\Gamma\,(\alpha)}\,\,,
\end{equation}
at integer $\alpha\,=\,n\,+\,1$. (If this looks unfamiliar to you, take
$\alpha$ integer and work out the residue of the integrand at
$\eta\,=\,0$.) Therefore
\renewcommand{\theequation}{1.38}
\begin{equation}
\label{bform}
b_{f}\,(s)\,=\,\sum\limits_{n\,=\,0}^{\infty}\,\frac{a_{n}}{n!}\,s^{n}\,\,.
\end{equation}
We could have simply {\em defined} $b_{f}$ by this series, but since we
are going to need the concept of Borel transform for objects more
general than the series (\ref{fasymp}), we have chosen to derive it here
from the more general integral transform (\ref{borel}).

\qquad
Thus we have manufactured a series with much faster-decreasing
coefficients than (\ref{fasymp}), and if the $a_{n}$ coefficients grow
no faster than (\ref{largen}) at large $n$, then $b_{f}$ is actually
something much better than a mere asymptotic expansion of $b\,(s)$,
namely a truly convergent expansion: it is given by the generalized
geometric series
\renewcommand{\theequation}{1.39}
\begin{equation}
\label{geoser}
c\,\sum\limits_{n\,=\,0}^{\infty}\,\frac{\Gamma\,(n\,+\,1\,+\,B)}{\Gamma\,(n\,+\,1)}\,\left(\frac{s}{A}\right)^{n}\,=\,
c\,\Gamma\,(B\,+\,1)\,\left(1\,-\,\frac{s}{A}\right)^{-(B\,+\,1)}\,\,,
\end{equation}
plus a faster-convergent remainder series having a least the same
convergence radius, and is therefore convergent in the circle
\renewcommand{\theequation}{1.40}
\begin{equation}
\label{scirc}
|s|\,<\,|A|\,\,,
\end{equation}
where it defines $b\,(s)$ {\em constructively}. This is not yet all of
$b\,(s)$; we shall need to perform an {\em analytic-continuation
process} ${\cal{A}}$ to continue $b$ from the circle (\ref{scirc}) into
as much as possible of the $s$ plane, and in particular along the real
positive $s$ axis if we are to recover the original $f\,(z)$ uniquely at
the real positive values of $z$ in which we are primarily interested. It
is no surprise that the conditions of theorem (\ref{unique}) also
guarantee that this can be done -- a statement known as the
\cite{SOK}
\renewcommand{\theequation}{1.41}
\begin{equation}
\label{sokal}
\left\|\quad
{\parbox{10cm}{
{\bf Nevanlinna-Sokal theorem}: If $f\,(z)$ fulfills the
conditions
of theorem (\ref{unique}), then its Borel transform $b\,(s)$ is
analytic be
yond the circle (\ref{scirc}) in the striplike region $|s\,-\,{\rm
l\!R}^{+}|\,<\,|A|$
on both sides of the real $s$ axis l\!R$^{+}$, as illustrated in
Fig.
1.6.
Moreover, $b\,(s)$ is exponentially bounded there by const.
$\exp$
$\left(\frac{|s|}{R}\right)$, and $f\,(z)$ is uniquely given by
the Laplace integral
(\ref{laplace})
below for all $z$ in its region of analyticity Re
$\frac{1}{z}\,>\,\frac{1}{R}$.}}
\quad\right\|
\end{equation}
We can see from (\ref{geoser}) that (at least for $B\,>\,-1)$ the sum of
the $b_{f}$ series is singular at $s\,=\,+|A|$ for $A\,>\,0$, and at
$s\,=\,-|A|$ for $A\,<\,0$. Thus under the conditions of the
Nevanlinna-Sokal
theorem, the $A$ in (\ref{largen}) can only be negative: the large-$n$
terms of (\ref{fasymp}) must have {\em alternating} signs. We
emphasize that (\ref{sokal}) again gives {\em sufficient} conditions;
the desirable behavior of $b\,(s)$ on the positive $s$ axis may still
obtain in other cases with suitable modified hypotheses.

\begin{center}
\epsfig{file=fig17.eps,width=12cm}\\
$\begin{array}{ll}
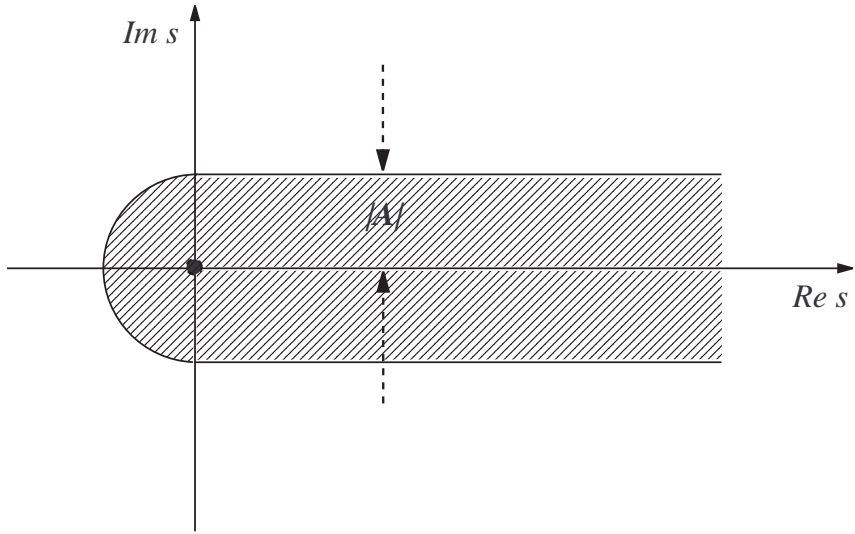
\text{\bf Fig. 1.6\quad} & \text{Region of analyticity in the $s$ plane
of the Borel}\\
& \text{transform $b\,(s)$ in the Nevanlinna-Sokal theorem}
\end{array}$
\end{center}

\qquad
With such behavior of $b\,(s)$ secured, we can finally reconstruct from
it the function $f\,(z)$ in its region of analyticity by inverting the
transformation ${\cal{B}}$ of (\ref{borel}): the inversion theorem in
the theory of Laplace transforms \cite{GRY} states that
\renewcommand{\theequation}{1.42}
\begin{equation}
\label{laptilde}
\tilde{f}\,(\zeta)\,=\,\int\limits_{0}^{\infty}\,d\,s\
e^{-\zeta\,s}\,b\,(s)\,\,.
\end{equation}
[If you feel you need proof of this, write eq. (\ref{borel}), with
$\zeta$ parameterized as $\rho\,+\,i\,\sigma$, in the form of a Fourier
transform,
\renewcommand{\theequation}{1.43}
\begin{equation}
\label{futb}
\E^{-\rho\,s}\,b\,(s)\,=\,\frac{1}{2\,\pi}\,\int\limits_{-\infty}^{+\infty}\,
d\,\sigma\,\E^{i\,s\,\sigma}\,\tilde{f}\,(\rho\,+\,i\,\sigma)\,\,,
\end{equation}
and evaluate the Fourier transform of the product
\renewcommand{\theequation}{1.44}
\begin{equation}
\label{thetb}
\Theta\,(s)\,[\E^{-\rho\,s}\,b\,(s)]\,\,,
\end{equation}
by the Fourier convolution theorem plus contour closing, using
$[i\,(\omega\,-\,i\,\varepsilon)]^{-1}$ as the Fourier transform of the
unit-step function $\Theta\,(s)$].
Therefore we recover our
function $f\,(z)$ through what for our purposes we shall call the {\em
Laplace transform},
\renewcommand{\theequation}{1.45}
\begin{equation}
\label{laplace}
f\,(z)\,=\,{\cal{L}}\,[b;\,z):\,=\,\frac{1}{z}\,
\int\limits_{0}^{\infty}\,d\,s\ e^{-\frac{s}{z}}\,b\,(s)\,\,,
\end{equation}
for values of $z$ such that the integral converges.
Under the conditions of the Nevanlinna-Sokal theorem (\ref{sokal})
above,
this is obviously the case for Re $\frac{1}{z}\,>\,\frac{1}{R}$, i. e.
in the entire original region of analyticity of $f$. Under the stronger
Phragmén-Lindelöf or Watson conditions, {\em that region does not become
any larger}, but $b\,(s)$ then has in addition a "sectorial"
region
of analyticity in the $s$ plane, with opening $\gamma\,-\,\pi$, that may
be useful for some purposes. For example, the integration path in the
variable $t\,=\,\frac{s}{z}$, namely, the ray
\renewcommand{\theequation}{1.46}
\begin{equation}
\label{teeray}
t\,=\,\frac{1}{z}\,\cdot\,(0\,.....\,\infty)\,\,,
\end{equation}
can in this case be rotated back to the real axis without changing the
value of the integral provided $|{\rm
arg}\,z|\,<\,\frac{1}{2}\,(\gamma\,-\,\pi)$, and (\ref{laplace}) is then
equivalent to
\renewcommand{\theequation}{1.47}
\begin{equation}
\label{tlapla}
f\,(z)\,=\,\int\limits_{0}^{\infty}\,d\,t\ e^{-t}\,b\,(t\,z)\,\,,
\end{equation}
which otherwise can be justified only for real, positive $z$. This
form lends itself more readily to generalization to more than
one variable \cite{STSH}, and incidentally is best suited to
performing the final check on our reconstruction procedure: from
(\ref{tlapla}), the Taylor coefficients of $f$ from the right follow by
differentiations under the integral sign and use of (\ref{bform}),
\renewcommand{\theequation}{1.48}
\begin{equation}
\label{taylap}
\frac{1}{n!}\,\left(\frac{d^{n}\,f}{d\,z^{n}}\right)_{z\to 0+}\,=\,
\int\limits_{0}^{\infty} d\,t\ e^{-t}\,\left[\frac{1}{n!}\,\left(
t\frac{d}{d\,s}\right)^{n}\,b\,(s)\right]_{s=0}\,=\,
\frac{a_{n}}{n!}\,\int\limits_{0}^{\infty} d\,t\ e^{-t}\,t^{n}\,\,,
\end{equation}
and since
\renewcommand{\theequation}{1.49}
\begin{equation}
\label{gamma}
\int\limits_{0}^{\infty}\,d\,t\
e^{-t}\,t^{\beta}\,=\,\Gamma\,(\beta\,+\,1)\,\,,
\end{equation}
this becomes equal to $a_{n}$, so (\ref{laplace}) indeed gives the
function $f\,(z)$ that has the series (\ref{fasymp}) as its asymptotic
expansion.

\qquad
A remark on conventions: our definition of the Laplace transform, eq.
(\ref{laplace}), is adapted to the study of semiconvergent expansions
{\em at small} $z$, and has been chosen so as to agree with the
convention used in the majority of the field-theory textbooks that
discuss the subject \cite{WEI} \cite{ZNJ} \cite{ITZ}.
It is also the convention used by Watson \cite{WAT} and by Sokal
\cite{SOK}. Most of
the mathematical literature on resurgent analysis (such as references
\cite{ECAL},\,\cite{CNP}, and parts of \cite{STSH}) studies expansions
at
a {\em large} argument $\zeta$ and therefore needs to be translated to
our context through $z\,=\,\frac{1}{\zeta}$; moreover it uses a
different definition, namely eq. (\ref{laptilde}), for the Laplace
transform. This makes for simpler appearance of a number of formulas
(for example, the convolution theorem, line 10 in Table 1 below, gets
rid
of the extra $z$ factor on its l. h. s.) but on the other hand requires
the separate introduction of a unit element, the delta distribution
$\delta\,(s)$, for the algebra of formal Borel transforms $b_{f}\,(s)$,
since (\ref{laptilde}) fails to map $b\,\equiv\,1$ onto
$\tilde{f}\,\equiv\,1$. To further complicate matters, a recent QFT
textbook \cite{SIEG} also uses this "mathematical" definition of the
Laplace transform, while reference \cite{STSH} occasionally switches
between small-argument and large-argument considerations. So be prepared
to spend some time climbing around between the various conventions.

\qquad
To summarize, the BL reconstruction proceeds in three steps:

{\bf (1)} " Borelize" the given asymptotic series:
\renewcommand{\theequation}{1.50}
\begin{equation}
\label{borstep}
f_{f}\,(z)\ \ ({\rm eq.} (\ref{fasymp}))\quad
\stackrel{\displaystyle {\cal{B}}}{\longrightarrow}\quad
b_{f}\,(s)\ \ ({\rm eq.} (\ref{bform}))\,\,.
\end{equation}
{\bf (2)} Continue $b_{f}$ to $b$ analytically along the positive real
$s$ axis:
\renewcommand{\theequation}{1.51}
\begin{equation}
\label{constep}
b_{f}\,(s)\quad \stackrel{\displaystyle{\cal{A}}}{\longrightarrow}\quad
b\,(s)\,\,.
\end{equation}
{\bf (3)} Obtain $f\,(z)$ from $b\,(s)$ through Laplace transform:
\renewcommand{\theequation}{1.52}
\begin{equation}
\label{lapstep}
b\,(s)\quad \stackrel{\displaystyle{\cal{L}}}{\longrightarrow}\quad
f\,(z)\,\,.
\end{equation}
The importance of the second step, the analytic continuation
${\cal{A}}$, can hardly be overemphasized: without it we would make no
progress at all. Indeed if we forget that step and simply plug expansion
(\ref{bform}) into  the integral of (\ref{tlapla}) -- i.\,e. if we use
(\ref{bform}) beyond its radius of convergence -- then obviously we do
nothing but repeat the calculation of eqs. (\ref{taylap}) and
(\ref{gamma}), and are led back to our original semiconvergent series
(\ref{fasymp}) without having come any closer to recovering the "true"
function $f\,(z)$. (Actually, one standard way in which semiconvergence
arises is as a punishment for your having used a series in an
integration outside its circle of convergence.) Needless to say the
simplest method of continuing (\ref{bform}) analytically, namely to sum
the series in closed form as we did in (\ref{geoser}), is hardly ever
applicable in realistic cases, where the $a_{n}$ coefficients result
from complicated perturbative calculations, and often are only available
numerically for a limited number of orders $n$. The practical methods
that have been used -- particularly in studies of critical phenomena
based on the superrenormalizable $\phi^{4}$ theory in $D\,=\,3$ -- are
surveyed in chapter 37 of \cite{ZNJ}. Here I would like to mention just
one method which, while no more the most refined one for the specific
purpose of calculating universal quantities in phase-transition theory,
is still widely used in statistical mechanics because of its simplicity
and flexibility. This is again the method of Padé
approximants, whose direct application to the original asymptotic series
(\ref{fasymp}) we already mentioned in sect. 1.1, but which is usually
applied more profitably to the better-converging "borelized" series
(\ref{bform}), where its convergence in the lower orders as a rule is
less erratic. Thus one approximates
\renewcommand{\theequation}{1.53}
\begin{equation}
\label{bpade}
b\,(s)\,\approx\,b_{(K,\,L)}\,(s)\,=\,
\frac{P_{K}\,(s)}{Q_{L}\,(s)}
\end{equation}
with polynomials $P_{k}$ and $Q_{L}$ of orders $K$ and $L$ respectively,
and with $Q_{L}\,(0)$ normalized to 1 without loss of generality. The
$K\,+\,L\,+\,1$ coefficients of this fraction are then determined either
by matching the coefficients of
$s^{0},\,s^{1},\,s^{2},\,...\,s^{K\,+\,L}$ in its expansion around
$s\,=\,0$ to those of $b_{f}\,(s)$ (Pade-I method), or by matching the
entire fraction at $K\,+\,L\,+\,1$ points $s_{i}$ within the circle of
convergence of $b_{f}$ to the numerically summed $b_{f}$ (Pade-II
method). The $L$ roots $s_{l}$ of the denominator polynomial,
\renewcommand{\theequation}{1.54}
\begin{equation}
\label{qfact}
Q_{L}\,(s)\,=\,\prod\limits_{l\,=\,1}^{L}\,(s\,-\,s_{l})\,\,,
\end{equation}
then approximate the singularities of $b\,(s)$ -- it is an advantage of
the method over, e.\,g., conformal-mapping methods that one does not
have to know the positions of these singularities in advance. Nor, in
principle, does one need to know much about the large-$|s|$ behavior of
$b\,(s)$, provided one is willing to work through the entire
"Padé table", i.\,e. the entire set of possible $(K,\,L)$ pairs at
given $K\,+\,L\,+\,1$. If the true $b\,(s)$ bears some resemblance to
(\ref{geoser}) at large $s$, then the approximants with $K\,<\,L$ will
converge best, and one may write a partial-fraction decomposition,
\renewcommand{\theequation}{1.55}
\begin{equation}
b_{(K,\,L)}\,(s)\,=\,\sum\limits_{l\,=\,1}^{L}\,\frac{r_{l}}{s\,-\,s_{l}}\,\,,\qquad
r_{l}\,=\,\frac{P_{K}\,(s_{l})}{Q_{L}^{'}\,(s_{l})}
\end{equation}
which allows the Laplace transform (\ref{laplace}) to be carried out
analytically according to line 11 of table 1 below:
\renewcommand{\theequation}{1.56}
\begin{equation}
\label{fklei}
f\,(z)\,\approx\,f_{(K,\,L)}\,=\,-\frac{1}{z}\,\sum\limits_{l\,=\,0}^{L}\,r_{l}\,
e^{-\frac{s_{l}}{z}}\,Ei\,\left(\frac{s_{l}}{z}\right)\,\,.
\end{equation}
If $f$ is Borel summable and $f_{f}$ has real coefficients, the $s_{l}$
must all be off the positive real $s$ axis, and thus come either at
negative real $s$ or in complex-conjugate pairs, so all the $Ei$
functions in (\ref{fklei}) are unambiguous.

\qquad
For the last step (\ref{lapstep}) of the reconstruction scheme, it is
often convenient to have a little toolkit of basic BL transforms
available, so Table 1 offers such a list. All entries in this table are straightforward to verify by elementary
calculation. (Many lines carry some clauses or restrictions that are
usually obvious and not mentioned explicitly.) The first line, for
$\nu\,=\,n$ an integer, repeats what we
have already used to derive (\ref{bform}), but it also shows that it is
just as easy to "Borelize" more general formal series with some
noninteger power in front:
\renewcommand{\theequation}{1.57}
\begin{equation}
\label{nonint}
{\cal{B}}\,\Biggl[z^{\beta}\,\cdot\,\sum\limits_{n\,=\,0}^{\infty}\,a_{n}\,z^{n};\,s\Biggr)\,=\,
s^{\beta}\,\sum\limits_{n\,=\,0}^{\infty}\,\frac{a_{n}}{\Gamma\,(n\,+\,\beta\,+\,1)}\,s^{n}
\qquad (\beta\,>\,-1)\,\,.
\end{equation}
The fourth line is interesting because it states that ${\cal{B}}$ and
${\cal{L}}$ commute with the infinitesimal generator of dilatations, so
that $f\,(\lambda\,z)$ and $b\,(\lambda\,s)$ form a transform pair if
$f\,(z)$ and $b\,(s)$ do. The tenth line expresses the
well-known {\em convolution theorem} of Laplace-transform theory, with
the convolution $b_{1}\,*\,b_{2}$ of two functions being defined by the
r. h. s.

Lines 11 and 12 feature a simple higher transcendental function studied
already by Euler in the 18th century, the exponential-integral function
$Ei\,(z)$ defined \cite{GRY} by the integral
\renewcommand{\theequation}{1.58}
\begin{equation}
\label{eidef}
Ei\,(z)\,=\,\int\limits_{-\infty}^{z}\,d\,u\,\frac{\E^{u}}{u}\qquad
(z\,\not=\,0)\,\,.
\end{equation}
\newpage
\begin{center}
{\bf Table 1}: Some Borel-Laplace transform pairs
\end{center}
\begin{center}
\begin{tabular}{|c|c|c|}\hline\hline
\rule[-5mm]{0mm}{11mm}
\mbox{} & $f\,(z)\,=\,{\cal{L}}\,[b;\,z)$ &
$b\,(s)\,=\,{\cal{B}}\,[f;\,s)$ \\ \hline\hline
1. &
$z^{\nu}\quad (\nu\,>\,-1)$ &
\rule[-5mm]{0mm}{12mm}
$\displaystyle{\frac{1}{\Gamma\,(\nu\,+\,1)}\,s^{\nu}}$ \\ \hline
2. &
$z\,\cdot\,f\,(z)$ &
\rule[-7mm]{0mm}{17mm}
$\displaystyle{\int\limits_{0}^{s}\,d\,t\ b\,(t)}$ \\ \hline
\rule[-5mm]{0mm}{12mm}
3. &
$\displaystyle{\frac{1}{z}\,\cdot\,f\,(z)}$ &
$\displaystyle{\Bigl[\delta\,(s)\,+\,\frac{d}{d\,s}\Bigr]\,b\,(s)}$ \\
\hline
\rule[-5mm]{0mm}{12mm}
4. &
$\displaystyle{z\,\cdot\,\frac{d\,f\,(z)}{d\,z}}$ &
$\displaystyle{s\,\cdot\,\frac{d\,b\,(s)}{d\,s}}$ \\ \hline
\rule[-5mm]{0mm}{12mm}
5. &
$\displaystyle{\frac{d\,f\,(z)}{d\,z}}$ &
$\displaystyle{\Bigl(\frac{d}{d\,s}\,s\,\frac{d}{d\,s}\Bigr)\,b\,(s)}$
\\ \hline
\rule[-5mm]{0mm}{12mm}
6. &
$\displaystyle{z\,\frac{d}{d\,z}\,[z\,f\,(z)]}$ &
$s\,\cdot\,b\,(s)$ \\ \hline
\rule[-7mm]{0mm}{17mm}
7. &
$\displaystyle{\frac{1}{z}\,\int\limits_{0}^{z}\,d\,w\ f\,(w)}$ &
$\displaystyle{\frac{1}{s}\,\int\limits_{0}^{s}\,d\,t\ b\,(t)}$ \\
\hline
\rule[-5mm]{0mm}{12mm}
8. &
$\displaystyle{\frac{f\,(z)\,-\,f\,(0)}{z}}$ &
$\displaystyle{\frac{d\,b\,(s)}{d\,s}}$ \\ \hline
\rule[-5mm]{0mm}{12mm}
9. & $\displaystyle{\frac{1}{1\,-\,\omega\,z}\,f\,
\left(\frac{z}{1\,-\,\omega\,z}\right)}$ & $e^{\omega\,s}\,b\,(s)$\\
\hline
10. &
$z\,\cdot\,f_{1}\,(z)\,\cdot\,f_{2}\,(z)$ &
$\displaystyle{(b_{1}\,*\,b_{2})\,(s)\,=\,\int\limits_{0}^{s}\,d\,t\
b_{1}\,(t)\,b_{2}\,(s\,-\,t)}$ \\ \hline
\rule[-5mm]{0mm}{12mm}
11. &
$\displaystyle{-\frac{1}{z}\,e^{-\frac{s_{0}}{z}}\,Ei\,
\Bigl(\frac{s_{0}}{z}\Bigr)}$ &
$\displaystyle{\frac{1}{s\,-\,s_{0}}}$ \\ \hline
\rule[-5mm]{0mm}{12mm}
12. &
$\displaystyle{-\frac{1}{z}\,Ei\,\Bigl(-\frac{s_{0}}{z}\Bigr)}$ &
$\displaystyle{\frac{1}{s}\,\Theta\,(s\,-\,s_{0})}$ \\ \hline
\rule[-5mm]{0mm}{12mm}
13. &
$\displaystyle{e^{-\frac{s_{0}}{z}}\,f\,(z)}$ &
$\displaystyle{\Theta\,(s\,-\,s_{0})\,b\,(s\,-\,s_{0})}$ \\ \hline
\end{tabular}
\end{center}

The equivalent spectral representation $(w\,=\,z\,-\,u)$
\renewcommand{\theequation}{1.59}
\begin{equation}
\label{eispec}
Ei\,(z)\,=\,\E^{z}\,\int\limits_{0}^{\infty}\,d\,w\,\frac{\E^{-w}}{z\,-\,w}
\end{equation}
shows that the function is analytic in the $z$ plane except for a cut
along the positive real axis, where its discontinuity is
\renewcommand{\theequation}{1.60}
\begin{equation}
\label{eidisc}
Ei\,(x\,+\,i\,0)\,-\,Ei\,(x\,-\,i\,0)\,=\,-2\,\pi\,i\,\Theta\,(x)\qquad
(x\,\,\,\text{real})\,\,.
\end{equation}
The cut is thus logarithmic, as shown explicitly by the alternative
representation
\renewcommand{\theequation}{1.61}
\begin{equation}
\label{eilog}
Ei\,(z)\,=\,\gamma_{E}\,+\,\ln\,(-z)\,+\,
\int\limits_{0}^{z}\,d\,u\,\frac{\E^{u}\,-\,1}{u}\,\,,
\end{equation}
in which the integral is an entire function. Formal expansion of the
denominator in (\ref{eispec}) and integration beyond the convergence
circle results in the divergent series
\renewcommand{\theequation}{1.62}
\begin{equation}
\label{eilarg}
Ei\,(z)\,=\,\frac{\E^{z}}{z}\,\sum\limits_{p\,=\,0}^{\infty}\,p!\,\frac{1}{z^{p}}\,\,,
\end{equation}
so that
\renewcommand{\theequation}{1.63}
\begin{equation}
\label{eiasy}
\frac{a}{z}\,\E^{-\frac{a}{z}}\,Ei\,\left(\frac{a}{z}\right)\,=\,
\sum\limits_{p\,=\,0}^{\infty}\,(a^{-p}\,p!)\,z^{p}
\end{equation}
is a simple example of a factorially divergent series of type
(\ref{largep}); line 11 of Table 1 just describes the BL reconstruction
from this series. Line 12, on the other hand, follows from the
application of line 3 to the Laplace-transform representation
\renewcommand{\theequation}{1.64}
\begin{equation}
\label{eilap}
-Ei\,\left(-\frac{a}{z}\right)\,=\,\frac{1}{z}\,\int\limits_{0}^{\infty}\,
d\,s\,\E^{-\frac{s}{z}}\,\left[\Theta\,(s\,-\,a)\,\ln\,\left(\frac{s}{a}\right)\right]\,\,,
\end{equation}
where $a$ is real and positive. This representation embodies the
modern-day significance of the $Ei$ function as the simplest among a
family
of elementary building blocks for resurgent functions, the so-called
resurgence monomials \cite{ECAL}.

The last two lines of Table 1 actually go beyond the framework
we
discussed so far: by considering Borel-transform functions with jumps,
they remind us of the fact that for the Laplace integral (\ref{laplace})
to be well-defined, it is quite sufficient that $b\,(s)$ should only be
{\em piecewise continuous} on the real $s$ axis, rather than fully
analytic in a region along it. In this they point ahead to chapter 2,
where we will exploit this greater generality.

\qquad
As a matter of etiquette, we must acknowledge that the three-step scheme
of (\ref{borstep}) to (\ref{lapstep}) should actually be preceded by a
zeroth step:
\renewcommand{\theequation}{1.65}
\begin{equation}
\label{prestep}
\text{{\bf{(0)}}\,\,Check conditions for summability, i.\,e., uniqueness
of $f\,(z)$}\,\,.
\end{equation}

\qquad
But you know that real life isn´t like that -- outside the arcadian
realm of the superrenormalizable theories, physicists hardly ever have
enough information to do such a check. So as a rule they will proceed
heuristically, apply the BL machinery to the limited information they
have, and expect to be punished by some singularities on the real Borel
axis should they have committed some invisible offense. If you feel bad
about this state of things, you may find solace in the words of the
great
Poincaré, who clearly had a heart for the poor physicists when (on the
limits of "pragmatism" in dealing with asymptotic series) he wrote
\cite{POIN}: "Les astronomes ne les connaissent pas toujours d´une
facon bien précise, mais ils les franchissent rarement, ... d´ailleurs
leur instinct les guide et, s´il les trompait, le contrôle de
l´observation les avertirait promptement de leur erreur".

\renewcommand{\thesection}{2}
\section{Strictly Renormalizable Theories: Resurgence}
\subsection{Coupling Renormalization and Lambda Scale}

\qquad
When passing from the superrenormalizable theories discussed in sect.
1.2 to a "realistic", strictly renormalizable theory, such as
$(\phi^{4})_{D\,=\,4}$, (Yukawa)$_{D\,=\,4}$, or a nonabelian gauge
theory in $D\,=\,4$, we encounter no changes in the form of the
corresponding Lagrangians, or in the symmetries displayed by these
Lagrangians. Also, the renormalizations related to the {\em bilinear}
parts of the Lagrangians -- field renormalizations and, if bare-mass
terms are present, mass renormalizations -- remain ultraviolet
divergent, as they were already in the superrenormalizable cases, the
only change being that they are now divergent in {\em all} orders of
perturbation theory, rather than in a few low orders only. The only
{\em fundamentally} new element in the definition of the theory is the
divergent coupling renormalization (\ref{recoup}). It becomes necessary
because the three-point and four-point vertex functions that evolve
through quantum effects from the trilinear and quadrilinear Lagrangian
interaction terms now become superficially divergent, which amounts to
saying that the bare coupling $\overline{g}_{0}$ becomes dimensionless
as $D\,\to\,4$.

\qquad
The information about this new constitutive element, which will turn out
to have a strong effect on all the correlation functions of the theory,
is originally contained in the coupling-renormalization constant
$Z_{\alpha}$ of
(\ref{recoup}), which perturbatively takes the form of a Laurent
expansion in the dimensional regulator
$\varepsilon\,=\,(4\,-\,D)\,/\,2$,
\renewcommand{\theequation}{2.1}
\begin{equation}
\label{zalaur}
Z_{\alpha}\,(g^{2},\,\varepsilon)\,=\,1\,+\,\zeta_{1}\,(g^{2})\,\frac{1}{\varepsilon}\,+\,
\zeta_{2}\,(g^{2})\,\frac{1}{\varepsilon^{2}}\,+\,...\,\,.
\end{equation}
Each $\zeta_{n}$ is in turn a (semiconvergent) series in $g^{2}$
starting at order $(g^{2})^{n}$, since ultraviolet divergence of type
$\varepsilon^{-n}$ only comes from diagrams with at least $n$ loops:
\renewcommand{\theequation}{2.2}
\begin{equation}
\label{zetnfu}
\zeta_{n}\,(g^{2})\,=\,\zeta_{n,\,n}\,\left(\frac{g}{4\,\pi}\right)^{2\,n}\,+\,
\zeta_{n,\,n\,+\,1}\,\left(\frac{g}{4\,\pi}\right)^{2\,n\,+\,2}\,...\qquad
(n\,\geq\,1)\,\,.
\end{equation}
The set of all coefficients $\zeta_{n,\,k}\,(n\,\geq\,1,\,k\,\geq\,n)$
perturbatively represents the complete information about the coupling
renormalization. However there is a much more compact and convenient way
of encoding the same information, suggested by renormalization-group
(RG) analysis, namely, in terms of the RG beta function,
\renewcommand{\theequation}{2.3}
\begin{equation}
\label{betafu}
\beta\,(g\,(\mu))\,=\,\mu\,\frac{d\,g\,(\mu)}{d\,\mu}\,\,.
\end{equation}
This function is UV finite (the renormalized coupling $g\,(\mu)$ being
UV finite by definition), and its dependence on the regulator is
therefore much simpler. Since this is an important point in all that
follows, let me briefly repeat how these two encodings are connected
\cite{THO7}. We can in principle determine $\beta\,(g)$ from
$Z_{\alpha}$ by applying the operation $\mu\,d\,/\,d\,\mu$ on both sides
of (\ref{recoup}) and using the fact that $\overline{g}_{0}$, a bare
quantity, knows nothing about the sliding scale $\mu$, so
$d\,\overline{g}_{0}\,/\,d\,\mu\,=\,0$. We find
\renewcommand{\theequation}{2.4}
\begin{equation}
\label{betafrac}
\beta\,(g,\,\varepsilon)\,=\,
\frac{-\varepsilon\,g}{1\,+\,g^{2}\,\frac{\partial}{\partial\,g^{2}}\,(\ln\,Z_{\alpha})}\,\,.
\end{equation}
Upon inserting expansion (\ref{zalaur}) this becomes
\renewcommand{\theequation}{2.5}
\begin{equation}
\label{belaur}
\beta\,(g)\,=\,-\varepsilon\,g\,+\,g^{3}\,\zeta_{1}^{'}\,+\,
\frac{g^{3}}{\varepsilon}\,\left[\zeta_{2}^{'}\,-\,\zeta_{1}\,\zeta_{1}^{'}\,-\,
g^{2}\,(\zeta^{'}_{1})^{2}\right]\,+\,{\cal{O}}\,\left(\frac{1}{\varepsilon^{2}}\right)
\end{equation}
Here the primes denote differentiation by $g^{2}$. Since $\beta\,(g)$
must be finite as $\varepsilon\,\to\,0$ we conclude, first, that all
terms with powers of $\frac{1}{\varepsilon}$ on the r. h. s. must vanish
separately. This gives a sequence of conditions on the $\zeta_{n}$ which
generally determine $\zeta^{'}_{n}$ for $n\,\geq\,2$ in terms of lower
$\zeta$´s. For example, the $\frac{1}{\varepsilon}$ term displayed in
(\ref{belaur}) gives
\renewcommand{\theequation}{2.6}
\begin{equation}
\label{zet2prim}
\zeta^{'}_{2}\,(g^{2})\,=\,\left[\zeta_{1}\,(g^{2})\,+\,g^{2}\,
\zeta^{'}_{1}\,(g^{2})\right]\,\zeta_{1}^{'}\,(g^{2})\,\,.
\end{equation}
These relations are interesting in themselves; they may be exploited
e.\,g. to derive a succession of partially resummed forms for the
perturbative expansion of $Z_{\alpha}$, but since such forms are more
easily derived from the closed-form result of eq. (\ref{zaclos}) below,
we do not dwell on them here, noting only that all the $\zeta_{n}$ with
$n\,\geq\,2$ are in principle determined by the function
$\zeta_{1}\,(g^{2})$ alone, the residue of $Z_{\alpha}$ at
$\varepsilon\,=\,0$. Thus (\ref{zalaur}) is actually a highly redundant
form of conveying the information on coupling renormalization.

\qquad
Second, comparison of the finite parts in eq. (\ref{belaur}) now gives
\renewcommand{\theequation}{2.7}
\begin{equation}
\label{betaclos}
\beta\,(g,\,\varepsilon)\,=\,-\varepsilon\,g\,+\,g^{3}\,
\frac{d\,\zeta_{1}\,(g^{2})}{d\,g^{2}}
\end{equation}
and therefore a (semiconvergent) perturbative expansion
\renewcommand{\theequation}{2.8}
\begin{equation}
\label{betaex}
\beta\,(g,\,\varepsilon)\,=\,-g\,\left[
\varepsilon\,+\,\beta_{0}\,\left(\frac{g}{4\,\pi}\right)^{2}\,+\,
\beta_{1}\,\left(\frac{g}{4\,\pi}\right)^{4}\,+\,...\,\right]\,\,,
\end{equation}
\renewcommand{\theequation}{2.9}
\begin{equation}
\label{betacof}
\beta_{k}\,=\,-(k\,+\,1)\,\zeta_{1,\,k\,+\,1}\qquad
(k\,=\,0,\,1,\,2,\,...\,)\,\,.
\end{equation}
You will remember the particular significance of $\beta_{0}$, the
leading beta-function coefficient at $\varepsilon\,=\,0$ -- the theory
is asymptotically free, i. e. its coupling decreases like an inverse
logarithm at large scales $\mu$, if and only if $\beta_{0}\,>\,0$. Also,
let me recall without comment that the first two coefficients,
$\beta_{0}$ and $\beta_{1}$, are independent of the renormalization
scheme $R$ in a large class of schemes that includes the dimensional
schemes \cite{WEI}.

\qquad
Having expressed, in eq. (\ref{betaclos}), the beta function in terms of
the residue at $\varepsilon\,=\,0$ of $Z_{\alpha}$, we may now invert
the procedure and express $Z_{\alpha}$ in terms of the beta function.
From (\ref{betafrac}), we have
\renewcommand{\theequation}{2.10}
\begin{equation}
\label{zaloder}
(4\,\pi)^{2}\,\frac{\partial}{\partial\,g^{2}}\,\ln\,Z_{\alpha}\,=\,
\frac{1}{\kappa\,+\,\varepsilon\,\chi\,(\kappa)}\,\,,
\end{equation}
where we abbreviated $(g\,/\,4\,\pi)^{2}\,=\,\kappa$, and where
\renewcommand{\theequation}{2.11}
\begin{equation}
\label{chifu}
\chi\,(\kappa)\,=\,-\frac{1}{(4\,\pi)^{2}\,\zeta_{1}^{'}\,(g^{2})}\,=\,
\frac{1}{\beta_{0}\,+\,\beta_{1}\,\kappa\,+\,
\beta_{2}\,\kappa^{2}\,+\,...}
\end{equation}
contains the information of the beta function at $\varepsilon\,=\,0$.
Integrating with respect to $g^{2}$ under the initial condition
$Z_{\alpha}\,(0,\,\varepsilon)\,=\,1$ then gives the integral
representation \cite{THO7}
\renewcommand{\theequation}{2.12}
\begin{equation}
\label{zaclos}
Z_{\alpha}\,(g^{2},\,\varepsilon)\,=\,\exp\,\Biggl\{
-\int\limits_{0}^{(g\,/\,4\,\pi)^{2}}
\frac{d\,\kappa^{'}}{\kappa^{'}\,+\,\varepsilon\,\chi\,(\kappa^{'})}\Biggr\}\,\,.
\end{equation}
This exact expression shows the well-known double face of
renormalization constants: when expanding the integrand
"perturbatively", i. e. in powers of $\kappa$ with
$\beta_{0}\,/\,\varepsilon$ as a leading term, and integrating termwise,
you get back the perturbation expansion of (\ref{zalaur}) and
(\ref{zetnfu}) in which every term is ultraviolet divergent as
$\varepsilon\,\to\,0$ -- though with all coefficients now expressed in
terms of those of $\beta\,(g)$ or $\zeta_{1}\,(g^{2})$, and with the
relations for the higher $\zeta_{n}$´s, like (\ref{zet2prim}), therefore
built in:
\renewcommand{\theequation}{2.13}
\begin{equation}
\label{zaexp}
Z_{\alpha}\,(g,\,\varepsilon)\,=\,1\,-\,\beta_{0}\,\left(
\frac{g}{4\,\pi}\right)^{2}\,\frac{1}{\varepsilon}\,+\,
{\cal{O}}\,(g^{4})\,\,.
\end{equation}
On the other hand if you ask about the behavior as $\varepsilon\,\to\,0$
of the full, resummed expression (\ref{zaclos}), you find
$\exp\,(-\infty)\,=\,0\,!$ To see exactly how the function vanishes
requires a minor detour in order to get rid of the lower-limit
divergence of the $\kappa$ integral at $\varepsilon\,=\,0$: write
\renewcommand{\theequation}{2.14}
\begin{equation}
\frac{1}{\kappa\,+\,\varepsilon\,\chi\,(\kappa)}\,=\,
\frac{1}{1\,+\,\varepsilon\,\chi^{'}\,(\kappa)}\,
\frac{d}{d\,\kappa}\,\ln\,[\kappa\,+\,\varepsilon\,\chi\,(\kappa)]\,\,,
\end{equation}
where the prime is now a differentiation with respect to $\kappa$, and
perform a partial integration. You get
\begin{equation*}
Z_{\alpha}\,(g,\,\varepsilon)\,=\,
\frac{\varepsilon}{\beta_{0}\,\left[\left(\frac{g}{4\,\pi}\right)^{2}\,+\,
\varepsilon\,\chi\,\left(\left(\frac{g}{4\,\pi}\right)^{2}\right)\right]}\,\,\times
\end{equation*}
\renewcommand{\theequation}{2.15}
\begin{equation}
\label{zapart}
\times\,\,\exp\,\Biggl\{\varepsilon\,\Biggl[
\frac{\chi^{'}\,(\kappa)\,\ln\,(\kappa\,+\,\varepsilon\,\chi\,(\kappa))}{
1\,+\,\varepsilon\,\chi^{'}\,(\kappa)}\,-\,
\frac{\chi^{'}\,(0)\,\ln\,\left(\frac{\varepsilon}{\beta_{0}}\right)}{1\,+\,\varepsilon\,\chi^{'}\,(0)}
\end{equation}
\begin{equation*}
-\int\limits_{0}^{\kappa}\,d\,\kappa^{'}\,
\frac{\chi^{''}\,(\kappa^{'})}{[1\,+\,\varepsilon\,\chi^{'}\,(\kappa^{'})]^{2}}\,\ln\,
[\kappa^{'}\,+\,\varepsilon\,\chi\,(\kappa^{'})]\Biggr]\Biggr\}_{\kappa\,=\,\left(\frac{g}{4\,\pi}\right)^{2}}\,\,.
\end{equation*}
The remaining integral is finite at $\varepsilon\,=\,0$; the exponent
has terms of order $\varepsilon$ and $\varepsilon\,\ln\,\varepsilon$
only in that limit; so
\renewcommand{\theequation}{2.16}
\begin{equation}
\label{zalim}
Z_{\alpha}\,(g^{2},\,\varepsilon)\,=\,\frac{\varepsilon}{\beta_{0}\,\left(\frac{g}{4\,\pi}\right)^{2}}\,
[1\,+\,{\cal{O}}\,(\varepsilon,\,\varepsilon\,\ln\,\varepsilon)\,(g^{2}\,(\mu))]
\end{equation}
is an exact statement.
(The writing emphasizes that the additional terms still depend on
$g^{2}\,(\mu)$ in a complicated way). This will not surprise lattice
theorists, since when inserted into (\ref{recoup}) it says that in the
removal-of-regulator limit of a {\em nonperturbative} calculation, the
bare coupling should be tuned to zero according to
\renewcommand{\theequation}{2.17}
\begin{equation}
\label{bartu}
\left(\frac{\overline{g}_{0}\,(\varepsilon)}{4\,\pi}\right)^{2}\,=\,
\frac{\varepsilon}{\beta_{0}}\,\mu^{2\,\varepsilon}\,[1\,+\,{\cal{O}}\,
(\varepsilon,\,\varepsilon\,\ln\,\varepsilon)\,(g^{2}\,(\mu))]\,\,,
\end{equation}
and for them this is entirely familiar if they replace $\varepsilon$
with something proportional to their regulator $a$, the lattice
constant.

\qquad
There is yet a third way of encoding the information on coupling
renormalization, which in a sense is the most physical since it relates
to the representation of dimensionful observables. In the strictly
renormalizable theories the coupling becomes dimensionless as
$D\,\to\,4$, and if the theory has no explicit mass terms in its
Lagrangian (or, as in QCD with only light quarks, masses that are orders
of magnitude too small to explain anything observed), then there seems
to be at first sight no dimensionful quantity available to set the scale
for dimensionful observables. For example, a pure Yang-Mills
theory can still generate nonzero glueball masses, or nonzero scattering
lengths (low-energy scattering amplitudes) between its gauge bosons. The
solution to the paradox \cite{GNV} is that the theory creates its own
mass scale "spontaneously" through quantum effects -- and essentially
the same quantum effects (i. e., loops) that are responsible for the
necessity of coupling renormalization. Since observable quantities are
to be expressed as multiples of powers of this scale, it cannot simply
be the arbitrary sliding scale $\mu$, nor can it depend on arbitrary
gauge-fixing parameters. In an otherwise massless theory, this scale,
called $\Lambda$, must therefore depend on $\mu$ and on $g\,(\mu)$ in
such a way as to be an {\em RG invariant}, i. e. it must be a solution
of
\renewcommand{\theequation}{2.18}
\begin{equation}
\label{ladgl}
\mu\,\frac{d}{d\,\mu}\,\Lambda\,(g\,(\mu),\,\mu)\,=\,
\left[\mu\,\frac{\partial}{\partial\,\mu}\,+\,\
\beta\,(g\,(\mu),\,\varepsilon)\,\frac{\partial}{\partial\,g}\right]\,
\Lambda\,(g,\,\mu)\,=\,0\,\,,
\end{equation}
the simplest RG equation conceivable. Mass dimension requires $\Lambda$
to be of the form $\mu\,\cdot\,{\cal{E}}_{\varepsilon}\,(g\,(\mu))$,
with
the dimensionless function ${\cal{E}}_{\varepsilon}\,(g)$ fulfilling
\renewcommand{\theequation}{2.19}
\begin{equation}
\left[1\,+\,\beta\,(g,\,\varepsilon)\,\frac{\partial}{\partial\,g}\right]\,
{\cal{E}}_{\varepsilon}\,(g)\,=\,0\,\,,
\end{equation}
so separation of variables gives the representation
\renewcommand{\theequation}{2.20}
\begin{equation}
\label{lambdae}
\Lambda_{\varepsilon}\,(g\,(\mu),\,\mu)\,=\,\mu\,\exp\,\left\{
-\int\limits_{g_{1}}^{g\,(\mu)}\frac{d\,g^{'}}{\beta\,(g^{'},\,\varepsilon)}\right\}
\end{equation}
with an integration constant $g_{1}$, a trivial RG invariant. A simpler
but somewhat more formal way of seeing the existence of $\Lambda$ ist to
write eq. (\ref{betafu}) as
\renewcommand{\theequation}{2.21}
\begin{equation}
\frac{1}{\mu}\,d\,\mu\,=\,\frac{1}{\beta\,(g,\,\varepsilon)}\,d\,g
\end{equation}
and integrate between any two values $\mu_{1},\,\mu_{2}$ of the sliding
scale, which gives
\renewcommand{\theequation}{2.22}
\begin{equation}
\label{psirel}
\mu_{1}\,\E^{-\Psi\,(g\,(\mu_{1}),\,\varepsilon)}\,=\,
\mu_{2}\,\E^{-\Psi\,(g\,(\mu_{2}),\,\varepsilon)}\,\,,
\end{equation}
with $\Psi\,(g,\,\varepsilon)$ denoting a primitive of
$1\,/\,\beta\,(g,\,\varepsilon)$; this says that the combination
(\ref{lambdae}) is $\mu$-independent. Thus a divergent coupling
renormalization, as expressed in the RG beta function (\ref{betaclos}),
{\em directly} translates into the "spontaneous" emergence of a
$\Lambda$ scale.

\qquad
The integration constant $g_{1}$, like the higher beta-function
coefficients (\ref{betacof}) with $k\,\geq\,2$, depends on the
renormalization scheme adopted, and in fact is part of its
specification. Changing $g_{1}$ to $g_{1}^{'}$ while keeping the
higher $\beta_{k}$'s unchanged (a special change of scheme) gives
another scale
$\Lambda^{'}$ differing from $\Lambda$ by a finite di\-mensionless
factor:
\renewcommand{\theequation}{2.23}
\begin{equation}
\label{lambtraf}
\frac{\Lambda_{\varepsilon}^{'}}{\Lambda_{\varepsilon}}\,=\,
\exp\,\left[\Psi\,(g_{1}^{'})\,-\,\Psi\,(g_{1})\right]\,\,.
\end{equation}
Thus the $\Lambda$ scale, while an invariant within one renormalization
scheme, is still different for different schemes, with the $\Lambda$´s
for two
\newpage
schemes differing by a pure-number factor. In that sense it is
not yet on the same level of theoretical nobility as a physical mass,
which must in addition be scheme-independent, but it is a necessary and
important intermediary between the arbitrary $\mu$ and the measurable
masses and lengths.

\qquad
From the point of view of coupling dependence, the essential point about
the $\Lambda$ scale emerges in the limit $\varepsilon\,\to\,0$: by
expanding perturbatively,
\renewcommand{\theequation}{2.24}
\begin{equation}
\label{invbet}
\frac{1}{\beta\,(g,\,0)}\,=\,-\frac{(4\,\pi)^{2}}{\beta_{0}\,g^{3}}\,
\left[1\,-\,\left(\frac{\beta_{1}}{\beta_{0}}\right)\,\left(
\frac{g}{4\,\pi}\right)^{2}\,+\,{\cal{O}}\,(g^{4})\right]\,\,,
\end{equation}
where the terms ${\cal{O}}\,(g^{4})$ again contain the scheme-dependent
higher beta coefficients, we have
\renewcommand{\theequation}{2.25}
\begin{equation}
\label{lambnull}
\Lambda_{0}\,=\,\mu\,\E^{-\frac{(4\,\pi)^{2}}{2\,\beta_{0}\,g\,(\mu)^{2}}}
\left[\beta_{0}\left(\frac{g}{4\,\pi}\right]^{2}\right]^{-\frac{\beta_{1}}{2\,\beta_{0}^{2}}}
\left\{1\,-\left(\frac{\beta^{2}_{1}\,-\,\beta_{0}\,\beta_{2}}{2\,\beta_{0}^{3}}\right)\left(\frac{g}{4\,\pi}\right)^{2}+\,
{\cal{O}}\,(g^{4})\right\}_{R}
\end{equation}
where the (semiconvergent) series in curly brackets depends on the
renormalization scheme $R$, and where for the class of schemes we use we
have chosen $g_{1}$ such that, in the notation of (\ref{psirel}),
\renewcommand{\theequation}{2.26}
\begin{equation}
\label{g1conv}
\Psi\,(g_{1},\,0)\,=\,-\frac{\beta_{1}}{2\,\beta_{0}^{2}}\,\ln\,
\left[\frac{\beta_{0}}{(4\,\pi)^{2}}\right]\,\,.
\end{equation}
Thus $\Lambda$ at $D\,=\,4$ is the perfect example on an "escaping"
function of the form of (\ref{exsupp}), a quantity that will never show
up in a perturbation expansion in $g^{2}$. In making this
identification, it is clearly essential that $\beta_{0}\,>\,0$, i. e.
that {\em the theory should be asymptotically free}. In asymptotically
non-free theories the scale $\Lambda_{0}$, rather than being
exponentially suppressed, would be {\em exploding} as $g^{2}\,\to\,0+$,
and none of the things we are going to discuss in the following would be
applicable.

\qquad
Moreover, one should not lose sight of the fact that when talking about
"behavior as $g^{2}\,\to\,0+$" here, we are actually talking about a
rather fictitious, unphysical limit -- a limit in which we imagine
$g^{2}$ being tuned to zero while $\mu$ is being kept constant. Only in
this purely formal limit can we speak of $\Lambda_{0}$ being
"exponentially suppressed" -- in the physical world there is no way of
getting $\Lambda_{0}$ to vanish, since the only parameter there which we
can "tune" at will is $\mu$, and $\Lambda_{0}$ by construction just
does {\em not} change with $\mu$.

\qquad
Now eq. (\ref{lambnull}) would not create an essentially new situation
if the only way
$\Lambda_{0}$ occurred in correlation functions were the trivial one:
one may of course invert eq. (\ref{lambdae}) to express $g$ in terms of
$\Lambda_{\varepsilon}\,/\,\mu$, obtaining at $\varepsilon\,=\,0$ the
well-known expression (\cite{WEI})
\renewcommand{\theequation}{2.27}
\begin{equation}
\label{alfmu}
\frac{\alpha\,(\mu)}{4\,\pi}\,=\,
\frac{1}{\beta_{0}\,\lambda_{0}}\,\left\{1\,-\,\frac{2\,\beta_{1}}{\beta^{2}_{0}}\,
\left(\frac{\ln\,\lambda_{0}}{\lambda_{0}}\right)\,+\,{\cal{O}}\,
\left(\frac{1}{\lambda_{0}^{2}},\,\frac{\ln\,\lambda_{0}}{\lambda_{0}^{2}}\right)\right\}
\end{equation}
where
\renewcommand{\theequation}{2.28}
\begin{equation}
\label{logmu}
\lambda_{0}\,=\,\ln\,\left(\frac{\mu^{2}}{\Lambda_{0}^{2}}\right)\,\,.
\end{equation}
One may then parameterize the perturbation series in terms of
the inverse of this logarithm rather than in terms of $\alpha$, but this
will not introduce any new information beyond the perturbative one, and
this remains true even if we go one step further and invoke the RG
scaling properties of vertex functions to rewrite the expansion in terms
of a coupling $\overline{\alpha}$ that "runs" with $Q^{2}$, some
representative combination of the function´s momenta, rather than with
$\mu^{2}$ \cite{WEI}. (This would be the {\em same} function as
(\ref{alfmu}) with $\mu^{2}$ simply replaced by $Q^{2}$.) Since
\renewcommand{\theequation}{2.29}
\begin{equation}
\label{alfbar}
\overline{\alpha}\,(Q^{2})\,=\,\alpha\,\left\{
1\,+\,\left[\text{power series
in}\,\frac{\alpha}{4\,\pi}\,\beta_{0}\,\ln\,
\left(\frac{Q^{2}}{\mu^{2}}\right)\right]\right\}
\end{equation}
(as follows by combining eqs. (\ref{lambdae}) at scales $\mu$ and
$\sqrt{Q^{2}}$), this amounts to a resummation of the perturbation
series that is useful for many purposes, but again creates no new
content beyond that series. But then if your QFT text mentions no other
application of the lambda scale than eq. (\ref{alfmu}), you should
protest -- they are hiding the best from you! For example, an $s$-wave
scattering amplitude in a massless theory at low center-of-mass energies
$\sqrt{s}$ would look like
\renewcommand{\theequation}{2.30}
\begin{equation}
\label{scattle}
f_{0}\,(s)\,=\,\frac{1}{d_{0}\,\Lambda_{0}\,+\,
d_{1}\,\sqrt{s}\,+\,d_{2}\,\frac{s}{\Lambda_{0}}\,+\,...}\,\,,
\end{equation}
with $d_{0}\,\Lambda_{0}$ the inverse scattering length, and with the
dimensionless coefficients $d_{i}$ possibly having some perturbative,
power-series dependence on the coupling (\ref{alfbar}). Or a bound-state
pole in a four-point amplitude $T$ of a massless theory would appear as
\renewcommand{\theequation}{2.31}
\begin{equation}
\label{bopole}
T\,(p^{'},\,P,\,p;\,\alpha)\,=\,
\frac{\alpha\,\Phi\,(p^{'}\,/\,\Lambda_{0};\,\alpha)\,\cdot\,
\alpha\,\overline{\Phi}\,(p\,/\,\Lambda_{0};\,\alpha)}{P^{2}\,+\,
c\,\Lambda_{0}^{2}}\,+\,[\text{regular terms}]\,\,,
\end{equation}
(where the notation indicates that the factorizing residue, in contrast
to the purely nonperturbative denominator, is a mixed
perturbative-nonperturbative quantitiy beginning with at least an
overall factor of $\alpha^{2}$). In other words, the truly interesting
quantities are powers and inverse powers, polynomials and inverse
polynomials, series and inverse series in
$\Lambda_{0}$, and in these,
unlike in (\ref{alfmu}), $\Lambda_{0}$ appears in its full, non-analytic
nastiness, eq. (\ref{lambnull}). We will, for the purpose of these
seminars, refer to such quantities as {\em strongly nonanalytic} at
$\alpha\,=\,0$ or {\em strongly nonperturbative}, as opposed to mere
resummations of a perturbation series, where $\Lambda_{0}$ only appears
in inverse logarithms.
\subsection{Operator-Product Expansion and Lambda}

\qquad
Our digression into the Operator-Product Expansion (OPE) will be brief
and highly selective, serving only the narrow purpose of reminding you
of a long established way \cite{WIL} of asymptotically representing the
strongly nonperturbative coupling dependence. For a systematic
development of the OPE, refer to \cite{WEI}; this reference also
contains, in the special context of renormalons, a discussion equivalent
to the one below. For a terse account of Zimmermann´s perturbative
renormalization \cite{ZIM} of the OPE, you may consult \cite{KUG}.

\qquad
Again we only look at one simple example that will exhibit all features
of interest -- the Euclidean vacuum expectation of a single product of
two generic elementary fields with naive mass dimension one at variable
separation $x$,
\renewcommand{\theequation}{2.32}
\begin{equation}
\label{twox}
D\,(x)\,=\,\braket{0|\varphi\,\left(\frac{1}{2}\,x\right)\,\varphi\,\left(-\frac{1}{2}\,x\right)}{0}\,\,,
\end{equation}
in the context of an asymptotically free theory $(\beta_{0}\,>\,0)$
without Lagrangian masses ($m\,=\,m_{0}\,=\,0)$. (General OPE treatments
would consider correlation functions with this pair of operators
inserted into an arbitrary product
$\varphi\,(y_{1})\,\varphi\,(y_{2})\,...\,\varphi\,(y_{n})$ of fields at
fixed, distinct, nonzero points $y_{i}$). Loosely stated, the OPE is the
postulate that the field product permits a small-$x$ expansion of the
form
\renewcommand{\theequation}{2.33}
\begin{equation}
\label{opef}
\varphi\,\left(\frac{1}{2}\,x\right)\,\varphi\,\left(-\frac{1}{2}\,x\right)\,=\,
\sum\limits_{n\,=\,0}^{\infty}\,\sum\limits_{i\,=\,1}^{l\,(n)}\,W_{n,\,i}\,(x)\,{\cal{O}}_{2\,n,\,i}\,(0)\,\,,
\end{equation}
where the ${\cal{O}}_{2\,n,\,i}\,(z)$ are the full set of local
operators of mass dimension $2\,n$,
elementary or composite, that bear the conserved quantum numbers of the
field product. Thus at the core of the OPE is a {\em completeness
hypothesis} which, to my knowledge, has nowhere been proven with full
mathematical rigor but is nevertheless highly plausible: the full set of
local operators at $x$ forms a basis for the expansion of operator
products clustering around $x$.

\qquad
These operators are enumerated here according to their naive mass
dimensions $2\,n\,=\,0,\,2,\,4,\,...$ and to a counting label $i$
ranging over the finite number $l\,(n)$ of such operators with given
$n$, the enumeration beginning at $n\,=\,0$ with $l\,(0)\,=\,1$ and the
dimensionless unit operator,
\renewcommand{\theequation}{2.34}
\begin{equation}
\label{unit}
{\cal{O}}_{0,\,1}\,=\,{\rm 1\!l}\,\,.
\end{equation}
To make up for the increasing mass dimensions, the $c$-number functions
$W_{n,\,i}\,(x)$, the "Wilson coefficients", consist of powers
$(x^{2})^{n-1}$ times a perturbatively calculable series in the coupling
that contains powers of $\ln\,x^{2}$.

\qquad
More precisely, what is meant \cite{WIL} by (\ref{opef}) is that this
expansion should hold in the sense of "weak operator convergence", i.
e., not with respect to any operator norm, but only when both sides are
sandwiched between normalizable states $\ket{\Psi_{1}}$ and
$\bra{\Psi_{2}}$ of the quantized-field system. The two-point function
(\ref{twox}) is then the simplest sandwich of this kind, where
$\ket{\Psi_{1}}\,=\,\ket{\Psi_{2}}\,=\,\ket{0}$, the physical vacuum
state. In a renormalizable theory, the resulting matrix elements, the
"vacuum condensates"
\renewcommand{\theequation}{2.35}
\begin{equation}
\label{vacond}
\braket{0|{\cal{O}}_{2\,n,\,i}\,(0)}{0}\qquad
(n\,=\,0,\,1,\,2,\,...\,\,,\quad i\,=\,1,\,2,\,...\,l\,(n))\,\,,
\end{equation}
should finally appear in a renormalized form, referring to a definite
renormalization scheme $R$ and to a scale $\mu$ within that scheme, but
since the detailed way the renormalization works in an expansion like
(\ref{opef}) has some aspects of interest for our later discussion, we
will begin by
considering a not-yet-completely-renormalized version, parameterized by
$\alpha\,(\mu)$ but still at a positive value of the dimensional
regulator $\varepsilon\,=\,(4\,-\,D)\,/\,2$. When taking Fourier
transforms, such as
\renewcommand{\theequation}{2.36}
\begin{equation}
\label{ftwilc}
W_{n,\,i}\,(x)\,=\,\int\,d^{4\,-\,2\,\varepsilon}\,x\,\tilde{W}_{n,\,i}\,(k^{2};\,\alpha;\,\frac{1}{\varepsilon})\,
\E^{i\,k\,\cdot\,x}\,\,,
\end{equation}
the above mentioned structure of the Wilson coefficients is mirrored in
the momentum-space structure
\renewcommand{\theequation}{2.37}
\begin{equation}
\label{wtilde}
\tilde{W}_{n,\,i}\,=\,\left(\frac{1}{k^{2}}\right)^{n\,+\,1}\,\times\,
\{\text{ perturbatively calculable series in}\,\,\alpha\}\,\,,
\end{equation}
with the coefficients of the series containing, in general, both
$\frac{1}{\varepsilon}$ ultraviolet divergences and the typical
logarithmic modifications $\ln\,(k^{2}\,/\,\mu^{2})$. Therefore when
passing to the two-point vertex function $\Gamma_{2}\,=\,-D^{-1}$ the
expansion will look like
\renewcommand{\theequation}{2.38}
\begin{equation}
\label{opem}
-\Gamma_{2}\,(k^{2};\,\alpha;\,\varepsilon)\,=\,k^{2}\,
\sum\limits_{n\,=\,0}^{\infty}\,\sum\limits_{i\,=\,1}^{l\,(n)}\,
V_{n,\,i}\,\left(\alpha;\,\ln\,\frac{k^{2}}{\mu^{2}};\,\frac{1}{\varepsilon}\right)\,
\frac{\braket{0|{\cal{O}}_{2\,n,\,i}}{0}_{\alpha,\,\varepsilon}}{(k^{2})^{n}}\,\,,
\end{equation}
with perturbatively calculable coefficients,
\renewcommand{\theequation}{2.39}
\begin{equation}
\label{vwilc}
V_{n,\,i}\,\left(\alpha;\,\ln\,\frac{k^{2}}{\mu^{2}};\,\frac{1}{\varepsilon}\right)\,=\,
\sum\limits_{p\,=\,0}^{\infty}\,V_{n,\,i,\,p}\,\left(\ln\,\frac{k^{2}}{\mu^{2}};\,\frac{1}{\varepsilon}\right)\,\alpha\,(\mu)^{p}\,\,.
\end{equation}
As for the vacuum condensates, which at this stage have their own
ultraviolet divergences at $\varepsilon\,\to\,0$, let us start by
considering the expansion purely within the context of perturbation
theory. There, we have no known way of "renormalizing" composite
operators; i. e. redefining them so as to have finite matrix elements,
other than defining them as free-field normal products
:${\cal{O}}_{2\,n,\,i}$:, but since those all have zero expectation
values in the perturbative vacuum, all condensate terms with
$n\,\geq\,1$ in (\ref{opem}) would then "renormalize" to zero. Thus we
conclude that the $n\,=\,0$ term, involving the trivial operator
(\ref{unit}), is the perturbative portion of $D^{-1}$,
\renewcommand{\theequation}{2.40}
\begin{equation}
\label{opert}
-\Gamma^{pert}_{2}\,(k^{2};\,\alpha;\,\varepsilon)\,=\,k^{2}\,V_{0\,1}\,\left(
\alpha;\,\ln\,\frac{k^{2}}{\mu^{2}};\,\frac{1}{\varepsilon}\right)\,\,,
\end{equation}
and, since the zeroth-order perturbative part is simply $k^{2}$,
\renewcommand{\theequation}{2.41}
\begin{equation}
\label{vnull}
V_{0\,1}\,\left(\alpha;\,\ln\,\frac{k^{2}}{\mu^{2}};\,\frac{1}{\varepsilon}\right)\,=\,1\,+\,{\cal{O}}\,(\alpha)\,\,.
\end{equation}
The situation would be different, of course, for the more general
correlation-with-insertion functions on which most discussions of the
OPE focus: there one encounters matrix elements of the free-field normal
product between non-vacuum states, which in general have nonvanishing
perturbative limits. Pure vacuum expectations like (\ref{twox}), which
in a sense are more vulnerable to perturbative mutilation, are
particularly suited to our purpose because they compel us to make a
crucial point: as originally envisaged by its inventor \cite{WIL},
expansion (\ref{opem}) is to be understood here in a more general,
nonperturbative sense, where vacuum condensates for $2\,n\,\geq\,2$ can
be nonzero even though the OPE by itself offers no immediate recipe for
calculating them in a given renormalization scheme. This indeed is the
standpoint taken in the successful semi-empirical applications of the
OPE through QCD sum rules \cite{SVZ}, another wide and interesting field
on which these seminars will be unable to enter \cite{NAR}.

\qquad
Even without calculating the condensates we may, to some extent, discuss
their renormalization formally to see more precisely what is
"incalculable" about them. Under renormalization, composite operators
of the same mass dimension are known to mix linearly:
\renewcommand{\theequation}{2.42}
\begin{equation}
\label{comix}
\braket{0|{\cal{O}}_{2\,n;\,i}\,(0)}{0}_{\alpha,\,\varepsilon}\,=\,
\sum\limits_{j\,=\,1}^{l\,(n)}\,Z_{i\,j}^{(2\,n)}\,\left(\alpha,\,\frac{1}{\varepsilon}\right)\,
\langle{\cal{O}}_{2\,n,\,j}\rangle_{R,\,\mu}\,\,.
\end{equation}
The renormalization matrices $Z^{(2\,n)}$ are of the general form
\renewcommand{\theequation}{2.43}
\begin{equation}
\label{zmat}
Z^{(2\,n)}_{i\,j}\,\left(\alpha,\,\frac{1}{\varepsilon}\right)\,=\,
\delta_{i\,j}\,+\,{\cal{O}}\,(\alpha)\,\,,
\end{equation}
with all terms ${\cal{O}}\,(\alpha)$ ultraviolet divergent, while the
renormalized condensates $\langle{\cal{O}}_{2\,n,\,j}\rangle_{R,\,\mu}$
are by definition ultraviolet finite but in general dependent on $\mu$
and on the the scheme $R$. The right-hand side of (\ref{opem}) becomes,
\renewcommand{\theequation}{2.44}
\begin{equation}
\label{opdiv}
k^{2}\,V_{0,\,1}\,\left(\alpha;\,\ln\,\frac{k^{2}}{\mu^{2}};\,\frac{1}{\varepsilon}\right)\,+\,
k^{2}\,\sum\limits_{n\,=\,1}^{\infty}\,\sum\limits^{l\,(n)}_{j\,=\,1}\,
\left\{\,
\sum\limits_{i\,=\,1}^{l\,(n)} V_{n,\,i}\,Z^{(2\,n)}_{i\,j}\right\}\,
\frac{\langle{\cal{O}}_{2\,n,\,j}\rangle_{R,\,\mu}}{(k^{2})^{n}}\,\,.
\end{equation}
For such an expansion to remain within the realm of renormalizable QFT
in the presently accepted sense of the term, one of the minimal
prerequisites is that nonlocal ultraviolet divergences of the form
\renewcommand{\theequation}{2.45}
\begin{equation}
\label{nonpow}
\left(\frac{1}{\varepsilon}\right)^{m}\,\left(\frac{1}{k^{2}}\right)^{n}\,\,,
\end{equation}
which would require counterterms of the same form in the Lagrangian,
should be absent. Therefore the curly-bracketed quantities in
(\ref{opdiv}) must all be finite for at least $n\,\geq\,2$; the Wilson
coefficients must be rendered finite by the composite-operator
renormalizations {\em separately} for each $n$:
\renewcommand{\theequation}{2.46}
\begin{equation}
\label{vren}
\sum\limits_{i\,=\,1}^{l\,(n)}\,V_{n,\,i}\left(
\alpha;\,\ln\,\frac{k^{2}}{\mu^{2}};\,\frac{1}{\varepsilon}\right)Z^{(2\,n)}_{i\,j}\left(\alpha,\,\frac{1}{\varepsilon}\right)=
V_{n,\,j}^{(R)}\left(\alpha;\,\ln\,\frac{k^{2}}{\mu^{2}}\right),\,
\text{finite as}\,\,\,\varepsilon\,\to\,0\,\,,
\end{equation}
where, because of (\ref{zmat}),
\renewcommand{\theequation}{2.47}
\begin{equation}
\label{vrenex}
V_{n,\,j}^{(R)}\,=\,V_{n,\,j}\,(\alpha\,=\,0)\,+\,{\cal{O}}\,(\alpha)\,\,.
\end{equation}
This "internal" renormalization at each $n\,\geq\,2$ is indeed what
happens in the purely perturbative calculations alluded to before, for
quantities where our condensates are replaced by more general matrix
elements having nonzero perturbative limits. It is also implicit
(though, curiously, never discussed explicitly) in all cases where one-
or two-loop Wilson coefficients have been calculated under
nonperturbative assumptions allowing for nonvanishing vacuum condensates
\cite{NAR} \cite{RRY} \cite{LAV}.
This is analogous to the well-known result of renormalization theory
that nonlocal divergences of type
\renewcommand{\theequation}{2.48}
\begin{equation}
\label{nonlog}
\left(\frac{1}{\varepsilon}\right)^{m}\,\left(\ln\,\frac{k^{2}}{\mu^{2}}\right)^{p}
\end{equation}
cancel separately in each order of the {\em perturbative} expansion.
Here, since we lack the time to delve into explicit examples, we
will simply accept this general statement as plausible. (Incidentally,
one of the uses of the techniques described in chapter 3 may be to
provide a more precise framework for such demonstrations.)

\qquad
In the $n\,=\,1$ term, the counterterm required for an ultraviolet
divergence would be constant in momentum, of the form of a mass
renormalization, and therefore not excluded by renormalizability, but it
turns out to be really necessary only in the case of $m\,\not=\,0$, a
nonzero Lagrangian mass; for a massless theory, where this term
represents a contribution to dynamical-mass formation not visible in the
Lagrangian, the finiteness of eq. (\ref{vren}) will persist at
$n\,=\,1$.

\qquad
On the other hand for the perturbative term with $n\,=\,0$, associated
with the trivial operator (\ref{unit}), there is no internal
renormalization to bring ultraviolet finiteness -- the unit operator
"doesn´t volunteer" to tame its Wilson coefficient, and that
coefficient alone -- that is, the perturbative series -- must therefore
be supplied with the usual "external" renormalization through a
field-renormalization factor $Z_{3}$. Thus the general scheme, in
the presence of the nonperturbative OPE terms, is no more that of
multiplication with one overall $Z_{3}$ for the entire vertex
function, but rather
\renewcommand{\theequation}{2.49}
\begin{equation}
\label{opren}
(D^{-1})_{R,\,\mu}\,=\,\underbrace{Z_{3}\,(D^{-1})^{pert}_{\varepsilon}}_{\text{finite}}\,+\,
\sum\limits_{n,\,j}\,\underbrace{V_{n,\,j}^{(R)}}_{\text{finite}}\,
\frac{\langle{\cal{O}}_{2\,n,\,j}\rangle_{R,\,\mu}}{(k^{2})^{n\,-\,1}}\,\,.
\end{equation}
I have been dwelling on this simple but important point because the OPE
literature, vast as it is, does not seem to state it clearly, and
because it means that in the presence of the nonperturbative terms,
multiplicative renormalizability in the usual simple sense is no more a
property of QFT amplitudes. We will encounter this peculiar
"two-track" renormalization pattern again, in the context of a more
dynamical approach, in ch. 3. below.

\qquad
In eq. (\ref{opren}) we have not yet separated the perturbative and
strongly nonperturbative coupling dependences completely. To achieve
such a separation, we need to note that in a massless theory, the
renormalized condensates at $\varepsilon\,=\,0$ must be of the form
\renewcommand{\theequation}{2.50}
\begin{equation}
\label{coren}
\langle{\cal{O}}_{2\,n,\,j}\rangle_{R;\,\mu}\,=\,C_{n,\,j}\,(\alpha\,(\mu))_{R}\,(\Lambda_{0}^{2})_{R}^{\,\,\,n}\,\,,
\end{equation}
since $\Lambda_{0}$ is then the only invariant mass scale available for
this purpose. The coefficients $C_{n,\,j}$, power series in
$\alpha\,(\mu)$ and in general semiconvergent,
\renewcommand{\theequation}{2.51}
\begin{equation}
\label{coser}
C_{n,\,j}\,(\alpha)\,=\,\sum\limits_{q\,=\,0}^{\infty}\,c_{n,\,j}^{(q)}\,\alpha^{q}\,\,,
\end{equation}
are what gives the vacuum condensates their residual $\mu$ dependence
within the scheme $R$ -- "residual" because the terms without explicit
powers of $\alpha$,
\renewcommand{\theequation}{2.52}
\begin{equation}
\label{conull}
c_{n,\,j}^{(0)}\,(\Lambda_{0}^{2})^{n}\,\,,
\end{equation}
are RG invariants within $R$. It is these terms, with no perturbative
power of $\alpha$ but with the strongly nonanalytic $\alpha$ dependence
in $\Lambda_{0}$ still present, that form the "incalculable" cores of
the condensates in the context of the OPE. One of the exhilarating
experiences to which the formalism of chapter 3 will invite you is that
these cores become calculable there as multiples of powers of
$\Lambda_{0}$ in a given (dimensional) scheme, with coefficients
$c_{n,\,j}^{(0)}$ given by a systematic sequence of approximations,
which in turn can be determined self-consistently from the
equations of motion of the basic vertex functions.

\qquad
Putting things together, and introducing modified Wilson coefficients
\renewcommand{\theequation}{2.53}
\begin{equation}
\label{uwilc}
U_{n}^{(R)}\,\left(\ln\,\frac{k^{2}}{\mu^{2}};\,\alpha\,(\mu)\right)\,=\,
\sum\limits_{j\,=\,1}^{l\,(n)}\,V_{n,\,j}^{(R)}\,\left(
\ln\,\frac{k^{2}}{\mu^{2}};\,\alpha\right)\,C_{n,\,j}\,(\alpha)
\end{equation}
that refer only to the mass dimension $2\,n$ but no more to individual
condensates, we then have a more transparent form of the OPE,
\renewcommand{\theequation}{2.54}
\begin{equation}
\label{opela}
\begin{array}{c}
\displaystyle{
[D^{-1}\,(k^{2};\,\alpha;\,\mu)]_{R}\,=\,k^{2}\,\Biggl\{
U_{0}^{(R)}\,\left(\ln\,\frac{k^{2}}{\mu^{2}};\,\alpha\right)}\\
\mbox{}\\
\displaystyle{
+\,\sum\limits_{n\,=\,1}^{\infty}\,U_{n}^{(R)}\,\left(\ln\,\frac{k^{2}}{\mu^{2}};\,\alpha\right)\,
\left[\frac{(\Lambda_{0}^{2})_{R}}{k^{2}}\right]^{n}\Biggr\}\,\,,}
\end{array}
\end{equation}
where
\renewcommand{\theequation}{2.55}
\begin{equation}
\label{rpert}
U_{0}^{(R)}\,\left(\ln\,\frac{k^{2}}{\mu^{2}};\,\alpha\right)\,=\,
\left\{Z_{3}\,\left(\alpha;\,\frac{1}{\varepsilon}\right)\,V_{0,\,1}\,\left(
\ln\,\frac{k^{2}}{\mu^{2}};\,\alpha;\,\frac{1}{\varepsilon}\right)\right\}_{\varepsilon\,=\,0}
\end{equation}
denotes the renormalized perturbation series. Again, the terms without
powers of $\alpha$ but with $\Lambda_{0}$ kept nonzero,
\renewcommand{\theequation}{2.56}
\begin{equation}
\label{invdnull}
(D^{-1})^{(0)}_{R}\,=\,k^{2}\,\left\{
1\,+\,\sum\limits_{n\,=\,1}^{\infty}\,U_{n}^{(R)}\,(0,\,0)\,\left(
\frac{\Lambda_{0}^{2}}{k^{2}}\right)^{n}\right\}\,\,,
\end{equation}
with
\renewcommand{\theequation}{2.57}
\begin{equation}
\label{unnull}
U_{n}^{(R)}\,(0,\,0)\,(\Lambda^{2}_{0})^{n}\,=\,
\sum\limits_{j\,=\,1}^{l\,(n)}\,V_{n,\,j}\,(\alpha\,=\,0)\,\left[
c_{n,\,j}^{(0)}\,(\Lambda^{2}_{0})^{n}\right]\,\,,
\end{equation}
form an RG-invariant "core" for the two-point function, to which the
$U_{n}$ power series add $\mu$-dependent, perturbative
corrections.

\qquad
As rearranged in the form of eq. (\ref{opela}), the OPE tells us three
things. First, the genuinely nonperturbative coupling dependence of a
correlation function can be represented by a series in {\em integer}
powers of $\Lambda_{0}$, each power being accompanied by its
own series of perturbation corrections with the typical logarithmic
momentum
dependences. This structure has been confirmed in all cases where an OPE
could be studied to some degree of explicitness in renormalizable model
theories with a lambda scale \cite{DAV} \cite{BBK}, and may nowadays be
regarded as well established, but it was not always so clear in the
early years when attempts to calculate condensates by instanton methods
seemed to indicate the necessity of terms with noninteger powers of
$\Lambda$ \cite{THO8}. Work by G. Münster \cite{MUN} later established
that these noninteger powers were really an artefact of the
"dilute-instanton-gas" approximation used in most instanton
calculations, and that by taking account of the denseness of the
instanton gas one is led back to integer powers of $\Lambda$ (see
t´Hooft´s introduction for the 1994 reissue of his 1980 lectures, p.
457 of ref. \cite{THO3}).

\qquad
Second, since $\Lambda_{0}$ is dimensionful as opposed to the
dimensionless $\alpha$, such an expansion is {\em automatically} a
large-momentum expansion in powers of $\Lambda_{0}^{2}\,/\,k^{2}$. This
automatic linkage between coupling and momentum dependences is much
tighter and more direct than in the perturbative series, where
rearrangement in terms of a momentum-dependent running coupling
$\overline{\alpha}\,(Q^{2})$ is a useful option for physics but not a
mathematical necessity.

\qquad
Third, we have learned that each of the strongly nonperturbative
$(n\,\geq\,1)$ terms
must establish itself in an ultraviolet-{\em finite} manner -- a
requirement which obviously places nontrivial restrictions on any
dynamical theory seeking to calculate those terms.

\qquad
The OPE, since it does not by itself produce a prescription to calculate
its own vacuum condensates, remains a deep {\em structural} statement
but does not represent a completely known starting point for an {\em ab
initio} dynamical theory. It is nevertheless
-- even apart from the successful applications already mentioned -- of
great value for clarifying the coupling dependence of amplitudes in the
strictly renormalizable theories with a $\Lambda$ scale. To see this
dependence more explicitly, we may once more rewrite our result
(\ref{opela}) by using the structure (\ref{lambnull}) of $\Lambda_{0}$.
Introducing the perturbative expansions
\renewcommand{\theequation}{2.58}
\begin{equation}
\label{urser}
\begin{array}{c}
\displaystyle{
U_{n}^{(R)}\,\left(\ln\,\frac{k^{2}}{\mu^{2}};\,\alpha\right)\,\cdot\,
\left\{1\,-\,\left(\frac{\beta_{1}^{2}\,-\,\beta_{0}\,\beta_{2}}{2\,\beta_{0}^{3}}\right)\,
\left(\frac{g}{4\,\pi}\right)^{2}\,+\,O\,(g^{4})\right\}^{2\,n}_{R}}\\
\mbox{}\\
\displaystyle{
=\,\sum\limits_{p\,=\,0}^{\infty}\,\tilde{u}_{n,\,p}^{(R)}\,\left(
\ln\,\frac{k^{2}}{\mu^{2}}\right)\,\alpha^{p}\,\,,}
\end{array}
\end{equation}
we then have for our two-point function the form
\renewcommand{\theequation}{2.59}
\begin{equation}
\label{oprsymb}
(D^{-1})_{R}\,=\,k^{2}\,\sum\limits_{n\,=\,0}^{\infty}\,\left(
\frac{\mu^{2}}{k^{2}}\right)^{n}\,\left(\E^{-\frac{4\,\pi}{\beta_{0}\,\alpha}}\right)^{n}\,
\left\{\left(\beta_{0}\,\frac{\alpha}{4\,\pi}\right)^{-\frac{n\,\beta_{1}}{\beta_{0}^{2}}}\,
\sum\limits_{p\,=\,0}^{\infty}\,\tilde{u}_{n,\,p}^{(R)}\,\alpha^{p}\right\}\,\,,
\end{equation}
a form that stands directly at the entrance door to the theory of
resurgent functions.
\vspace*{6mm}

\subsection{Resurgent Symbols and Functions}

\qquad
You will remember that at the end of sect. 1.1 we made a point of the
fact that the "escaping" functions of a semiconvergent expansion
typically appear again as small-$z$ expansions, eq. (\ref{exsupp}),
whose power-series factors $\{1\,+\,O\,(z)\}$ in general are again
semiconvergent. This observation may be taken as a starting point for
the following heuristic consideration.

\qquad
Assume we are dealing with a function $f\,(z)$ having an asymptotic
expansion
\renewcommand{\theequation}{2.60}
\begin{equation}
\label{f0ser}
f\,(z)\,\sim\,\sum\limits_{p\,=\,0}^{\infty}\,a_{0,\,p}\,z^{p}\,\,,
\end{equation}
but not sufficient analyticity to be reconstructed uniquely from it as
in chapter 1. We may take some representative $f_{0}\,(z)$ from the
class of functions having this expansion and write
\renewcommand{\theequation}{2.61}
\begin{equation}
\label{fplusg}
f\,(z)\,=\,f_{0}\,(z)\,+\,g_{0}\,(z)\,\,,
\end{equation}
with $g_{0}\,(z)$ the "escaping" remainder. Assume further that
$f\,(z)$ has bounding constants of the form of eq. (\ref{gevrey}) so the
bound of eq. (\ref{bndmin}) holds, and that $g_{0}\,(z)$ actually has
terms, as e.\,g. in (\ref{exsupp}), that exhaust the bound:
\renewcommand{\theequation}{2.62}
\begin{equation}
\label{g0ser}
g_{0}\,(z)\,=\,e^{-\frac{s_{1}}{z}}\,F_{1}\,(z)\,\,;\qquad
F_{1}\,(z)\,\sim\,z^{-B_{1}}\,\sum\limits_{p\,=\,0}^{\infty}\,a_{1,\,p}\,z^{p}
\qquad (Re\,s_{1}\,>\,0)\,\,.
\end{equation}
Of course, $g_{0}$ may have other terms, of faster exponential or even
super-exponential decrease, but the series in the $F_{1}$ factor, an
object as formal as the original series (\ref{f0ser}), has enough
freedom of interpretation to accommodate all of these. In the simplest
case, that series will turn out to have a finite radius of convergence
and thus to determine uniquely  an analytic function $F_{1}\,(z)$, so
that the $\sim$ sign in (\ref{g0ser}) turns into an equals sign within
that convergence radius. In this case, the combination of eqs.
(\ref{fplusg}) and (\ref{g0ser}) is a complete and exact representation
of our function $f\,(z)$.

\qquad
But a much more interesting case, and the one we concentrate on in all
that follows, is for the situation of eq. (\ref{f0ser}) to repeat
itself: assume $F_{1}\,(z)$ again has a small-$z$ expansion that is only
semiconvergent, and "escaping" terms led by an exponentially
suppressed one:
\renewcommand{\theequation}{2.63}
\begin{equation}
\label{f1g1}
F_{1}\,(z)\,=\,f_{1}\,(z)\,+\,g_{1}\,(z)\,\,,\qquad
g_{1}\,(z)\,=\,\E^{-\frac{r_{2}}{z}}\,F_{2}\,(z)\qquad
(Re\,r_{2}\,\geq\,0)\,\,,
\end{equation}
where $f_{1}$ is some representative of the class of functions having
the second member of (\ref{g0ser}) as asymptotic expansion. Again
$g_{1}\,(z)$ may have other terms, of faster exponential or even
super-exponential decrease, but we are again free to stuff these into
the "escaping" terms left undetermined by the formal series expansion
\renewcommand{\theequation}{2.64}
\begin{equation}
F_{2}\,(z)\,\sim\,z^{-B_{2}}\,\sum\limits_{p\,=\,0}^{\infty}\,a_{2,\,p}\,z^{p}\,\,.
\end{equation}
With that understanding we may then write
\renewcommand{\theequation}{2.65}
\begin{equation}
\label{ftwex}
f\,(z)\,=\,f_{0}\,(z)\,+\,\E^{-\frac{s_{1}}{z}}\,f_{1}\,(z)\,+\,
\E^{-\frac{s_{2}}{z}}\,F_{2}\,(z)\qquad
(s_{2}\,=\,s_{1}\,+\,r_{2})\,\,.
\end{equation}
Imagine now that this pattern keeps repeating itself, or "resurging",
so that at every stage the escaping functions turn out to be led by an
exponentially suppressed term. Then upon putting conventionally
\renewcommand{\theequation}{2.66}
\begin{equation}
s_{0}\,=\,B_{0}\,=\,0\,\,,
\end{equation}
we arrive at a formal representation
\renewcommand{\theequation}{2.67}
\begin{equation}
\label{suppser}
f\,(z)\,\sim\,\sum\limits_{n\,=\,0}^{\infty}\,\E^{-\frac{s_{n}}{z}}\,f_{n}\,(z)\,\,
\end{equation}
where
\renewcommand{\theequation}{2.68}
\begin{equation}
\label{fndef}
\begin{array}{l}
f_{n}\,(z):\,=\,\text{a representative of the class of functions}\\
\text{having asymptotic
expansion}\,\,z^{-B_{n}}\,\sum\limits_{p\,=\,0}^{\infty}\,a_{n,\,p}\,z^{p}\,\,,
\end{array}
\end{equation}
and where
\renewcommand{\theequation}{2.69}
\begin{equation}
\label{asequ}
0\,=\,Re\,s_{0}\,<\,Re\,s_{1}\,\leq\,Re\,s_{2}\,\leq\,Re\,s_{3}\,...
\end{equation}
The representation may terminate if at some $n$ the formal series for
the function $F_{n}\,(z)$ turns out to have a finite radius of
convergence and determines uniquely an analytic function $F_{n}\,(z)$,
which is then equal to $f_{n}\,(z)$ without a nonanalytic remainder.
(This case actually occurs \cite{STSH} in some of the classical
transcendental functions obtained as solutions to linear, second-order
differential equations.)

\qquad
However if the decomposition process does not stop and new, nonzero,
suppression constants $r_{n}$ (or $s_{n}\,=\,s_{n\,-\,1}\,+\,r_{n}$)
keep emerging indefinitely, then the representation (\ref{suppser})
remains highly formal in that it lacks proper definition at two fronts.
First, we have said nothing as yet about the convergence or divergence
of the
$n$ series, nor about its resummation in the case of divergence. Second,
while it appears plausible that by going far enough in the expansion
(\ref{suppser}) we will eventually account for all the {\em
exponentially} suppressed terms, we have completely lost track of the
possible terms with super-exponential suppression, which at every step
we swept under the rug of the next $F_{n\,+\,1}\,(z)$. Thus at every
{\em finite} stage of the decomposition, such as eq. (\ref{ftwex}), they
were still accounted for by the simple device of the last term having an
$F_{n}$, instead of $f_{n}$, that still had room for such terms -- just
as in an ordinary semiconvergent series {\em all} suppressed terms were
still present in the exact remainder term of the finite Taylor expansion
(\ref{remtayl}). We lost them the moment we discarded this subtle
difference and jumped ahead to the infinite formal series
(\ref{suppser}) -- just as in earlier times we lost {\em all} terms with
faster-than-powers suppression in jumping from (\ref{remtayl}) to the
formal series (\ref{formser}). The first and second deficiencies are of
course related -- by carrying the analogy one step further, we can guess
that the super-exponentially vanishing terms, if any, must be hiding
behind a divergence of the $n$ series, and that accounting for such
terms must be tantamount to giving a prescription for resumming that
series.

\qquad
Making bad things worse, we now forget even about our list of specific
choices of representatives $f_{n}\,(z)$ for the sums of the various
semiconvergent $p$ series, and simply write
\renewcommand{\theequation}{2.70}
\begin{equation}
\label{resymb}
f\,(z)\,\sim\,\sum\limits_{n\,=\,0}^{\infty}\,\E^{-\frac{s_{n}}{z}}\,\left\{
z^{-B_{n}}\,\sum\limits_{p\,=\,0}^{\infty}\,a_{n,\,p}\,z^{p}\right\}\,\,,
\end{equation}
a representation still more formal than (\ref{suppser}) in that it now
cries out for answers, not only to the above-mentioned and related two
questions, but also to the question of the meaning of each of the $p$
sums. This formal double expansion is called a general {\em resurgent
symbol} in the mathematical literature \cite{ECAL}, and the successive
terms of its $n$ sum -- each carrying a definite exponential-suppression
factor $\exp\,(-s_{n}\,/\,z)$, except for the $n\,=\,0$ term -- are
referred to as the {\em elementary resurgent symbols}. Because of its
highly formal character, it would at first seem like an exotic way of
making one´s life complicated; yet it is precisely such a double
expansion that arises in several areas of both mathematics and physics
-- the perturbation theory of quantum-mechanical energy levels, the JWKB
semi-classical expansion, the thermodynamic functions of the ideal Fermi
gas, the perturbation expansions of celestial mechanics, or, to mention
a purely mathematical problem that forms the subject of vol. II of ref.
\cite{ECAL}, the classification of germs of analytic functions around
the origin through a study of their iterations. Most important for us, a
glimpse at eq. (\ref{oprsymb}) tells us that the OPE, in a strictly
renormalizable and asymptotically free theory, is also of this form: in
a QFT
with divergent coupling renormalization (\ref{recoup}) and therefore a
coupling-nonanalytic mass scale (\ref{lambdae}), the {\em
operator-product expansion is a resurgent-symbol formal representation
with respect to the coupling} of the correlation functions. This
representation in terms of a formal double expansion is the new paradigm
that replaces the formal perturbation series (\ref{pertex}) in the realm
of the strictly renormalizable theories.

\qquad
You should realize that eq. (\ref{resymb}) is a highly peculiar kind of
expansion: it looks as if one and the same functional dependence on a
single variable $z$ were to be expanded in two sets of basis functions
simultaneously. If somebody would suggest this to you in the context of
convergent expansions, you would surely protest against the uncontrollable
double counting. What makes the
resurgent-symbol expansion possible is the specific, and matching,
incompleteness in both types of "basis functions": the $z^{p}$
expansion can never account for an $\exp\,(-\,\text{const.}\,/\,z)$, and
$\exp\,(-\,\text{const.}\,/\,z)$, being an infinite Laurent series with
only negative powers of $z$, can never account for positive powers
$z^{p}$.

\qquad
The possibilities of reconstructing analytic functions from such
resurgent symbols form an important part of the theory of {\em resurgent
functions}. As a preliminary definition -- which will not satisfy the
mathematician, but conveys enough of the agenda to be helpful to the
consumer -- we may thus formulate:
\renewcommand{\theequation}{2.71}
\begin{equation}
\label{predef}
\left\|\quad
{\parbox{10cm}{
A resurgent function is an analytic function that can be
reconstructed from its formal small-$z$ expansion if the
latter is given at least "to exponential accuracy", i. e. in
the form of a resurgent symbol (\ref{resymb}).}}
\quad \right\|
\end{equation}
In dealing with such functions, the physicist must revise his or her
ingrained habit of neglecting exponentially suppressed terms in the
small-$z$ expansion "because they are smaller than any power". While
this may be a useful approximation {\em numerically} at sufficiently
small $z$ and under favorable circumstances (namely, if the leading
coefficients $a_{n,\,0},\,n\,\geq\,1$, are not too large), it is
completely irrelevant when it comes to knowing the "true" function: in
reconstructing the latter, one either keeps the exponentially suppressed
terms or gets nonsense, {\em tertium non datur}.

\qquad
Our above heuristic argument emphasized the peculiar property that gave
resurgent
functions their name -- the fact that semiconvergent expansions with the
special properties of "factorial" error bounding in $N$, and of the
ensuing exponential error suppression in $z$, "resurge" at every level
of refinement of the small-$z$ expansion. Yet it does not lead to the
most general type of resurgent symbol. The more systematic mathematical
approach \cite{ECAL} proceeds by firmly placing the problem from the
outset in the
framework of the most powerful and versatile reconstruction method, the
Borel-Laplace process of sect. 1.3. The starting point there is a
reformulation of the basic Laplace-transform representation
(\ref{laplace}): one introduces a modified Borel-transform function
$B\,(s)$, called the {\em major} of $f$ as opposed to the {\em minor}
$b\,(s)$, such that $B\,(s)$ has as its sole singularity a branch
cut
along the positive real $s$ axis with discontinuity equal to the minor
$b\,(s)$:
\renewcommand{\theequation}{2.72}
\begin{equation}
\label{madisco}
B\,(s\,+\,i\,\varepsilon)\,-\,B\,(s\,-\,i\,\varepsilon)\,=\,\Theta\,(s)\,b\,(s)\,\,.
\end{equation}
For a $b\,(s)$ sufficiently decreasing as $s\,\to\,+\infty$, such a
function may simply be constructed as a spectral integral,
\renewcommand{\theequation}{2.73}
\begin{equation}
\label{smaj}
B\,(s)\,=\,\frac{1}{2\,\pi\,i}\,\int\limits_{0}^{\infty}\,d\,s^{'}\,
\frac{b\,(s^{'})}{s^{'}\,-\,s}\,\,,
\end{equation}
which by virtue of the well-known distribution relation
\renewcommand{\theequation}{2.74}
\begin{equation}
\label{deldiff}
\frac{1}{s^{'}\,-\,(s\,+\,i\,\varepsilon)}\,-\,
\frac{1}{s^{'}\,-\,(s\,-\,i\,\varepsilon)}\,=\,
2\,i\,\pi\,\delta\,(s^{'}\,-\,s)
\end{equation}
fulfills condition (\ref{madisco}). (Such functions have occasionally
been used in physics work too \cite{DAV}.) If $b\,(s^{'})$ does not
decrease fast enough, or even increases, so that the simple spectral
integral fails to exist, one may use various "subtracted" or
"divided" forms of the spectral representation: for example, if under
Nevanlinna-Sokal conditions $b\,(s)$ turns out to be unfriendly enough
to exhaust its exponential-growth restriction $|b\,(s)|\,\leq\,{\rm
const.}\,\exp\,\left(\frac{s}{R}\right)$ at $s\,\geq\,0$, one may set up
a spectral representation for the function
$\exp\,\left(-\frac{s}{R^{'}}\right)\,\cdot\,B\,(s)$ with $R^{'}\,<\,R$,
which retains all analytic properties of $B$ since the exponential is an
entire function, and obtain
\renewcommand{\theequation}{2.75}
\begin{equation}
\label{dmaj}
B_{e}\,(s)\,=\,\E^{\left(\frac{s}{R^{'}}\right)}\,\frac{1}{2\,\pi\,i}\,
\int\limits_{0}^{\infty}\,d\,s^{'}\,
\frac{\E^{-\frac{s^{'}}{R^{'}}}\,b\,(s^{'})}{s^{'}\,-\,s}\,\,,
\end{equation}
where the integral now converges. Of course, this superconvergence
device may also be applied in cases where it is not strictly necessary;
for example, when the pure-and-simple Cauchy integral (\ref{smaj})
already exists, we obviously have
\renewcommand{\theequation}{2.76}
\begin{equation}
\label{comaj}
B_{e}\,(s)\,=\,B\,(s)\,+\,
\frac{1}{2\,\pi\,i}\,\int\limits_{0}^{\infty}\,d\,s^{'}\,\frac{
\E^{\frac{s\,-\,s^{'}}{R^{'}}}\,-\,1}{s^{'}\,-\,s}\,b\,(s^{'})\,\,,
\end{equation}
where the second term is an entire function since it is now analytic
both off
and on the positive $s$ axis. This example illustrates the general truth
that an infinite family of functions with the prescribed right-hand
discontinuity (\ref{madisco}) can be constructed, which all differ from
each other by the addition of entire functions. The notion of major
therefore really stands for such a whole equivalence class of functions,
known to mathematicians as a {\em hyperfunction}. In a QFT context we
are usually dealing with minors $b\,(s)$ sufficiently decreasing as
$s\,\to\,+\infty$ -- the toy example of eq. (\ref{fullap}) below being
typical in that respect -- and therefore will usually be able to get
along with the "canonical" representatives of eq. (\ref{smaj}) without
really using this generality of hyperfunctions.

\qquad
For an ordinary asymptotic series summable to a function $f\,(z)$ by the
Borel-Laplace procedure, the Laplace representation (\ref{laplace}), by
virtue of eq. (\ref{madisco}), can now be rewritten in terms of the
major $B\,(s)$ as
\renewcommand{\theequation}{2.77}
\begin{equation}
\label{lagaplus}
f\,(z)\,=\,\frac{1}{z}\,\int\limits_{\Gamma_{+}}\,d\,s\
\E^{-\frac{s}{z}}\,B\,(s)\,\,,
\end{equation}
the integral being taken along the contour $\Gamma_{+}$ on both sides of
the positive $s$ axis as shown in Fig. 2.1. In view of the analyticity
of $B\,(s)$, this contour can in turn be opened up, as indicated in
Fig. 2.1, to turn into the contour $\Gamma_{0}$ parallel to the
imaginary $s$ axis. Thus as an alternative to (\ref{laplace}) we have an
integral transform along the path $\Gamma_{0}$ -- that is, in essence a
Fourier transform -- of a major function $B\,(s)$ whose sole
singularity is a branch point at $s_{0}\,=\,0$.
\begin{center}
\epsfig{file=fig21.eps,height=8.5cm}\\
$\begin{array}{ll}
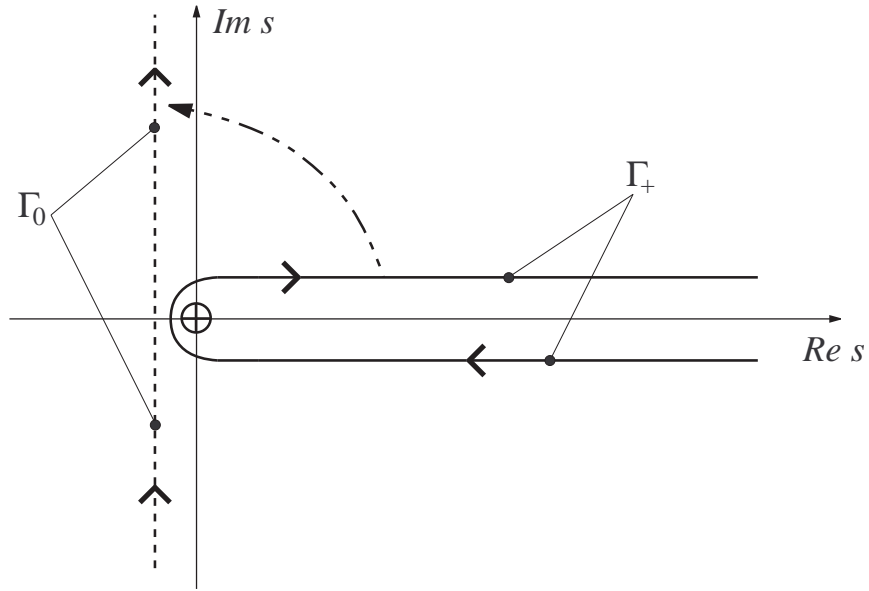
\text{\bf Fig. 2.1} \quad & \text{Integration contour $\Gamma_{+}$ in
major-based}\\
& \text{BL transform and its deformation into $\Gamma_{0}$}
\end{array}$
\epsfig{file=fig21a.eps,height=8.5cm}\\
$\begin{array}{ll}
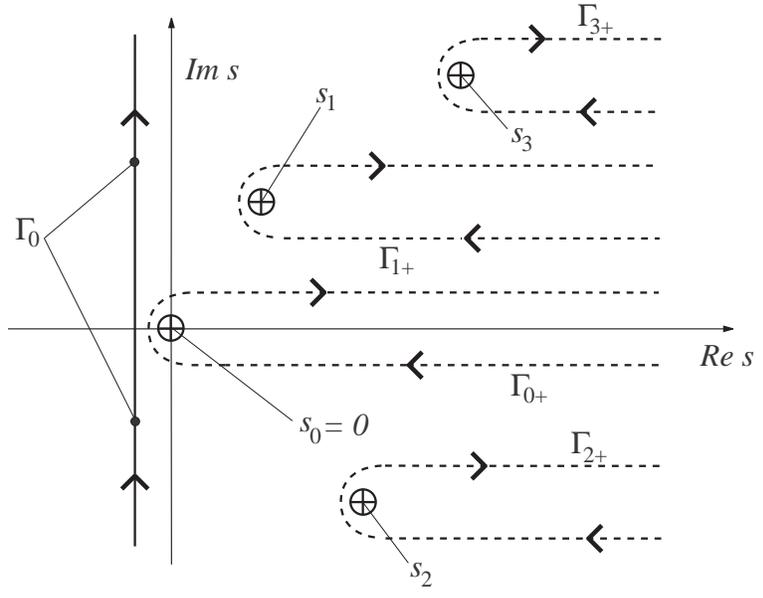
\text{\bf Fig. 2.2} \quad & \text{Integration contours in BL integral
for}\\
 & \text{resurgent function with support points $s_{n}$}
\end{array}$
\end{center}

\qquad
It is this alternative Laplace integral that forms a convenient point of
departure for the generalization to resurgent functions. Assume we are
now given a function $B\,(s)$ which instead of a single branch point at
$s_{0}\,=\,0$ has a denumerable set of singularity points $s_{n}$,
either branch points or poles, in the right half plane $Re\,s\,>\,0$,
possibly including one at $s_{0}\,=\,0$, and ordered as in eq.
(\ref{asequ}). See Fig. 2.2. We label $B\,(s)$, somewhat summarily, by
this set
$\{s_{0},\,s_{1},\,s_{2},\,...\,\}$ of singularities, which in the
mathematical literature is referred to as the {\em support} of a
resurgent function or symbol (or both). Assume that $B\,(s)$ is such
that the Laplace integral with contour $\Gamma_{0}$,
\renewcommand{\theequation}{2.78}
\begin{equation}
\label{resnull}
f\,(z)\,=\,\frac{1}{z}\,\int\limits_{\Gamma_{0}}\,d\,s\
\E^{-\frac{s}{z}}\,B_{\{s_{0},\,s_{1},\,s_{2},\,...\,\}}\,(s)\,\,,
\end{equation}
exists in at least some "anisotropic", right-hand $z$-plane domain.
This is the basic representation of a general resurgent function, {\em
provided} one further condition is fulfilled: in order to obtain a
formula suitable for calculating $f\,(z)$ at small, positive $z$, we
must be able to bend the contour $\Gamma_{0}$ back to the right as in
Fig. 2.2 so as to turn it into a union of horizontal back-and-forth
contours $\Gamma_{n+},\,n\,=\,0,\,1,\,2,\,...\,,$ each slung around one
of the singularity points $s_{n}$, so that
\renewcommand{\theequation}{2.79}
\begin{equation}
\label{resdec}
f\,(z)\,=\,\sum\limits_{n}\,\frac{1}{z}\,\int\limits_{\Gamma_{n+}}\,d\,s\,
\E^{-\frac{s}{z}}\,B_{\{s_{0},\,s_{1},\,s_{2},\,...\,\}}\,(s)\,\,.
\end{equation}
By using integration variables $s^{'}\,=\,s\,-\,s_{n}$ in the individual
terms, this can be written as
\renewcommand{\theequation}{2.80}
\begin{equation}
\label{resasy}
f\,(z)\,=\,\sum\limits_{n}\,\E^{-\frac{s_{n}}{z}}\,\left\{
\frac{1}{z}\,\int\limits_{0}^{\infty}\,d\,s^{'}\,\E^{-\frac{s^{'}}{z}}\,b_{n}\,(s^{'})\right\}
\end{equation}
with a set of discontinuity functions, or Borel minors, given by
\renewcommand{\theequation}{2.81}
\begin{equation}
\label{bn}
b_{n}\,(s^{'})\,=\,B\,(s_{n}\,+\,s^{'}\,+\,i\,0)\,-\,
B\,(s_{n}\,+\,s^{'}\,-\,i\,0)\,\,,\qquad
n\,=\,0,\,1,\,2,\,...\,\,,
\end{equation}
where $s^{'}$ is real and positive. The representation (\ref{resasy})
is precisely of the form encountered in eq. (\ref{suppser}) above, but
with the $f_{n}\,(z)$ factors now being given a more concrete meaning
through the bracketed Laplace transforms. The final step towards a
small-$z$ expansion is then to perform Taylor expansions at $z\,=\,0+$
-- in general semiconvergent -- of these transforms. For the more common
types of singularities of the major $B\,(s)$, this step is
straightforward: in particular, if $B\,(s)$, locally around each of the
$s_{n}$, is of the logarithmically singular form
\renewcommand{\theequation}{2.82}
\begin{equation}
\label{logmaj}
\left[-\frac{1}{2\,i\,\pi}\,\ln\,(s_{n}\,-\,s)\right]\,\phi_{n}\,(s\,-\,s_{n})\,+\,
\{\text{regular at}\,\,s_{n}\}\,\,,
\end{equation}
with factors $\phi_{n}$ analytic at zero argument and thus having
expansions
\renewcommand{\theequation}{2.83}
\begin{equation}
\label{phin}
\phi_{n}\,(v)\,=\,\sum\limits_{q\,=\,0}^{\infty}\,b_{n,\,q}\,v^{q}
\end{equation}
with finite convergence radii, then since
\renewcommand{\theequation}{2.84}
\begin{equation}
\label{lndisc}
\ln\,(-s^{'}\,-\,i\,0)\,-\,\ln\,(-s^{'}\,+\,i\,0)\,=\,
-2\,i\,\pi\,\Theta\,(s^{'})
\end{equation}
at real $s^{'}$, the $n$-th Borel minor simply becomes
\renewcommand{\theequation}{2.85}
\begin{equation}
\label{simpser}
b_{n}\,(s^{'})\,=\,\phi_{n}\,(s^{'})\,=\,
\sum\limits_{q\,=\,0}^{\infty}\,b_{n,\,q}\,s^{'\,q} \qquad
(s^{'}\,>\,0)\,\,,
\end{equation}
and the small-$z$ expansion is obtained in the form of eq.
(\ref{resymb}) above, with all $B_{n}$´s equal to zero. A function
$f\,(z)$ with such an "all-logarithmic" Borel major is called a
"resurgent function with simple singularities", and is the kind of
function most frequently considered in the mathematical literature on
resurgence. In the QFT context, where in view of the form
(\ref{oprsymb}) of the Operator-Product Expansion we expect an expansion
with non-integer $B_{n}$ exponents, only a minor extension of this
concept is necessary as long as $0\,<\,B_{n}\,<\,1$: we assume $B\,(s)$
to be locally of the form (\ref{logmaj}) only near $s\,=\,s_{0}\,=\,0$,
but of the form
\renewcommand{\theequation}{2.86}
\begin{equation}
\label{pomaj}
\frac{(s_{n}\,-\,s)^{-B_{n}}\,-\,1}{2\,i\,\sin\,(B_{n}\,\pi)}\,\phi_{n}\,
(s\,-\,s_{n})\,+\,\{\text{regular at}\,\,s_{n}\}\qquad
(n\,=\,1,\,2,\,...\,)
\end{equation}
near the other $s_{n}$, with $B_{n}$ non-integer. (The numerator term of
minus unity serves no purpose other than to make the limit
$B_{n}\,\to\,0$ exist and agree with (\ref{logmaj})). Then
\renewcommand{\theequation}{2.87}
\begin{equation}
\label{bnser}
b_{n}\,(s^{'})\,=\,s^{'\,-\,B_{n}}\,\phi_{n}\,(s^{'})\,=\,
\sum\limits_{q\,=\,0}^{\infty}\,b_{n,\,q}\,(s^{'})^{q\,-\,B_{n}}\qquad
(0\,<\,B_{n}\,<\,1)\,\,,
\end{equation}
and the general form of eq. (\ref{resymb}) is obtained. We see how a
resurgent symbol arises as a small-$z$ expansion "to exponential
accuracy" of a resurgent function given by eq. (\ref{resnull}).

\qquad
It has been essential, in deriving the resurgent symbol, that the
original contour $\Gamma_{0}$ could be deformed into the set of
$\Gamma_{n+}$´s without changing the value of the integral. For this it
is necessary, of course, that the growth (if any) of the major $B\,(s)$,
or at least of its discontinuities (\ref{bn}), as $Re\,s\,\to\,+\infty$
be sufficiently limited to allow the Laplace transforms of eq.
(\ref{resasy}) to converge. But there is also a more subtle
precondition, which in the presence of an {\em infinite} set of
singularities $s_{n}$ as allowed by our definition is no more trivial:
we tacitly assumed that the arcs-at-infinity that are invisible in Fig.
2.2, but necessary to turn the union of (minus) the old $\Gamma_{0}$ and
the new $\Gamma_{n+}$´s into a closed total contour and thus to
establish the equivalence, can always be supplied even if the
singularities occur at arbitrarily large distances $|s_{n}|$. In other
words the set $\{s_{n}\}$, while possibly spreading out to infinity in
the right half of the $s$ plane, must not be allowed to accumulate
anywhere in such a way as to completely obstruct, for some direction or
sector of directions, analytic continuation of $B\,(s)$ to infinity
along paths between them. The mathematical literature on resurgence
calls such functions {\em endlessly continuable}. We are thus in a
position to state the mathematical definition of resurgent functions:
\renewcommand{\theequation}{2.88}
\begin{equation}
\label{findef}
\left\| \quad
{\parbox{10cm}{
A resurgent function is an analytic function $f\,(z)$
permitting a
Laplace-transform representation (\ref{resnull}) /
(\ref{resdec}) with an
endlessly continuable Borel major $B\,(s)$.}}
\quad \right\|
\end{equation}
It is clear that this definition is extremely general -- it places
almost no restrictions on the detailed nature of the singularities of
the major at the points $s_{n}$, or on the detailed structure of the
analytic factors $\phi_{n}\,(s\,-\,s_{n})$ associated with them, except
that they should not destroy the possibility of a Laplace-transform
representation, and that they should remain sufficiently isolated to
permit endless continuation in between them.
It is no more necessary, as our introductory heuristic argument seemed
to
suggest, that each $r_{n}\,=\,s_{n}\,-\,s_{n-1}$ be identifiable as the
exponential-suppression constant in the series of the preceding,
$(n\,-\,1)$-th term.
That
argument, leading to eqs. (\ref{suppser}) and (\ref{resymb}), really
dealt
with a special case, where one not only encountered the same type of
singularity, namely (\ref{pomaj}), for all $n$, but also the same type
of semiconvergent series $F_{n}\,(z)$ (namely, with factorial error
bounding), and therefore the same type of convergence in the analytic
factors $\phi_{n}$, so that in fact the complete local behavior of
the major near any
$s_{n}$ gets revived near all the others. It is this
special but also especially interesting case which gave the entire class
of resurgent functions their name. In Écalle´s words \cite{ECAL}, these
are functions "possessing a built-in mechanism that counteracts the
usual "erosion" of local properties under analytic continuation and
provides instead for their spotwise resurgence".

\qquad
Even within this special class, you should still be prepared for all
sorts of nasty little complications, when actually trying to reconstruct
a resurgent function through, say, the series (\ref{resasy}) of partial
Laplace transforms. Our definitions required the expansions (\ref{phin})
of the analytic factors $\phi_{n}\,(v)$ to have finite convergence
radii, but this does not prevent them from having the singularities that
limit their convergence circles sit maliciously at real positive $v$. If
these singularities are poles, then the corresponding partial Borel
minors, obtained by analytic continuation of (\ref{bnser}), have poles
at real positive $s^{'}$, and the associated Laplace integrals in
(\ref{resasy}) have no unique meanings any more -- they depend on the
way one evades the pole when integrating, and therefore become
two-valued functions. The complete resurgent function is then, in
general, multi-valued.

\qquad
The mathematical definition (\ref{findef}), apart from the weak
restriction of endless continuability, allows for very general forms of
the support $\{s_{n}\}$, the set of singularities of the Borel major,
which via eq. (\ref{resasy}) become the set of exponential-suppression
constants. In many cases -- particularly when considering nonlinear
equations where products of a function with itself or its derivatives
must be dealt with -- one may want to move with\-in sets of resurgent
symbols that are closed under formal series multiplication (form a
multiplicative algebra), and mathematical theory therefore prefers to
deal with support sets that form {\em additive semigroups}. Of special
importance among these is
one particular kind of
support, where all $s_{n}$ sit equidistantly on the nonnegative real
axis,
\renewcommand{\theequation}{2.89}
\begin{equation}
\label{equidi}
s_{n}\,=\,n\,s_{1}\,\,;\qquad s_{1}\,\,\text{real and}\,\,>\,0\,\,;
\qquad n\,=\,0,\,1,\,2,\,...\,\,,
\end{equation}
and which not only plays a prominent role in resurgence theory (in
particular,
it is ubiquitous in Écalle´s treatment of resurgent solutions to
nonlinear differential equations) but, as it happens, is also present in
the resurgent symbols of OPE type relevant to QFT coupling dependence.
These, as illustrated by our two-point example of eq. (\ref{oprsymb})
above, feature the constants
\renewcommand{\theequation}{2.90}
\begin{equation}
\label{opsn}
s_{n}\,=\,n\,s_{1}\,\,,;\qquad
s_{1}\,=\,\frac{4\,\pi}{\beta_{0}}\,>\,0\,\,;\qquad
n\,=\,0,\,1,\,2,\,...
\end{equation}
\renewcommand{\theequation}{2.91}
\begin{equation}
\label{opBn}
B_{n}\,=\,n\,B_{1}\,\,;\qquad
B_{1}\,=\,\frac{\beta_{1}}{\beta_{0}^{2}}\qquad ;\qquad
n\,=\,0,\,1,\,2,\,...
\end{equation}
In this case, the resurgent symbol (\ref{resymb}) assumes the special
form of a double formal power series:
\renewcommand{\theequation}{2.92}
\begin{equation}
\label{posymb}
f\,(\alpha)\,\sim\,\sum\limits_{n\,=\,0}^{\infty}\,u^{n}\,\left(
\alpha^{-B_{1}}\,\E^{-\frac{s_{1}}{\alpha}}\right)^{n}\,\sum\limits_{p\,=\,0}^{\infty}\,a_{n,\,p}\,\alpha^{p}\,\,.
\end{equation}
The occurrence of an $n$-th power of a dimensionless "counting
parameter" $u$ for the $n$ summation, where in our case
\renewcommand{\theequation}{2.93}
\begin{equation}
\label{uparm}
u\,=\,\frac{\mu^{2}}{k^{2}}\,\,,
\end{equation}
is a typical feature of such symbols. Since all $s_{n}$ are now on the
real $s$ axis, the representation of eq. (\ref{resdec}) can
alternatively be written as a single Laplace integral,
\renewcommand{\theequation}{2.94}
\begin{equation}
\label{reslap}
f\,(\alpha)\,=\,\frac{1}{\alpha}\,\int\limits_{0}^{\infty}\,d\,s\
\E^{-\frac{s}{\alpha}}\,b\,(s)\,\,,
\end{equation}
with a {\em combined Borel minor} given, according to line 13 of Table
1, by
\renewcommand{\theequation}{2.95}
\begin{equation}
\label{combo}
b\,(s)\,=\,\sum\limits_{n\,=\,0}^{\infty}\,\Theta\,(s\,-\,n\,s_{1})\,b_{n}\,(s\,-\,n\,s_{1})\,\,.
\end{equation}
It is no more analytic along the positive $s$ axis, but has finite
discontinuities in the support points. For the individual minors
$b_{n}\,(s)$, the class of functions (\ref{bnser}) considered up to now
must
be enlarged here to include distribution-valued terms, since $B_{n}$,
by eq. (\ref{opBn}), becomes larger than one as $n$ increases so the
first line of Table 1 can no more be used. (Recall typical values such
as
$B_{1}\,=\,\frac{32}{81}$ or $B_{1}\,=\,\frac{13}{49}$ for $SU\,(3)$
gauge theory with 3 or 6 fermion species respectively, for which
$B_{n}\,>\,1$ at $n\,\geq\,3$ or $n\,\geq\,4$.)
For the terms in eq. (\ref{posymb}) where $p\,-\,n\,B_{1}\,=\,-k$ is a
negative integer, the Borel transform is a delta distribution
$\delta\,(s)$ when $k\,=\,1$ and a derivative of a delta distribution,
$\delta^{(k\,-\,1)}\,(s\,-\,s_{n})$, when $k\,\geq\,2$:
\renewcommand{\theequation}{2.96}
\begin{equation}
\label{delap}
\alpha^{-k}\,\E^{-\frac{s_{n}}{\alpha}}\,=\,{\cal{L}}\,[\delta^{(k\,-\,1)}\,(s\,-\,s_{n});\,\alpha)\qquad
(k\,=\,1,\,2,\,3,\,....\,)\,\,.
\end{equation}
This is verified immediately by using the definition
\renewcommand{\theequation}{2.97}
\begin{equation}
\label{deridel}
\int\,d\,s\ \delta^{(k\,-\,1)}\,(s)\,g\,(s)\,=\,
(-)^{k\,-\,1}\,\left(\frac{d^{k\,-\,1}\,g}{d\,s^{k\,-\,1}}\right)_{s\,=\,0}
\end{equation}
in the Laplace integral (\ref{reslap}). For the terms with noninteger
$p\,-\,n\,B_{1}\,=\,-\beta\,<\,-1$, one even needs to introduce
fractional derivatives of delta distributions, functionals that perform
the minor miracle of the mapping
\renewcommand{\theequation}{2.98}
\begin{equation}
\label{nonidel}
\frac{1}{z}\,\int\limits_{0}^{\infty}\,d\,s\
\E^{-\frac{s}{z}}\,\delta^{(\beta\,-\,1)}\,(s\,-\,s_{n})\,=\,z^{-\beta}
\,\E^{-\frac{s_{n}}{z}}
\qquad
(\beta\,\,\text{non-integer})\,\,.
\end{equation}
I can hear your oophs -- that is quite some machinery to deal with an
object as simple as $\alpha^{-n\,B_{1}}$! Indeed the combined minor of
eq. (\ref{combo}), having now not only the finite step-function jumps
but also all kinds of distribution-valued terms at the support points
$s\,=\,s_{n}$, becomes complicated in a way that appears somewhat out of
proportion for the relatively simple structure of the symbol
(\ref{posymb}). While the Borel-Laplace framework remains essential for
the {\em definition} and basic theory of resurgence, it may therefore
well turn out to be a clumsy tool when it comes to the practical
recovery of a resurgent $f\,(z)$ from its symbol, and it may become
legitimate, for that purpose, to look around for simpler alternatives --
we emphasized earlier that Borelization, for all its power, has no
monopoly.

\qquad
On the other hand these observations show clearly that
distribution-valued terms of delta and derivative-of-delta type are
natural ingredients for Borel transforms of resurgent functions -- as
indeed we might have guessed already from line 3 our Table 1. They do
not create new problems, in the sense that they do not infringe on
either the existence or the univaluedness of the Laplace transform. This
is the main reason why in the following we will not be concerned with
the so-called "instanton singularities" of Borel transforms that arise
in
the semiclassical treatment of QFT's with instanton solutions. These, as
discussed by t´Hooft \cite{THO8}, are simply integer-order derivatives
of $\delta\,(s\,-\,s_{n})$, which produce discrete additions to the
Laplace transform by the pattern of eq. (\ref{delap}) above.

\qquad
When does a resurgent symbol (\ref{resymb}), or the more special
double-power-series form (\ref{posymb}), determine uniquely an analytic
function $f\,(\alpha)$ whose small-$\alpha$ expansion it represents?
(In this case the symbol is again referred to as {\em summable}.)
Here comes a big disappointment: to the best of my knowledge, no
sufficient conditions are presently known that would come in as handy
for a realistic, asymptotically free QFT as the Nevanlinna-Sokal
criterion (\ref{unique}) does for the superrenormalizable theories. Such
conditions, if they were to be of practical value, would as a minimum
have to allow for the special analyticity situation in the complex
$\alpha$ plane governed by the so-called t´Hooft singularities
\cite{THO8}, at which we had a glance in Fig. 1.1 c) and whose origins
we will explain in more detail in sect. 2.4 below -- singularities that
preclude
even the minimalistic assumption of a small-angle
"sectorial" analyticity of $f\,(\alpha)$ that is routinely made in the
mathematical treatments of resurgence. The modest contribution of sect.
2.4 will be to prove, by explicit counterexample, that these
singularities by themselves need not sabotage unique reconstruction. But
a
compact and convenient positive criterion accommodating them seems to be
lacking. It is clear that if {\em all} the Borel minors obtained by
analytic continuation of the series (\ref{bnser}) have sufficient
smoothness and limited growth to make the partial Laplace transforms in
(\ref{resasy}) well-defined, {\em and} if the $n$ series in
(\ref{resasy}), with these transforms as weights, converges on at least
some segment of the positive $\alpha$ axis including the point
$\alpha\,=\,0$, then at least on that segment the resurgent function
$f\,(\alpha)$ is uniquely recoverable. But this set of conditions is so
restrictive as to be practically useless. For example it has been
plausibly conjectured for realistic QFT amplitudes \cite{DAV}, and will
be strikingly apparent again in sect. 2.4, that the interesting case is
for all $b_{n}$´s individually to have positive-$s$ poles, but for
these poles to cancel in the combined minor $b\,(s)$ of eq.
(\ref{combo}) so the total $f\,(\alpha)$ is nevertheless univalued.
(This scenario seems also to be known \cite{CNP} in mathematical
theory.) The
resurgent symbol is then obviously an awkward way of decomposing such an
amplitude, but we still need to rack our brains about it because physics
makes us run straight into such awkward decompositions.

\qquad
In this (for the time being) unsatisfactory situation, the best one can
do perhaps is to ask what the "escaping" parts, the parts undetermined
by the resurgent symbol, would look like if $f\,(\alpha)$ were {\em not}
to be uniquely recoverable. We already noted in passing that just as the
escaping terms for an ordinary semiconvergent expansion must vanish
faster than any power as $\alpha\,\to\,0+$, the escaping parts for the
more comprehensive resurgent symbol must be what resurgence theory calls
{\em rapidly decreasing functions}, functions vanishing faster than any
exponential $\exp\,(-\,{\rm const.}\,/\,\alpha$). To be more precise we
must, as earlier, adopt more specific hypotheses: let us therefore focus
on the special case of (\ref{posymb}) relevant for QFT, and let us
assume a behavior of the $a_{n,\,p}$ coefficients with respect to $n$
such that the $n$ sum at fixed $p$,
\renewcommand{\theequation}{2.99}
\begin{equation}
\label{xsum}
\sum\limits_{n\,=\,0}^{\infty}\,a_{n,\,p}\,X^{n}\,\,,\qquad
\text{where}\qquad
X\,=\,u\,\alpha^{-B_{1}}\,\E^{-\,\frac{s_{1}}{\alpha}}\,\,,
\end{equation}
is no worse than semiconvergent with factorial error bounding,
analogous to eq. (\ref{gevrey}). We may then repeat verbatim the
error-minimization argument with respect to $N$ of sect. 1.1, and
conclude that the sum of (\ref{xsum}) is determined up to escaping
terms with the small-$\alpha$ expansion
\renewcommand{\theequation}{2.100}
\begin{equation}
\label{xescap}
\E^{-\frac{K_{p}}{X}}\,X^{-L_{p}}\,\{1\,+\,{\cal{O}}\,(X)\}_{p}\,\,,
\end{equation}
with constants $K_{p}\,(>\,0)$ and $L_{p}$ (real), and with the
bracketed power series in $X$ possibly being again semiconvergent. This
formula is interesting in that it shows that -- under these more special
but still entirely reasonable assumptions -- the super-exponential
suppression of the "rapidly decreasing" functions turns out to really
be as fast as $\exp\,(-\exp\,1/\alpha)$, or in other words, faster than
$\exp\,(-c\,/\,\alpha^{k})$ for any $c$ {\em and} $k$. For the
semiconvergent
bracket, arguing heuristically, we may repeat our chain of argument at
the beginning of this section, and conclude that the escaping terms form
the nontrivial part of a resurgent symbol in $X$,
\renewcommand{\theequation}{2.101}
\begin{equation}
\label{xresym}
\sum\limits_{q\,=\,0}^{\infty}\,\E^{-\frac{K_{p,\,q}}{X}}\,X^{-L_{p,\,q}}\,\{
1\,+\,{\cal{O}}\,(X)\}_{p,\,q}\,\,.
\end{equation}
By resupplying the $p$ summation omitted up to now and reordering the
entire expansion by increasing $K$´s (here of course, is another tacit
assumption -- that there is a smallest $K$), we find that the terms left
undetermined by the symbol (\ref{posymb}), if any, should themselves be
representable as a small-$\alpha$ expansion of the form
\renewcommand{\theequation}{2.102}
\begin{equation}
\label{symbesc}
\sum\limits_{m\,=\,0}^{\infty}\,\E^{-K_{m}\,(\alpha^{B_{1}}\,\E^{\frac{s_{1}}{\alpha}})}\,
(\alpha^{B_{1}}\,\E^{\frac{s_{1}}{\alpha}})^{L_{m}}\,R_{m}\,(\alpha)\,\,,
\end{equation}
where $R_{m}\,(\alpha)$ is some resurgent symbol in $\alpha$.

\qquad
In the physics context, clarity may be helped by rewriting the symbol
(\ref{posymb}) as a double expansion in powers of $\Lambda_{0}$ and
$\alpha$ as in (\ref{opela}), i. e. by undoing the series multiplication
of (\ref{urser}). Call $d_{\Gamma}$ the mass dimension, and $Q^{2}$ some
typical, Euclidean squared momentum of the vertex function $\Gamma$
under consideration; all other invariant momentum arguments of $\Gamma$
may then be replaced by their dimensionless ratios, $x_{i}$, with
$Q^{2}$. The heuristic argument just given may then be summed up in the
small-$\alpha$ expansion (we mostly omit dependences on the
uninteresting set $\{x_{i}\}$ for leaner writing):
\renewcommand{\theequation}{2.103}
\begin{align}
\label{hiera}
\Gamma\,(Q^{2},\,\{x_{i}\};\,\mu^{2};\,\alpha) &=
(Q^{2})^{\frac{1}{2}\,d_{\Gamma}}\,\Biggl\{\sum\limits_{p\,=\,0}^{\infty}\,a_{0,\,0,\,p}\,
\left(\frac{Q^{2}}{\mu^{2}}\right)\,\alpha^{p}\nonumber\\
&+
\sum\limits_{n\,=\,1}^{\infty}\,\left(\frac{\Lambda^{2}_{0}}{Q^{2}}\right)^{n}\,\left[
\sum\limits_{p\,=\,0}^{\infty}\,a_{0,\,n,\,p}\,\left(
\frac{Q^{2}}{\mu^{2}}\right)\,\alpha^{p}\right]\nonumber\\
&+
\sum\limits_{m\,=\,1}^{\infty}\,\E^{-K_{m}\,\left(\frac{Q^{2}}{\Lambda_{0}^{2}}\right)}\,
\left[\sum\limits_{n\,=\,1}^{\infty}\sum\limits_{p\,=\,1}^{\infty}\,
a_{m,\,n,\,p}\,\left(\frac{Q^{2}}{\mu^{2}}\right)\,
\left(\frac{\Lambda_{0}^{2}}{Q^{2}}\right)^{n}\,\alpha^{p}\right]\nonumber\\
&+ .................................\,\Biggr\}\,\,.
\end{align}
The new escaping terms are characterized by a {\em Gaussian} decay at
large momentum, with $\Lambda_{0}$ providing the decay length. In view
of their exotic coupling dependence displayed in eq. (\ref{symbesc}),
they are probably outside the range of what can usefully be treated by a
Borel-Laplace process with respect to $\alpha$.

\qquad
We should now pause and reflect on the fact that here, at a point where
we have hardly done more about resurgence theory than quote a few
general concepts and basic formulas, we have already been touched by its
great organizing power: where in the matter of coupling dependence we
originally expected a horrible mess, we now see an orderly {\em
hierarchy of small-$\alpha$ suppressions} emerging, with progressively
stronger suppression at small coupling accompanied by progressively
faster decrease at large momenta. This insight remains valuable even
while we have no complete certainty about the presence or absence of the
third line (and possibly fourth and further lines) in eq. (\ref{hiera}).
If we find ways of treating the dynamical effects leading to the second
line of (\ref{hiera}), i. e. of constructing vertex functions whose
small-$\alpha$ expansions are resurgent symbols rather than perturbation
series, we will have made definite progress and omitted at most terms
with a definite, rather exotic large-momentum behavior. Moreover, we
will have gained a framework for adequately treating an impressive array
of nontrivial dynamical problems. Take nonperturbative mass generation
and dynamical symmetry breaking: the conversion of a massless
perturbative into a massive nonperturbative propagator, described
schematically by
\renewcommand{\theequation}{2.104}
\begin{equation}
\label{massi}
k^{2}\,\to\,k^{2}\,+\,c\,\Lambda_{0}^{2}\,+\,...\,=\,
k^{2}\,\left[1\,+\,c\,\left(\frac{\Lambda_{0}^{2}}{k^{2}}\right)\,+\,...\,\right]
\end{equation}
involves power, not exponential, corrections in
$\Lambda_{0}^{2}\,/\,k^{2}$. The breaking of dilatation symmetries by
renormalization involves trace anomalies that are powers, not
exponentials, of $\Lambda_{0}^{2}$.

\qquad
In my opinion, the view is even brighter, since everything in present
knowledge favors the assumption that for the strictly renormalizable and
asymptotically free theories, eq. (\ref{hiera}) is complete after its
second line, just as for the superrenormalizable theories it was
complete after the first line. (Here "complete" refers to unique
recoverability from a small-$\alpha$ expansion.) Let me list a few
arguments (there may be more) in favor of this assumption.
\begin{itemize}
\item[$\bullet$]
A purely qualitative but nevertheless strong argument is that the
completeness postulate underlying the OPE is very probably correct. In a
theory based on a finite number of local fields as basic degrees of
freedom, it is difficult to see what basic operators should be available
for expanding products near a point other than the set of all local
composites at that point. Therefore the OPE, very probably, does {\em
not} miss further escaping terms as in the third line of eq.
(\ref{hiera}), and represents a {\em summable} resurgent-symbol
expansion.
\item[$\bullet$]
Although the question of convergence for the OPE is not settled,
the $n$ summation at fixed $p$ in the OPE, eq. (\ref{xsum}), in contrast
to the perturbative $p$ sum at fixed $n$, is very probably {\em not}
semiconvergent but has a finite radius of convergence, so that in fact
it leaves no room for escaping terms of type (\ref{xescap}). This is
because the growth of the coefficients $a_{n,\,p}$ with $n$ is
controlled by the growth in the number $l\,(n)$ of local operators with
mass dimension $n$ and vacuum quantum numbers, and this number, in
contrast to the factorial growth of the number of Feynman diagrams in
the $p$ direction, increases only exponentially. Consider a theory with
$B$
bosonic fields $\phi_{b}\,(x)\,(b\,=\,1\,...\,B)$ of unit mass dimension
in $D$ spacetime dimensions, and regard a local operator product of mass
dimension $n$ as an array of $n$ slots. Each slot can be filled, either
with one of the $\phi_{b}$, or (apart from the rightmost slot) by one of
the $D$ spacetime derivatives $\partial^{\mu}$. (We adopt the convention
of having each $\partial^{\mu}$ acting only on the next $\phi$ field to
the right, since the action of a $\partial^{\mu}$ on two or more fields
gives terms like
\renewcommand{\theequation}{2.105}
\begin{equation}
\partial^{\mu}\,(\phi_{a}\,\phi_{b})\,=\,
(\partial^{\mu}\,\phi_{a})\,\phi_{b}\,+\,\phi_{a}\,(\partial^{\mu}\,\phi_{b})
\end{equation}
and therefore nothing new). Thus the number of local composites of mass
dimension $n$ is
\renewcommand{\theequation}{2.106}
\begin{equation}
\label{nnum}
B\,\cdot\,(B\,+\,D)^{n\,-\,1}\,\,.
\end{equation}
The number we called $l\,(n)$ above is distinctly lower, since only the
operators contracted to vacuum quantum numbers will give nonzero vacuum
condensates, but in any case we have the exponential estimate
\renewcommand{\theequation}{2.107}
\begin{equation}
\label{llim}
l\,(n)\,<\,(B\,+\,D)^{n}\,\,.
\end{equation}
It is true that coincidence of the $n^{'}\,\phi$-field factors, where
$n^{'}\,\leq\,n$, at the same spacetime point $x$ leads in momentum
space to $n^{'}$ loop integrations in a vacuum condensate, but this
increase in the number of loops is already present in the perturbation
sum in the $p$ direction, where it is known to contribute only to the
power-and-exponential growth in eq. (\ref{largep}), not to the factorial
one (the $p$-loop integrals can usually be bounded from above by a
$p\,-\,th$ power of some one-loop integral). Thus
\renewcommand{\theequation}{2.108}
\begin{equation}
a_{n,\,p}\,\to\,({\rm
const.})_{p}\,n^{V}\,u^{n}\,\left[1\,+\,{\cal{O}}\,
\left(\frac{1}{n}\right)\right]\qquad (n\,\to\,\infty)
\end{equation}
is the likely behavior. (Well, yes, as I hear some initiates among you
saying, there are still the "renormalon" type sequences of $p$-bubble
diagrams that we will also meet in sect. 2.4 and that generate their own
$p\,!$ growth in the $p$ direction in the manner exemplified by eq.
(\ref{xpint}) below -- but remember we are dealing here with behavior
{\em at fixed} $p$. In the $n$ direction, what we have diagrammatically
is growth in the number of lines entering and leaving a single composite
vertex, at constant number of interaction vertices in the remainder
diagram, and this does not generate bubble chains). Thus the series of
(\ref{xsum}) is likely to have the finite convergence circle (in $X$) of
a generalized geometric series, such as (\ref{geoser}), and the question
of escaping terms (\ref{xescap}) does not arise.
\item[$\bullet$]
Dimensionful observables in lattice-QFT calculations, in the
removal-of-regulator limit, scale like a power of $\Lambda_{0}$ as
functions of the coupling, but not like
\renewcommand{\theequation}{2.109}
\begin{equation}
({\rm power\ of}\,\,\Lambda_{0})\,\times\,\left\{
c_{0}\,+\,c_{1}\,\E^{-\frac{{\rm
const.}_{1}}{\Lambda_{0}^{2}}}\,+\,...\,\right\}
\end{equation}
(I am, of course, simplifying a little: lattice calculations are
parametrized in terms of a bare coupling $g_{0}\,(a)$ at lattice spacing
$a$ rather than a renormalized one, but the dependence  of
$(\Lambda_{0})_{lattice}\,\cdot\,a$ on
$g_{0}$ is still the same as for our $\Lambda_{}\,/\,\mu$
in eq. (\ref{lambnull}) on $g$). Indeed this behavior is routinely used
by lattice theorists as an indicator that they are sufficiently close to
the continuum limit. This is a special case of the general statement
that momentum or coupling dependences as in the third line of
(\ref{hiera}) have nowhere been encountered in any of the
nonperturbative approaches applied so far to QFT.

\qquad
There are therefore sound reasons for believing that strictly
renormalizable and asymptotically free field theories will be about
summable resurgent symbols, just as the superrenormalizable theories are
about summable perturbation series. The methods described in Chapter 3
are tuned to this exceptation.
\end{itemize}

\subsection{A Resurgent Model Amplitude}

\qquad
Several features of amplitudes having resurgent coupling dependence
through the lambda scale, such as their peculiar singularity structure
with respect to the coupling, and the tricky things that can happen when
such an amplitude is approached through decomposition into its
semiconvergent perturbation series and a nonperturbative remainder, are
well illustrated by a simple model function that can be analyzed
completely in both the $\alpha$ and Borel planes. This model depends,
other than on the coupling $\alpha$, on a single scalar Euclidean
momentum
variable $k^{2}$, as does a two-point function in QFT, and explicitly on
the sliding scale $\mu$, but since it is taken to be dimensionless, it
can depend on the latter two arguments only through the dimensionless
ratio $k^{2}\,/\,\mu^{2}$. It is given by the spectral representation
\renewcommand{\theequation}{2.110}
\begin{equation}
\label{model}
\Gamma\,\left(\frac{k^{2}}{\mu^{2}};\,\alpha\right)\,=\,1\,-\,
\int\limits_{0}^{\infty}\,d\,(q^{2})\,\frac{\rho\,\left(
\frac{q^{2}}{\mu^{2}};\,\alpha\right)}{k^{2}\,+\,q^{2}}
\end{equation}
with the deceptively simple spectral function,
\renewcommand{\theequation}{2.111}
\begin{equation}
\label{specfun}
\rho\,\left(\frac{q^{2}}{\mu^{2}};\,\alpha\right)\,=\,
\Theta\,(q^{2}\,-\,\Lambda^{2}\,(\mu,\,\alpha))\,
\frac{\alpha}{1\,+\,\alpha\,\frac{\eta}{2}\,\ln\,
\left(\frac{q^{2}}{\mu^{2}}\right)}\,\,,
\end{equation}
where as a caricature of a QFT lambda scale we adopt
\renewcommand{\theequation}{2.112}
\begin{equation}
\label{lambmod}
\Lambda^{2}\,(\mu,\,\alpha)\,=\,\mu^{2}\,\E^{-\frac{1}{\alpha}}\quad
\Leftrightarrow\quad \alpha\,(\mu)\,=\,\left[
\ln\,\left(\frac{\mu^{2}}{\Lambda^{2}}\right)\right]^{-1}\,\,.
\end{equation}
The first, "nonperturbative" factor in (\ref{specfun}), which simply
limits the spectral integration to $q^{2}\,\geq\,\Lambda^{2}$,
schematically
represents the assumption that the intermediate states contributing to
$\Gamma$ have through nonperturbative mass generation acquired invariant
masses of at least $\Lambda$. The second, "perturbative" factor
mimicks a typical resummation of a perturbative subseries; it
displays the logarithmic
$q^{2}$ dependence typical of perturbative
amplitudes, and contains a positive constant $\eta$ characteristic of
its (imagined) building-block diagram. We will always restrict $\eta$ to
\renewcommand{\theequation}{2.113}
\begin{equation}
\label{rangeta}
0\,<\,\eta\,<\,2\,\,,
\end{equation}
since at $\eta\,=\,2$ the amplitude would cease to exist, even in its
renormalized form of eq. (\ref{subtrac}) below, due to a divergence of
the integral at the lower limit $q^{2}\,=\,\Lambda^{2}$. (This is merely
a consequence of the fact that in order to minimize kinematical
complications, we have avoided building correct threshold behavior into
the spectral weight (\ref{specfun}).
If we were more fussy about threshold behavior, we would include in the
$\rho$ of eq. (\ref{specfun}) an extra factor
\renewcommand{\theequation}{2.114}
\begin{equation}
\label{thresh}
\frac{\sqrt{q^{2}\,(q^{2}\,-\,\Lambda^{2})}}{q^{2}\,+\,c\,\Lambda^{2}}
\end{equation}
that would preserve integrability at $q^{2}\,=\,\Lambda^{2}$ even for
$\eta\,=\,2$, but this would also burden us with an extra parameter, and
would deprive the model of the useful generating-function property
mentioned after eq. (\ref{modest}) below, thus rendering the whole
exercise much less transparent and instructive).

\qquad
Since (\ref{specfun}) decreases too slowly at large $q^{2}$ to make the
spectral integral convergence at the upper limit, we apply a caricature
of renormalization, rewriting the amplitude as
\renewcommand{\theequation}{2.115}
\begin{equation}
\label{subtrac}
\Gamma\,\left(\frac{k^{2}}{\mu^{2}};\,\alpha\right)\,=\,\Gamma\,(1;\,\alpha)\,+\,
\alpha\,\int\limits_{\mu^{2}}^{k^{2}}\,\frac{d\,(p^{2})}{p^{2}}\,F\,
\left(\frac{p^{2}}{\mu^{2}};\,\alpha\right)
\end{equation}
in terms of the derivative function
\renewcommand{\theequation}{2.116}
\begin{equation}
\label{efamp}
F\,\left(\frac{k^{2}}{\mu^{2}};\,\alpha\right)\,=\,
\frac{k^{2}}{\alpha}\,\frac{\partial}{\partial\,(k^{2})}\,\Gamma\,
\left(\frac{k^{2}}{\mu^{2}};\,\alpha\right)\,=\,\frac{k^{2}}{\alpha}\,
\int\limits_{0}^{\infty}\,d\,(q^{2})\,
\frac{\rho\,\left(\frac{q^{2}}{\mu^{2}};\,\alpha\right)}{(k^{2}\,+\,q^{2})^{2}}\,\,,
\end{equation}
while absorbing the ultraviolet divergence in the constant
\renewcommand{\theequation}{2.117}
\begin{equation}
\label{divconst}
\Gamma\,(1;\,\alpha)\,=\,1\,-\,\int\limits_{0}^{\infty}\,d\,(q^{2})\,\frac{
\rho\,\left(\frac{q^{2}}{\mu^{2}};\,\alpha\right)}{q^{2}\,+\,\mu^{2}}\,\,.
\end{equation}
It is the finite amplitude $F$ of eq. (\ref{efamp}), rather than the
original quantity of eq. (\ref{subtrac}) which serves more or less as a
pretext, that will be the main object of our study.

We first take a look at the {\em singularity structure in $\alpha$} of
(\ref{model}) or (\ref{efamp}) with the spectral weight (\ref{specfun}).
From the "perturbative" factor of the latter, the amplitude receives a
$k^{2}$-independent singularity at
\renewcommand{\theequation}{2.118}
\begin{equation}
\label{pertsing}
1\,+\,\alpha\,\frac{\eta}{2}\,\ln\,\left(\frac{q^{2}}{\mu^{2}}\right)\,=\,0\,\,,
\end{equation}
while from the spectral denominator we have a $k^{2}$-dependent
singularity at
\renewcommand{\theequation}{2.119}
\begin{equation}
\label{nonpsing}
q^{2}\,+\,k^{2}\,=\,0\,\,,
\end{equation}
which when varying $k^{2}$ produces a branch cut in the $k^{2}$ plane at
"Minkowskian" values $k^{2}\,=\,-\Lambda^{2}\,...\,-\infty$, the
standard pattern in QFT two-point functions. Here, however, we are after
the $\alpha$ plane: in both conditions $q^{2}$ ranges from $\Lambda^{2}$
to $+\infty$ so we may parameterize
\renewcommand{\theequation}{2.120}
\begin{equation}
\label{tparam}
q^{2}\,=\,\Lambda^{2}\,\E^{t}\,=\,\mu^{2}\,\E^{\left(t\,-\,\frac{1}{\alpha}\right)}\,\,,
\qquad t\,=\,0\,...\,\infty\,\,.
\end{equation}
Then since $\alpha\,\ln\,(q^{2}\,/\,\mu^{2})\,=\,\alpha\,t\,-\,1$,
condition (\ref{pertsing}) gives a "perturbative" coupling-plane
singularity line at
\renewcommand{\theequation}{2.121}
\begin{equation}
\alpha\,=\,-\frac{\left(\frac{2}{\eta}\right)\,-\,1}{t}\,\,,\qquad
t\,=\,0\,...\,\infty\,\,,
\end{equation}
i. e., a branch cut on the negative real axis at
\renewcommand{\theequation}{2.122}
\begin{equation}
\label{pertcut}
\alpha\,=\,-\infty\,...\,0\,\,.
\end{equation}
If this were the only singularity with respect to the coupling, we would
again
be in the enviable situation of having Phragmén-Lindelöf analyticity in
a sector of infinite radius and opening angle $2\,(\pi\,-\,\delta)$ with
infinitesimal $\delta$. The physics behind this kind of branch cut --
which, incidentally, is also inherent in Dyson´s QED argument
\cite{DYS5} -- is
generally that the QFT´s underlying an amplitude such as (\ref{model})
become unstable at real negative $\alpha$ and produce complex,
two-valued Euclidean correlation functions, whose imaginary parts are
related to the decay width of the unstable vacuum state.

\qquad
The spoiler of this simple picture is the singularity arising from
(\ref{nonpsing}), which upon using (\ref{tparam}) reads
\renewcommand{\theequation}{2.123}
\begin{equation}
\E^{\left(t\,-\,\frac{1}{\alpha}\right)}\,=\,-
\frac{k^{2}}{\mu^{2}}
\end{equation}
or, since $k^{2}$ is Euclidean and therefore positive,
\renewcommand{\theequation}{2.124}
\begin{equation}
\label{neglog}
t\,-\,\frac{1}{\alpha}\,-\,\ln\,\left(\frac{k^{2}}{\mu^{2}}\right)\,=\,\ln\,(-1)\,=\,\pm\,i\,\pi\,(2\,n\,+\,1)\,\,;\qquad
n\,=\,0,\,1,\,2,\,...
\end{equation}
At a given Euclidean $k^{2}$, we therefore have a whole infinite
sequence of singularity curves in the $\alpha$ plane,
\renewcommand{\theequation}{2.125}
\begin{equation}
\label{nonpcut}
\alpha\,(t)\,=\,\frac{1}{\left(t\,-\,\ln\,\left(\frac{k^{2}}{\mu^{2}}\right)\,\mp\,
i\,\pi\,(2\,n\,+\,1)\right)}\,\,;\qquad
t\,=\,0\,...\,\infty\,\,;\qquad n\,=\,0,\,1,\,2,\,...
\end{equation}
By taking the real and imaginary parts of this equation and eliminating
the parameter $t$ between the two, one finds
\renewcommand{\theequation}{2.126}
\begin{equation}
\label{circular}
(Re\,\alpha)^{2}\,+\,(Im\,\alpha\,\mp\,r_{n})^{2}\,=\,r_{n}^{2}\,\,,
\end{equation}
i. e., the equations of two sequences of nested circular arcs in the
$\alpha$ plane, with radii
\renewcommand{\theequation}{2.127}
\begin{equation}
\label{radii}
r_{n}\,=\,\frac{1}{2\,(2\,n\,+\,1)\,\pi}\,=\,
\frac{1}{2\,\pi},\,\frac{1}{6\,\pi},\,\frac{1}{10\,\pi},\,...\,\,,
\end{equation}
and tangent to the real axis from above and below at $\alpha\,=\,0+$
(see Fig. 2.3). The $k^{2}$ dependence is in the lengths of these arcs,
which are more or less than those of semicircles according to whether
$k^{2}$ is more or less than $\mu^{2}$. (The situation depicted in Fig.
2.3 corresponds to the first, that of Fig. 1.1 (c) to the second case).

\begin{center}
\epsfig{file=fig22.eps,width=12.5cm}\\
$\begin{array}{ll}
{\bf Fig. 2.3}\quad &
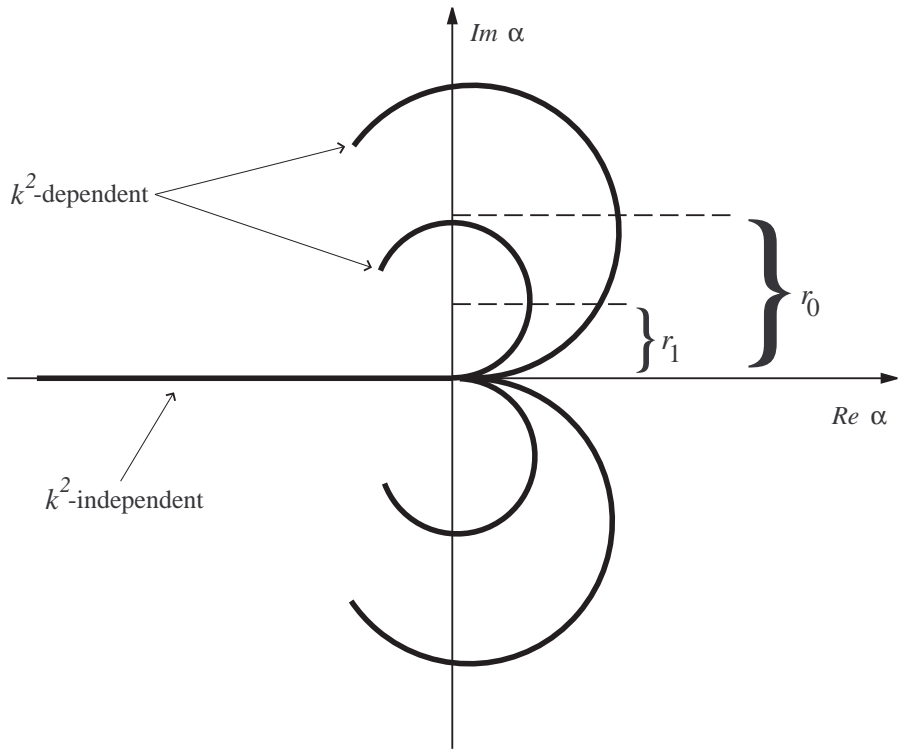
\text{Singularity lines of the model amplitude in the
complex $\alpha$ plane}\\
& \text{at Euclidean $k^{2}$. In addition to the
$k^{2}$-independent branch line}\\
& \text{on the real negative $\alpha$ axis,
only the first two pairs from the}\\
& \text{infinite sequence of nested,
$k^{2}$-dependent branch lines are}\\
& \text{shown.}
\end{array}$
\end{center}

\qquad
Obviously, this "nonperturbative" kind of singularity is very general:
it arises solely from the fact that the scale of the spectral
integration, or of the $k^{2}$ branch cut in the Minkowskian, is set by
a threshold proportional to the strongly nonperturbative $\Lambda^{2}$.
This is why t´Hooft, in discovering this peculiar family of
singularities \cite{THO8}, did not even need an explicit amplitude
function. It is clear that this set of singularities precludes not only
the use of theorem (\ref{unique}) to ascertain uniqueness of
reconstruction from the perturbative expansion, but also any form of
sectorial analyticity -- the "horn-shaped" region of analyticity
defined by the two largest arcs allows for no finite sectorial opening
angle along the positive real axis. Although we have emphasized that the
conditions of theorem (\ref{unique}) are sufficient but not neccessary
conditions, this has generally been taken as a plausibility argument
against the possibility of unique reconstruction, and the t´Hooft
singularities ever since have hovered as a kind of vague threat over the
entire subject of coupling dependence.
This still should not
deter us from proceeding with the analysis of (\ref{efamp}), since after
all at physical $\alpha$, which are real and positive, that amplitude is
{\em manifestly} well-defined and single-valued, and since we know by
now that there is more to physics than the perturbative expansion.

\qquad
To examine the perturbation series of the amplitude (\ref{efamp}), it is
convenient to use
\renewcommand{\theequation}{2.128}
\begin{equation}
\label{xlavar}
x\,=\,\frac{1}{2}\,\ln\,\left(\frac{q^{2}}{\mu^{2}}\right)\,\,,\qquad
\lambda\,(k^{2})\,=\,\frac{1}{2}\,\ln\,\left(\frac{k^{2}}{\mu^{2}}\right)
\end{equation}
as integration and external variables respectively. Then
\renewcommand{\theequation}{2.129}
\begin{equation}
\label{fxint}
F\,\left(\frac{k^{2}}{\mu^{2}};\,\alpha\right)\,=\,
\frac{1}{2}\,\int\limits_{-\frac{1}{2\,\alpha}}^{\infty}\,d\,x\,
\frac{1}{\cosh^{2}\,(x\,-\,\lambda)}\,
\frac{1}{1\,+\,\alpha\,\eta\,x}\,\,.
\end{equation}
Look first at the Taylor coefficients of $F$ with respect to $\alpha$ at
$\alpha\,=\,0+$: you quickly realize that the nonperturbative $\alpha$
dependence through the lower integration limit does not contribute here.
For example,
\renewcommand{\theequation}{2.130}
\begin{align}
\label{fprime}
\frac{\partial\,F}{\partial\,\alpha} &=
\frac{1}{2}\,\int\limits_{-\frac{1}{2\,\alpha}}^{\infty}\,d\,x\,
\frac{1}{\cosh^{2}\,(x\,-\,\lambda)}\,\left[\frac{\partial}{\partial\,\alpha}\,\left(
\frac{1}{1\,+\,\alpha\,\eta\,x}\right)\right]\nonumber\\
&-
\frac{1}{2}\,\left[\frac{d}{d\,\alpha}\,\left(-\frac{1}{2\,\alpha}\right)\right]\,\left\{
\frac{1}{\cosh^{2}\,(x\,-\,\lambda)}\,\frac{1}{1\,+\,\alpha\,\eta\,x}\right\}_{x\,=\,-\frac{1}{2\,\alpha}}\,\,.
\end{align}
As $\alpha\,\to\,0+$, the second term vanishes exponentially, and so do
all terms in the higher derivatives
$\partial^{n}\,F\,/\,\partial\,\alpha^{n}$ except those in which all $n$
derivations act only on the "perturbative" factor of the integrand:
the lower limit in (\ref{fxint}) only creates the exponentially
suppressed, and therefore "escaping", $\alpha$ dependence we described
in (\ref{exsupp}). Thus the perturbation series is identical with the
formal expansion around $\alpha\,=\,0$ of the integral
\renewcommand{\theequation}{2.131}
\begin{equation}
\label{fpert}
F^{pert}\,\left(\frac{k^{2}}{\mu^{2}};\,\alpha\right)\,=\,
\frac{1}{2}\,\int\limits_{-\infty}^{+\infty}\,d\,x
\,\frac{1}{\cosh^{2}\,(x\,-\,\lambda)}\,\frac{1}{1\,+\,\alpha\,\eta\,x}\,\,.
\end{equation}
Accordingly, we would call the remainder integral
\renewcommand{\theequation}{2.132}
\begin{equation}
\label{fnonp}
F^{nonp}\,\left(\frac{k^{2}}{\mu^{2}};\,\alpha\right)\,=\,
-\frac{1}{2}\,\int\limits_{-\infty}^{-\frac{1}{2\,\alpha}}\,d\,x
\,\frac{1}{\cosh^{2}\,(x\,-\,\lambda)}\,
\frac{1}{1\,+\,\alpha\,\eta\,x}
\end{equation}
the nonperturbative part, so that
\renewcommand{\theequation}{2.133}
\begin{equation}
F\,\left(\frac{k^{2}}{\mu^{2}};\,\alpha\right)\,=\,F^{pert}\,+\,F^{nonp}\,\,.
\end{equation}
But something weird has happened here: the prescription to let
$\alpha\,\to\,0+$ in the Taylor coefficients -- that is, the very {\em
definition} of perturbation theory -- has forced us to take the lower
integration limit in (\ref{fpert}) to minus infinity, so that now we not
only integrate beyond the circle of convergence,
$|x|\,<\,\frac{1}{\eta\,\alpha}$, of the perturbative integrand --
a sin for which we expect punishment in the form of a semiconvergent
perturbation series --, but inevitably also over its pole at
$x\,=\,-1\,/\,\eta\,\alpha$, thus creating a branch cut and the
associated multi-valuedness in $\alpha$ on the {\em entire} real
$\alpha$ axis in the $F^{pert}$ and the {\em positive} $\alpha$ axis in
the
$F^{nonp}$
portion. (On the other hand it is obvious that in $F^{pert}$, t´Hoofts
"circular" $\alpha$-plane singularities of Fig. 2.2 have disappeared:
with $\Lambda^{2}$ no more setting the scale of the $q^{2}$ integration
as in (\ref{tparam}), condition (\ref{nonpsing}) no more translates into
a singularity condition in the $\alpha$-plane.) Clearly this problem is
entirely spurious, since in the full amplitude (\ref{fxint}) at physical
positive $\alpha$, the integration starts only at
$x\,=\,-\frac{1}{2\,\alpha}$ and never touches that pole. Thus if we
want to interpret (\ref{fpert}) as the sum of perturbation theory, we
will
have to specify a prescription of how to approach the spurious branch
cut, and be careful to use the same prescription in the nonperturbative
piece (\ref{fnonp}) so the two nuisances can ultimately cancel. Looking
back at (\ref{xlavar}), we see that $x\,=\,-\infty$ corresponds to
having $q^{2}\,=\,0$, rather than the "true" $q^{2}\,=\,\Lambda^{2}$, as
the lower limit in the spectral integral of (\ref{model}) or
(\ref{efamp}): perturbation theory has created for us a {\em spurious
infrared problem} by forcing us to integrate over massless intermediate
states that are plainly absent in the full theory. Again it is
immediately clear that this problem is very general, and of significance
far beyond the present simple model: the full theory has the spontaneous
$\Lambda$ scale in almost all of its thresholds (the only exception
being electroweak amplitudes involving the massless photon), but
perturbation theory, by "losing" that scale, gets us involved in false
infrared problems that will manifest themselves in false nonsummability.

\qquad
For the moment, we may ignore this problem, since we are only interested
in the {\em formal} power-series expansion of (\ref{fpert}),
\renewcommand{\theequation}{2.134}
\begin{equation}
\label{fpser}
F^{pert}_{f}\,\left(\frac{k^{2}}{\mu^{2}};\,\alpha\right)\,=\,
\sum\limits_{p\,=\,0}^{\infty}\,\alpha^{p}\,F^{(p)\,pert}\,(\lambda)\,\,,
\end{equation}
whose individual Taylor coefficients are well defined:
\renewcommand{\theequation}{2.135}
\begin{equation}
\label{fpcint}
F^{(p)\,pert}\,(\lambda)\,=\,\frac{1}{2}\,(-\eta)^{p}\,\int\limits_{-\infty}^{+\infty}\,
d\,x^{'}\,\frac{(x^{'}\,+\,\lambda)^{p}}{\cosh^{2}\,x^{'}}\,\,.
\end{equation}
Here we used $x^{'}\,=\,x\,-\,\lambda$. The evaluation is facilitated by
the observation that $F^{pert}$ possesses a scaling property of the
renormalization-group type: by writing in (\ref{fpert})
\renewcommand{\theequation}{2.136}
\begin{equation}
\frac{1}{1\,+\,\alpha\,\eta\,(x^{'}\,+\,\lambda)}\,=\,
\left[\frac{\left(\frac{\alpha}{1\,+\,\alpha\,\eta\,\lambda}\right)}{\alpha}\right]\,
\frac{1}{1\,+\,\left(\frac{\alpha}{1\,+\,\alpha\,\eta\,\lambda}\right)\,\eta\,x^{'}}\,\,,
\end{equation}
one sees that apart from the square-bracketed prefactor, $F^{pert}$ at
general $k^{2}$ looks like the one at $k^{2}\,=\,\mu^{2}$, i. e.
at $\lambda\,=\,0$, but with a $k^{2}$-dependent effective coupling:
\renewcommand{\theequation}{2.137}
\begin{equation}
\label{fpscal}
F^{pert}\,\left(\frac{k^{2}}{\mu^{2}};\,\alpha\right)\,=\,
\left[\frac{\overline{\overline{\alpha}}\,(k^{2})}{\alpha}\right]\,F^{pert}\,(1;\,
\overline{\overline{\alpha}}\,(k^{2}))\,\,;
\end{equation}
\renewcommand{\theequation}{2.138}
\begin{equation}
\label{bbcoup}
\overline{\overline{\alpha}}\,(k^{2})\,=\,\frac{\alpha}{1\,+\,\alpha\,\frac{\eta}{2}\,\ln\,
\left(\frac{k^{2}}{\mu^{2}}\right)}\,\,.
\end{equation}
It is therefore sufficient to evaluate the coefficients (\ref{fpcint})
at $\lambda\,=\,0$. This can be done with the help of the formula
\cite{GRY}
\renewcommand{\theequation}{2.139}
\begin{equation}
\label{xpint}
\int\limits_{0}^{\infty}\,d\,x\,\frac{x^{p}}{\cosh^{2}\,x}\,=\,
\frac{1}{2^{p\,-\,1}}\,\left(1\,-\,\frac{1}{2^{p\,-\,1}}\right)\,\zeta\,(p)\,\Gamma\,(p\,+\,1)\qquad
(p\,\not=\,1)
\end{equation}
and of the symmetry properties of the integrand, giving
\renewcommand{\theequation}{2.140}
\begin{equation}
\label{fpnull}
F^{(p)\,pert}\,(\lambda\,=\,0)\,=\,[(-)^{p}\,+\,1]\,
\left(1\,-\,\frac{1}{2^{p\,-\,1}}\right)\,\zeta\,(p)\,\left(\frac{\eta}{2}\right)^{p}\,p!
\end{equation}
Sine $\zeta\,(p)\,\to\,1$ as $p\,\to\,\infty$, this indeed displays the
factorial-growth pattern of (\ref{largep}), with constants
$c_{0}\,=\,1,\,B_{0}\,=\,0$, and $A_{0}\,=\,\frac{2}{\eta}$. Due to the
symmetric range of the $x$ integration, "alternating" and
"monotonic"
contributions are present in equal mixture so only even-$p$
coefficients  are nonzero -- an oversimplified feature of the model.
This feature will disappear anyway when calculating the full expansion
(\ref{fpser}) at $k^{2}\,\not=\,\mu^{2}$ from the scaling relation
(\ref{fpscal}): by expanding the last factor in each term of
\renewcommand{\theequation}{2.141}
\begin{equation}
F^{pert}_{f}\,\left(\frac{k^{2}}{\mu^{2}};\,\alpha\right)\,=\,
\sum\limits_{p\,=\,0}^{\infty}\,F^{(p)\,pert}\,(\lambda\,=\,0)\,\alpha^{p}\,[
1\,+\,\alpha\,(\eta\,\lambda)]^{-(p\,+\,1)}
\end{equation}
in powers of $\alpha$, and collecting powers of $\alpha$, one finds the
series (\ref{fpser}) with
\renewcommand{\theequation}{2.142}
\begin{equation}
\label{fpcoeff}
F^{(p)\,pert}\,(\lambda)\,=\,(-)^{p}\,\left(\frac{\eta}{2}\right)^{p}\,
\left[\sum\limits_{q\,=\,0}^{p}\,f^{(p)}_{q}\,(2\,\lambda)^{q}\right]\,p!\,\,,
\end{equation}
\renewcommand{\theequation}{2.143}
\begin{equation}
\label{pqcoeff}
f^{(p)}_{q}\,=\,[1\,+\,(-)^{p\,-\,q}]\,\frac{1}{q!}\,\left(
1\,-\,\frac{1}{2^{p\,-\,q\,-\,1}}\right)\,\zeta\,(p\,-\,q)\,\,.
\end{equation}
These coefficients, for all the simplicity of the model, again display
typical properties of QFT perturbation series: the $k^{2}$ dependence
through logarithms, which in this case are infrared singular at
$k^{2}\,=\,0$ due to the spurious infrared problem we produced, and the
fact that the $p$ - $th$ term contains a $p$ - $th$ degree
polynomial in these logarithms. The "leading-logarithms" (LL)
subseries,
comprising the terms with $q\,=\,p$, has $f^{(p)}_{p}\,\cdot\,p!\,=\,1$
(use $\zeta\,(0)\,=\,-\frac{1}{2}$), and is therefore, as usual, a
purely geometric series,
\renewcommand{\theequation}{2.144}
\begin{equation}
\label{fpll}
F^{pert}_{(LL)}\,\left(\frac{k^{2}}{\mu^{2}};\,\alpha\right)\,=\,
\sum\limits_{p\,=\,0}^{\infty}\,(-\alpha\,\eta\,\lambda)^{p}\,=\,
\frac{1}{1\,+\,\alpha\,(\eta\,\lambda)}\,\,,
\end{equation}
which is identical with the leading term in
$\overline{\overline{\alpha}}$ (the prefactor) of relation
(\ref{fpscal}) and, moreover, with the complete perturbative factor of
our spectral function (\ref{specfun}) taken at $k^{2}$ instead of
$q^{2}$ -- in LL approximation, that factor seems to sail unchanged
through the folding operation of (\ref{fpert}). It is clear from a look
at (\ref{fpcoeff}), however, that such an approximation barely scratches
the surface of the resummation problem: as mentioned already in chapter
0, the summation of subseries with finite convergence radius usually
evades the difficulty of the $p!$ growth that resides here in the
$q\,=\,0$ (i. e. $\lambda\,=\,0$) terms.

\qquad
For later use we note a different, still {\em formal} representation of
$F^{pert}$ that follows upon using the property
\renewcommand{\theequation}{2.145}
\begin{equation}
\label{zetarel}
\left(1\,-\,\frac{1}{2^{p\,-\,1}}\right)\,\zeta\,(p)\,=\,
\sum\limits_{n\,=\,1}^{\infty}\,\frac{(-)^{n\,-\,1}}{n^{p}}
\end{equation}
of the zeta function \cite{GRY}. In the ensuing, double formal series
\renewcommand{\theequation}{2.146}
\begin{equation}
\label{fpnex}
F_{f}^{pert}\,(1;\,\alpha)\,=\,-\sum\limits_{n\,=\,1}^{\infty}\,(-)^{n}\,
\left\{\sum\limits_{p\,=\,0}^{\infty}\,[(-)^{p}\,+\,1]\,p!\,\left(
\frac{\eta\,\alpha}{2\,n}\right)^{p}\right\}\,\,,
\end{equation}
we would be tempted to identify the inner series as the asymptotic
expansion of a combination of $E\,i$ functions according to
(\ref{eiasy}):
\renewcommand{\theequation}{2.147}
\begin{equation}
\label{fpeiser}
F_{f}^{pert}\,(1;\,\alpha)\,=\,-\frac{1}{\eta\,\alpha}\,
\sum\limits_{n\,=\,1}^{\infty}\,(-)^{n}\,2\,n\,\left[
\E^{-\frac{2\,n}{\eta\,\alpha}}\,E\,i\,\left(\frac{2\,n}{\eta\,\alpha}\right)\,-\,
\E^{\frac{2\,n}{\eta\,\alpha}}\,E\,i\,\left(-\frac{2\,n}{\eta\,\alpha}\right)\right]
\end{equation}
but even this partial resummation would confront us with the necessity
of supplying a prescription for approaching the spurious branch cut,
which at real and positive values of $\alpha$ would reside in the second
$Ei$ function of (\ref{fpeiser}) at each $n$. (The function at
$k^{2}\,/\,\mu^{2}\,\not=\,1$ would then again follow from the scaling
relation (\ref{fpscal}), i. e. by making the substitution
\renewcommand{\theequation}{2.148}
\begin{equation}
\label{bbsub}
\frac{1}{\alpha}\quad \longrightarrow\quad
\frac{1}{\alpha}\,+\,\eta\,\lambda
\end{equation}
and supplying the prefactor). Exploiting a representation of type
(\ref{fpnex}) will therefore be much easier for the Borel transform.

\qquad
{\em Borel-Laplace analysis} of the $F^{pert}$ amplitude begins with the
observation that the right-hand side of the scaling relation
(\ref{fpscal}) is precisely of the form appearing in the 9th line of our
Table 1 (page 30) on the left-hand side, with
$\omega\,=\,-\eta\,\lambda$. Thus
\renewcommand{\theequation}{2.149}
\begin{equation}
\label{bpscal}
b^{pert}\,\left(\frac{k^{2}}{\mu^{2}};\,s\right)\,=\,
\E^{-(\eta\,\lambda)\,s}\,b^{pert}\,(1;\,s)
\end{equation}
is the Borel image of the scaling relation (\ref{fpscal}). The
"Borelized" perturbation expansion at $k^{2}\,/\,\mu^{2}\,=\,1$ has
coefficients (\ref{fpnull}) without the $p$! factor:
\renewcommand{\theequation}{2.150}
\begin{equation}
\label{mbser}
b_{f}^{pert}\,(1;\,s)\,=\,\sum\limits_{p\,=\,0}^{\infty}\,
\left\{[(-)^{p}\,+\,1]\,\left(1\,-\,
\frac{1}{2^{p\,-\,1}}\right)\,\zeta\,(p)\,\left(\frac{\eta}{2}\right)^{p}\right\}\,s^{p}\,\,.
\end{equation}
Its circle of convergence is $|s|\,<\,2\,/\,\eta$. Analytic continuation
beyond the circle is, in this case, possible by summing the series in
closed form. One may, for example, apply relation (\ref{xpint}) in
reverse after having divided it by $p!$\ :
\renewcommand{\theequation}{2.151}
\begin{align}
\label{mbirep}
b^{pert}\,(1;\,s) &= \frac{1}{2}\,\int\limits_{-\infty}^{+\infty}\,
\frac{d\,x}{\cosh^{2}\,x}\,\left[\sum\limits_{p\,=\,0}^{\infty}\,\frac{1}{p!}\,(\eta\,x\,s)^{p}\right]\nonumber\\
&= \frac{1}{2}\,\int\limits_{-\infty}^{+\infty}\,d\,x\,
\frac{\E^{\eta\,s\,x}}{\cosh^{2}\,x}\,=\,\frac{1}{2}\,
\int\limits_{-\infty}^{+\infty}\,d\,x\,
\frac{\cosh\,(\eta\,s\,x)}{\cosh^{2}\,x}\,\,,
\end{align}
where we used the symmetry of the integrand. Use of the integral
formula \cite{GRY}
\renewcommand{\theequation}{2.152}
\begin{equation}
\label{chint}
\int\limits_{0}^{\infty}\,d\,x\,
\frac{\cosh\,(y\,x)}{\cosh^{2}\,x}\,=\,
B\,\left(1\,+\,\frac{y}{2},\,1\,-\,\frac{y}{2}\right)\,=\,
\frac{\Gamma\,\left(1\,+\,\frac{y}{2}\right)\,\Gamma\,\left(1\,-\,\frac{y}{2}\right)}{\Gamma\,(2)}
\end{equation}
and of a well-known property of the gamma function \cite{MOS},
\renewcommand{\theequation}{2.153}
\begin{equation}
\Gamma\,(1\,+\,z)\,\Gamma\,(1\,-\,z)\,=\,\frac{\pi\,z}{\sin\,(\pi\,z)}\,\,,
\end{equation}
gives finally
\renewcommand{\theequation}{2.154}
\begin{equation}
\label{mbp1}
b^{pert}\,(1;\,s)\,=\,\frac{\frac{\pi}{2}\,\eta\,s}{\sin\,\left(
\frac{\pi}{2}\,\eta\,s\right)}
\end{equation}
and, therefore,
\renewcommand{\theequation}{2.155}
\begin{equation}
\label{mbpfull}
b^{pert}\,\left(\frac{k^{2}}{\mu^{2}};\,s\right)\,=\,\left(\frac{\mu^{2}}{k^{2}}\right)^{\frac{\eta}{2}\,s}\,
\frac{\frac{\pi}{2}\,\eta\,s}{\sin\,\left(\frac{\pi}{2}\,\eta\,s\right)}\,\,.
\end{equation}
The consequences of our having spuriously integrated over the
perturbative pole in (\ref{fpert}) are now manifestly on us: this Borel
transform has poles on the real $s$ axis at
\renewcommand{\theequation}{2.156}
\begin{equation}
\label{spoles}
s\,=\,\pm\,\frac{2}{\eta}\,n\,\,,\qquad n\,=\,1,\,2,\,3,\,...\,\,;
\end{equation}
and while those at negative $s$ do not obstruct unique Laplace
reconstruction (\ref{laplace}) of $F^{pert}$ at physical values of
$\alpha$, those at positive $s$ do. The perturbative expansion of $F$ is
not Borel summable. We already pointed out that this property has
at best indirectly to do with the presence of t´Hooft´s set of
circular-arc
singularities in Fig. 2.3, since in $F^{pert}$ these are altogether
absent; instead it arises from our having created, through the very
definition of the perturbative expansion, a spurious branch cut in
$F^{pert}$ on the real positive $\alpha$ axis -- that is, from our
having {\em lost} the $\Lambda^{2}$ threshold that gave rise to the
t´Hooft arcs in the first place.
I am emphasizing this
in order to bring more precision to the vague notion that the
perturbative nonsummability
in, for example, QCD amplitudes is somehow connected to the t´Hooft
singularities.

\qquad
It is useful for later purposes to make the poles of $b^{pert}$ more
explicit, either by looking up the Mittag-Leffler expansion of the
inverse sine in (\ref{mbpfull}), or by using the representation
analogous to (\ref{fpnex}), with $\alpha$ replaced by $s$ and $p!$
omitted, and with its inner series therefore summing to
\renewcommand{\theequation}{2.157}
\begin{equation}
\frac{1}{1\,+\,\left(\frac{\eta\,s}{2\,n}\right)}\,+\,
\frac{1}{1\,-\,\left(\frac{\eta\,s}{2\,n}\right)}\,\,.
\end{equation}
Either route leads to the representation
\renewcommand{\theequation}{2.158}
\begin{equation}
\label{mbpolre}
b^{pert}\,\left(\frac{k^{2}}{\mu^{2}};\,s\right)\,=\,\E^{-(\eta\,\lambda)\,s}\,\frac{2}{\eta}\,
\sum\limits_{n\,=\,1}^{\infty}\,(-)^{n}\,\left[
\frac{-n}{s\,+\,\left(\frac{2}{\eta}\right)\,n}\,+\,
\frac{+n}{s\,-\,\left(\frac{2}{\eta}\right)\,n}\right]\,\,.
\end{equation}
It consists of a meromorphic part,
\renewcommand{\theequation}{2.159}
\begin{equation}
\label{mbmero}
\left[b^{pert}\,\left(\frac{k^{2}}{\mu^{2}};\,s\right)\right]_{poles}\,=\,
\sum\limits_{n\,=\,1}^{\infty}\,\left[
\frac{-\left(\frac{2}{\eta}\right)\,n\,(-\E^{2\,\lambda})^{n}}{s\,+\,
\left(\frac{2}{\eta}\right)\,n}\,+\,\frac{+\left(\frac{2}{\eta}\right)\,n\,(-\E^{-2\,\lambda})^{n}}{s\,-\,
\left(\frac{2}{\eta}\right)\,n}\right]\,\,,
\end{equation}
and an entire remainder function. (Remember
$\E^{2\,\lambda}\,=\,k^{2}\,/\,\mu^{2}$). In trying to give a meaning to
the Laplace integral
\renewcommand{\theequation}{2.160}
\begin{equation}
\label{plapla}
F^{pert}\,\left(\frac{k^{2}}{\mu^{2}};\,\alpha\right)\,=\,\frac{1}{\alpha}\,
\int\limits_{0}^{\infty}\,d\,s\,\E^{-\frac{s}{\alpha}}\,b^{pert}\,\left(
\frac{k^{2}}{\mu^{2}};\,s\right)
\end{equation}
we are therefore forced \cite{ECAL} to specify a path of integration
around {\em each} of the infinite sequence of positive-$s$ poles in the
second term of (\ref{mbmero}): at each of these we may deform the path
of integration along the positive real $s$ axis to run slightly above
($s\,\to\,s\,+\,i\,\varepsilon)$ or slightly below
($s\,\to\,s\,-\,i\,\varepsilon)$ the pole. We thus face an infinitely
multivalued function of $\alpha$, whose branches may be indexed as
\renewcommand{\theequation}{2.161}
\begin{equation}
\label{fpmulti}
F^{pert}_{\{\sigma_{1},\,\sigma_{2},\,\sigma_{3},\,...\}}\,\left(
\frac{k^{2}}{\mu^{2}};\,\alpha\right)\,\,;\qquad
\sigma_{n}\,=\,+1\,\,\text{or}\,\,-1\quad (\text{all}\,\,n)
\end{equation}
and are computed by replacing the second term in (\ref{mbmero}) by
\renewcommand{\theequation}{2.162}
\begin{equation}
\sum\limits_{n\,=\,1}^{\infty}\,\left(-\frac{\mu^{2}}{k^{2}}\right)^{n}\,
\frac{\left(\frac{2}{\eta}\right)\,n}{s\,+\,i\,\varepsilon\,\sigma_{n}\,-\,
\left(\frac{2}{\eta}\right)\,n}\,\,.
\end{equation}
Among these, the two branches where all $\sigma_{n}$´s are +1 or -1 may
be characterized by a single overall shift,
$s\,\to\,s\,+\,i\,\varepsilon$ or $s\,\to\,s\,-\,i\,\varepsilon$, of
the Borel variable, which by virtue of (\ref{tlapla}) may in turn be
transferred to the variable $\alpha$:
\renewcommand{\theequation}{2.163}
\begin{equation}
\label{epsmig}
b\,((t\,\pm\,i\,\varepsilon)\,\alpha)\quad \longrightarrow\quad
b\,(t\,\alpha\,\pm\,i\,\varepsilon^{'})\quad \longrightarrow\quad
b\,(t\,(\alpha\,\pm\,i\,\varepsilon^{''}))\,\,.
\end{equation}
So these two branches meet at the branch cut on the positive real
$\alpha$
axis of what by definition is the {\em physical sheet with respect to
$\alpha$} of the function $F^{pert}$:
\renewcommand{\theequation}{2.164}
\begin{equation}
\label{fpplus}
F^{pert}\,\left(\frac{k^{2}}{\mu^{2}};\,\alpha\,+\,i\,0\right)\,=\,
F^{pert}_{\{+,\,+,\,+,\,...\}}\,\left(\frac{k^{2}}{\mu^{2}};\,\alpha\right)\,\,,
\end{equation}
\renewcommand{\theequation}{2.165}
\begin{equation}
\label{fpminus}
F^{pert}\,\left(\frac{k^{2}}{\mu^{2}};\,\alpha\,-\,i\,0\right)\,=\,
F^{pert}_{\{-,\,-,\,-,\,...\}}\,\left(\frac{k^{2}}{\mu^{2}};\,\alpha\right)\,\,.
\end{equation}

\qquad
With this clarification of branch-point structure, both the perturbative
amplitude $F^{pert}$ in its two main branches (\ref{fpplus}) and
(\ref{fpminus}) and its Borel transform may now be given in well-defined
closed forms involving known transcendental functions. The function
\cite{MOS}
\renewcommand{\theequation}{2.166}
\begin{equation}
\label{gfunc}
G\,(z)\,=\,\psi\,\left(\frac{z\,+\,1}{2}\right)\,-\,\psi\,\left(\frac{z}{2}\right)\,=\,
2\,\sum\limits_{n\,=\,0}^{\infty}\,\frac{(-)^{n}}{n\,+\,z}
\end{equation}
and its property
\renewcommand{\theequation}{2.167}
\begin{equation}
\label{grecur}
G\,(z\,+\,1)\,=\,-G\,(z)\,+\,\frac{2}{z}
\end{equation}
may be used to express
\renewcommand{\theequation}{2.168}
\begin{equation}
\label{mbgfu}
b^{pert}\,\left(\frac{k^{2}}{\mu^{2}};\,s\,\pm\,i\,0\right)\,=\,
\E^{-(\eta\,\lambda)\,s}\,\frac{\eta\,s}{4}\,\left[
G\,\left(\frac{\eta}{2}\,s\right)\,+\,G\,\left(1\,-\,\frac{\eta}{2}\,s\,\mp\,i\,0\right)\right]\,\,,
\end{equation}
where the first term in the square bracket contains the contribution
from "harmless", negative-$s$ poles, while the second term brings in
the undesirable poles at positive $s$. Analogously, the integral
representation (\ref{fpert}), now given a precise meaning as a
representation of the two main branches through the
$\alpha\,\to\,\alpha\,\pm\,i\,0$ prescription, may be written (at
$k^{2}\,=\,\mu^{2})$
\renewcommand{\theequation}{2.169}
\begin{equation}
\label{fpbrs}
F^{pert}\,(1;\,\alpha\,\pm\,i\,0)\,=\,
\frac{1}{2}\,\int\limits_{0}^{\infty}\,
\frac{d\,x}{\cosh^{2}\,x}\,\left[
\frac{1}{1\,+\,\alpha\,\eta\,x}\,+\,
\frac{1}{1\,-\,(\alpha\,\pm\,i\,0)\,\eta\,x}\right]
\end{equation}
and compared to an integral representation of the function
$\psi^{'}\,(z)\,=\,d^{2}\,\ln\,\Gamma\,(z)\,/\,d\,z^{2}$ \cite{MOS}
along its branch line $Re\,z\,=\,\frac{1}{2}$,
\renewcommand{\theequation}{2.170}
\begin{equation}
\label{psiprime}
\frac{i}{\pi\,y}\,\psi^{'}\,\left(
\frac{1}{2}\,\pm\,0\,+\,\frac{i}{\pi\,y}\right)=1+
\frac{1}{2}\,\int\limits_{0}^{\infty}\,
\frac{d\,x}{\cosh^{2}\,x}\,\left[
\frac{1}{1\,+\,y\,x}+\frac{1}{1\,-\,(y\,x\,\pm\,i\,0)}\right]\,,
\end{equation}
to give
\renewcommand{\theequation}{2.171}
\begin{equation}
\label{fpclos}
F^{pert}\,(1;\,\alpha\,\pm\,i\,0)\,=\,
\frac{i}{\pi\,\eta\,\alpha}\,\psi^{'}\,\left(\frac{1}{2}\,\pm\,0\,+\,\frac{i}{\pi\,\eta\,\alpha}\right)\,-\,1\,\,.
\end{equation}
These are evidently unphysical constructs, since an Euclidean amplitude
at real momenta and physical couplings must be real and single-valued.

\qquad
Enough of the perturbative portion -- there is more to life! We turn to
the nonperturbative remainder (\ref{fnonp}), at first treated formally
since again we wish to derive only a formal resurgent-symbol expansion.
In the variable
\renewcommand{\theequation}{2.172}
\begin{equation}
y\,=\,-\left(2\,x\,+\,\frac{1}{\alpha}\right)\,=\,-\left(
\frac{1}{\alpha}\,+\,\ln\,\frac{q^{2}}{\mu^{2}}\right)
\end{equation}
we have
\renewcommand{\theequation}{2.173}
\begin{equation}
\label{fnint}
F^{nonp}\,\left(\frac{k^{2}}{\mu^{2}};\,\alpha\right)\,=\,-\frac{1}{4}\,
\int\limits_{0}^{\infty}\,d\,y\,\frac{1}{
\cosh^{2}\,\left(\frac{1}{2}\,\left(y\,+\,\frac{1}{\alpha}\right)\,+\,\lambda\right)}\,
\frac{1}{\left(1\,-\,\frac{\eta}{2}\right)\,-\,\frac{\eta}{2}\,\alpha\,y}\,\,.
\end{equation}
We expand
\renewcommand{\theequation}{2.174}
\begin{equation}
\label{chexp}
\frac{1}{4\,\cosh^{2}\,\left(\frac{1}{2}\,y\,+\,\left(\frac{1}{2\,\alpha}\,+\,\lambda\right)\right)}\,=\,
\sum\limits_{n\,=\,1}^{\infty}\,(-)^{n\,-\,1}\,\left[
\E^{-\left(\frac{1}{\alpha}\,+\,2\,\lambda\right)}\right]^{n}\,\E^{-n\,y}
\end{equation}
and obtain $F^{nonp}$ in the form (\ref{resymb}) of a resurgent symbol,
\renewcommand{\theequation}{2.175}
\begin{equation}
\label{fnsymb}
F^{nonp}\,\left(\frac{k^{2}}{\mu^{2}};\,\alpha\right)\,=\,
\sum\limits_{n\,=\,1}^{\infty}\,\left(
\frac{\mu^{2}}{k^{2}}\right)^{n}\,\E^{-\frac{n}{\alpha}}\,
f_{n,\,f}\,(\alpha)\,\,,
\end{equation}
with support at the points $s_{n}\,=\,n,\,n\,=\,1,\,2,\,3,\,...\,,$ and
with the formal series resulting from the expansion of
\renewcommand{\theequation}{2.176}
\begin{equation}
\label{fnser}
f_{n}\,(\alpha)\,=\,(-)^{n}\,n\,\int\limits_{0}^{\infty}\,d\,y
\,\frac{\E^{-n\,y}}{\left(1\,-\,\frac{\eta}{2}\right)\,-\,\left(
\frac{\eta}{2}\right)\,\alpha\,y}
\end{equation}
as coefficients. By using $t\,=\,n\,y$ as an integration variable, we
obtain
\renewcommand{\theequation}{2.177}
\begin{equation}
\label{phidila}
f_{n}\,(\alpha)\,=\,(-)^{n\,-\,1}\,f_{1}\,\left(\frac{\alpha}{n}\right)\,\,;\qquad
n\,=\,1,\,2,\,3,\,...\,\,,
\end{equation}
where
\renewcommand{\theequation}{2.178}
\begin{equation}
\label{phi1int}
f_{1}\,(\alpha)\,=\,-\left(\frac{2}{2\,-\,\eta}\right)\,\int\limits_{0}^{\infty}\,d\,t
\,\frac{\E^{-t}}{1\,-\,\left(\frac{\eta}{2\,-\,\eta}\right)\,\alpha\,t}
\end{equation}
has the formal expansion
\renewcommand{\theequation}{2.179}
\begin{equation}
\label{phi1ser}
f_{1,\,f}\,(\alpha)\,=\,-\left(\frac{2}{2\,-\,\eta}\right)\,
\sum\limits_{p\,=\,0}^{\infty}\,\left(
\frac{\eta}{2\,-\,\eta}\right)^{p}\,p!\,\alpha^{p}\,\,.
\end{equation}
The series $f_{n,\,f}$ therefore all exhibit the factorial-divergence
pattern of eq. (\ref{largep}), with constants
\renewcommand{\theequation}{2.180}
\begin{equation}
\label{phi1lap}
c_{n}\,=\,(-)^{n}\,\frac{2}{2\,-\,\eta}\,\,,\quad
B_{0}\,=\,0\,\,,\quad
A_{n}\,=\,n\,\left(\frac{2\,-\,\eta}{\eta}\right)\,\,,
\end{equation}
and, thererfore, also a pattern similar to the perturbative  expansion,
whose coefficients (at $k^{2}\,=\,\mu^{2})$ were given by
(\ref{fpnull}). For higher $n$´s they all follow from the series
$f_{1,\,f}$ by simple dilatations, $\alpha\,\to\,\frac{\alpha}{n}$.
This is a special, simple form of the phenomenon of resurgence described
in sect. 2.3. With a look back at eq. (\ref{eispec}), $f_{n}$ may
alternatively be expressed as
\renewcommand{\theequation}{2.181}
\begin{equation}
\label{fnei}
f_{n}\,(\alpha)\,=\,(-)^{n}\,\frac{2}{\eta}\,\frac{n}{\alpha}\,
\E^{-\frac{(2\,-\,\eta)\,n}{\eta\,\alpha}}\,Ei\,\left(
\frac{(2\,-\,\eta)\,n}{\eta\,\alpha}\right)\,\,.
\end{equation}

\qquad
It is one of the oversimplified and unrealistic features of our model
that at $k^{2}\,\not=\,\mu^{2}$, the $f_{n,\,f}$ series for $n\,\geq\,1$
do not develop the perturbative logarithms,
$2\,\lambda\,=\,\ln\,(k^{2}\,/\,\mu^{2})$, that were present in eq.
(\ref{fpcoeff}) for the perturbative series
\renewcommand{\theequation}{2.182}
\begin{equation}
\label{phinull}
f_{0,\,f}\,(\alpha)\,:=\,F^{pert}_{f}\,\left(\frac{k^{2}}{\mu^{2}};\,\alpha\right)\,\,.
\end{equation}
This is essentially a consequence of the simple factorizing structure of
eq. (\ref{specfun}). The only $k^{2}$ dependence in (\ref{fnsymb}) is
therefore the purely nonperturbative one of "power corrections"
through the appearance of $n$ - $th$ powers of the dimensionless
parameter,
\renewcommand{\theequation}{2.183}
\begin{equation}
\label{kappa}
u\,=\,\E^{-2\,\lambda}\,=\,\left(\frac{\mu^{2}}{k^{2}}\right)\,\,,
\end{equation}
just as in the generic form of eq. (\ref{posymb}). In view of the
lambda-scale definition (\ref{lambmod}), this gives to the complete
resurgent-symbol representation of
$F$,
\renewcommand{\theequation}{2.184}
\begin{equation}
\label{fresym}
F_{f}\,\left(\frac{k^{2}}{\mu^{2}};\,\alpha\right)\,=\,
\sum\limits_{n\,=\,0}^{\infty}\,\left(\frac{\Lambda^{2}}{k^{2}}\right)^{n}\,f_{n,\,f}\,(\lambda;\,\alpha)\,\,,
\end{equation}
the form of an operator-product expansion as in (\ref{opela}).

\qquad
It is easy to find the "Borelized" $f_{n}$ series,
\renewcommand{\theequation}{2.185}
\begin{equation}
\label{mbuser}
b_{n,\,f}\,(s)\,=\,(-)^{n}\,\left(\frac{2}{2\,-\,\eta}\right)\,\sum\limits_{p\,=\,0}^{\infty}\,
\left(\frac{\eta}{2\,-\,\eta}\right)^{p}\,\left(\frac{s}{n}\right)^{p}\qquad
(n\,\geq\,1)\,\,,
\end{equation}
and their analytic continuations through resummation,
\renewcommand{\theequation}{2.186}
\begin{equation}
\label{mbn}
b_{n}\,(s)\,=\,(-)^{n\,-\,1}\,n\,\left(\frac{2}{\eta}\right)\,
\frac{1}{s\,+\,n\,-\,\left(\frac{2}{\eta}\right)\,n}\,\,.
\end{equation}
Each has a pole on the positive real $s$ axis, but the interesting
observation is where these poles are when we form the combined Borel
transform (\ref{bndmin}) of the nonperturbative part (\ref{fnsymb}) by
using the last line of Table 1:
\renewcommand{\theequation}{2.187}
\begin{equation}
\label{mbnfull}
b^{nonp}\,\left(\frac{k^{2}}{\mu^{2}};\,s\right)\,=\,
-\sum\limits_{n\,=\,1}^{\infty}\,\Theta\,(s\,-\,n)\,
\frac{\left(\frac{2}{\eta}\right)\,n\,\left(-\E^{-2\,\lambda}\right)^{n}}
{s\,-\,\left(\frac{2}{\eta}\right)\,n}\,\,.
\end{equation}
The poles, because of (\ref{rangeta}), are at points where the step
functions are unity, so in order to isolate them we write
\renewcommand{\theequation}{2.188}
\begin{equation}
\label{stepfu}
\Theta\,(s\,-\,n)\,=\,1\,-\,\Theta\,(n\,-\,s)
\end{equation}
and have
\renewcommand{\theequation}{2.189}
\begin{equation}
\label{mbnd}
b^{nonp}\,\left(\frac{k^{2}}{\mu^{2}};\,s\right)\,=\,
\sum\limits_{n\,=\,1}^{\infty}\,\Theta\,(n\,-\,s)\,
\frac{n\,\left(-\E^{-2\,\lambda}\right)^{n}}{\left(\frac{\eta}{2}\right)\,s\,-\,n}\,-\,
\sum\limits_{n\,=\,1}^{\infty}\,\frac{n\,\left(-\E^{-2\,\lambda}\right)^{n}}
{\left(\frac{\eta}{2}\right)\,s\,-\,n}\,\,.
\end{equation}
What we suspected all along is now manifest: the second term, which sums
up the poles on the positive Borel axis, {\em exactly} cancels the
corresponding second term in the meromorphic part (\ref{mbmero}) of
$b^{pert}$. The full Borel transform
\renewcommand{\theequation}{2.190}
\begin{equation}
\label{mbfull}
b\,\left(\frac{k^{2}}{\mu^{2}};\,s\right)\,=\,b^{pert}\,+\,b^{nonp}
\end{equation}
is only piecewise smooth on the positive $s$ axis due to the first
term in
(\ref{mbnd}), but otherwise perfectly well-behaved in the Laplace
integral, which therefore yields a well-defined function having the
resurgent symbol (\ref{fresym}) as its formal small-$\alpha$ expansion.

\qquad
It only remains for us to check whether the Laplace reconstruction
indeed gives back the original function (\ref{fxint}). We use
(\ref{mbgfu}) and (\ref{mbnd}) to write
\renewcommand{\theequation}{2.191}
\begin{equation}
\label{fullap}
F\,\left(\frac{k^{2}}{\mu^{2}};\,\alpha\right)\,=\,\frac{1}{\alpha}\,
\int\limits_{0}^{\infty}\,d\,s\,\E^{-\frac{s}{\alpha}}\,\left\{
\E^{-\eta\,\lambda\,s}\,\left[\frac{\eta\,s}{4}\,G\,\left(\frac{\eta\,s}{2}\right)\,-\,
\frac{1}{2}\right]\,+\,R\,\left(\frac{\eta\,s}{2},\,\lambda\right)\right\}\,\,,
\end{equation}
where $R$ collects the pieces with smoothed-out, positive-$s$ poles:
\renewcommand{\theequation}{2.192}
\begin{equation}
\label{rfunc}
R\,\left(\frac{\eta}{2}\,s,\,\lambda\right)\,=\,
\sum\limits_{n\,=\,1}^{\infty}\,n\,\left(-\E^{-2\,\lambda}\right)^{n}\,
\frac{\left[\E^{-\lambda\,(\eta\,s\,-\,2\,n)}\,-\,1\right]\,+\,
\Theta\,(n\,-\,s)}{\frac{\eta}{2}\,s\,-\,n}\,\,.
\end{equation}
For the first term in (\ref{fullap}), use of relation (\ref{grecur}) and
of the integral representation \cite{GRY}
\renewcommand{\theequation}{2.193}
\begin{equation}
\label{gint}
G\,\left(\frac{z\,+\,1}{2}\right)\,=\,2\,\int\limits_{0}^{\infty}\,d\,x\,
\frac{\E^{-z\,x}}{\cosh\,x}
\end{equation}
gives, after a few lines of calculation, the contribution
\renewcommand{\theequation}{2.194}
\begin{equation}
\label{fullef1}
\frac{1}{2}\,\int\limits_{0}^{\infty}\,\frac{d\,x}{\cosh^{2}\,x}\,
\frac{1}{1\,+\,\eta\,\alpha\,(x\,+\,\lambda)}
\end{equation}
to the Laplace integral. Note the lower limit of {\em zero}, rather than
of minus infinity as in $F^{pert}$. In the second, $R$ part of
(\ref{fullap}), the only minor trick you may need is to observe that in
the term coming from the step functions, which may be written as
\renewcommand{\theequation}{2.195}
\begin{equation}
\label{stepef2}
\frac{1}{\alpha}\,\int\limits_{0}^{1}\,d\,y\,\frac{1}{1\,-\,\left(
\frac{\eta}{2}\right)\,y}\,\left[\sum\limits_{n\,=\,1}^{\infty}\,(-)^{n\,-\,1}\,n\,\E^{-n\,\left(\frac{y}{\alpha}\,+\,2\,\lambda\right)}\right]
\end{equation}
with integration variables $y\,=\,s\,/\,n$, the convergent series in
square
brackets can be resummed in the manner of eq. (\ref{chexp}) above. The
rest is straightforward and the contribution from the $R$ term becomes
\renewcommand{\theequation}{2.196}
\begin{equation}
\label{fullef2}
\frac{1}{2}\,\int\limits_{-\frac{1}{2\,\alpha}\,-\,\lambda}^{0}\,
\frac{d\,x}{\cosh^{2}\,x}\,\frac{1}{1\,+\,\eta\,\alpha\,(x\,+\,\lambda)}\,\,.
\end{equation}
The two contributions together indeed reproduce the function
(\ref{fxint}). The Borel-Laplace process at physical values of the
coupling $\alpha$ has worked perfectly, unobstructed by either the
t´Hooft set of singularities, or by the spurious poles on the positive
Borel axis introduced by the artificial
perturbative-plus-nonperturbative decomposition. It has worked because
we based it on the full resurgent symbol as the adequate asymptotic
representation of a resurgent amplitude, rather than on the inadequate
perturbation series.

\qquad
The spurious right-hand poles in the {\em perturbative} Borel transform
(second term of eq. (\ref{mbmero})) are of course, as some of you
already noted, a not-so-unrealistic model of the so-called {\em infrared
renormalons}, just as those on the left-hand Borel axis (first term of
eq. (\ref{mbmero})), which are not spurious at all but remain a true and
important feature of the full resurgent amplitude, represent the
so-called {\em ultraviolet renormalons} \cite{GNV} \cite{WEI}
\cite{ZNJ}. (The completely symmetric manner in which both arise here
from eq. (\ref{mbp1}) is presumably among the oversimplified features of
our model.) The renormalons -- another specialty on which we will be
unable to dwell in any detail \cite{BEN} -- generally arise through one
of the perturbative partial resummations with finite
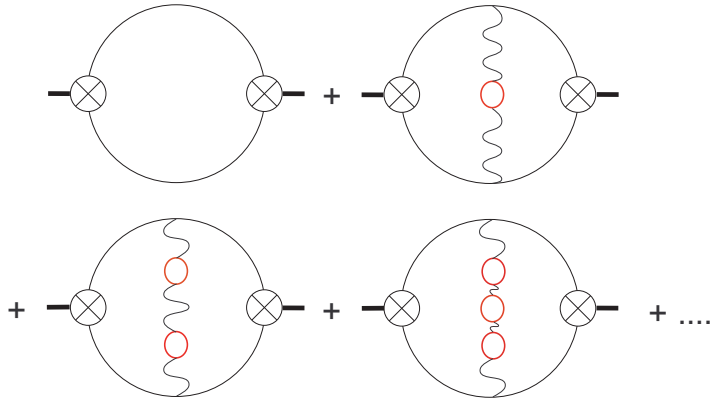
\begin{center}
\epsfig{file=fig24.eps,width=10cm}\\
$\begin{array}{ll}
{\bf Fig. 2.4}\quad &
\text{Example of a family of diagrams whose summation in perturbation}\\
& \text{theory
gives rise to renormalons. The diagrams represent loop contri-}\\
& \text{butions to
the two-point function of an elementary or composite opera-}\\
& \text{tor denoted
by a cross.}
\end{array}$
\end{center}
convergence
radius mentioned in the introduction: in this case, the geometric-series
summation of chains of two-point or four-point bubble graphs inserted
into a larger loop for some two-point correlation (of elementary or
composite fields), as illustrated in Fig. 2.4. Although in such a
diagram family there is only one chain diagram per loop order $p$ --
namely, the one with $p\,-\,2$ bubbles --, their insertion into the
larger loop generates $p!$ growth by the mathematics expressed in
formula (\ref{xpint}), so that here one has a mechanism producing
factorial divergence independent of the "standard" mechanism of
growth-in-number-of-diagrams. That these infrared renormalons must get
cancelled, on the level of the combined Borel minor (\ref{combo}), by
opposite contributions from what we would now call the higher elementary
resurgent symbols, seems to have been inferred first, in the context of
an OPE analysis of the nonlinear sigma model, by David \cite{DAV}, who
also conjectured the cancellation would persist in realistic theories.
What the present model study can add to his (exact but somewhat
indirect) argument is to show, with the greatest simplicity and
explicitness, how the problem arises directly from the definition of
perturbation theory, and how it disappears when the resurgent coupling
dependence of an amplitude is viewed and treated as a whole.

\qquad
What, then, are the morals to be drawn from this model? It seems to me
there are at least two. First, there is absolutely no reason to be
afraid of t´Hooft´s $\alpha$-plane singularities: in the perturbative
portion, they are missing anyway; in the full resurgent function, they
do not impede BL reconstruction. In my opinion they are actually a sign
of health -- they indicate that the nonperturbative thresholds
proportional to $\Lambda_{0}$ have been left intact, and have not been
perturbatively mutilated.

\qquad
Second, as for the recoverability of the amplitude from its resurgent
symbol, it is clearly imperative to keep together the perturbative and
strongly nonperturbative portions as much as possible,
since only the sum of both leads (at positive $\alpha$) to an univalent
function. {\em As much as possible?} Since we were dealing here with a
simple model amplitude having an explicit integral representation, we
were able to "sum the whole thing", i.\,e. to perform Laplace
reconstruction from the combined Borel minor in the form of eqs.
(\ref{fullef1}) and (\ref{fullef2}) -- but that route is clearly not
viable in the real world where all we may know about an amplitude is
that it should be (part of) a coupling-resurgent solution to some
complicated Dyson-Schwinger hierarchy. But here a simple idea almost
suggests itself: if we cannot hope to sum everything in one blow, can´t
we at least establish what one might call a {\em quasi-perturbative
series}, i.\,e. an expansion in which the $n$ summation in the
"nonperturbative direction" has been performed, either completely or
in some systematic approximation, at each fixed $p$? In the physically
relevant case of the real-and-equidistant support, eq. (\ref{posymb}),
we would then have something like
\renewcommand{\theequation}{2.197}
\begin{equation}
\label{quapex}
f\,(\alpha)\,\sim\,\sum\limits_{p\,=\,0}^{\infty}\,c_{p}\,(\Lambda_{0}^{2})
\,\alpha^{p}\,\,,
\end{equation}
a semiconvergent series in $\alpha$ in which the coefficients $c_{p}$
carry the strongly nonperturbative $\alpha$ dependence through
$\Lambda_{0}\,\propto\,\exp\,(-s_{1}\,/\,\alpha)$. Since each $c_{p}$
keeps the perturbative $(n\,=\,0)$ and nonperturbative $(n\,\geq\,1)$
terms together, and since we know the full function is ultimately
univalued at real positive $\alpha$, we may be bold and conjecture that
{\em this series should be Borel summable as a formal series in}
$\alpha$, i.\,e. if we treat the $\Lambda_{0}^{2}$ dependence in the
$c_{p}$ as a dependence on an extra parameter unrelated to $\alpha$.
It will be interesting, and yield a little harvest of extra insight into
the resurgent structure, to check this out for our model amplitude.

\qquad
We write the complete resurgent symbol $F_{f}$ of eq. (\ref{fresym}) in
the form
\renewcommand{\theequation}{2.198}
\begin{equation}
\label{fdoser}
F_{f}\,=\,\sum\limits_{n\,=\,0}^{\infty}\,\sum\limits_{p\,=\,0}^{\infty}\,
a_{n,\,p}\,\left(\frac{\Lambda^{2}}{k^{2}}\right)^{n}\,\alpha^{p}\,\,,
\end{equation}
where the coefficients $a_{n,\,p}$ can be assembled from eqs.
(\ref{fpcoeff}) and (\ref{phi1ser}):
\renewcommand{\theequation}{2.199}
\begin{align}
\label{fdocoeff}
a_{n,\,p}
&=
\Biggl\{\delta_{n,\,0}\,(-)^{p}\,\left(\frac{\eta}{2}\right)^{p}\,
\Biggl[\sum\limits_{q\,=\,0}^{p}\,f_{q}^{(p)}\,(2\,\lambda)^{q}\Biggr]\nonumber\\
&+
(1\,-\,\delta_{n,\,0})\,(-)^{n}\,\frac{\left(\frac{\eta}{2}\right)^{p}}{\left(1\,-\,\frac{\eta}{2}\right)^{p\,+\,1}}\,
\frac{1}{n^{p}}\Biggr\}\,p!
\end{align}
Performing the $n$ sum at fixed $p$ then gives the form of eq.
(\ref{quapex}), with coefficients
\renewcommand{\theequation}{2.200}
\begin{equation}
\label{quacp}
c_{p}\,(\Lambda^{2})\,=\,a_{0,\,p}\,+\,
\sum\limits_{n\,=\,1}^{\infty}\,a_{n,\,p}\,\left(
\frac{\Lambda^{2}}{k^{2}}\right)^{n}\,\,.
\end{equation}
Here, the sum of the $n\,\geq\,1$ terms,
\renewcommand{\theequation}{2.201}
\begin{equation}
\label{npsum}
\frac{\left(\frac{\eta}{2}\right)^{p}}{\left(
1\,-\,\frac{\eta}{2}\right)^{p\,+\,1}}\,p!\,
\sum\limits_{n\,=\,1}^{\infty}\,\frac{1}{n^{p}}\,\left(-
\frac{\Lambda^{2}}{k^{2}}\right)^{n}\,\,,
\end{equation}
converges in the domain
\renewcommand{\theequation}{2.202}
\begin{equation}
\label{nconc}
\left|\frac{\Lambda^{2}}{k^{2}}\right|\,<\,1\,\,,
\end{equation}
which illustrates our general expectation that the $n$ summation should
not be semiconvergent but have a finite radius of convergence. Within
that radius, it can be performed in terms of a known class of
transcendental functions, the polylogarithms,
\renewcommand{\theequation}{2.203}
\begin{equation}
\label{polylog}
Li_{p}\,(z)\,=\,\sum\limits_{n\,=\,1}^{\infty}\,\frac{z^{n}}{n^{p}}\qquad
(|z|\,<\,1)\,\,,
\end{equation}
whose $p\,=\,2$ member you may have met in calculations of two-loop (or
three-point one-loop) Feynman graphs; the $p\,=\,0$ and $p\,=\,1$
members are actually elementary functions:
\renewcommand{\theequation}{2.204}
\begin{equation}
\label{01log}
Li_{0}\,(z)\,=\,\frac{z}{1\,-\,z}\,\,;\qquad
Li_{1}\,(z)\,=\,-\ln\,(1\,-\,z)\,\,.
\end{equation}
The analytic continuation beyond (\ref{nconc}) is provided by the
integral representation
\renewcommand{\theequation}{2.205}
\begin{equation}
\label{polyint}
Li_{p}\,(z)\,=\,\frac{1}{p!}\,\int\limits_{0}^{\infty}\,d\,x\
x^{p}\,\frac{z\,\E^{-x}}{(1\,-\,z\,\E^{-x})^{2}}\,\,,
\end{equation}
which again encompasses the two elementary cases (\ref{01log}). (We have
rewritten the usual integral definition of $Li_{p}$ found in tables
\cite{DEV} through a change of variable and a couple of partial
integrations.) Thus,
\renewcommand{\theequation}{2.206}
\begin{equation}
\label{modcp}
c_{p}\,(\Lambda^{2})=p!\left(\frac{\eta}{2}\right)^{p}\left\{
(-)^{p}\sum\limits_{q=0}^{p}\,f^{(p)}_{q}\left[\ln\left(\frac{k^{2}}{\mu^{2}}\right)\right]^{q}
+\frac{1}{\left(1-\frac{\eta}{2}\right)^{p\,+\,1}}\,
Li_{p}\left(-\frac{\Lambda^{2}}{k^{2}}\right)\right\}\,.
\end{equation}
Now for Borel summability we do not seem to have made much progress --
in particular, the telltale alternating signs that would indicate
banishment of Borel singularities to the left-hand, real Borel axis are
not visible. But wait! Up to now we failed to express both parts of eq.
(\ref{modcp}) in the same variable. Maybe in order to really fuse them
into one quantity, what we should do is to reexpress the perturbative
portion of $F_{f}$,
\renewcommand{\theequation}{2.207}
\begin{equation}
\label{fpmn}
F_{f}^{(pert)}\,=\,\sum\limits_{p\,=\,0}^{\infty}\,a_{0,\,p}\,\left(
\ln\,\frac{k^{2}}{\mu^{2}}\right)\,\cdot\,\alpha^{p}\,\,,
\end{equation}
in terms of the variable $k^{2}\,/\,\Lambda^{2}$, rather than
$k^{2}\,/\,\mu^{2}$, by using
\renewcommand{\theequation}{2.208}
\begin{equation}
\label{logrel}
\ln\,\left(\frac{k^{2}}{\mu^{2}}\right)\,=\,\ln\,
\left(\frac{k^{2}}{\Lambda^{2}}\right)\,-\,\frac{1}{\alpha}
\end{equation}
(remember eq. (\ref{lambmod})) and reshuffling the perturbative $\alpha$
series. This requires a page of algebra but leads to a simple result,
expressible as another RG-type scaling relation:
\renewcommand{\theequation}{2.209}
\begin{equation}
\label{fprew}
F_{f}^{(pert)}\,\left(\frac{k^{2}}{\mu^{2}};\,\alpha\right)\,=\,
\frac{1}{1\,-\,\frac{\eta}{2}}\,F_{f}^{(pert)}\,\left(
\frac{k^{2}}{\Lambda^{2}};\,\frac{\alpha}{1\,-\,\frac{\eta}{2}}\right)\,\,.
\end{equation}
(Note that the $\mu$ dependence hidden in $\alpha\,=\,\alpha\,(\mu)$ is
still present on the r. h. s.; we merely changed the scaling of the
momentum variable $k^{2}$.) In other words we now have for
$F^{(pert)}$ a rearranged expansion
\renewcommand{\theequation}{2.210}
\begin{equation}
\label{fpla}
F_{f}^{(pert)}\,=\,\sum\limits_{p\,=\,0}^{\infty}\,\overline{a}_{0,\,p}\,\left(
\ln\,\frac{k^{2}}{\Lambda^{2}}\right)\,\alpha^{p}\,\,,
\end{equation}
\renewcommand{\theequation}{2.211}
\begin{equation}
\label{apla}
\overline{a}_{0,\,p}\,=\,(-)^{p}\,p!\,\frac{\left(\frac{\eta}{2}\right)^{p}}{\left(
1\,-\,\frac{\eta}{2}\right)^{p\,+\,1}}\,\sum\limits_{r\,=\,0}^{p}\,f^{(p)}_{r}\,
\left[\ln\,\left(\frac{k^{2}}{\Lambda^{2}}\right)\right]^{r}\,\,,
\end{equation}
which is no more the perturbation expansion in the strict sense -- the
Taylor series of $F$ at $\alpha\,=\,0+$ -- but has the formal advantage
of being expressed in the same variables as the nonperturbative terms.
(Let me once more emphasize it is an oversimplified feature of our model
that this reshuffling of logarithms should be necessary only in the
perturbative part. More realistic amplitudes always have dependence on
$\ln\,(k^{2}\,/\,\mu^{2})$ also in the modified Wilson coefficients
$U_{n}^{(R)},\,n\,\geq\,1$, of eq. (\ref{opela}), and will need
regrouping of logarithms in these quantities as well). It is then
plausible that we should redefine our quasi-perturbative expansion to
mean
\renewcommand{\theequation}{2.212}
\begin{equation}
\label{quapert}
F\,=\,\sum\limits_{p\,=\,0}^{\infty}\,\overline{c}_{p}\,\left(
\frac{\Lambda^{2}}{k^{2}}\right)\,\alpha^{p}\,=\,\overline{F}\,
\left(\frac{k^{2}}{\Lambda^{2}};\,\alpha\right)\,\,,
\end{equation}
with coefficients now given by
\renewcommand{\theequation}{2.213}
\begin{equation}
\overline{c}_{p}\,=\,\overline{a}_{0,\,p}\,+\,\sum\limits_{n\,=\,1}^{\infty}\,a_{n,\,p}\,
\left(\frac{\Lambda^{2}}{k^{2}}\right)^{n}\,\,,
\end{equation}
rather than by eq. (\ref{quacp}). Since the structure (\ref{pqcoeff}) of
the $f_{r}^{(p)}$ coefficients enforces $(-)^{r}\,=\,(-)^{p}$, we have
\renewcommand{\theequation}{2.214}
\begin{equation}
\label{cpbar}
\overline{c}_{p}\,\left(\frac{\Lambda^{2}}{k^{2}}\right)\,=\,
\frac{(\eta\,/\,2)^{p}}{[1\,-\,(\eta\,/\,2)]^{p\,+\,1}}\,p!\,
\Biggl\{\sum\limits_{r\,=\,0}^{p}\,f^{(p)}_{r}\,\left[\ln\,\left(
\frac{\Lambda^{2}}{k^{2}}\right)\right]^{r}\,+\,Li_{p}\,\left(-\frac{\Lambda^{2}}{k^{2}}\right)\Biggr\}\,\,.
\end{equation}
This redefinition now pays off in a surprising way: what we have in the
curly bracket is the r. h. s. of a known "duality relation" for the
polylogarithms \cite{DEV}, a relation connecting $Li_{p}$´s at large and
small arguments:
\renewcommand{\theequation}{2.215}
\begin{equation}
\label{lidual}
(-)^{p\,+\,1}\,Li_{p}\,(z)\,=\,\sum\limits_{r\,=\,0}^{p}\,f^{(p)}_{r}\,\left[
\ln\,\left(-\frac{1}{z}\right)\right]^{r}\,+\,Li_{p}\,\left(\frac{1}{z}\right)\,\,.
\end{equation}
Under this new definition we can therefore indeed merge the perturbative
and nonperturbative parts at a fixed $p$ into a new quasi-perturbative
object given, up to factors, by a single polylogarithm:
\renewcommand{\theequation}{2.216}
\begin{equation}
\label{cpmerg}
\overline{c}_{p}\,\left(\frac{\Lambda^{2}}{k^{2}}\right)\,=\,
(-)^{p}\,p!\,\frac{(\eta\,/\,2)^{p}}{[1\,-\,(\eta\,/\,2)]^{p\,+\,1}}\,
\left[-Li_{p}\,\left(-\frac{k^{2}}{\Lambda^{2}}\right)\right]\,\,.
\end{equation}
From eq. (\ref{polyint}), it is clear that $-Li_{p}\,(-z)$ at real,
positive $z$ is positive, so the alternating signs $(-)^{p}$ we were
looking for are now present. That the series (\ref{quapert}) with
coefficients (\ref{cpmerg}), and with $\Lambda^{2}\,/\,k^{2}$ treated as
a parameter unrelated to $\alpha$, is indeed Borel summable with respect
to $\alpha$ can be seen by yet another rewriting of the integral
representation (\ref{fxint}), using as integration variable
\renewcommand{\theequation}{2.217}
\begin{equation}
\label{teevar}
t\,=\,2\,x\,+\,\frac{1}{\alpha}\,=\,-y\,\,,
\end{equation}
which is the negative of the one employed in (\ref{fnint}) for the
nonperturbative part alone. We get
\renewcommand{\theequation}{2.218}
\begin{equation}
\label{ftint}
F\left(\frac{k^{2}}{\mu^{2}};\,\alpha\right)=\overline{F}\left(\frac{k^{2}}{\Lambda^{2}};\,\alpha\right)=
\frac{1}{2\,(2-\eta)}\int\limits_{0}^{\infty}d\,t\,
\frac{1}{\cosh^{2}\left(\frac{t-\kappa}{2}\right)\left[1+\left(
\frac{\eta}{2-\eta}\,\alpha\right)t\right]}\,,
\end{equation}
where
\renewcommand{\theequation}{2.219}
\begin{equation}
\label{kapvar}
\kappa\,=\,2\,\lambda\,+\,\frac{1}{\alpha}\,=\,\ln\,\left(
\frac{k^{2}}{\Lambda^{2}}\right)\,\,.
\end{equation}
Due to the strong convergence provided by the $\cosh^{-2}$ factor, the
function $\overline{F}\,(e^{\kappa};\,\alpha)$ at fixed $\kappa$ is
obviously analytic with respect to $\alpha$ in $Re\,\alpha\,>\,0$. Note
that the t´Hooft singularities do not now disturb this conclusion; they
would arise from the "hidden"  $\alpha$ dependence in
$e^{\kappa}\,=\,k^{2}\,/\,\Lambda^{2}$ which for the present purpose we
have eliminated by treating $\kappa$ as a fixed parameter. Moreover, one
easily evaluates
\renewcommand{\theequation}{2.220}
\begin{equation}
\label{derfbar}
\frac{\partial^{p}\,\overline{F}}{\partial\,\alpha^{p}}\,=\,
\frac{(-\eta)^{p}\,p!}{2\,(2\,-\,\eta)^{p\,+\,1}}
\int\limits_{0}^{\infty} d\,t\,
\frac{t^{p}}{\cosh^{2}\,\left(\frac{t\,-\,\kappa}{2}\right)\,\left[
1\,+\,\left(\frac{\eta}{2\,-\,\eta}\,\alpha\right)\,t\right]^{p\,+\,1}}\,\,,
\end{equation}
and by estimating
\renewcommand{\theequation}{2.221}
\begin{equation}
\label{estalf}
\left|\left[1\,+\,\left(\frac{\eta}{2\,-\,\eta}\,\alpha\right)\,t\right]^{-(p\,+\,1)}\right|\,\leq\,1 \qquad
(Re\,\alpha\,>\,0)\,\,,
\end{equation}
as well as
$\cosh^{-2}\,\left(\frac{t\,-\,\kappa}{2}\right)\,\leq\,4\,e^{\kappa\,-\,t}$, one establishes a bound of type (\ref{derbnd}), similar to
what we had for superrenormalizable amplitudes in (\ref{derest}):
\renewcommand{\theequation}{2.222}
\begin{equation}
\label{modest}
\left|\frac{\partial^{p}\,\overline{F}}{\partial\,\alpha^{p}}\right|\,\leq\,
\frac{2}{2\,-\,\eta}\,\left(\frac{k^{2}}{\Lambda^{2}}\right)\,
\left(\frac{2\,-\,\eta}{\eta}\right)^{-p}\,(p!)^{2}\,\,.
\end{equation}
Therefore we have Nevanlinna-Sokal conditions satisfied in the
half-plane $Re\,\alpha\,>\,0$, and $\overline{F}$ is Borel summable with
respect to $\alpha$. (As a purely mathematical aside, we have shown the
function $\overline{F}$ of eq. (\ref{ftint}) at $\eta\,=\,1$ to be a
{\em generating function} for the functions
$2\,(-)^{p\,+\,1}\,p!\,Li_{p}\,(-e^{\kappa})$, i.\,e. to generate them
as coefficients when expanded with respect to $\alpha$.)

\qquad
The quasi-perturbative expansion (\ref{quapert}), in this respect, is
better behaved than the perturbative one, but we should not lose sight
of the fact that there is now no easy way back from the former to the
latter. The beautiful duality formula (\ref{lidual}) became applicable
only after we replaced the perturbation expansion in the strict sense by
the rearranged series (\ref{fpla}), which is no more the Taylor
expansion at $\alpha\,=\,0+$ of $F$. The rearrangement is infinite, in
the sense that one quasi-perturbative term collects contributions from
infinitely many perturbative terms and vice versa. As a result, in
writing (\ref{quapert}) we have "lost the perturbative limit": the
formal limit of $\Lambda\,\to\,0$ with $\alpha^{p}$´s remaining finite,
which previously could be obtained by simply dropping the last term in
each of the coefficients $c_{p}$ of (\ref{modcp}), cannot now be
recovered any more from one or a finite number of quasi-perturbative
terms. In the {\em zeroth quasi-perturbative order}, the analog of the
OPE object (\ref{invdnull}) which sums up the terms without powers of
$\alpha$ but with all powers of $\Lambda^{2}$, we have
\renewcommand{\theequation}{2.223}
\begin{equation}
\label{fbarnull}
\overline{F}\,\left(\frac{k^{2}}{\Lambda^{2}};\,0\right)\,=\,
\overline{c}_{0}\,\left(\frac{k^{2}}{\Lambda^{2}}\right)\,=\,
\frac{1}{1\,-\,\left(\frac{\eta}{2}\right)}\,
\frac{k^{2}}{k^{2}\,+\,\Lambda^{2}}\,\,,
\end{equation}
and although the limit $\Lambda\,\to\,0$ here exists, it differs from
the perturbative $F^{(0)\,pert}\,=\,1$ by the factor
$\left(1\,-\,\frac{\eta}{2}\right)^{-1}$, which sums a geometric
series coming from {\em all} perturbative terms. In the higher
quasi-perturbative orders, the $\Lambda^{2}\,\to\,0$ limit does not even
exist termwise; we could retrieve it only when undoing the rearrangement
of eq. (\ref{fpla}) by applying relation (\ref{logrel}) backwards --
i.\,e. by acknowledging that $\Lambda^{2}$ and $\alpha$ are, in fact,
{\em not} unrelated. {\em In a calculation where we can determine only a
finite number of quasi-perturbative terms, we therefore have no easy way
of making contact with the perturbative series}, and of checking whether
it comes out as a limiting case. One might take the view, why bother at
all about a perturbative limit in this sense? After all, the very point
of an RG-invariant scale is that it cannot be switched off in any way,
and since the interesting regime we would like to explore for
nonperturbative phenomena, in both Quantum Chromodynamics and
electroweak theory, is the regime
$k^{2}\,\begin{array}{c} <\vspace*{-3mm}\\
\sim\end{array}\,\Lambda^{2}$, we
could dismiss the $\Lambda\,\to\,0$ limit as uninteresting. Still, the
difficulty of staying in touch with the perturbative expansion remains a
nuisance in some respects. Take, for example, the RG results on
large-momentum behavior: they stipulate that the {\em perturbative}
zeroth-order vertex functions should be approached, with logarithmic
corrections otainable through {\em perturbative} leading-log summations,
and since these are exact limiting results one would like to preserve,
one feels unhappy about losing the easy way back to their verification.

\qquad
The loss is offset, to be sure, by considerable gain in the description
of the small-$k^{2}$ regime. By eq. (\ref{polylog}), we now have the
correct behavior $F\,\to\,0$ as $k^{2}\,\to\,0$ in each
quasi-perturbative term. What is more, due to the analytic properties of
the $Li_{p}$ functions -- for $p\,\geq\,1$, they all have branch cuts at
real argument larger than unity --, each term now also reflects the
correct "spectral" singularities of $F$ at Minkowskian values
$k^{2}\,\leq\,-\Lambda^{2}$, including the pole at
$k^{2}\,=\,-\Lambda^{2}$ carried by the $p\,=\,0$ term (\ref{fbarnull})
and the physical cut at $k^{2}\,<\,-\Lambda^{2}$ in the $p\,\geq\,1$
quasi-perturbative corrections. Thanks to the tight linkage between the
$\Lambda^{2}$ and $k^{2}$ dependences, the quasi-perturbative
resummation of $(\Lambda^{2})^{n}$ powers in the resurgent symbol has
automatically furnished  a very satisfactory {\em analytic continuation
in the $k^{2}$ plane}, which resummations of perturbation theory alone,
even disregarding its false infrared singularities, could not provide.
It would then seem that the quasi-perturbative series deserves a try in
the "real" problem, and chap. 3 will sketch some steps in that
direction.

\renewcommand{\thesection}{3}
\section{Construction of Resurgent Amplitudes}
\subsection{The Approximating Sequence}

\qquad
We have seen that the operator-product expansion, in a strictly
renormalizable and asymptotically free QFT, is of the form of a
resurgent symbol in the coupling constant $\alpha$. As such, it asserts
that amplitudes in those QFT´s are resurgent functions of $\alpha$. We
have also seen that there are good reasons for assuming that this symbol
is summable, that is, the small-$\alpha$ expansion of a unique analytic
function of $\alpha$. In physics terms, there are good reasons for
expecting the OPE to be complete, i. e. to leave no room for further
terms that would escape the resurgent-symbol representation by having
stronger-than-exponential suppression as $\alpha\,\to\,0+$.

\qquad
The implications of these two insights are quite strong. If we find
methods of constructing systematically, or at least in a systematic
approximation, the unique analytic coupling functions waiting behind the
resurgent symbols, we will have accounted not just for one special type
(as an easy misunderstanding might suggest) but for {\em the totality of
the strongly nonanalytic coupling dependence}. It therefore seems
worthwhile to look for and try out such methods. The method described in
this, our last chapter, builds on the plausible hypothesis of
summability of the resurgent symbols, and attempts to systematically
reconstruct the strongly nonperturbative coupling dependence of
correlation functions -- the dependence manifesting itself in the
resurgent symbol through powers of the RG-invariant scale $\Lambda$ --
from their equations of motion.

\qquad
In developing such a method we will find ourselves compelled to deviate
somewhat from standard strategy. Standard procedure relies on the
function being a solution to some differential or
differential-cum-functional equation. Its first step is then to obtain
the function´s resurgent-symbol expansion by purely formal manipulation,
inserting the symbol into the governing equation and comparing terms.
(If the equation is nonlinear, it becomes important at this point that
the set of elementary resurgent symbols form an algebra under
multiplication.) In mathematical theory, completion of this step is
usually taken for granted; it belongs, so to speak, to a realm of
pre-mathematical heuristics. The procedure then continues in the manner
we illustrated in sect. 2.4 with our model amplitude: one "borelizes"
the individual power series, separately  for each $n$, and performs
analytic continuation to obtain the Borel minors $b_{n}\,(s)$. If these
are sufficiently nonsingular (integrable  singularities, as pointed out
by Sokal \cite{SOK}, can be tolerated) and restricted in growth on the
positive Borel axis, one may proceed to evaluating their Laplace
transforms that appear in eq. (\ref{resasy}). Finally, one confronts the
$n$ series of that equation: if it has nonzero convergence radius, it
may require another analytic-continuation step; if it is summably
divergent, it will call for another BL resummation. If all goes well,
one finally reaches the holy grail, a globally valid representation of
the full resurgent function.

\qquad
For physicists dealing with QFT, while this standard procedure should be
in the back of their minds, there are cogent reasons for proceeding in
different order. As we noted in studying our model, it is a consequence
of the very definition of perturbation theory that individual $b_{n}$´s
are in general afflicted with spurious Borel nonsummability, and thus
force us to cope with spurious multivaluedness at all intermediate
stages of the reconstruction, while we are convinced that the full
function must ultimately be univalued at physical, positive couplings.
We saw that the easiest way to avoid this is to keep perturbative and
nonperturbative contributions tightly together by using a
quasi-perturbative expansion, that is, by first performing the $n$
summation in the symbol´s nonperturbative direction at fixed $p$. You
might choose to call this a "resummation of the OPE in the $n$
direction", but you know by now why I am reluctant to use that term --
it sounds as if the OPE, for realistic amplitudes, were something fully
calculable {\em ab initio} that one could simply take and subject
to some process of continuation. In truth, as we have emphasized,
the
OPE is not known ab initio, since is does not by itself determine its
own strongly nonperturbative ingredients, the vacuum condensates. We
will therefore be forced to recover our strongly nonperturbative
contributions from exact equations of motion, which supply the dynamical
principle the OPE is lacking. In the process, the vacuum condensates
will lose some of their central role in describing the strongly
nonperturbative effects, that role being taken over by {\em
dimensionless} nonperturbative parameters more easily accessible through
the equations of motion (and from which one may later recover the
condensates if one so desires).

\qquad
Of course, with realistic amplitudes there is again no way of performing
the entire $n$ sum at fixed $p$ in one blow, as we did for our model in
writing eqs. (\ref{modcp}) or (\ref{cpbar}). But we may attempt to do it
in terms of a suitable systematic sequence of approximations, amounting
to resummations of the $(\Lambda^{2})^{n}$ power series that grow
progressively more comprehensive with increasing approximation order.
These resummations, as again we learned in our model, will automatically
provide a sequence of analytic continuations, increasingly accurate, to
the momentum region of $Q^{2}\,\begin{array}{c}
<\vspace*{-3mm}\\
\sim\end{array}\,\Lambda^{2}$.

\qquad
As we stressed already at the end of sect. 1.2, it is not compulsory for
us to use "Borelization" in doing this -- all the more as we trust the
$n$ summation at fixed $p$ to already have nonzero radius of
convergence. Indeed we shall find other approximating sequences better
adapted  to the task.
Enumerating
successive approximants (we´ll discuss their detailed form in a moment)
by an index $r\,=\,0,\,1,\,2,\,...$, we denote
\renewcommand{\theequation}{3.1}
\begin{equation}
\label{genapp}
\Gamma^{[r,\,p)}\,(\Lambda^{2})\,=\,r{\rm -}t\,h\quad \text{approximant
to}\quad
\sum\limits_{n\,=\,0}^{\infty}\,c_{n,\,p}\,\left(\frac{\Lambda^{2}}{Q^{2}}\right)^{n}
\end{equation}
at a fixed quasi-perturbative order $p$.

\qquad
There is, to be sure, a subtle drawback to this necessity of employing
approximants to the $n$ sum. At finite level $r$, we can no more expect
a quasi-perturbative expansion based on such approximants to achieve the
complete cancellation of spurious poles in the combined Borel minor that
we were entitled to expect from the full $n$ summation. This is not much
of a problem as long as we
calculate amplitudes at an
energy scale $\mu$ where the coupling $\alpha\,(\mu)$ is small enough to
ensure good semi-convergence for the sequence of partial $p$ sums,
\renewcommand{\theequation}{3.2}
\begin{equation}
\label{partp}
\Gamma^{[r,\,p]}\,=\,\sum\limits_{q\,=\,0}^{p}\,\Gamma^{[r,\,q)}\,(\Lambda^{2})\,
\alpha\,(\mu)^{q}\qquad
(r,\,p\,=\,0,\,1,\,2,\,...\,)\,\,,
\end{equation}
as we do in most of present-day electroweak theory: then these sums may
be taken directly as approximants to the amplitude, without ever
confronting the Borel plane. But when $\alpha\,(\mu)$ is not small, as
in the hadronic region of QCD, we will have to rely on a BL process for
the
$p$ summation, and at low approximation levels $r$ we will then be able
to avoid only a small number, increasing with $r$, of the "nearest",
i.\,e. next-to-the-origin, spurious Borel poles. Since nearest poles in
the $s$ plane dominate large-$\alpha$ behavior in the $\alpha$-plane,
and since in an asymptotically free theory the large-$\alpha$ regime is
also the low-momentum regime, we may qualitatively expect our
approximation to have its largest errors at low momenta and, on
increasing $r$, to gradually work its way down from the large-momentum
regime. Nevertheless, the crude analytic continuations it provides at
low $r$ to the low-momentum region will have finite error and be usable
even down to $Q^{2}\,=\,0$, where the error of the original resurgent
symbol, an expansion in terms of $\Lambda^{2}\,/\,Q^{2}$, would become
infinite.

\qquad
If we then start our reconstruction scheme by inserting such a
continuation-through-resummation, rather than the original OPE-type
expansion, into the equations of motion, we reap several benefits.
First, it turns out it is really only for $p\,=\,0$ that we need to
specify an explicit sequence of approximating functions: the {\em
nonperturbatively modified, zeroth quasi-perturbative orders}
$\Gamma^{[r,\,0)}$, whose exact counterparts we met in the form of eq.
(\ref{invdnull}) for the OPE, and in the form of eq. (\ref{fbarnull})
for our model amplitude. The higher quasi-perturbative corrections,
$\Gamma^{[r,\,p)}$ with $p\,\geq\,1$, will then be generated for us by
the equations of motion themselves. Now choosing approximants for
$\Gamma^{[r,\,0)}$ is a lot easier than for general $p$, since they do
not need to be saddled with typical complications that only the
higher-$p$ corrections are supposed to bring -- in particular, with
logarithmic corrections to large-momentum behavior, and with the
associated dependence on $\mu$ (compare eq. (\ref{uwilc}) in our
discussion of the OPE). Drawing on the body of experience we alluded to
in connection with formulas (\ref{fpade}) and (\ref{bpade}), we
therefore again choose {\em rational} approximants, but this time with
respect to $\Lambda$ and not to $\alpha$ -- i.\,e. expressions of the
form
\renewcommand{\theequation}{3.3}
\begin{equation}
\label{lambfrac}
\frac{\sum\limits_{m\,=\,0}^{l^{'}}\,f_{m}\,(\{k\})\,(\Lambda^{2})^{m}}
{\sum\limits_{n\,=\,0}^{l}\,g_{n}\,(\{k\})\,(\Lambda^{2})^{n}}
\end{equation}
(for bosonic amplitudes). We seem to need {\em two} indices, the
polynomial degrees $l^{'}$ and $l$, to label such an approximant, but
here we reap our next benefit, since asymptotic freedom greatly
simplifies the matter. Recall that the RG leads to exact asymptotic
formulas for vertex functions when the entire set $\{k\}$ of momentum
arguments are scaled up uniformly, and that these formulas simply
feature the zeroth-order perturbative expression, $\Gamma^{(0)\,pert}$,
multiplied by certain logarithmic corrections. But the logarithmic
factors arise from (resummation of) perturbative corrections with
$p\,\geq\,1$, and in our context are therefore for the higher
$\Gamma^{[r,\,p)}$´s to bring; they have no business with
$\Gamma^{[r,\,0)}$. So of the latter we must demand {\em naive}
asymptotic freedom:
\renewcommand{\theequation}{3.4}
\begin{equation}
\label{klarge}
\Gamma_{N}^{[r,\,0)}\,(\{k\},\,\Lambda)\quad \longrightarrow\quad
\Gamma_{N}^{(0)\,pert}\,(\{k\})\qquad(\text{all}\,\,k\,\gg\,\Lambda)\,\,.
\end{equation}
In exploiting this condition, it is best to proceed in steps and deal
first with the simplest case of a two-point vertex function $(N\,=\,2)$,
or negative-inverse propagator, depending on a single scalar Euclidean
$k^{2}$; vertices with $N\,\geq\,3$ that require more momentum variables
will be taken up later. Since $\Lambda^{2}$ is dimensionful, and we
assume no Lagrangian masses present, the approximant (\ref{lambfrac}) is
then automatically rational with respect to $k^{2}$ too, with
\renewcommand{\theequation}{3.5}
\begin{equation}
\label{kpow}
f_{m}\,=\,\zeta_{m}\,(k^{2})^{l^{'}\,-\,m}\,\,,\qquad
g_{n}\,=\,\eta_{n}\,(k^{2})^{l\,-\,n}\,\,,
\end{equation}
and with real constants $\zeta_{m}$ and $\eta_{n}$ since the complete
approximant should be real at real Euclidean $k^{2}$. (At this point,
things would seem to become substantially more complicated when
Lagrangian masses $\hat{m}$ are present, since the $\zeta_{m}$ and
$\eta_{n}$ could then still depend on the dimensionless ratios
$\hat{m}^{2}\,/\,k^{2}$ -- but let us continue to concentrate on the
massless case.) Now the
$\Gamma^{(0)\,pert}$ in condition (\ref{klarge}), which are none other
than the ordinary Feynman rules, are polynomial in $k$, and in
particular for $N\,=\,2$ and a massless scalar field,
\renewcommand{\theequation}{3.6}
\begin{equation}
\label{ga2pert}
-\Gamma^{(0)\,pert}_{2}\,(\{k\})\,=\,k^{2}\,\,.
\end{equation}
Thus condition (\ref{klarge}) fixes the {\em relative} degrees of our
approximant in advance at $l^{'}\,=\,l\,+\,1$. We do not face the
prospect, as one usually does in working with rational approximants, of
having to work through the entire Padé table at given total degree
$l^{'}\,+\,l$; we need to explore only the next-to-diagonal line
$l^{'}\,-\,l\,=\,1$ of it. So in fact we need only one labeling index,
and
we will from now on adopt the {\em convention of using the denominator
degree $l$ as the classifying index $r$ for our approximants to
irreducible vertex functions}:
\renewcommand{\theequation}{3.7}
\begin{equation}
\label{lisr}
r\,=\,l\,=\,l^{'}\,-\,1\qquad (N\,=\,2)
\end{equation}
(For
{\em propagators}, or connected two-point functions,
\renewcommand{\theequation}{3.8}
\begin{equation}
\label{g2conn}
D\,\equiv\,G_{2}^{(conn)}\,=\,-(\Gamma_{2})^{-1}\,\,,
\end{equation}
and the index $r$ is then also the {\em numerator} degree of rational
propagator approximants.) Since without any loss of generality we may
choose
\renewcommand{\theequation}{3.9}
\begin{equation}
\label{etnorm}
\eta_{0}\,=\,1\,\,,\qquad
g_{0}\,(k)\,=\,(k^{2})^{r}\,\,,
\end{equation}
the level-$r$ approximant is now characterized by a set of $2\,r\,+\,2$
dimensionless, real parameters
\renewcommand{\theequation}{3.10}
\begin{equation}
\label{ga2par}
\zeta_{0}^{[r]},\,\zeta_{1}^{[r]},\,...\,\zeta_{r\,+\,1}^{[r]}\,\,;\qquad
\eta_{1}^{[r]},\,\eta_{2}^{[r]},\,...\,\eta_{r}^{[r]}
\end{equation}
and assumes the explicit form
\renewcommand{\theequation}{3.11}
\begin{equation}
\label{ga2r0}
-\Gamma^{[r,\,0)}_{2}\,(k^{2})\,=\,
\frac{N_{2}^{[r]}\,(k^{2},\,\Lambda^{2})}
{\Delta_{2}^{[r]}\,(k^{2},\,\Lambda^{2})}
\end{equation}
for a scalar two-point vertex, where
\renewcommand{\theequation}{3.12}
\begin{equation}
\label{n2r}
N_{2}^{[r]}\,(k^{2},\,\Lambda^{2})\,=\,
\sum\limits_{m\,=\,0}^{r\,+\,1}\,\zeta_{m}^{[r]}\,(k^{2})^{r\,+\,1\,-\,m}\,(\Lambda^{2})^{m}\,\,,
\end{equation}
\renewcommand{\theequation}{3.13}
\begin{equation}
\label{de2r}
\Delta_{2}^{[r]}\,(k^{2},\,\Lambda^{2})\,=\,
\sum\limits_{n\,=\,0}^{r}\,\eta_{n}^{[r]}\,(k^{2})^{r\,-\,n}\,
(\Lambda^{2})^{n}\qquad (\eta_{0}^{[r]}\,=\,1)\,\,.
\end{equation}
(Of course, if $\Gamma_{2}$ has Lorentz-tensor or Dirac-matrix
structure, as do the fields in a gauge theory, such a formula applies
separately to each of the Lorentz-scalar "invariant functions" in a
decomposition of the vertex into basis tensors or basis matrices. For
example, an $SU\,(N)$ gauge-field propagator $D^{\mu\,\nu}_{a\,b}\,(k)$,
with Euclidean vector indices $\mu,\,\nu\,=\,1\,...\,4$ and color
indices $a,\,b\,=\,1\,...\,N^{2}\,-\,1$, would decompose into transverse
and longitudinal pieces with respect to $k$,
\renewcommand{\theequation}{3.14}
\begin{equation}
\label{provec}
D^{\mu\,\nu}_{a\,b}\,(k)\,=\,
\delta_{a\,b}\,[t^{\mu\,\nu}\,(k)\,D_{T}\,(k^{2})\,+\,
l^{\mu\,\nu}\,(k)\,D_{L}\,(k^{2})]\,\,,
\end{equation}
with projectors $t^{\mu\,\nu}$ and $l^{\mu\,\nu}$ given by
\renewcommand{\theequation}{3.15}
\begin{equation}
\label{tralo}
t^{\mu\,\nu}\,(k)\,=\,\delta^{\mu\,\nu}\,-\,
\frac{k^{\mu}\,k^{\nu}}{k^{2}}\,=\,
\delta^{\mu\,\nu}\,-\,l^{\mu\,\nu}\,(k)\,\,,
\end{equation}
and the functions $[D_{T}\,(k^{2})]^{-1}$, $[D_{L}\,(k^{2})]^{-1}$ would
each have an approximant of the form (\ref{ga2r0})). The ubiquitous
indices $[r]$ are cumbersome but necessary at this stage to remind us
that the coefficients of a rational approximant must be recalculated at
every new order $r$.

\qquad
You may ask at this point why we did not exploit condition
(\ref{klarge}) / (\ref{ga2pert}) more fully -- up to now, we only
ensured
\renewcommand{\theequation}{3.16}
\begin{equation}
\label{zetlim}
-\Gamma^{[r,\,0)}_{2}\,(k^{2})\quad \longrightarrow\quad
\zeta_{0}^{[r]}\,k^{2}\qquad
(k^{2}\,\gg\,\Lambda^{2})\,\,,
\end{equation}
so why don´t we impose $\zeta_{0}\,=\,1$, i.\,e.
$f_{0}\,(k)\,=\,(k^{2})^{r\,+\,1}$? Here I have to remind you of the
experience with our model amplitude, and ask you to cast a glance at eq.
(\ref{fbarnull}) above. If we insist on establishing a
quasi-perturbative expansion in the sense originally envisaged, and
exemplified by eqs. (\ref{quapex}) and (\ref{modcp}), then indeed we
should require $\zeta_{0}\,=\,1$. (For the inverse of the function
$D_{L}\,(k^{2})$ defined by eq. (\ref{provec}), the correct analog would
be
\renewcommand{\theequation}{3.17}
\begin{equation}
\label{xilim}
[D_{L}\,(k^{2})]^{-1}\quad \longrightarrow\quad
\frac{1}{\xi}\,k^{2}\qquad (k^{2}\,\gg\,\Lambda^{2})\,\,,
\end{equation}
with $\xi$ the gauge-fixing parameter, in the linear covariant gauge
fixing usually employed.) But if we allow for the possibility that our
approximation might turn out to be quasi-perturbative in the redefined
sense of eq. (\ref{quapert}) and (\ref{cpmerg}), we must allow for a
constant factor $\zeta_{0}\,\not=\,1$, analogous to the
$\left(1\,-\,\frac{\eta}{2}\right)^{-1}$ of eq. (\ref{fbarnull}), which
amounts to a finite renormalization. As long as we have not begun to
determine our approximant (\ref{ga2r0}) from a dynamical equation, we
cannot know whether that equation will reveal a tendency to prefer one
or the other of the two expansions, or something in between. So caution
would advise us to keep the option of $\zeta_{0}\,\not=\,1$ (or
$\not=\,1\,/\,\xi$) at this time.

\qquad
It is trivial to recover from eq. (\ref{ga2r0}) the expansion
corresponding to what we met in the OPE context as eq. (\ref{invdnull}):
by expanding (\ref{ga2r0}) in powers of $\Lambda^{2}\,/\,k^{2}$ for
$k^{2}\,\gg\,\Lambda^{2}$, we have
\renewcommand{\theequation}{3.18}
\begin{align}
\label{ga2rlak}
-\Gamma^{[r,\,0)}_{2}\,(k^{2}) &= k^{2}\,\Big\{
\zeta_{0}^{[r]}\,+\,\left(\zeta_{1}^{[r]}\,-\,\eta_{1}^{[r]}\,\zeta_{0}^{[r]}\right)\,\frac{\Lambda^{2}}{k^{2}}\nonumber\\
&+
\left[(\zeta_{2}^{[r]}\,-\,\eta_{1}^{[r]}\,\zeta_{1}^{[r]}\,+\,
\zeta_{0}^{[r]}\,\left(\eta_{1}^{[r]^{2}}\,-\,\eta_{2}^{[r]}\right)\right]\left(\frac{\Lambda^{2}}{k^{2}}\right)^{2}\,+\,...\,\Big\}\,\,,
\end{align}
so if we imposed $\zeta^{[r]}_{0}\,=\,1$ we could identify, at level
$r$,
\renewcommand{\theequation}{3.19}
\begin{equation}
\begin{array}{l}
\label{capus}
U_{1}^{(R)}\,(0,\,0)\,=\,\zeta_{1}^{[r]}\,-\,\eta_{1}^{[r]}\,\,,\\
\mbox{}\\
U_{2}^{(R)}\,(0,\,0)\,=\,\zeta_{2}^{[r]}\,-\,
\eta_{1}^{[r]}\,\left(\zeta_{1}^{[r]}\,-\,\eta_{1}^{[r]}\right)\,-\,
\eta_{2}^{[r]}\,\,,
\end{array}
\end{equation}
etc. Indeed it would be nothing but standard Padé procedure to determine
the $\zeta_{m}$, $\eta_{n}$ coefficients from the first  $2\,r\,+\,1$ of
these equations if the $U_{n}\,(0,\,0)$ were known quantities. They
aren´t, and so eqs. (\ref{capus}) are useful mainly as a kind of
sum-rule check on the vacuum-condensate combinations
$U_{n}^{(R)}\,(0,\,0)\,(\Lambda^{2}_{0})^{n}$ of eq. (\ref{unnull})
that
can in principle be computed once the $\zeta$´s and $\eta$´s have been
determined from dynamics.

\qquad
More interesting, particularly with a view toward generalizations to
higher vertex functions, is another simple rewriting of (\ref{ga2r0}),
namely the partial-fraction decomposition of its inverse (the propagator
approximant) with respect to the variable $k^{2}$. Decompose the
numerator polynomial (\ref{n2r}) into its root factors,
\renewcommand{\theequation}{3.20}
\begin{equation}
\label{profac}
N_{2}^{[r]}\,(k^{2},\,\Lambda^{2})\,=\,\zeta_{0}\,
\prod\limits_{l\,=\,1}^{r\,+\,1}\,\left(k^{2}\,+\,\sigma_{l}^{[r]}\,\Lambda^{2}\right)
\end{equation}
then
\renewcommand{\theequation}{3.21}
\begin{equation}
\label{propol}
D^{[r,\,0)}\,(k^{2})\,=\,
\sum\limits_{l\,=\,1}^{r\,+\,1}\,\frac{\rho_{l}^{[r]}}{
k^{2}\,+\,\sigma^{[r]}_{l}\,\Lambda^{2}}\,\,,
\end{equation}
with residues,
\renewcommand{\theequation}{3.22}
\begin{equation}
\label{prores}
\rho_{l}^{[r]}\,=\,
\frac{\sum\limits_{n\,=\,0}^{r}\,\eta_{n}^{[r]}\,(-\sigma_{l}^{[r]})^{r\,-\,n}}{
\zeta_{0}\,\prod\limits^{r\,+\,1}_{\begin{array}{c}
\scriptstyle{k\,=\,1}\vspace*{-2mm}\\
\scriptstyle{(k\,\not=\,l)}
\end{array}}\,\left(\sigma_{k}^{[r]}\,-\,\sigma_{l}^{[r]}\right)}\qquad
(\eta_{0}^{[r]}\,=\,1)\,\,.
\end{equation}
This may be viewed as arising from the Euclidean Källén-Lehmann
representation,
\renewcommand{\theequation}{3.23}
\begin{equation}
\label{kaelle}
D\,(k^{2})\,=\,\int\limits_{0}^{\infty}\,d\,s\,
\frac{\rho_{2}\,(s)}{k^{2}\,+\,s\,\Lambda^{2}}\,\,,
\end{equation}
by a discrete approximation of the spectral function,
\renewcommand{\theequation}{3.24}
\begin{equation}
\label{disc2}
\rho_{2}\,(s)\quad \longrightarrow\quad
\rho_{2}^{[r,\,0)}\,(s)\,=\,\sum\limits_{l\,=\,1}^{r\,+\,1}\,\rho_{l}^{[r]}\,
\delta\,\left(s\,-\,\sigma_{l}^{[r]}\right)\,\,.
\end{equation}
Since the spectral function is positive and has its support in the
Minkowskian, we expect a sensible approximant to have all its
$\rho_{l}$´s and $\sigma_{l}$´s non-negative; for the residues,
comparison of the $k^{2}\,\gg\,\Lambda^{2}$ limits of (\ref{propol}) and
(\ref{ga2r0}) gives the sum rule
\renewcommand{\theequation}{3.25}
\begin{equation}
\label{sumrho}
\zeta_{0}\,\sum\limits_{l\,=\,1}^{r\,+\,1}\,\rho^{[r]}_{l}\,=\,1\,\,.
\end{equation}
The $2\,r\,+\,2$ dimensionless quantities
\renewcommand{\theequation}{3.26}
\begin{equation}
\label{propar}
\sigma_{1}^{[r]},\,\sigma^{[r]}_{2},\,...\,
\sigma_{r\,+\,1}^{[r]}\,\,;\qquad
\rho_{1}^{[r]},\,\rho^{[r]}_{2},\,...\,\rho^{[r]}_{r\,+\,1}
\end{equation}
represent a parametrization equivalent to (\ref{ga2par}); when
$\zeta_{0}\,=\,1$, equation (\ref{sumrho}) turns into a relation among
the $\rho_{l}$´s that reduces the number of independent parameters by
one. For later convenience, we also note the corresponding decomposition
of the vertex function (\ref{ga2r0}) itself, which proceeds from the
root-factor decomposition of the denominator polynomial (\ref{de2r}),
\renewcommand{\theequation}{3.27}
\begin{equation}
\label{delfac}
\Delta_{2}^{[r]}\,(k^{2},\,\Lambda^{2})\,=\,
\prod\limits_{s\,=\,1}^{r}\,\left(k^{2}\,+\,u_{2\,s}^{[r]}\,\Lambda^{2}\right)\,\,.
\end{equation}
Since in this case the numerator polynomial is of higher degree, the
decomposition now takes a slightly different form,
\renewcommand{\theequation}{3.28}
\begin{equation}
\label{ga2pol}
-\Gamma^{[r,\,0)}_{2}\,=\,\zeta_{0}^{[r]}\,\left[k^{2}\,+\,u_{1}^{[r]}\,
\Lambda^{2}\,+\,\sum\limits_{s\,=\,1}^{r}\,
\frac{u_{2\,s\,+\,1}^{[r]}\,\Lambda^{4}}{k^{2}\,+\,u_{2\,s}^{[r]}\,\Lambda^{2}}\right]\,\,,
\end{equation}
with the sum now representing a discrete approximation to the branch cut
of the self-energy that mirrors the Källén-Lehmann propagator cut. By
comparison with eq. (\ref{ga2rlak}), its parameters obey
\renewcommand{\theequation}{3.29}
\begin{equation}
\label{uodd}
\begin{array}{c}
\displaystyle{
u_{1}^{[r]}\,=\,\left(\frac{\zeta_{1}}{\zeta_{0}}\,-\,\eta_{1}\right)^{[r]}\,\,,}\\
\mbox{}\\
\displaystyle{
\sum\limits_{s\,=\,1}^{r}\,u_{2\,s\,+\,1}^{[r]}\,=\,
\left(\frac{\zeta_{2}}{\zeta_{0}}\,-\,\eta_{1}\,\frac{\zeta_{1}}{\zeta_{0}}\,+\,
\eta_{1}^{2}\,-\,\eta_{2}\right)^{[r]}\,\,,}
\end{array}
\end{equation}
etc.

\qquad
If the field whose two-point correlation $D$ represents had exactly one
particle associated with it (as is usually assumed without discussion),
then $\sqrt{\sigma_{1}}\,\Lambda$ and $\rho_{1}$ would approximate its
mass and field-renormalization constant, and the higher $\rho_{l}$´s and
$\sigma_{l}$´s would provide a discrete approximation to its dressing
cut. But in a nonperturbative context, where the zeroth-order
perturbative spectrum may suffer not only small shifts but also
qualitative changes, it is quite possible for a quantized field to give
rise to more than one particle, or none at all, and in such cases we
encounter a practical disadvantage of our rational approximation: at low
degrees $r$, there is no simple way of deciding whether a given pole
term in (\ref{propol}) approximates a genuine particle pole, or is
merely part of the approximation to the dressing cut. The answer can
only be found indirectly by observing the approximation as $r$
increases: all experience with rational approximants indicates that
poles approximating "true" poles then display reasonable convergence
of their positions and residues, whereas the poles approximating a cut
keep "hopping", and at the same time spread to ever higher values of
$s$, since at each successive $r$ the approximation invents a new set of
positions and strengths to achieve an improved scan of the continuous
part of the spectral function.

\qquad
We should again pause, since some of you may have taken offense at this
point with our apparent lightheartedness in talking about such
treacherous objects as particle masses and field renormalizations.
Haven´t we all learned that these things are ultraviolet divergent, and
can at best be traded for something experimental but not really
computed? Here you should recall that at this stage we are dealing with
a {\em zeroth quasi-perturbative order}, and at zeroth order in powers
of $\alpha$, renormalization constants are not yet divergent. The
ultraviolet divergences will catch up with us soon enough once we start
computing quasi-perturbative corrections, but at the zeroth order we are
only into reconstruction of the quantity that in the OPE context
appeared as eq. (\ref{invdnull}) -- a quantity which we noted is
RG-invariant and {\em finite}. At {\em this} level, things like the
nonperturbative mass shift of eq. (\ref{massi}) do indeed have finite
values. Even when it comes to computing quasi-perturbative corrections,
you should still remember the pattern we noted in eq. (\ref{opren}) --
it is only for the perturbative portion that we expect to have to
cancel net ultraviolet divergences. For the nonperturbative remainder,
what we are aiming at is a direct continuation-through-resummation of
the $n$ sum of eq. (\ref{opren}) with its ultraviolet-finite
$V_{n,\,j}^{(R)}$ coefficients -- we do not attempt to separately deal
with vacuum condensates and Wilson coefficients, whose ultraviolet
divergences then cancel by eq. (\ref{vren}). That the dynamical
equations
should provide a mechanism allowing for all these features is of course
not trivial, but nevertheless true, as the next section will show.

\qquad
Let us return to eqs. (\ref{propol}) and (\ref{ga2pol}) where we
exploited the tight linkage, inherent in the dimensionful character of
$\Lambda^{2}$, between the $\Lambda^{2}$ and $k^{2}$ dependences to
reinterpret our original $\Lambda^{2}$ approximation as an approximation
to the expected $k^{2}$-plane analyticity structure of our propagator or
self-energy. It is this reinterpretation that provides the best guide
for our next and slightly more difficult step: choosing
zeroth-order approximants for
higher-than-two-point vertex functions, consistent  with the two-point
form of eq. (\ref{ga2r0}). Here again we start with the simplest and
most important case of a single scalar three-point vertex function with
Bose symmetry, depending on the three squared Euclidean momenta
\renewcommand{\theequation}{3.30}
\begin{equation}
\label{3mom}
k_{1}^{2},\quad
k^{2}_{2},\quad
k^{2}_{3}
\end{equation}
of its three external legs, and being otherwise dimensionless with unit
perturbative limit,
\renewcommand{\theequation}{3.31}
\begin{equation}
\label{ga3pert}
\Gamma_{3}^{(0)\,pert}\,(\{k\})\,=\,1\,\,.
\end{equation}
(We have tacitly adopted the widespread convention of defining the
three-point vertex without a coupling factor $\overline{g}_{0}$, so what
actually enters into diagrams is $\overline{g}_{0}\,\Gamma_{3}$ rather
than $\Gamma_{3}$ alone.) This again is not precisely what occurs in a
realistic gauge theory -- a vertex for three $SU\,(N)$
gauge-vector-fields, for example, has mass dimension plus one rather
than zero, and that dimension is carried by the Lorentz basis tensors in
a decomposition of the full tensorial vertex. In the index notation
adopted for eq. (\ref{provec}) above, the beginnings of such a
decomposition would read like
\renewcommand{\theequation}{3.32}
\begin{equation}
\label{3tens}
\begin{array}{l}
\Big\{\Gamma_{3\,V,\,a\,b\,c}^{\mu\,\lambda\,\nu}\,(k_{1},\,k_{2},\,k_{3})\Big\}_{k_{1}\,+\,k_{2}\,+\,k_{3}\,=\,0}\\
\mbox{}\\
=\,f_{a\,b\,c}\,\Big\{\delta^{\lambda\,\nu}\,(k_{2}\,-\,k_{3})^{\mu}\,F_{0}^{(f)}\,
(k_{2}^{2},\,k_{3}^{2};\,k_{1}^{2})\\
\mbox{}\\
\hspace*{1.5cm}+\,\delta^{\nu\,\mu}\,(k_{3}\,-\,k_{1})^{\lambda}\,F^{(f)}_{0}\,
(k_{3}^{2},\,k_{1}^{2};\,k_{2}^{2})\\
\mbox{}\\
\hspace*{1.5cm}+\,\delta^{\mu\,\lambda}\,(k_{1}\,-\,k_{2})^{\nu}\,F_{0}^{(f)}\,
(k_{1}^{2},\,k_{2}^{2};\,k_{3}^{2})\,+\,...\,\Big\}\\
\mbox{}\\
+\,d_{a\,b\,c}\,\Big\{\delta^{\lambda\nu}\,(k_{2}\,-\,k_{3})\,
F^{(d)}_{0}\,(k_{2}^{2},\,k_{3}^{2};\,k_{1}^{2})\\
\mbox{}\\
\hspace*{1.5cm}+\,\delta^{\nu\,\mu}\,(k_{3}\,-\,k_{1})^{\lambda}\,
F^{(d)}_{0}\,(k^{2}_{3},\,k_{1}^{2};\,k_{2}^{2})\\
\mbox{}\\
\hspace*{1.5cm}+\,\delta^{\mu\,\lambda}\,(k_{1}\,-\,k_{2})^{\nu}\,F^{(d)}_{0}\,
(k_{1}^{2},\,k_{2}^{2};\,k_{3}^{2})\,+\,...\,\Big\}\,\,,
\end{array}
\end{equation}
with $f_{a\,b\,c}$ and $d_{a\,b\,c}$ denoting respectively the
antisymmetric and symmetric $SU\,(N)$ structure constants, which enter
here in their capacity as Clebsch-Gordans for the two possible ways of
coupling three adjoint-representation objects to a (global) gauge
singlet. What is dimensionless is then the $F_{0}^{(f)}$ and
$F_{0}^{(d)}$ invariant functions, but they will of course display only
the reduced Bose symmetry required to match the partial symmetry of
their associated tensor structures, and thus be symmetric or
antisymmetric only
with respect to the first two of their arguments. The Feynman rule
$\Gamma_{3\,V}^{(0)\,pert}$, as you will remember, has only the
$f_{a\,b\,c}$ terms of eq. (\ref{3tens}) with $F^{(f)}_{0}$´s replaced
by unity, so we have the perturbative limits
\renewcommand{\theequation}{3.33}
\begin{equation}
\label{f3pert}
\left(F_{0}^{(f)}\right)^{(0)\,pert}\,=\,1\,\,,\qquad
\left(F^{(d)}_{0}\right)^{(0)\,pert}\,=\,0\,\,,
\end{equation}
as $\Lambda\,\to\,0$.

\qquad
The dots in eq. (\ref{3tens}) in each case stand for a finite but
lengthy
list of other independent Lorentz-tensor structures (actually, no less
than eleven of them), forming triplets or singlets under the group of
permutations of the three vector-field legs, and each accompanied by its
own invariant function. It will emerge only much further down the road
(and be commented upon in our closing section) that in the zeroth
quasi-perturbative order we are considering here, this cumbersome
$(2\,\times\,14)$-term tensorial structure fortunately simplifies --
most of the Lorentz tensors not made explicit in eq. (\ref{3tens}) are
tensors involving three powers of momentum rather than one, and being
therefore accompanied by invariant functions of mass dimension minus two
rather than zero, and in the zeroth quasi-perturbative order these turn
out to be excluded by considerations of renormalizability. Here,
however, we should not let ourselves be sidetracked too much by these
purely kinematical and enumerative aspects of tensor structure, so you
will allow me to refer you rather summarily to the standard reference
\cite{BCU} on these aspects. For the issues we wish to address,
consideration of a purely schematic, scalar three-variable function
characterized by eq. (\ref{ga3pert}), or equivalently restriction to the
$F^{(f)}_{0}$ function of eq. (\ref{3tens}) with limit as in the first
of eqs. (\ref{f3pert}), will be quite adequate.

\qquad
For such a function, a rational approximant with respect to
$\Lambda^{2}$, of the form of eq. (\ref{lambfrac}), must have
$l^{'}\,=\,l$ by virtue of condition (\ref{klarge}), so again we need
only one labelling index. But otherwise, eq. (\ref{lambfrac}) alone is
now much too unspecific to represent a computationally useful
approximant. (All we can say at this stage about its $f_{n}$ and $g_{n}$
coefficients is that each must be of the form $(Q_{n}^{2})^{l\,-\,n}$,
where $Q_{n}^{2}$ is some dimension-two combination of the variables
(\ref{3mom}), times some dimensionless function of two independent
dimensionless combinations.)
It might be sufficient if the dynamical equations to be invoked were
differential or integral equations {\em with respect to the coupling},
which would correspond to the situation called "equational resurgence"
in mathematical theory \cite{ECAL} \cite{STSH}. But here the
equations at our disposal will be integral equations with respect to
{\em momentum} variables, and for an approximant to be capable of
exploiting these equations, it must then detail the momentum dependence
in a manner comparable to eq. (\ref{ga2r0}) for the one-variable case.
In
addition to the resurgent coupling structure, which we are resumming at
the $p\,=\,0$ level through functions (\ref{lambfrac}), we need some
additional input here. What can help us across this gap are
considerations of analyticity. From the general properties of QFT
amplitudes, we expect $\Gamma_{3}$ to be analytic (in the usual loose
sense, i.\,e. except for a denumerable set of poles and cuts) in the
complex plane of each of its variables (\ref{3mom}) at fixed values of
the other two, in a manner analogous to the single-$k^{2}$ analyticity
embodied in the spectral representation (\ref{kaelle}). It therefore
seems sensible, and not unduly restrictive, to postulate that our
$\Gamma_{3}$, {\em at least at the zeroth quasi-perturbative order},
fits into the form of a triple Cauchy-integral representation,
\renewcommand{\theequation}{3.34}
\begin{equation}
\label{trip}
\Gamma_{3}^{(p\,=\,0)}\,
(k_{1}^{2},\,k_{2}^{2},\,k_{3}^{3};\,\Lambda^{2})=\!
\int\limits_{{\cal{C}}_{1}}\!\!ds_{1}\!
\int\limits_{{\cal{C}}_{2}}\!\!ds_{2}\!
\int\limits_{{\cal{C}}_{3}}\!\!ds_{3}\!
\frac{\rho_{3}^{(0)}(s_{1},\,s_{2},\,s_{3})}{
(k_{1}^{2}+s_{1}\,\Lambda^{2})
(k^{2}_{2}+s_{2}\,\Lambda^{2})
(k^{2}_{3}+s_{3}\,\Lambda^{2})}.
\end{equation}
For a Bose-symmetric function, the spectral weight $\rho_{3}$, which may
include distribution-valued terms, should be symmetric in its three
variables, and the contours ${\cal{C}}_{i}$ in the three $s_{i}$
planes should be identical to one ${\cal{C}}$. Also the full function
must be real at real Euclidean momenta, so if ${\cal{C}}$ comprises any
sections away from the real $s$ axis (which for off-shell variables like
our $k_{i}^{2}$ cannot be excluded), these, as well as the associated
portions of $\rho_{3}^{(0)}$,
should only appear in complex-conjugate
pairs. We will always understand the form (\ref{trip}) to allow for
terms depending on less than three variables, by interpreting the
compact notation $\rho^{(0)}_{3}$ as
\renewcommand{\theequation}{3.35}
\begin{align}
\label{3spec}
\rho_{3}^{(0)}
&=
\xi_{0}\,\prod\limits_{i\,=\,1}^{3}\,[k_{i}^{2}\,\delta\,(s_{i})]\,+\,
\Lambda^{2}\,\sum\limits_{i\,<\,j}\,[k_{i}^{2}\,\delta\,(s_{i})][k_{j}^{2}\,
\delta\,(s_{j})]\,\rho_{3,\,1}\,(s_{k})\nonumber\\
&+
\Lambda^{4}\,\sum\limits_{k\,=\,1}^{3}\,[k_{k}^{2}\,\delta\,(s_{k})]\,\rho_{3,\,2}\,(s_{i},\,s_{j})\,+\,
\Lambda^{6}\,\rho_{3,\,3}\,(s_{1},\,s_{2},\,s_{3})\,\,,
\end{align}
with a positive constant $\xi_{0}$, and with $(i,\,j,\,k)$ cyclic in the
second and third terms. This function then has large-momentum limit
\renewcommand{\theequation}{3.36}
\begin{equation}
\label{xi0lim}
\Gamma^{(p\,=\,0)}_{3}\,\to\,\xi_{0}\qquad
(\text{all}\,\,k_{i}^{2}\,\gg\,\Lambda^{2})\,\,,
\end{equation}
and all our earlier comments on the $\zeta_{0}^{[r]}$ of eq.
(\ref{zetlim}) are again applicable.

\qquad
The main consequences of postulate (\ref{trip}), for our purposes, are
that if we now generate an approximant to $\Gamma^{(0)}_{3}$ by adopting
a discrete approximation analogous to (\ref{disc2}),
\renewcommand{\theequation}{3.37}
\begin{equation}
\label{disc3}
\begin{array}{c}
\rho_{3}^{(0)}\,(s_{1},\,s_{2},\,s_{3})\quad \longrightarrow\quad
\rho^{[r,\,0)}_{3}\,(s_{1},\,s_{2},\,s_{3})\,=\,
\sum\limits_{n_{1},\,n_{2},\,n_{3}\,=\,0}^{r}
\xi^{[r]}_{n_{1}\,n_{2}\,n_{3}}\\
\mbox{}\\
\times\,\,
\left[\delta_{n_{1}0}\,k^{2}_{1}+(1-\delta_{n_{1}0})\,\Lambda^{2}\right]\!\!\left[
\delta_{n_{2}0}\,k^{2}_{2}+(1-\delta_{n_{2}0})\,\Lambda^{2}\right]\!\!\left[
\delta_{n_{3}0}\,k^{2}_{3}+(1-\delta_{n_{3}0})\,\Lambda^{2}\right]\\
\mbox{}\\
\times\,\,
\delta\,\left(s_{1}\,-\,\tau_{n_{1}}^{[r]}\right)\,\delta\,\left(s_{2}\,-\,\tau_{n_{2}}^{[r]}\right)\,
\delta\,\left(s_{3}\,-\,\tau_{n_{3}}^{[r]}\right)\,\,,
\end{array}
\end{equation}
then automatically that approximant will (i) be rational not only with
respect to $\Lambda^{2}$ but also to the $k^{2}_{i}$, and (ii) possess a
{\em denominator factorizing} in the three $k_{i}^{2}$ which mirrors the
product structure of the Cauchy-integral denominator. (To accommodate
the structure (\ref{3spec}) we add the convention that
\renewcommand{\theequation}{3.38}
\begin{equation}
\label{taunull}
\tau_{0}^{[r]}\,=\,0\qquad
(\text{all}\,\,r)
\end{equation}
in eq. (\ref{disc3}).) For full Bose symmetry we should have
\renewcommand{\theequation}{3.39}
\begin{equation}
\label{xisym}
\xi_{n_{1}\,n_{2}\,n_{3}}^{[r]}\quad \text{completely symmetric
in}\quad (n_{1},\,n_{2},\,n_{3})\,\,.
\end{equation}
Therefore if we put the entire approximant over a common denominator we
arrive at the form
\renewcommand{\theequation}{3.40}
\begin{equation}
\label{ga3r0}
\Gamma^{[r,\,0)}_{3}\,(\{k_{i}^{2}\};\,\Lambda^{2})\,=\,
\frac{N_{3}^{[r]}\,(\{k_{i}^{2}\};\,\Lambda^{2})}
{\Delta_{3}^{[r]}\,(k_{1}^{2};\,\Lambda^{2})\,\Delta^{[r]}_{3}\,(k_{2}^{2};\,\Lambda^{2})\,
\Delta_{3}^{[r]}\,(k_{3}^{2};\,\Lambda^{2})}
\end{equation}
with a numerator polynomial of degree $3\,r$ in $\Lambda^{2}$  and of
degree $r$ in each of the squared momenta,
\renewcommand{\theequation}{3.41}
\begin{align}
\label{n3r}
N_{3}^{[r]}\,(\{k_{i}^{2}\};\,\Lambda^{2})
&=
\sum\limits_{\lambda\,=\,0}^{3\,r}\,\sum\limits_{m_{1},\,m_{2},\,m_{3}\,=\,0}^{r} \delta_{\lambda\,+\,m_{1}\,+\,m_{2}\,+\,m_{3},\,3\,r}\,
x_{m_{1}\,m_{2}\,m_{3}}^{[r]}\nonumber\\
&\times
(k_{1}^{2})^{m_{1}}\,(k_{2}^{2})^{m_{2}}\,(k_{3}^{2})^{m_{3}}\,(\Lambda^{2})^{\lambda}\,\,,
\end{align}
\renewcommand{\theequation}{3.42}
\begin{equation}
\label{ixsym}
x_{m_{1}\,m_{2}\,m_{3}}^{[r]}\quad \text{completely symmetric in}
\quad (m_{1},\,m_{2},\,m_{3})\,\,,
\end{equation}
\renewcommand{\theequation}{3.43}
\begin{equation}
\label{ixrrr}
x_{rrr}^{[r]}\,=\,\xi^{[r]}_{0}
\end{equation}
and with a denominator factorizing into three degree-$r$ polynomials
similar to (\ref{de2r}),
\renewcommand{\theequation}{3.44}
\begin{equation}
\label{de3r}
\Delta_{3}^{[r]}\,(k^{2};\,\Lambda^{2})\,=\,
\prod\limits_{s\,=\,1}^{r}\,(k^{2}\,+\,\tau_{s}^{[r]}\,\Lambda^{2})\,\,.
\end{equation}
In the notation of our original eq. (\ref{lambfrac}) for the rational
approximants, eq. (\ref{ga3r0}) therefore has
\renewcommand{\theequation}{3.45}
\begin{equation}
\label{lis3r}
l^{'}\,=\,l\,=\,3\,r\qquad (N\,=\,3)\,\,,
\end{equation}
rather than $l\,=\,r$ as in eq. (\ref{ga2r0}). Indeed it will turn out
that on this point analyticity has given us good guidance, since it will
be just this combination -- of $l\,=\,r$ for $\Gamma_{2}$ with
$l\,=\,3\,r$ for $\Gamma_{3}$, and more matching $l$´s for the
$\Gamma_{4}$ vertex --
that will prove itself capable of self-consistency in the
equations of motion.

\qquad
When all $k_{i}^{2}\,\gg\,\Lambda^{2}$, eq. (\ref{ga3r0}) gives an
expansion
\renewcommand{\theequation}{3.46}
\begin{equation}
\label{ga3rlak}
\begin{array}{c}
\Gamma^{[r,\,0)}_{3}\,(\{k_{i}^{2}\};\,\Lambda^{2})\,=\,\xi_{0}^{[r]}\\
\mbox{}\\
\displaystyle{
\times\,\left\{1\,+\,\left[\frac{x_{r,\,r,\,r-1}^{[r]}}{\xi_{0}^{[r]}}\,-\,
\left(\sum\limits_{s\,=\,1}^{r}\,\tau_{s}^{[r]}\right)\right]\,
\left(\frac{\Lambda^{2}}{k_{1}^{2}}\,+\,
\frac{\Lambda^{2}}{k_{2}^{2}}\,+\,
\frac{\Lambda^{2}}{k_{3}^{2}}\right)\,+\,...\right\}}
\end{array}
\end{equation}
that may be regarded as a three-point counterpart of eq.
(\ref{invdnull}) -- a three-variable OPE, restricted here to terms of
zeroth perturbation order with respect to $\alpha$.

\qquad
The foregoing discussion of $\Gamma^{[r,\,0)}_{2}$ and
$\Gamma_{3}^{[r,\,0)}$ approximants has, I hope, provided sufficient
illustration of the guidelines and technicalities that go into a
construction of zeroth quasi-perturbative orders as a systematic
sequence. You will therefore allow me to skip the corresponding
construction for a four-boson vertex -- the highest function for which
such a construction will be found necessary. With no less than six
independent kinematic invariants, and with a plethora of possible tensor
structures, this function requires a much lengthier but not much more
illuminating construction. Only much further along the way -- that is,
after having learned some lessons from trying to get these constructions
self-consistent in the equations of motion -- will it emerge that {\em
for the zeroth quasi-perturbative order}, $\Gamma_{4}^{[r,\,0)}$,
tensorial structures and functional dependences are strongly restricted,
so that the final approximants are still manageable. For the moment, it
may be more important to address your feelings that my remark about this
being the highest construct necessary sounds completely obscure. The
equations for correlation functions, in whatever variant, are always
hierarchically coupled -- so in order to establish a four-point
amplitude, aren´t we going to need a five-point one, and so on forever?
In other words, mustn´t we live in fear that unlike their perturbative
zeroth-order counterparts -- the Feynman rules $\Gamma_{N}^{(0)\,pert}$
-- our new "rules" will endlessly {\em proliferate}? It seems nontrivial
(at least to me) that such fears in fact should turn out to be
unfounded.
\subsection{Self-Consistency of the Generalized Feynman Rules}
\qquad
The best of the benefits from our $n$-resummed, quasi-perturbative
reconstruction strategy are yet to be reaped. They emerge once we insert
the matched set of zeroth-order approximants, as constructed in the last
section, into the interaction (loop) terms of exact equations of motion.
As announced, and as expected, these loop terms then start producing the
quasi-radiative corrections $\Gamma^{[r,\,p)}$, $p\,\geq\,1$, in what
is basically the first step in an iteration process around a
nonperturbatively improved starting point. But more importantly, we find
four pleasant surprises. First, despite the strongly nonperturbative
character of the $\Gamma^{[r,\,0)}$´s with their ubiquitous
$\Lambda$-scale terms, the loops turn out to be readily {\em calculable}
by the known methods of loop computation and regularization, simply due
to their rational structure in momenta. Second, {\em in addition} to
producing $p\,\geq\,1$ corrections, the loop terms prove themselves
capable of reproducing the nonperturbative parts of our zeroth-order
input, thus conferring true {\em self-consistency} on the improved
starting point. Third, while the $p\,\geq\,1$ corrections continue to
carry the ultraviolet loop divergences, the mechanism operative in the
self-reproduction sees to it that the nonperturbative zeroth-order
terms, the quantities $\Gamma^{[r,\,0)}$ -- $\Gamma^{(0)\,pert}$ at all
levels $r$, establish themselves in a {\em finite} manner, as one would
expect and demand from our consideration of the OPE. Fourth, the same
mechanism relieves us from our nightmare of proliferation of Feynman
rules, by {\em strictly} limiting the formation of zeroth-order proper
vertices to a small finite set of these functions. It is these features
together that seem to justify designation of this finite and
self-consistent set of $\Gamma^{[r,\,0)}$´s as {\em generalized Feynman
rules}.

\qquad
The system of exact equations of motion we are going to exploit is the
oldest known such system, the integral equations first established by
Dyson \cite{DYS4}, who derived them by Feynman-graph summation, and by
Schwinger \cite{SCHWI}, whose functional derivation made it plain that
they are in fact independent of any perturbative framework.
They are
referred to as the DS equations. Although they still were fundamental
field-theory tools for such researchers as Schwinger, Bogolyubov, and
Symanzik, they are rarely discussed in textbooks nowadays (two 1980´s
texts that give them more than passing mention are \cite{ITZ} and
\cite{RIV}), but I hope you may have had at least a superficial
encounter with them and will recognize a DS system (from its typical,
hierarchical structure) when you see one. For the DS equations of
non-abelian gauge theories, the standard original reference is
\cite{EIF}; condensed statements of their derivation and main properties
can also be found in the review articles \cite{ROW} and \cite{AVS},
which concentrate on exact numerical solutions -- another line of
research which our focus on coupling dependence will not allow us to
pursue in any detail.

\qquad
To collect the minimum of formal relations we will need to refer to, we
temporarily ignore the fact that we actually have to deal with
non-abelian gauge-theories (these being the only ones for which our
basic premise of asymptotic freedom is correct in the physical dimension
$D\,=\,4$) and write simplified equations for some generic field
$\varphi\,(x)$, using notations as in eq. (\ref{pathint}) above. The
generating functional of Euclidean correlation functions,
\renewcommand{\theequation}{3.47}
\begin{equation}
\label{genfu}
Z_{E}\,[J]\,=\,\frac{1}{{\cal{N}}}\,\int\,{\cal{D}}\,[\varphi]\,
\E^{-S_{E}\,[\varphi]\,+\,(J,\,\varphi)}\,\,,
\end{equation}
with normalization ${\cal{N}}$ such that $Z_{E}\,[0]\,=\,1$, is
postulated to be invariant under simultaneous
infinitesimal translations of all its integration variables,
\renewcommand{\theequation}{3.48}
\begin{equation}
\label{fitra}
\varphi\,(x)\quad \longrightarrow\quad
\varphi^{'}\,(x)\,=\,\varphi\,(x)\,+\,\delta\,\varphi\,(x)\qquad
(\text{all}\,\,x)\,\,,
\end{equation}
in the sense that ${\cal{D}}\,[\varphi^{'}]\,=\,{\cal{D}}\,[\varphi]$.
(For the mathematically minded, it is not unproblematic to use
translation invariance for an infinite-dimensional measure, but since
the procedure leads quickly and elegantly to equations of motion that
can also be establish\-ed independently by operator techniques, we will
accept it here as a formal generating tool.) This gives the statement
that
\renewcommand{\theequation}{3.49}
\begin{equation}
\label{divan}
\frac{1}{{\cal{N}}}\,\int\,{\cal{D}}\,[\varphi]\,
\frac{\delta}{\delta\,\varphi\,(x)}\,\left\{
\E^{-S_{E}\,[\varphi]\,+\,(J,\,\varphi)}\right\}\,=\,0\,\,,
\end{equation}
which may be written as
\renewcommand{\theequation}{3.50}
\begin{equation}
\label{exDS}
\Big\langle
-\frac{\delta\,S_{E}\,[\varphi]}{\delta\,\varphi\,(x)}\,+\,
J\,(x)\Big\rangle_{J}\,=\,0\,\,,
\end{equation}
using the expectation-value notation
\renewcommand{\theequation}{3.51}
\begin{equation}
\label{apex}
\langle A\rangle_{J}\,=\,\frac{1}{{\cal{N}}}\,
\int\,{\cal{D}}\,[\varphi]\,A\,[\varphi]\,\E^{-S_{E}\,[\varphi]\,+\,(J,\,\varphi)}
\end{equation}
for functionals $A$ of $\varphi$. Further rewriting along familiar lines
puts this into the form
\renewcommand{\theequation}{3.52}
\begin{equation}
\label{fuDS}
\left\{\left(\frac{\delta\,S_{E}\,[\varphi]}{\delta\,\varphi\,(x)}\right)_{\varphi\,\to\,\frac{\delta}{\delta\,J}}\,-\,
J\,(x)\,\right\}\,Z_{E}\,[J]\,=\,0\,\,,
\end{equation}
a relation sometimes called the master DS equation. The equations for
the 1-, 2-, 3- ... point correlation functions then follow from this
relation the same way that the functions themselves follow from
(\ref{genfu}): by functionally differentiating 0, 1, 2, ... times with
respect to $J$ and putting $J\,=\,0$. For example, after one
differentiation with respect to $J\,(y)$ one arrives at
\renewcommand{\theequation}{3.53}
\begin{equation}
\label{2peq}
\Big\langle\varphi\,(y)\,\frac{\delta\,S_{E}\,[\varphi]}{\delta\,\varphi\,(x)}\Big\rangle_{0}\,=\,
\delta^{D}\,(y\,-\,x)\,\,,
\end{equation}
and if $S_{E}$, as usual, consists of a bilinear kinetic term
\renewcommand{\theequation}{3.54}
\begin{equation}
\label{sfree}
\varphi\,(x)\,K_{x}\,\varphi\,(x)
\end{equation}
with symmetric, positive-definite differential operator $K_{x}$, plus
tri- and quadrilinear interaction terms collected in a "potential"
$V\,(\varphi\,(x))$, then (\ref{2peq}) will be an integro-differential
equation for the two-point function $G_{2}\,(y,\,x)$ of eq.
(\ref{xcorr}) of the form
\renewcommand{\theequation}{3.55}
\begin{equation}
\label{2pdi}
K_{x}\,G_{2}\,(y,\,x)\,=\,\delta^{4}\,(y\,-\,x)\,-\,
\langle\varphi\,(y)\,V^{'}\,(\varphi\,(x))\rangle_{0}\,\,,
\end{equation}
coupling that function to three- and four-point functions of the special
forms
\renewcommand{\theequation}{3.56}
\begin{equation}
\label{dscomp}
\langle \varphi\,(y)\,(\varphi\,(x))^{2}\rangle_{0}\,\,,\qquad
\langle\varphi\,(y)\,(\varphi\,(x))^{3}\rangle_{0}\,\,.
\end{equation}
It is this characteristic coupling to the next higher functions that is
referred to as hier\-archical. Upon converting the equations from $x$
space to momentum space, the identifications of two or three points in
the functions (\ref{dscomp}) will turn into single or double
internal-momentum integrations (loops). Finally, one transcribes the
equations from full correlation functions $G_{N}$ to connected,
amputated, and 1 PI functions $\Gamma_{N}$ to give them their most
transparent form.

\qquad
In Fig. 3.1 I have attempted to write diagrammatically the three lowest
levels -- the equations for two-, three- and four-point vertices -- of
what might be called a generic DS system. (These three equations, for
our purpose, will actually turn out to bear a crucial intrinsic
distinction, being the equations for the superficially
ultraviolet-divergent functions). For each equation, Fig. 3.1 selects a
set of typical and recurrent features: in the first line of each, the r.
h. s. starts out with an "inhomogeneous" term representing a
zeroth-order perturbative vertex $\Gamma^{(0)\,pert}_{N}$ (terms 2\,A,
3\,A, 4\,A). Next on each r. h. s is a loop term containing no vertex
function higher than the one on the l. h. s., so that with such an
interaction term alone the equation would close upon itself (terms
2\,B, 3\,B, 4\,B). Spoiling this closure is, on the second line of each
equation, a loop term featuring the next higher vertex function, which
brings in the hierarchical coupling (terms 2\,C, 3\,C, 4\,C).

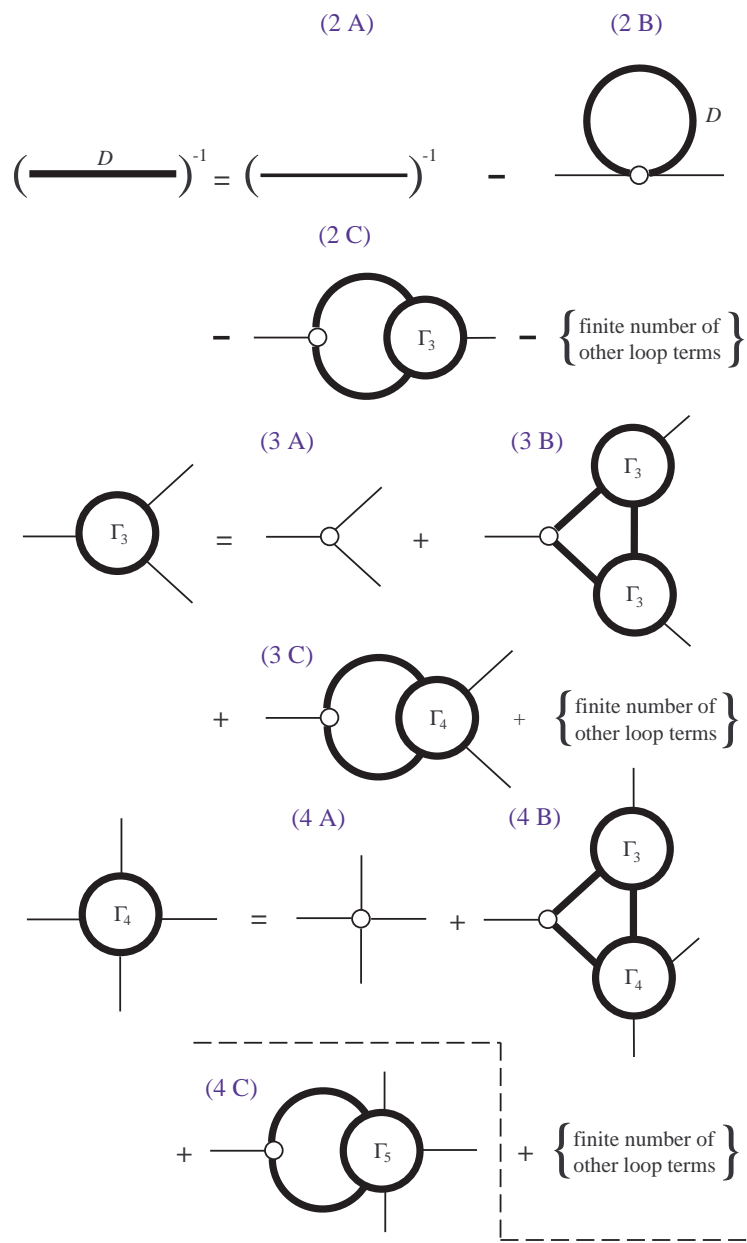
\begin{center}
\epsfig{file=fig31.eps,width=10.5cm}\\
$\begin{array}{ll}
\text{\bf Fig. 3.1} \quad &
\text{Beginnings of a schematic Dyson-Schwinger hierarchy in}\\
& \text{diagrammatic form. Small circles prepresent bare vertices;}\\
& \text{heavy lines, exact propagators; large circles, exact proper}\\
& \text{vertices as specified. Dashed line in $\Gamma_{4}$ equation
indicates}\\
& \text{limit beyond which loop terms do no more contribute to}\\
& \text{generation of zeroth quasi-perturbative order.}
\end{array}$
\end{center}

Also
present in most cases, but not made explicit, are a finite number of
other loop terms, which in the case of non-abelian gauge-boson
vertices will include two-loop terms arising from the second type of
function
in (\ref{dscomp}). By suitably interpreting the dressed propagators
(heavy lines) and dressed vertices (large circles) in their explicit
terms, and suitably filling in the non-explicit ones, these three
graphical equations may be viewed as encompassing the lower levels of
practically every known DS hierarchy.
\qquad
To guard against misunderstanding, I must interject a remark here.
Because of the hierarchical coupling, DS systems, like their
non-relativistic counterparts in statistical mechanics, are usually
solved by "decoupling" approximations whereby the next-higher vertices
in some equation are expressed in terms of lower ones, thus closing the
system.
Although these decoupling approximations are efficient devices for
generating partial resummations, and in my opinion do not deserve the
low esteem in which they seem to be held by most contemporary particle
theorists, it must be emphasized that what we are going to study in this
section -- the self-consistency of nonperturbatively modified zeroth
orders in a quasi-perturbative resurgence reconstruction -- has nothing
to do with decoupling approximations. The one "decoupling" that will
occur here will be both automatic and exact, and will arise from the
basic property of renormalizable theories of having only a limited
number of superficially divergent correlation functions. Decoupling
approximations may be useful at later stages, particularly when dealing
with the equations for superficially convergent higher amplitudes of
which our Fig. 3.1 stops short, but for the purpose at hand, they play
no role whatsoever.

\qquad
The common form under which all the equations of Fig. 3.1 may be
subsumed is
\renewcommand{\theequation}{3.57}
\begin{equation}
\label{phiDS}
\Gamma_{N}\,=\,\Gamma^{(0)\,pert}_{N}\,+\,
\overline{g}_{0}^{\,2}\,\Phi_{N}\,[\Gamma]\qquad
(N\,=\,2,\,3,\,4)\,\,,
\end{equation}
which makes explicit the fact that all interaction terms carry at least
one power of the squared bare coupling, $\overline{g}^{\,2}_{0}$. The DS
interaction functional, $\Phi_{N}$, consists of loop integrals over
integrands involving all vertices $\Gamma$ up to $\Gamma_{N\,+\,1}$ (or
$\Gamma_{N\,+\,2}$ in the presence of quartic interactions), and is
nonlinear, as is directly visible from the graphs of Fig. 3.1.
Straightforward iteration of eqs. (\ref{phiDS}) around the zeroth
perturbative orders is one way of generating the perturbation series
(\ref{pertex}), but in such an iteration, in which at every stage one
evaluates integrals over functions known from previous steps, one never
fully faces the true challenge of (\ref{phiDS}), which is a {\em
self-consistency} problem: find a set of $\Gamma_{N}$´s such that, when
inserted into the $\Phi_{N}$ functionals, they are capable of
reproducing themselves. Nonperturbative treatments, along whatever
lines, must address that challenge: to determine, in mathematical terms,
a fixed point of the complicated mapping defined by the r. h. s. of eqs.
(\ref{phiDS}). In our specific context, not the least part of that
challenge is to establish from eqs. (\ref{phiDS}) the nonperturbative
parts of the zeroth-order input -- the terms
\renewcommand{\theequation}{3.58}
\begin{equation}
\label{gnonp}
\Gamma_{N}^{[r,\,0)}\,-\,\Gamma^{(0)\,pert}_{N}\qquad
(r\,=\,0,\,1,\,2,\,...\,)
\end{equation}
which carry powers $\Lambda^{n}$ with $n\,\geq\,1$ but no powers
$(g^{2})^{p}$ with $p\,\geq\,1$ -- in spite of the fact that the
interaction terms always have at least one power of
$\overline{g}^{\,2}_{0}$ in front. For this to be possible, the
interaction functionals $\Phi_{N}$ must obviously be capable of
producing terms with factors of $1\,/\,g^{2}$ from sufficiently
nonperturbative input, and indeed that is what they will be seen to be
doing below. (One might say that the basic "mistake" of perturbation
theory is to miss precisely this possibility.)

\qquad
Besides a multitude of aspects of DS systems that will have to go
unmentioned in the restricted time and space of these lectures (a few
will be taken up in our closing section), there is one that is so rarely
discussed in the literature, and so intricately connected to
renormalizable ultraviolet-divergence structure, that I feel I should
comment on it before plunging into the technicalities of
self-consistency. In a theory with up to quartic terms in its action
functional $S_{E}\,[\varphi]$, the "master DS equation" (\ref{fuDS}),
of
which our individual equations (\ref{phiDS}) are functional moments, is
a functional-differential equation of the third order for the generating
functional $Z_{E}\,[J]$. From the
analogy with differential equations, one would therefore expect that the
first three functional Taylor coefficients in the expansion
\renewcommand{\theequation}{3.59}
\begin{equation}
\label{futayl}
Z_{E}\,[J]\,=\,
\sum\limits_{N\,=\,0}^{\infty}\,\frac{1}{N!}\,\int\,d^{D}\,x_{1}\,...\,
\int\,d^{D}\,x_{N}\,G_{N}\,(x_{1}\,...\,x_{N})\,J\,(x_{1})\,...\,
J\,(x_{N})
\end{equation}
will be left undetermined by the equation and will have to be specified
as "initial conditions" (for the evolution of the functional away from
$J\,=\,0$). The $N\,=\,0$ coefficient, the vacuum Norm
$G_{0}\,=\,Z_{E}\,[0]\,=\,\braket{0}{0}$, is fixed at unity by
normalization, and the $N\,=\,1$ coefficient, the single-field
expectation $G_{1}\,=\,\braket{0|\varphi\,(x)}{0}$, vanishes in theories
without elementary scalar fields (or indeed remains an empirical
parameter in Higgs-field theories, but then that´s what we have been
accustomed to). But the idea that the $N\,=\,2$ coefficient, the
propagator $G_{2}\,(x_{1}\,-\,x_{2})$, should have to be put in as an
initial datum sounds positively bewildering. I must refer you to the
excellent discussion in sect. 1.4 of ref. \cite{RIV}, which uses the
so-called static ultra-local model of Caianiello and Scarpetta
\cite{CAI}, for insight into how the perturbation series seems to evade
this problem -- basically, it steers you tacitly into a particular
solution in which the initial datum $G_{2}$ is assigned a special value,
while displacing the freedom of choosing initial conditions into its
undetermined "escaping" terms. (Again, analogs of this have been known
in Écalle´s theory of resurgent solutions to nonlinear differential
equations, where the perturbation series is referred to as the "formal
solution"; see the "epilogue" of \cite{CNP}). In any truly
nonperturbative treatment, however, the initial-conditions problem is
bound to reappear, and indeed the numerical studies of propagator
equations described in \cite{AVS} have met with indeterminacies
presumably related to this question. In our context, the problem will
resurface in a very clear-cut and interesting way: while the standard DS
system will produce self-consistency conditions determining the
polynomials $N_{2}^{[r]}\,(k^{2},\,\Lambda^{2})$ of eq. (\ref{ga2r0})
above, which at each level $r$ encode the particle-pole positions of the
propagator and the mesh points for the discretization of its continuum
integral through their root-factor decompositions (\ref{profac}),
conditions that would determine the coefficients of the polynomials
$\Delta_{2}^{[r]}\,(k^{2},\,\Lambda^{2})$ of eqs. (\ref{de2r}) and
(\ref{delfac}) -- that is, the weights to be assigned to those poles and
mesh points -- will be found lacking.
When reflecting on the origins of this indeterminacy, both in DS systems
in general and in our more special problem of zeroth quasi-perturbative
orders, we are led back to looking at the OPE in e. g. the form of eq.
(\ref{opela}). We noted that the coefficients $U_{n}^{(R)}$ appearing
there contain information about renormalized vacuum expectations of
composite operators of mass dimension $n$ -- singular products of up to
$n$ elementary fields at the same space-time point. In the ordinary DS
system, the loops do represent {\em some} of these composites --
the quadratic and cubic composites of eq. (\ref{dscomp}) -- but not
more, while the OPE, when following it to higher $n$´s, obviously
carries more information about higher composites. The natural conjecture
then is that the indeterminacy revealed by the initial-conditions
argument is related to insufficient composite-operator information in
the ordinary DS equations. To supply the missing information in a way
that is compatible with, and forms a natural extension of, the ordinary
DS system, one may therefore think of using DS equations with increasing
numbers of coalescing points in $x$ space, which in momentum space would
correspond to integrating out increasing numbers of the independent
momenta. On the simplest level, for example, the $y\,\to\,x$ limit of
eq. (\ref{2peq}) would give
\renewcommand{\theequation}{3.60}
\begin{equation}
\label{2pcond}
\Big\langle \varphi\,(x)\,\frac{\delta\,S_{E}\,[\varphi]}{\delta\,
\varphi\,(x)}\Big\rangle_{0}\,=\,0
\end{equation}
(remembering that $\delta^{D}\,(0)\,=\,0$ in dimensional
regularization), which in momentum space will turn into an equation for
the integrated quantity
\renewcommand{\theequation}{3.61}
\begin{equation}
\label{2pint}
\int\,\frac{d^{D}\,k}{(2\,\pi)^{D}}\,\left[D^{(0)\,pert}\,(k)\right]^{-1}\,
D\,(k)\,\,,
\end{equation}
rather than for $D\,(k)$ itself. (This equation is a single-number
condition, and at the order $r\,=\,1$ of rational approximation, where
only a single coefficient $u_{2}^{[1]}\,=\,\eta_{1}^{[1]}$ needs to be
determined in the polynomial $\Delta^{[1]}_{2}$, it would already
suffice for completing the "initial" conditions). On the next level,
coalescence of two of the three spacetime points in a three-point
equation would lead to an equation for the quantity
\renewcommand{\theequation}{3.62}
\begin{equation}
\label{3pint}
\int
\frac{d^{D}\,q}{(2\,\pi)^{D}}\,\tilde{G}_{3}\,\left(-p,\,\frac{1}{2}\,p\,+\,q,\,\frac{1}{2}\,p\,-\,q\right)\,\,,
\end{equation}
which is still an equation for a function of $p$, and coalescing all
three points would give the integral of this equation over $p$, which
like the equation for (\ref{2pint}) is a single-number condition.
Equations of this kind have been termed "equation-of-motion
condensates" in some OPE studies \cite{AHL}, and have been used as
checks
on Wilson-coefficient calculations, but in our context they stand to
assume a more fundamental role. In ordinary integral-equations theory
with well-behaved functions and convergent integrals, once you have
solved an equation, the integration of that equation over its domain of
definition will not, of course, generate new information. In a theory
with renormalizable ultraviolet divergences, the situation is different
-- composite-operator divergences require new renormalization steps
beyond the usual field and parameter renormalizations, and the
above-described {\em momentum-integrated DS equations} therefore do
bring in new information. In these lectures we will take the view that
they are an adequate means of completing the initial conditions.

\qquad
When taking up self-consistency in earnest, it is again best to first
concentrate on a partial task in which the basic mechanisms can be seen
at work in an exemplary way. Let us therefore focus on the first,
inverse-propagator equation of Fig. 3.1, and let us single out, among
the interaction terms on its r. h. s., the one-DS-loop diagram (2\,C),
which is the simplest one to bring in the crucial hierarchical coupling.
We take the $D^{-1}$ on the l. h. s. to be a scalar function of $k^{2}$,
$k$ being the external momentum flowing through the equation, with
perturbative limit $(D^{(0)\,pert})^{-1}$ given by
(\ref{ga2pert}), and with zeroth quasi-perturbative order given, at
level $r$ of rational continuation, by (\ref{ga2r0}) or (\ref{ga2pol}).
The equation reads, in the notation of (\ref{phiDS}),
\renewcommand{\theequation}{3.63}
\begin{equation}
\label{DS2p}
D^{-1}\,(k^{2})\,=\,k^{2}\,-\,\overline{g}_{0}^{2}\,\left\{
\Phi_{2\,C}\,[\Gamma;\,k^{2})\,+\,[\text{other loop terms}]\right\}\,\,,
\end{equation}
where, remembering our parenthetical remark after eq. (\ref{ga3pert})
\renewcommand{\theequation}{3.64}
\begin{equation}
\label{phi2C}
\begin{array}{c}
\displaystyle{
\Phi_{2\,C}\,[\Gamma;\,k^{2})\,=\,\frac{1}{2}\,\int\,\frac{
d^{D}\,q}{(2\,\pi)^{D}}\,\Gamma^{(0)\,pert}\,(q,\,-k,\,k\,-\,q)}\\
\mbox{}\\
\times\,\,D\,(q^{2})\,D\,((k\,-\,q)^{2})\,\Gamma_{3}\,(q\,-\,k,\,k,\,-q)\,\,.
\end{array}
\end{equation}
In order to get mass dimensions (and the associated degrees of
divergence) right without having to delve again into the
paraphernalia
of gauge-theory tensor structure, we will patch up things a little and
supply both $\Gamma_{3}$ and $\Gamma^{(0)\,pert}_{3}$ with an extra unit
of momentum dimension in an {\em ad hoc} fashion, such that their
product develops an extra contraction factor
\renewcommand{\theequation}{3.65}
\begin{equation}
\label{adhoc}
[(k\,-\,q)\,-\,q]^{\mu}\,[-q\,-\,(q\,-\,k)]^{\mu}\,=\,
(2\,q\,-\,k)^{2}\,\,.
\end{equation}
(Never mind this little dose of patchwork -- it will do all the
essential things right. If you feel uncomfortable with it, you may do an
instructive piece of homework by setting up, say, the gluon-loop
contribution to the gluon´s transverse $D^{-1}_{T}$ as defined by
(\ref{provec}), with full tensor structure for $\Gamma_{3-gluon}$ as in
eq. (\ref{3tens}), working in Landau gauge to avoid mixing in $D_{L}$´s,
and grind out the tensor contractions. You will find that our ad hoc
factor does things qualitatively right for the ultraviolet-divergent
terms, which turn out to be what matters here).

\qquad
In a first iteration, we now substitute $D$ and $\Gamma_{3}$ by their
zeroth quasi-perturbative orders, using eq. (\ref{ga3r0}) at level $r$
of rational continuation, with perturbative limits given by
(\ref{ga2pert}) and (\ref{ga3pert}). That is, we decide -- in order to
keep all inessential complications out of the following discussion of
the self-consistency procedure -- to only look for solutions with
\renewcommand{\theequation}{3.66}
\begin{equation}
\label{unlim}
\zeta_{0}^{[r]}\,=\,1\,\,,\qquad
\xi^{[r]}_{0}\,=\,1\,\,,
\end{equation}
which represents a kind of boundary condition. Indeed we remember from
sect. 2.4 that there may be at least two different meanings, related by
finite rescaling, to the concept of quasi-perturbative zeroth-order, and
we expect that the DS functionals in principle can produce both (and
perhaps more), so a supplemental condition of some kind will be
necessary. Eq. (\ref{DS2p}) assumes the form
\renewcommand{\theequation}{3.67}
\begin{equation}
\label{DS2e}
\begin{array}{c}
[D^{-1}\,(k^{2})]^{[r,\,0)}\,+\,g^{2}\,[D^{-1}\,(k^{2})]^{[r,\,1)}\,+\,O\,(g^{4})\\
\mbox{}\\
=\,k^{2}\,-\,\overline{g}^{\,2}_{0}\,\left\{
\Phi_{2\,C}\,[\Gamma^{[r,\,0)};\,k^{2})\,+\,
[\text{other loop terms}]^{[r,\,0)}\,+\,O\,(g^{2})\right\}\,\,,
\end{array}
\end{equation}
where
\renewcommand{\theequation}{3.68}
\begin{equation}
\label{ph2Cr}
\begin{array}{c}
\displaystyle{
\Phi_{2\,C}\,[\Gamma^{[r,\,0)};\,k^{2})\,=\,\Biggl\{
\frac{1}{2}\int\frac{d^{D}\,q}{(2\,\pi)^{D}}\,(2\,q\,-\,k)^{2}\,
\frac{\Delta^{[r]}_{2}\,(q^{2},\,\Lambda^{2})\,\Delta^{[r]}_{2}\,((
k\,-\,q)^{2},\,\Lambda^{2})}{N_{2}^{[r]}\,(q^{2},\,\Lambda^{2})\,N_{2}^{[r]}\,
((k\,-\,q)^{2},\,\Lambda^{2})}}\\
\mbox{}\\
\displaystyle{
\times\,\,
\frac{N_{3}^{[r]}\,(q\,-\,k)^{2},\,k^{2},\,q^{2};\,\Lambda^{2})}
{\Delta_{3}^{[r]}\,((q\,-\,k)^{2},\,\Lambda^{2})\,\Delta_{3}^{[r]}\,(q^{2},\,\Lambda^{2})}\Biggr\}\,
\frac{1}{\Delta_{3}^{[r]}\,(k^{2},\,\Lambda^{2})}\,\,.}
\end{array}
\end{equation}
At this point, to shorten the discussion, we anticipate a simplifying
fact that will soon be obvious: the rational-approximant poles at
$k^{2}\,=\,-u^{[r]}_{2\,s}\,\Lambda^{2}\,(s\,=\,1\,...\,r$), which
according to eq. (\ref{ga2r0}) we are postulating for the l. h. s. of
eq. (\ref{DS2e}), can at zeroth quasi-perturbative order arise only
from the last factor $1\,/\,\Delta_{3}^{[r]}$ of (\ref{ph2Cr}) but not
from the bracketed integral. Therefore we immediately have a first set
of self-consistency conditions, which in the notation of eq.
(\ref{de3r}) read
\renewcommand{\theequation}{3.69}
\begin{equation}
\label{utau}
\tau_{s}^{[r]}\,=\,u^{[r]}_{2\,s}\qquad
(s\,=\,1,\,...\,r)\,\,,
\end{equation}
and which imply that quite simply
\renewcommand{\theequation}{3.70}
\begin{equation}
\label{d3d2}
\Delta_{3}^{[r]}\,(k^{2})\,=\,
\Delta_{2}^{[r]}\,(k^{2})\qquad ({\rm all}\,\,r)\,\,.
\end{equation}
We remark, without elaborating, that this situation will repeat itself
when considering the three-point equation, to the effect that
\renewcommand{\theequation}{3.71}
\begin{equation}
\label{d4d3}
\Delta_{4}^{[r]}\,(k^{2})\,=\,\Delta_{3}^{[r]}\,(k^{2})\qquad
({\rm all}\,\,r)\,\,,
\end{equation}
where $\Delta_{4}^{[r]}$ is the polynomial appearing in denominators of
level-$r$ approximants to the invariant functions of the 4-point vertex.
In other words, the vertex-denominator
(or propagator-numerator)
polynomials $\Delta_{2}^{[r]}$, and the vertex-function branch cuts
which they represent, turn out to be a common element in all
zeroth-order,
basic vertex functions. Depending on viewpoint, this element is
"handed up" or "handed down" the ladder of DS equations by the
hierarchical coupling, which is seen here playing a prominent role.

\qquad
For our immediate task, this brings substantial simplification since
$\Delta$´s now cancel in the integrand of (\ref{ph2Cr}). For the
remaining fraction, it is straightforward to write a partial-fraction
decomposition, first with respect to the loop variables $q^{2}$ and
$(k\,-\,q)^{2}$ and then also with respect to the external $k^{2}$. The
former step reduces (\ref{ph2Cr}) to a linear combination of the basis
integrals
\renewcommand{\theequation}{3.72}
\begin{equation}
\label{bint}
I^{[r]}_{m,\,n}\,(k^{2},\,\Lambda^{2})\,=\,\int\,\frac{d^{D}\,q}{(2\,\pi)^{D}}\,
\frac{(2\,q\,-\,k)^{2}}{[(q\,-\,k)^{2}\,+\,\sigma_{m}^{[r]}\,\Lambda^{2}]\,
[q^{2}\,+\,\sigma_{n}^{[r]}\,\lambda^{2}]}\qquad
(m,\,n\,=\,1\,...\,r\,+\,1)
\end{equation}
These may be evaluated in $D\,=\,4\,-\,2\,\varepsilon$ by the standard
techniques, the result being
\renewcommand{\theequation}{3.73}
\begin{equation}
\label{vint}
I_{m,\,n}^{[r]}\,(k^{2},\,\Lambda^{2})\,=\,-\frac{1}{(4\,\pi)^{2}}\,
\left[A_{m,\,n}^{[r]}\,(k^{2},\,\Lambda^{2};\,\varepsilon)\cdot
k^{2}\,+\,B_{m,\,n}^{[r]}\,(k^{2},\,\Lambda^{2};\,\varepsilon)\cdot\Lambda^{2}\right]\,\,.
\end{equation}
The functions $A_{m,\,n}^{[r]}$ and $B_{m,\,n}^{[r]}$ are given by the
Feynman-parameter integrals
\renewcommand{\theequation}{3.74}
\begin{equation}
\label{amn}
\begin{array}{c}
\displaystyle{
A_{m,\,n}^{[r]}\,(k^{2},\,\Lambda^{2};\,\varepsilon)
\,=\,
\int\limits_{0}^{1}\,d\,z\,\left[\Sigma_{m,\,n}^{[r]}\,(k^{2},\,\Lambda^{2};\,z)\right]^{-\varepsilon}}\\
\mbox{}\\
\displaystyle{
\times\,\,
\left\{[12\,z\,(1\,-\,z)\,-\,1]\,\left[\frac{1}{\varepsilon}\,-\,
\gamma\,+\,\ln\,4\,\pi\right]\,+\,4\,z\,(1\,-\,z)\,+\,O\,(\varepsilon)\right\}\,\,,}
\end{array}
\end{equation}
\renewcommand{\theequation}{3.75}
\begin{equation}
\label{bmn}
\begin{array}{c}
\displaystyle{
B_{m,\,n}^{[r]}\,(k^{2},\,\Lambda^{2};\,\varepsilon)\,=\,
\int\limits_{0}^{1}\,d\,z\,\left[\Sigma_{m,\,n}^{[r]}\,(k^{2},\,
\Lambda^{2};\,z)\right]^{-\varepsilon}}\\
\mbox{}\\
\displaystyle{
\times\,\,
\left\{8\,[z\,\sigma_{m}^{[r]}\,+\,(1\,-\,z)\,\sigma_{n}^{[r]}]\,\left[
\frac{1}{\varepsilon}\,-\,\gamma\,+\,\ln\,4\,\pi\,+\,\frac{1}{2}\right]\,+\,
O\,(\varepsilon)\right\}\,\,,}
\end{array}
\end{equation}
and near $D\,=\,4$ have weak, logarithmic dependences on $k^{2}$ and
$\Lambda^{2}$ through the $(-\varepsilon)$-th power of the quantity
\renewcommand{\theequation}{3.76}
\begin{equation}
\label{simn}
\Sigma_{m,\,n}^{[r]}\,(k^{2},\,\Lambda^{2};\,z)\,=\,z\,(1\,-\,z)\,k^{2}\,+\,
\left[z\,\sigma_{m}^{[r]}\,+\,(1\,-\,z)\,\sigma_{n}^{[r]}\right]\,\Lambda^{2}\,\,.
\end{equation}
(To maintain correct mass dimensions, we have stopped short of expanding
the non-integer power). After collecting the debris of the
partial-fraction decompositions, eq. (\ref{DS2e}) now assumes the form
\renewcommand{\theequation}{3.77}
\begin{equation}
\label{DS2r}
\begin{array}{l}
[D^{-1}\,(k^{2})]^{[r,\,0)}\,+\,g^{2}\,[D^{-1}\,(k^{2})]^{[r,\,1)}\,+\,O\,(g^{4})\\
\mbox{}\\
\displaystyle{
=\,k^{2}\,+\,\left(\frac{\overline{g}_{0}}{4\,\pi}\right)^{2}\,\Biggl\{
r_{0}^{[r]}\,(k^{2},\,\Lambda^{2};\,\varepsilon)\,\cdot\,k^{2}\,+\,
r_{1}^{[r]}\,(k^{2},\,\Lambda^{2};\,\varepsilon)\,\cdot\,\Lambda^{2}}\\
\mbox{}\\
\displaystyle{
+\,\sum\limits_{s\,=\,1}^{r}\,\frac{r_{2\,s\,+\,1}^{[r]}\,(k^{2},\,\Lambda^{2};\,\varepsilon)\,\Lambda^{4}}
{k^{2}\,+\,u_{2\,s}\,\Lambda^{2}}\,+\,[\text{other loop
terms}]^{[r,\,0)}\,+\,O\,(g^{2})\Biggr\}\,\,,}
\end{array}
\end{equation}
where the quantities $r_{i}^{[r]}$ have structure
\renewcommand{\theequation}{3.78}
\begin{equation}
\label{rico}
r_{i}^{[r]}\,=\,\sum\limits_{m,\,n\,=\,1}^{r\,+\,1}\,\left[
a_{i,\,m,\,n}^{[r]}\,\cdot\,A_{m,\,n}^{[r]}\,(k^{2},\,\Lambda^{2};\,
\varepsilon)\,+\,b_{i,\,m,\,n}^{[r]}\,\cdot\,
B_{m,\,n}^{[r]}\,(k^{2},\,\Lambda^{2};\,\varepsilon)\right]\,\,,
\end{equation}
and therefore weak $k^{2}$ and $\Lambda^{2}$ dependence near $D\,=\,4$.
About the dimensionless coefficients $a_{i,\,m,\,n}^{[r]}$ and
$b^{[r]}_{i,\,m,\,n}$, we need to know no more than that they depend
linearly on the level-$r$, three-point vertex coefficients $x^{[r]}$,
and in general nonlinearly on the set of parameters of the inverse
propagator itself (the $u_{i}^{[r]}$ of eq. (\ref{ga2pol}), or any
equivalent set). With eq. (\ref{DS2r}), we are ready for posing the
self-consistency question. It is, by the way, no later than at this
stage that one realizes that condition (\ref{d3d2}) is a necessity.

\qquad
This is, once more, a moment to pause -- two observations are worth
mentioning. First, while our integrand, with its $\Lambda^{2}$ terms all
over the place, is manifestly a strong\-ly nonperturbative object which
has no power-series expansion in the coupling at all, nothing has kept
us from computing the loop by the standard techniques. These techniques
are sometimes, misleadingly, referred to as "`perturbative", which
sounds as if they were applicable only in pure perturbation-theory
calculations, but in fact they depend, of course, only on the rational
structure of integrands with respect to momenta.

\qquad
Second, it is clear that from a purely technical standpoint, nothing
would have stood in the way of our using directly the resurgent symbols
-- that is, the operator-product expansions -- of $D$ and of
$\Gamma^{3}$ in the loop integral (\ref{phi2C}). Instead of (\ref{bint})
we would have encountered momentum integrals of type
\renewcommand{\theequation}{3.79}
\begin{equation}
\label{oploop}
\int\,\frac{d^{D}\,q}{(2\,\pi)^{D}}\,
\frac{\left(q\,-\,\frac{1}{2}\,k\right)^{2}}{[(q\,-\,k)^{2}]^{s}\,[q^{2}]^{t}}\,
\propto\,\left(\frac{1}{k^{2}}\right)^{s\,+\,t\,-\,3\,+\,\varepsilon}\,\,,
\end{equation}
which are increasingly
singular as $k^{2}\,\to\,0$, and we might have proceeded to matching
the series involving these integrals to the series of eq.
(\ref{invdnull}) at increasing orders $n$, thus determining in principle
the resurgent-symbol coefficients $U_{n}^{(R)}\,(0,\,0)$. But it would
nevertheless have been foolish to do so. On the mathematical side, the
coefficients of proportionality in (\ref{oploop}) would be
Feynman-parameter integrals increasingly divergent for $s$ or $t\,>\,2$,
and would force us to throw in a new set of rather obscure
regularizations that might (hopefully) go away in the symbols's final
resummation. Even if we were willing to adopt such an ugly formalism,
after all the effort we would still be around with our resurgent symbol,
a
large-$k^{2}$ asymptotic representation, and would have made no progress
at all toward knowing the function $(D^{-1})$ at intermediate and small
$k^{2}$ -- just as in Borel-Laplace reconstruction when we forget the
crucial continuation step (\ref{constep}).
On the physical side, we would have decomposed the well-defined physical
effect described by our diagram Fig. 3.1 (2\,C) -- the self-dressing of
a field quantum by virtual pairs of quanta of its own kind -- in a
totally unreasonable  way: into a sum of increasingly infrared-singular
terms with increasingly violent small-$k^{2}$ cancellations between
them. Finally, this procedure would have broken down completely for the
"tadpole" diagram of Fig. 3.1 (2\,B): by using the resurgent symbol
for its single-$D$ integrand, we would have decomposed it into a sum of
scale-free integrals which vanish in dimensional regularization, whereas
with any reasonable analytic continuation this term contributes a
nonzero multiple of $\Lambda^{2}$, in marked contrast to perturbation
theory. Thus it is abundantly clear that (as mentioned already in the
introduction) we cannot simply plug our resurgent symbols into the
dynamical equations in order to determine their parameters -- some
amount of analytic continuation is necessary {\em from the outset} to
deal with objects globally usable in momentum space, and it is such
objects that we are trying to establish.

\qquad
Returning to eq. (\ref{DS2r}), we note that although our task here is
more intricate than the standard task of matching a rational approximant
to an explicitly known function, the two problems seem to share the
property that there is in principle a continuum of matching
prescriptions for determining the approximant´s coefficients -- Padé I
or II, as mentioned after our eq. (\ref{bpade}) above, are just two
among many (and Padé II in itself already represents a continuum of
possibilities). But a moment´s reflection will show that in the present
problem there {\em is} a distinctly preferred matching strategy: we
should clearly see to it that we get the {\em poles} (at the zeroes
$k^{2}\,=\,-u_{2\,s}^{[r]}\,\Lambda^{2}$ of the $\Delta_{2}$
denominator) right, from the r. h. s. of (\ref{DS2r}), for the
postulated zeroth-order $(D^{-1})^{[r,\,0)}$. If we failed to ensure
this, the remainder in eq. (\ref{DS2r}), which is to give (the
(2\,C)-type contribution to) the first-order-quasi-perturbative
correction $g^{2}\,(D^{-1})^{[r,\,0)}$, would have its typical one-loop
analyticity properties contaminated by splinters from the zeroth-order
poles, and thus would be plagued by approximation errors that become
locally infinite in the $k^{2}$ plane. Having already ensured correct
pole positions through conditions (\ref{utau}), we therefore need to
match {\em residues} at those positions. Looking at eq. (\ref{ga2pol})
(with $\zeta^{[r]}_{0}\,=\,1$) above, we are led to imposing the
self-consistency conditions
\renewcommand{\theequation}{3.80}
\begin{equation}
\label{upcond}
u_{2\,s\,+\,1}^{[r]}\,=\,\left(\frac{\overline{g}_{0}}{4\,\pi}\right)^{2}\,
r_{2\,s\,+\,1}^{[r]}\,(-u_{2\,s}\,\Lambda^{2},\,\Lambda^{2};\,\varepsilon)\qquad
(s\,=\,1\,...\,r)\,\,.
\end{equation}
For the $r_{1}\,\Lambda^{2}$ term, there is no such obviously preferred
matching prescription, so here we may freely choose a matching point
$k^{2}\,=\,-\omega_{1}\,\Lambda^{2}$:
\renewcommand{\theequation}{3.81}
\begin{equation}
\label{u1cond}
u_{1}^{[r]}\,=\,\left(\frac{\overline{g}_{0}}{(4\,\pi)}\right)^{2}\,r_{1}^{[r]}\,
(-\omega_{1}\,\Lambda^{2},\,\Lambda^{2};\,\varepsilon)
\end{equation}
(This freedom of choice is actually part of the freedom of performing
finite renormalizations, since the $u_{1}$ term can be shifted around by
a {\em local} Lagrangian counterterm). In all these conditions, the
anxious question of how´s-this-going-to-give-a-zeroth-order, with the
explicit $\overline{g}_{0}^{2}$ factor in front, is still hanging over
us.

\qquad
But only apparently! With $k^{2}$, in these conditions, now nailed to
fixed multiples of $\Lambda^{2}$, the $A_{m,\,n}$ and $B_{m,\,n}$
functions entering in eq. (\ref{rico}) now appear with values
\renewcommand{\theequation}{3.82}
\begin{equation}
\label{asub}
A_{m,\,n}^{[r]}\,(-c\,\Lambda^{2},\,\Lambda^{2};\,\varepsilon)\,=\,
(\Lambda^{2})^{-\varepsilon}\,\frac{1}{\varepsilon}\,\{
1\,+\,[O\,(\varepsilon)]\,(c)\}\,\,,
\end{equation}
\renewcommand{\theequation}{3.83}
\begin{equation}
\label{bsub}
B_{m,\,n}^{[r]}\,(-c\,\Lambda^{2},\,\Lambda^{2};\,\varepsilon)\,=\,
(\Lambda^{2})^{-\varepsilon}\,\frac{1}{\varepsilon}\,\left\{
4\,(\sigma_{m}\,+\,\sigma_{n})^{[r]}\,+\,[O\,(\varepsilon)]\,(c)\right\}\,\,,
\end{equation}
so all the $r_{i}^{[r]}$ at matching points $-c\,\Lambda^{2}$ in
conditions (\ref{upcond}) and (\ref{u1cond}) are found to have factors
of $(\Lambda^{2})^{-\varepsilon}\,\frac{1}{\varepsilon}$ up front. Now
take a closer look at the quantity
\renewcommand{\theequation}{3.84}
\begin{equation}
\label{capi}
\left(\frac{\overline{g}_{0}}{4\,\pi}\right)^{2}\,(\Lambda^{2})^{-\varepsilon}\,
\frac{1}{\varepsilon}\,=\,\left(\frac{g\,(\mu)}{4\,\pi}\right)^{2}\,Z_{\alpha}\,
\left(\frac{\Lambda^{2}}{\mu^{2}}\right)^{-\varepsilon}\,\frac{1}{\varepsilon}\,\,,
\end{equation}
where relation (\ref{recoup}) has been used. Proceed the way you are
accustomed to, Laurent-expanding the last two
factors around $\varepsilon\,=\,0$. You have
\renewcommand{\theequation}{3.85}
\begin{equation}
\label{epseps}
\frac{1}{\varepsilon}\,\left(\frac{\Lambda^{2}}{\mu^{2}}\right)^{-\varepsilon}\,=\,
-\ln\,\left(\frac{\Lambda^{2}}{\mu^{2}}\right)\,+\,
\frac{1}{\varepsilon}\,+\,O\,(\varepsilon)\,\,.
\end{equation}
The logarithmic term survives the $\varepsilon\,\to\,0$ limit in the
same, familiar way as the well-known $\ln\,(k^{2}\,/\,\mu^{2})$´s in a
perturbative calculation, but it differs from them in one important
respect: it leads them by one order in the quasi-perturbative
$(g^{2})^{p}$ classification. From eq. (\ref{lambnull}),
\renewcommand{\theequation}{3.86}
\begin{equation}
\label{lola}
-\ln\,\left(\frac{\Lambda^{2}}{\mu^{2}}\right)\,=\,
\left(\frac{4\,\pi}{g\,(\mu)}\right)^{2}\,\frac{1}{\beta_{0}}\,
\left[1\,+\,O\,(g^{2}\,\ln\,g^{2})\right]
\end{equation}
at the one-loop level. This is precisely the $1\,/\,g^{2}$ factor needed
to "eat" the overall $g^{2}$ factor in (\ref{capi}) and make
zeroth-order-in-$g^{2}$ terms possible. There´s more: we now have
\renewcommand{\theequation}{3.87}
\begin{equation}
\label{exppi}
\left(\frac{\overline{g}_{0}}{4\,\pi}\right)^{2}\,(\Lambda^{2})^{-\varepsilon}\,\frac{1}{\varepsilon}\,=\,
\frac{1}{\beta_{0}}\,\left[1\,+\,\beta_{0}\,\left(\frac{g}{4\,\pi}\right)^{2}\,
\frac{1}{\varepsilon}\,+\,...\right]\,Z_{\alpha}\,\,,
\end{equation}
but $Z_{\alpha}$, to the same order, is given by eq. (\ref{zaexp}), so
that to the order at which we calculate here, the ultraviolet divergence
too gets "eaten" in the special quantities that enter our matching
conditions. From eq. (\ref{upcond}), this in particular means that no
nonlocal counterterms will be needed for the pole terms in eq.
(\ref{DS2r}): as we expected from our discussion of the OPE, they
establish themselves in a finite manner.

\qquad
Our simple one-loop-level argument has established the values
\renewcommand{\theequation}{3.88}
\begin{equation}
\label{aval}
A_{m,\,n}^{[r]}\,(-c\,\Lambda^{2},\,\Lambda^{2};\,0)\,=\,\frac{1}{\beta_{0}}\,\,,
\end{equation}
\renewcommand{\theequation}{3.89}
\begin{equation}
\label{bval}
B_{m,\,n}^{[r]}\,(-c\,\Lambda^{2},\,\Lambda^{2};\,0)\,=\,
\frac{4}{\beta_{0}}\,\cdot\,(\sigma_{m}\,+\,\sigma_{n})^{[r]}
\end{equation}
at $D\,=\,4$ only up to terms $O\,(g^{2}\,\ln\,g^{2})$, which still seem
to be a
potential nuisance in conditions (\ref{upcond}) through (\ref{u1cond}).
It is therefore useful to know that a more careful argument \cite{STI}
establishes these values {\em exactly} to all orders. This argument
avoids the use of the unrelated expansions (\ref{epseps}) and
(\ref{zaexp}) completely, and instead employs the exact integral
representations for the two middle factors on the r. h. s. of eq.
(\ref{capi}). For $Z_{\alpha}$, we already noted t´Hooft´s
representation, eq. (\ref{zaclos}) above. For
$(\Lambda_{\varepsilon}^{2})^{-\varepsilon}$ -- be careful here to
consider the $\varepsilon\,>\,0$ quantity formed with the full
$\varepsilon\,>\,0$ beta function (\ref{betaex}) -- it is a matter of a
few lines of calculation to deduce from definition (\ref{lambdae}) the
exact representation
\renewcommand{\theequation}{3.90}
\begin{equation}
\label{laee}
\left(\frac{\Lambda_{\varepsilon}^{2}}{\mu^{2}}\right)^{-\varepsilon}\,=\,
\frac{g_{1}^{2}}{g^{2}\,(\mu)}\,\exp\,\Biggl\{\,\,\,
\int\limits_{(g_{1}\,/\,4\,\pi)^{2}}^{(g\,/\,4\,\pi)^{2}}\,\frac{d\,\kappa^{'}}
{\kappa^{'}\,+\,\varepsilon\,\chi\,(\kappa^{'})}\Biggr\}\,\,,
\end{equation}
involving again the function $\chi\,(\kappa)$ defined by (\ref{chifu}).
Upon combining the two representations one sees that in the quantity
(\ref{capi}), the dependence on $g^{2}\,(\mu)$, and therefore on $\mu$,
cancels exactly (remember $g_{1}$ is a trivially RG-invariant
integration constant). Further analysis of the integral in
\renewcommand{\theequation}{3.91}
\begin{equation}
\label{intpi}
\left(\frac{\overline{g}_{0}}{4\,\pi}\right)^{2}\,(\Lambda^{2}_{\varepsilon})^{-\varepsilon}\,\frac{1}{\varepsilon}\,=\,
\frac{g^{2}_{1}}{(4\,\pi)^{2}\,\varepsilon}\,\exp\,\Biggl\{
-\int\limits_{0}^{(g_{1}\,/\,4\,\pi)^{2}}\,
\frac{d\,\kappa^{'}}{\kappa^{'}\,+\,\varepsilon\,\chi\,(\kappa^{'})}\Biggr\}
\end{equation}
by exactly the same technique as applied in eqs. (\ref{zapart}) and
(\ref{zalim}) gives, finally,
\renewcommand{\theequation}{3.92}
\begin{equation}
\label{piex}
\left(\frac{\overline{g}_{0}}{4\,\pi}\right)^{2}\,(\Lambda^{2}_{\varepsilon})^{-\varepsilon}\,
\frac{1}{\varepsilon}\,=\,\frac{1}{\beta_{0}}\,[1\,+\,O\,(\varepsilon,\,\varepsilon\,\ln\,\varepsilon)]\,\,,
\end{equation}
{\em rigorously independent of} $g^{2}\,(\mu)$. Despite the superficial
similarity, this should not be confused with eq. (\ref{zalim}); it says
something conceptually quite different. While (\ref{zalim}) is a
representation for $Z_{\alpha}$ with a $g^{2}\,(\mu)$ dependence of the
r. h. s. that is difficult to control as long as one does not know the
exact RG beta function, the result (\ref{piex}) states the presence of
an {\em exact RG invariant}, which, moreover, is {\em finite and
scheme-independent} at $\varepsilon\,=\,0$.
Its point is the exact contragredience of the $g^{2}$ dependences in the
quantities (\ref{zaclos}) and (\ref{laee}) {\em already at}
$\varepsilon\,\not=\,0$. Both a
$g^{2}$ factor and an
$\frac{1}{\varepsilon}$ divergence get "eaten" in this remarkable
combination.

\qquad
It is not necessary, in these seminars, for us to pursue the consequence
of
eqs. (\ref{aval}) and (\ref{bval}) through the $r_{i}^{[r]}$ functions
to the self-consistency conditions (\ref{upcond})/(\ref{u1cond}) in
every detail. It is sufficient
to state that the latter now assume, at $\varepsilon\,=\,0$, the general
form
\renewcommand{\theequation}{3.93}
\begin{equation}
\label{uself}
u_{2\,s\,+\,1}^{[r]}\,=\,\frac{1}{\beta_{0}}\,
R_{2\,s\,+\,1}\,\left(\{u\}^{[r]},\,\{x\}^{[r]}\right)\qquad
(s\,=\,0\,...\,r)\,\,,
\end{equation}
with functions $R_{2\,s\,+\,1}$ depending linearly on the set
$\{x\}^{[r]}$ of three-point vertex coefficients, and in general
nonlinearly on the set $\{u\}^{[r]}$ of the inverse-propagator
parameters $u_{i}$ themselves. These conditions allow one, remarkably,
to establish functions $(D^{-1})^{[r,\,0)}$ of zeroth quasi-perturbative
order from DS loops with a $\overline{g}^{2}_{0}$ prefactor, and to do
so in an ultraviolet-finite manner not requiring nonlocal counterterms.

\qquad
As we have anticipated, apart from equations such as (\ref{d3d2}) and
(\ref{d4d3}) that establish a common set of denominator coefficients
$u_{2\,s}^{[r]}$ for all vertex functions, we have not found conditions
that would really determine these coefficients. We have already sketched
how one may go about filling this gap by invoking higher degrees of
compositeness; setting up equations for the quantities (\ref{2pint}) and
(\ref{3pint}) one obtains the $r$ additional conditions needed. Apart
from the technical complication that such equations now have at least
two-loop terms on their right-hand sides, and involve
the trickery of extracting two-loop divergences, their treatment
requires no truly new
ideas; in particular, relation (\ref{piex}) again plays a central role.
We therefore skip details of these "integrated" conditions entirely.
For the same reasons, we do not enter into details of the next higher
vertex equations, which involve application of the same technology to
more variables and lengthier loop integrals. For example, the middle
(three-point) equation of Fig. 3.1, when analyzed for terms of zeroth
quasi-perturbative order, will produce conditions for the three-point
parameters $\{x\}^{[r]}$ of the general form
\renewcommand{\theequation}{3.94}
\begin{equation}
\label{xself}
x_{m_{1}\,m_{2}\,m_{3}}^{[r]}\,=\,
\frac{1}{\beta_{0}}\,S_{m_{1}\,m_{2}\,m_{3}}\,
\left(\{u\}^{[r]},\,\{x\}^{[r]},\,\{t\}^{[r]}\right)\,\,,
\end{equation}
with right-hand sides now involving the set $\{t\}^{[r]}$ of four-point
numerator coefficients. The structure of these conditions, which after
multiplication with enough denominators become a system of nonlinear
algebraic equations, mirrors quite closely the structure of the
underlying DS system with its nonlinearities and hierarchical couplings.

\qquad
As for the remainder terms in eq. (\ref{DS2r}) that we expect to yield
the first quasi-perturbative correction, we are now left with
\renewcommand{\theequation}{3.95}
\begin{equation}
\label{dinvr1}
\begin{array}{l}
\displaystyle{
g^{2}\,[D^{-1}\,(k^{2})]^{[r,\,1)}=
\left(\frac{\overline{g}_{0}}{4\,\pi}\right)^{2}\Biggl\{
r_{0}^{[r]}\,(k^{2},\Lambda^{2};\,\varepsilon)\,k^{2}\,+}\\
\mbox{}\\
\displaystyle{
\left[r_{1}^{[r]}\,(k^{2},\,\Lambda^{2};\,\varepsilon)\,-\,
r^{[r]}_{1}\,(-\omega_{1}\,\Lambda^{2},\,\Lambda^{2};\,\varepsilon)\right]\,\Lambda^{2}}\\
\mbox{}\\
\displaystyle{+\,
\sum\limits_{s\,=\,1}^{r}\,\frac{r_{2\,s\,+\,1}^{[r]}\,(k^{2},\,\Lambda^{2};\,\varepsilon)-
r_{2\,s\,+\,1}^{[r]}\,(-u_{2\,s}\,\Lambda^{2},\,\Lambda^{2};\,\varepsilon)}
{k^{2}\,+\,u_{2\,s}\,\Lambda^{2}}\,\Lambda^{4}}\\
\mbox{}\\
+\,[\text{contributions from other loop terms}]\Biggr\}
\end{array}
\end{equation}
The unsubtracted term $r_{0}^{[r]}\,k^{2}$ carries the
$\Lambda^{2}\,\to\,0$ "`perturbative limit" as it would come out from
the usual perturbative calculation of the loop, {\em including} the
perturbative $\frac{1}{\varepsilon}$ ultraviolet divergence. The other
terms, which the self-consistency process has left subtracted, are both
ultraviolet finite and (for the terms with $\Lambda^{4}$) regular at
$k^{2}\,=\,-u_{2s}\,\Lambda^{2}$, since by eqs. (\ref{rico}),
(\ref{amn}) and (\ref{bmn}) they are linear combinations of the
quantities
\renewcommand{\theequation}{3.96}
\begin{equation}
\label{sigsub}
\begin{array}{l}
\displaystyle{
\left[\Sigma^{[r]}_{m,\,n}\,(k^{2},\,\Lambda^{2};\,z)\right]^{-\varepsilon}\,-\,
\left[\Sigma^{[r]}_{m,\,n}\,(-c\,\Lambda^{2},\,\Lambda^{2};\,z)\right]^{-\varepsilon}\,=\,}\\
\mbox{}\\
\displaystyle{
-\varepsilon\,(k^{2}\,+\,c\,\Lambda^{2})
\,z(1-z)\int\limits_{0}^{1}\!dw\Bigl\{
wz(1-z)\,k^{2}}\\
\mbox{}\\
\displaystyle{
+\,\left[z\,\sigma_{m}^{[r]}\,+\,
(1\,-\,z)\,\sigma_{n}^{[r]}\,-\,(1\,-\,w)\,c\,z\,(1\,-\,z)\right]\,\Lambda^{2}\Bigr\}^{-(1\,+\,\varepsilon)}\,\,.}
\end{array}
\end{equation}
With respect to renormalization, we are therefore led into precisely the
situation we expected from our OPE discussion around eq. (\ref{opren}):
we have to follow the "two-track" procedure of supplying an explicit
counterterm, proportional to $g^{2}\,\frac{1}{\varepsilon}$, for the
perturbative portion, while the terms with powers of $\Lambda^{2}$
establish themselves in an ultraviolet-finite manner. As for the
$\overline{g}_{0}^{2}$ in expression (\ref{dinvr1}), we note that its
rigorously $g^{2}$-independent combination (\ref{piex}) with the
quantity $(\Lambda^{2})^{-\varepsilon}\,\frac{1}{\varepsilon}$ no more
occurs, and since we are working within a quasi-perturbative scheme, we
must therefore revert to the usual pattern of treating it as
$\mu^{2\,\varepsilon}\,g^{2}\,(\mu)\,+\,O\,(g^{4})$, the
$\mu^{2\,\varepsilon}$ serving as usual to render the
$(\Sigma^{[r]})^{-\varepsilon}$ terms in the quantities (\ref{amn}) and
(\ref{bmn}) dimensionless.

\qquad
I is clear that our condition (\ref{unlim}) of maintaining strict
perturbative limits bars us from "oversubtracting" our result
(\ref{DS2r}), in the sense of subtracting also the
$r_{0}^{[r]}$ term at some point $k^{2}\,=\,-\omega_{0}\,\Lambda^{2}$
and invoking the mechanism of eq. (\ref{piex}). Such a procedure would
inevitably lead us away from the perturbative limit $\xi_{0}\,=\,1$ and
into a different version of the quasiperturbative expansion, analogous
to the one we noted for our model amplitude in eqs. (\ref{quapert}) and
(\ref{cpbar}). We are not going to explore this route here, interesting
as it may be, since it would also force us to reorganize the entire
perturbative-renormalization procedure, and therefore would lead us too
far away from our stated focus.

\qquad
Several features of this self-consistency mechanism are worth commenting
on. First, the final self-consistency conditions at $D\,=\,4$ (i. e.
$\varepsilon\,=\,0$) are independent of the chosen subtraction points
$-u_{2\,s}\,\Lambda^{2}$ and $-\omega_{1}\,\Lambda^{2}$, since the
latter only enter into the terms $O\,(\varepsilon)$ in eqs. (\ref{asub})
and (\ref{bsub}). The subtraction points do, however, enter into the
first-order, quasi-perturbative correction $[D^{-1}]^{[r,\,1)}$ of eq.
(\ref{dinvr1}),
and are essential for keeping it uncontaminated by the zeroth-order
rational-approximation poles.

\qquad
Second, the conditions at $D\,=\,4$, while derived here in a
dimensional-regularization scheme, make no reference any more to a
special renormalization scheme, since $\beta_{0}$ is scheme-independent.
Now different schemes, as a noted in eq. (\ref{lambtraf}) above, have
$\Lambda$ scales differing by constant numerical factors. This leads us
to suspect that upon scaling $\Lambda$ by a factor, the conditions will
remain invariant when scaling all the nonperturbative $x$ and $u$
coefficients inversely to the $\Lambda^{2}$ powers with which they
are associated. This {\em scaling property} indeed turns out to be
present
in the system, and has the effect that the conditions in fact determine
only the ratios of all coefficients to one of them. The undetermined
coefficient, with its associated $\Lambda^{2}$ power, effectively
defines a modified scale $\tilde{\Lambda}$, which can be fixed only by
calculating and matching some dimensionful observable.

\qquad
Third, it is manifest that the entire mechanism is intimately {\em tied
to the presence of the ultraviolet loop divergences}: the crucial term
(\ref{lola}), and the ensuing terms (\ref{aval}) and (\ref{bval}), while
themselves finite, have survived the removal-of-regulator limit
$\varepsilon\,\to\,0$ only due to their association with a
$\frac{1}{\varepsilon}$ divergence. We have all known such an effect
from our experience with perturbative logarithms, but here it assumes a
new and much more prominent role. The self-reproduction of zeroth
quasi-perturbative orders is therefore not only a genuine quantum effect
(as it comes from loops) but also a genuine quantum-field effect (as it
comes from divergent loops).

\qquad
This has the immediate consequence that, fourth, the formation of zeroth
quasi-perturbative orders becomes rigorously restricted to the small
finite number of superficially divergent vertices --
by and large, the same functions that already possess a zeroth {\em
perturbative} order. You need no big calculation to see this. In a
superficially convergent amplitude, the one-loop contributions have
factors of $\overline{g}_{0}^{\,2}$ but are, by definition, genuinely
convergent, so they cannot trigger the mechanism of eq. (\ref{piex}) and
remain quasi-perturbative corrections of the first order. The two-loop
contributions, with factors of $\overline{g}_{0}^{\,4}$, may exhibit
ultraviolet divergences, but only in the form of one-loop subdivergences
producing a single $\frac{1}{\varepsilon}$ factor, and generally in the
higher $l$-loop terms the number of $\frac{1}{\varepsilon}$ factors
produced lags behind the number of overall $\overline{g}_{0}^{\,2}$
factors
by at least one unit. So the mechanism of eq. (\ref{piex}) can "eat"
at most $l\,-\,1$ of the $l$ factors of $\overline{g}^{2}_{0}$, and all
contributions remain $p\,\geq\,1$ quasi-perturbative corrections.

\qquad
The same consideration applies to terms such as (4\,C) of Fig. 3.1 that
couple the highest superficially divergent to the lowest superficially
convergent vertices. Since there is no $\Gamma_{5}^{(0)\,pert}$ in a
renormalizable action, the lowest contributions to term (4\,C) have
$\Gamma_{5}$ replaced by its convergent one-loop terms and two factors
of $\overline{g}_{0}^{\,2}$ in front, of which at most one can be eaten.
Thus {\em as far as self-reproduction of zeroth orders} is concerned,
the hierarchical coupling automatically stops acting after the
four-boson vertex equation. Once more we have a fine line (depicted in
the third equation of Fig. 3.1) marking a fundamental divide, this time
within the strictly renormalizable theory itself: between the equations
for superficially divergent and for superficially convergent functions.

\qquad
The self-consistency conditions, fifth, do nevertheless retain an
infinite-expansion aspect of a different kind, at least in the somewhat
conservative treatment we have afforded them here. We have been
classifying the contributions to the right-hand sides of the integral
equations in the time-honored way by their number of loops, or
independent momentum integrations, that arise upon iterating the
equations around the newly established, nonperturbative starting points
$\Gamma^{[r,\,0)}$. It is clear that this will translate into an
analogous loop classification of contributions to the right-hand sides
of the self-consistency conditions. In a superficially divergent
amplitude, there is no lagging of the number of $\overline{g}^{2}_{0}$
prefactors; an $l$-loop diagram here produces terms of type
\renewcommand{\theequation}{3.97}
\begin{equation}
\label{divpat}
(\overline{g}^{2}_{0})^{l}\,\cdot\,\left\{
1,\,\frac{1}{\varepsilon},\,\frac{1}{\varepsilon^{2}},\,...\,\frac{1}{\varepsilon^{l}}\right\}
\end{equation}
and through the mechanism of eq. (\ref{piex}) therefore gives
contributions to the quasi-perturbative series of the forms
\renewcommand{\theequation}{3.98}
\begin{equation}
\label{ordpat}
(g^{2})^{l},\quad
\frac{1}{\beta_{0}}\,(g^{2})^{l\,-\,1},\quad
\frac{1}{\beta_{0}^{2}}\,(g^{2})^{l\,-\,2},\quad
...\,\frac{1}{\beta_{0}^{l}}\,\,,
\end{equation}
all the way down to the zeroth order: a given loop number $l$ produces
not only terms of order $(g^{2})^{l}$ but also corrects all previous
orders $(g^{2})^{p}$ with $p\,<\,l$. That the tight coupling between $l$
and $p$ orders characteristic of a {\em perturbative} single-coupling
theory should get loosened is presumably something that happens in any
truly nonperturbative treatment. (It complicates diagrammatic argument,
in that a quasi-perturbative correction $\Gamma^{[r,\,p)}$ with
$p\,\geq\,1$ can no more simply be written as a set of $p$-loop
diagrams, but must include finite counterterms to account for the pieces
that go into orders $p^{'}\,<\,p$). But here it has the specific
consequence that the self-consistency condition for a generic
coefficient $x^{[r]}$ of a zeroth-order nonperturbative approximant in
principle gets corrections from all higher loop orders that form a
series in powers of $1\,/\,\beta_{0}$:
\renewcommand{\theequation}{3.99}
\begin{equation}
\label{secall}
x^{[r]}\,=\,\frac{1}{\beta_{0}}\,X_{(l\,=\,1)}^{[r]}\,+\,\frac{1}{\beta_{0}^{2}}\,X_{(l\,=\,2)}^{[r]}\,+\,
...
\end{equation}
This type of series will not create new summability problems of its
own, since from (\ref{divpat}) you see that it arises from the "most
strongly divergent terms", terms proportional to
$(1\,/\,\varepsilon)^{l}$, in all loop orders $l$, and these terms, like
the "leading logarithms" to which they are closely related, generally
do not exhibit factorial divergence but form a series with nonzero
convergence radius, behaving roughly like a geometric series. But I have
not yet had enough experience with this formalism to have a ready
suggestion for its practical resummation, which at this time remains a
(somewhat technical but not unimportant) research problem. In pure
$SU\,(N)$ Yang-Mills theory, where $\beta_{0}\,=\,\frac{11}{3}\,N$, eq.
(\ref{secall}) clearly is of the nature of an $\frac{1}{N}$ expansion
(though only formally, since $X_{l}^{[r]}$ may well contain
contributions comparable to $(\beta_{0})^{l}$ of its own),
and since general experience with $\frac{1}{N}$ expansions indicates
they tend to work much better than anyone has a right to expect, it
seems not unreasonable to start out by drawing conclusions from the
lowest orders of (\ref{secall}) only.

\qquad
Sixth and last for these comments, while we have stopped short of
technical detail for three-point and four-point vertex equations
(on that subject I must refer you to refs. \cite{DRI}),
it is
already clear from the inverse-propagator equation alone, in e. g. the
form (\ref{DS2r}), that the mechanism allows us, in the well-defined
approximation of a zeroth quasi-perturbative order, to watch the
spectacle of the {\em spontaneous generation of mass} for a field that
was massless on the Lagrangian level -- the process we referred to in
eq. (\ref{massi}). If the field decides to form a single massive
particle,
its squared mass at level $r$ will be given by the first root of the
propagator denominator (\ref{profac}),
\renewcommand{\theequation}{3.100}
\begin{equation}
\label{massr}
(m^{2})^{[r]}\,=\,\sigma_{1}^{[r]}\,\Lambda^{2}\,\,,
\end{equation}
with $\sigma_{1}^{[r]}$ calculable from the self-consistent $u_{i}$
coefficients. Because of the scaling property, solution of the $[r,\,0)$
self-consistency problem alone does not, of course, make this an
absolute
prediction, since at this stage you do not yet know which $\Lambda$
scale you are working with. But ratios of particle masses do in
principle become predictable. It is interesting to note that in some
sense this is already true on the $r\,=\,0$ level (which is not to be
confused with the perturbative limit). The $r\,=\,0$ approximation is so
crude that it can be accused of internal inconsistency -- as you see
from (\ref{n3r}), it keeps the three-point, and in fact also the
four-point, vertex functions at their perturbative forms
$\Gamma^{(0)\,pert}_{3,\,4}$, at least under our "boundary condition"
(\ref{unlim}) above, but it does allow the inverse propagator a
nonperturbative mass term proportional to $\Lambda^{2}$,
\renewcommand{\theequation}{3.101}
\begin{equation}
\label{ga200}
(D^{-1})^{[0,\,0)]}\,=\,k^{2}\,+\,u^{[0]}_{1}\,\Lambda^{2}\,\,,
\end{equation}
so that $\sigma_{1}^{[0]}\,=\,u_{1}^{[0]}$. With such a crudely
restricted scheme, you are not surprised when you work out the
self-consistency condition and find the form $u_{1}^{[0]}$ =
$\text{const.}\,\cdot\,u_{1}^{[0]}$ which allows only the trivial
solution
$u_{1}^{[0]}\,=\,0$, since
for realistic values of $\beta_{0}$ the constant refuses to be unity.
But it is interesting that for an asymptotically free theory with
$\beta_{0}\,>\,0$, the constant is positive, so that even at this
oversimplified level the system sends a muted signal of its tendency to
produce a spontaneous mass.
\subsection{Things done and not done}
\qquad
At the end of these seminars, let us take a stroll along the periphery
of the landscape
that we tried to begin to explore, casting glances at the few things
than have been done or attempted in the matter of construction of
coupling-resurgent amplitudes, and at the larger number of things that
have not been done but could be worthwhile taking up if you find these
questions interesting.

\qquad
The method described in the last two sections for constructing resurgent
vertex functions from quasi-perturbative series, based on a set of
generalized Feynman rules, has not at this time been tested to a degree
that one could call comprehensive, but there has been a reasonably
complete application to massless QCD at $r\,=\,1$, the lowest nontrivial
level of rational continuation, for which I must refer you to the
articles \cite{DRI}. This study employs Landau gauge fixing and
plausibly motivated restrictions on the oppressively complicated tensor
structure of the four-gluon vertex, and concentrates on the case of at
most two flavors of massless quarks. If finds that the two superficially
divergent vertices for Faddéev-Popov ghosts, {\em at the zeroth
quasi-perturbative order}, remain perturbative in Landau gauge fixing, a
conclusion I would like to see corroborated by independent studies. For
the zeroth-order self-consistency problem of the remaining five
superficially divergent functions (three for the gluon and two for the
fermion sector), it finds acceptable solutions with all-real
approximants, which after throwing out a few on physical grounds are
even
remarkably unique. Unlike the pure Yang-Mills case (no quarks) which is
also considered, the $r\,=\,1$ transverse-gluon propagator obtained is
not entirely satisfactory, since its zero at
$k^{2}\,=\,-u_{2}^{[1]}\,\Lambda^{2}$ slips a bit into the Euclidean
region, thus violating reflection positivity on a small Euclidean
interval -- an effect coming from a strong coupling to virtual
quark-antiquark pairs, which the approximation obviously exaggerates.
You
will tolerate my not entering into further discussion of the propagators
obtained in those articles -- they are interesting, but their
interpretation would drag us into issues of confinement far removed from
the focus of these talks. Here I am quoting refs. \cite{DRI} mainly as a
demonstration of technical feasibility.

\qquad
One problem for which refs. \cite{DRI} do offer a reasonably complete
solution, even for arbitrary $r$, looks at first like a purely technical
one. But it turns out to have a bearing on the fundamental issue of
unitarity (total-probability conservation), and therefore let me briefly
stop at it on our stroll. When we found we had to impose eq.
(\ref{d3d2}) as a self-consistency condition, you may have wondered in
passing what would have happened in the diagram of Fig. 3.1 (2\,C), or
its analytic expression (\ref{ph2Cr}) at level $r$, without that
condition. Then we would have had on each of the two internal lines, in
addition to the $r$-level propagator poles of eq. (\ref{propol}) that
provide a discretized Källén-Lehmann description of the propagation of
the virtual particle, $r$ additional poles from an $1\,/\,\Delta_{3}$
factor that would plainly contradict the postulated propagator
approximant. Condition (\ref{d3d2}) saved us from that embarrassment,
but only because in that diagram both internal lines end in a bare
vertex on the left-hand side. But now take a look at the second of the
DS equations of Fig. 3.1, the three-point vertex equation, and its loop
term (3\,B). In that loop, the right-hand part involving the vertical
internal line represents a dressed one-particle exchange of the form
\renewcommand{\theequation}{3.102}
\begin{equation}
\label{onex}
\Gamma_{3}\,(\,...\,,\,...\,,\,Q)\,D\,(Q^{2})\,\Gamma_{3}\,(Q,\,...\,,\,...\,)\,\,,
\end{equation}
with $Q$ the momentum of the vertical line. Since that line now extends
between two {\em dressed} vertices, only one of the two
$1\,/\,\Delta_{3}$´s from these vertices gets cancelled by the
$\Delta_{2}$ numerator of $D\,(Q^{2})$ by virtue of eq. (\ref{d3d2}),
and now the embarrassment of the "superfluous poles" seems to be
really on us: such poles will, by the diagram-cutting rules for
unitarity relations, produce contributions to these relations that do
not correspond to any accessible external states, so unitarity
would be endangered. It is a minor surprise
(at least to me) to see that
the formalism solves that problem automatically by its own devices: one
finds that the four-point vertex $\Gamma_{4}$ in diagram (3\,C)
unavoidably develops poles in its crossed channels that are the exact
opposites of the "superfluous" poles in (\ref{onex}). (That
conclusion, remarkably, follows from residue-taking operations in the
three-point and two-point equations alone, that is, from {\em lower}
equations in the hierarchy). Note that since the denominators in these
"compensating poles" are not propagators of any of the elementary
fields of the theory, their presence does not contradict the
one-particle-irreducible nature of $\Gamma_{4}$. Their effect in the
present context is to turn the vertical internal line described by
(\ref{onex}) into what refs. \cite{DRI} denote by a {\em dotted internal
line}, that is, a partial-fraction decomposition of expression
(\ref{onex}) in which only the "true" propagator poles have been kept.
Further analysis quickly reveals that these dotted lines really are
almost everywhere, being present on every internal line between two
dressed vertices. They are an important ingredient for keeping the
formalism in line with the unitarity requirement. One amusing
consequence of their presence that has not even begun to be exploited is
that they provide a natural discretization of Bethe-Salpeter bound-state
equations that refines itself with increasing $r$.

\qquad
What is entirely lacking at present -- and that brings us to the part of
the countryside where the things not done are lying around -- is tests
of the formalism at levels $r\,>\,1$ of rational continuation. (They are
necessary not only for numerical refinement but also because rational
approximants have a reputation for sometimes behaving erratically at low
orders). The reasons are, to a large extent, purely technical and have
little to do with physics: the nonlinear algebraic systems of
self-consistency equations turn out to be a surprisingly tough
assignment. Refs. \cite{DRI} were still able to get away with ad hoc
methods of solution, exploiting accidental simplifying features of their
Landau-gauge, $r\,=\,1$ system, but higher $r$´s call for more
systematic methods. From present experience, it seems fair to say that
no presently known algorithm or available routine is capable of locating
the rapidly growing number of solutions of realistic $r\,\geq\,2$
systems {\em completely} in tolerable amounts of computing time, and
most methods even have a hard time coping with $r\,=\,1$ systems. So
here the formalism also creates a new challenge for applied mathematics.
The provisional solution currently available is to rely on minimization
methods that do not guarantee to find all solutions, to let them search
a reasonable portion of the space of starting values as finely as one
can afford, and to rely on the experience that a fully real and
physically acceptable solution, as a rule, stands out quite
conspicuously and essentially uniquely.

\qquad
The correlation functions of QCD, by the way, may not represent the
optimal proving ground for first tests of the method described here --
or, for that matter, of any method you may invent for coupling-resurgent
reconstruction. The fact that the elementary QCD fields have no
asymptotically detectable particles associated with them tends to
intermix and obscure all questions of nonperturbative methodology, in
which we are interested here, with questions of the physical
interpretation of various features of the functions obtained --
particularly of the propagators. A much cleaner environment conceptually
is the $SU\,(2)$ gauge system of
electroweak theory, with coupling to fermions but without strong
interactions. Here, quasi-perturbative corrections are almost certainly
small at energies around the relevant $\Lambda$ scale, and resurgent
amplitudes of the zeroth quasi perturbative order should allow one in
principle to explore with some amount of systematics the question of
precisely how much can and cannot be done without the help of a Higgs
field, particularly with respect to dynamical mass generation. All
elementary fields here have asymptotically detectable particles, so the
requirements on and interpretation of propagators are clean and simple.

\qquad
The same environment may be well suited for studying the question of
when-and-how-and-how-much the coupling-resurgent amplitudes break
symmetries, particularly the local gauge symmetries. Here I may remind
you of the fact that quantum local symmetries place much more stringent
invariance requirements on the functional measure of the generating
functional (\ref{genfu}) than the simple translation invariance we used
for deriving the DS equation (\ref{fuDS}). The DS equations,
consequently, are more fundamental than the Slavnov-Taylor identities
that express local gauge invariance on the level of quantum correlation
functions; they reserve the right of producing symmetry-breaking
solutions. This possibility becomes particularly acute in connection
with nonperturbative solutions containing $\Lambda$ terms -- since the
self-consistency of the latter, as we saw in sect. 3.2, is a genuine
quantum effect arising from ultraviolet-divergent loops, they have in
principle the power to break classical symmetries, much like anomalies
in that respect. On the other hand the perturbative limits of
amplitudes, denoted $\Gamma^{(0)\,pert}_{N}$ in eq. (\ref{pertex}), do
preserve the quantum-symmetry identities, and that is all that is both
necessary and sufficient for maintaining the most important practical
consequence of gauge invariance -- the perturbative renormalizability
\cite{WEI}. I could therefore imagine (why not speculate a little on a
casual stroll like this) that a picture of "subasymptotic symmetry
breaking" might emerge in general, where perturbative
$\Gamma^{(0)\,pert}$'s
as large-momentum limits, with their proven Slavnov-Taylor compliance,
uphold renormalizability, while the $\Lambda$ terms coming into view one
by one on moving down the momentum scale increasingly deviate from
Slavnov-Taylor. In QCD, things are probably different -- the simple fact
that gluons obviously do not get saddled with $b$- or even $t$-quark
masses (through virtual quark-antiquark pairs) would seem to require
much stricter adherence to Slavnov-Taylor, and the fact that the
solutions of refs. \cite{DRI} are not doing very well in that respect is
another indicator that the approximation level there is still too crude.

\qquad
A purely methodological but definitely challenging cluster of
things-not-done centers around the question of whether the DS equations,
as employed here, really offer the best framework for the reconstruction
of resurgent amplitudes. Clever rearrangements of the pure-and-simple DS
system, as it emerges from eq. (\ref{fuDS}) and is sketched in Fig. 3.1,
have long been known. Already in his first paper on the subject
\cite{DYS4}, Dyson used what may be called a Bethe-Salpeter resummation
of the QED three-point vertex equation, which allows one to replace the
bare three-point vertices at the left ends of diagrams (3\,B) and
(3\,C) in Fig. 3.1 by dressed ones, while replacing the right-hand
portions of both diagrams by a Bethe-Salpeter kernel $K_{4}$ one- and
two-particle irreducible in the horizontal channel. Such a rearrangement
presumably shifts a substantial amount of physics from higher into lower
loop orders, and may make a crucial difference when studying problems of
symmetry breaking. It also provides partial, though not complete, relief
from a perennial nuisance in most DS equations with $N\,\geq\,3$ -- the
lack of manifest Bose or Fermi symmetry due to the unsymmetric
distinction of the leftmost leg that corresponds to the primary
functional differentiation in eq. (\ref{divan}). In a non-abelian gauge
theory, however, one pays for the progress with equally substantial new
complication: the information on the four-point amplitude, already by
far the most complicated of the superficially divergent ones, now enters
the system in two different ways, as $\Gamma_{4}$ and as $K_{4}$, whose
relation to each other must be stated as an additional dynamical
equation. This roughly doubles the effort required for dealing with the
superficially divergent subsystem. Still more sophisticated
rearrangements, originating in work by Symanzik \cite{SYM} and carried
out for non-abelian gauge theory by Baker and Lee \cite{BAL}, allow the
elimination of bare vertices in favor of dressed ones to be continued
down into the propagator equations (thus creating a system with the
remarkable property that its perturbative iteration does not produce
diagrams with overlapping ultraviolet divergences).
But it once more brings substantial additional complication in that
internal lines now carry momentum derivatives of propagators in addition
to the propagators themselves. In the search for improvements along
these
lines, just about everything seems to come at a heavy price. None of
these improvements touches the one feature of all DS-type dynamical
equations that the mathematician studying resurgence would find most
disappointing: that they make no direct reference, either differential
or integral or otherwise, to coupling dependence, which enters them only
in a somewhat indirect, parametric way. Equations like the
t´Hooft-Weinberg RG equations, which directly react to coupling
dependence through their trademark
$\beta\,(g)\,\frac{\partial}{\partial\,g}$ terms, would be much more
after the heart of the resurgence theorist, but while formulating
important restrictions on correlation functions, they are not statements
of their full dynamics. Equations combining the coupling sensitivity of
the t´Hooft-Weinberg with the full dynamical content of the DS equations
could perhaps be found among various versions of the exact RG
differential equations of Polchinski \cite{POL}; one example that I
happen to know of is ref. \cite{WIE}. Equations of this kind could be
well suited for studying nonperturbative structure of the resurgent
type. Once more the subject has not even begun to be explored.

\qquad
Still strolling on, we dimly perceive from a distance a set of problems
that lie at the borderline to mathematics, and may well be the deepest
and most long-term ones. I emphasized at the outset that I would be
appealing only to a few simple notions from what may be called the
surface of resurgence theory, not to its deeper constructs and methods.
But to bring these into play in the elucidation of nonperturbative
coupling structure may be fascinating in the short and unavoidable in
the long run. To mention one example, the ultraviolet-renormalon poles
on the left-hand, real Borel axis that we found to persist in our
resurgent model amplitude are presumably a common feature of most if not
all correlations in asymptotically free theories. In trying to
characterize quantitatively the features of coupling dependence
controlled by these singularities, and in bringing to the fore the
common traits of different correlations arising from them, Écalle´s
alien-derivative calculus, with its beautiful constructive tool of
resurgence monomials, may have a nontrivial contribution to make.
Another conjecture
brings us, once more, to QCD. There, the quasi-perturbative corrections
are not small enough at the scale $\Lambda$ to be treated by
"pragmatic" semiconvergence (as one would not hesitate to do in
electroweak theory). Accordingly, the ultimate methods of solution may
well be those that abandon completely the $n$-versus-$p$-direction
dichotomy still prevailing in our quasi-perturbative scheme, and perform
partial resummations that cut across the entire resurgent symbol. The
prominent role, aready noted, that resurgent symbols with support on the
non-negative integers play in Écalle´s treatment of differential
equations may have heuristic significance here: perhaps the resummations
needed may be differential approximants in the coupling, resurgent
solutions to a well-conceived sequence of differential equations,
characterizable by holomorphic invariants of presently unknown physical
meaning ...

\qquad
Our little stroll has led us on to flights of imagination. For here and
now, in my opinion, the greater value of resurgence theory lies not in
specific techniques, intriguing as they may be. It lies in providing a
framework for thinking, and in connecting disparate pieces of physical
knowledge. The study of QFT, to the extent that it has used continuum
methods, has been dominated conceptually by the perturbation expansion
to such an extent that "summing the series" was all but synonymous
with establishing the full solution for generations of physicists.
Resurgent functions provide a much wider yet precise framework, wide
enough presumably to cover the strongly nonperturbative coupling
dependence of asymptotically free theories. They quite literally open up
a second dimension. From this, I would expect, will arise their
longer-term impact.

\vspace*{2cm}
\begin{center}
{\bf\Large Acknowledgements}
\end{center}

I am indebted to my colleague Gernot Münster for a number of valuable
references, and to Daniel Ebbeler for expert te$\chi$ing and graphics.
Several of the students (they will forgive me for not listing them by
name) who listened to the shorter original versions of these lectures
asked good questions that hopefully have led to clarification of some
unclear points -- the many that undoubtedly remain are my
responsibility.


\end{document}